\documentclass[a4paper,12pt]{report}
\usepackage[toc,page]{appendix}

\usepackage[T1,OT1,OT2,T2A]{fontenc}
\usepackage[utf8]{inputenc}
\usepackage[english,russian]{babel}
\usepackage{setspace}
\usepackage{natbib}

\usepackage{graphicx}

\usepackage{epsfig}

\usepackage[figuresright]{rotating}

\usepackage{enumerate}
\usepackage{amssymb}
\usepackage{amsmath}
\usepackage{multirow}
\usepackage{array}
\usepackage{bm}
\usepackage{color}

\newcolumntype{L}[1]{>{\raggedright\let\newline\\\arraybackslash\hspace{0pt}}m{#1}}
\newcolumntype{C}[1]{>{\centering\let\newline\\\arraybackslash\hspace{0pt}}m{#1}}
\newcolumntype{R}[1]{>{\raggedleft\let\newline\\\arraybackslash\hspace{0pt}}m{#1}}

\newcommand{\beq}{\begin{eqnarray}}
\newcommand{\eeq}{\end{eqnarray}}

\begin{document}

%%%%%%%%%%%%%%%%%%%%%%%% The title page%%%%%%%%%%%%%%%%%%%%%%%%%%%%%%
\begin{titlepage}
\setstretch{1.0}

\begin{center}
\vspace*{2cm}
{\bf\large COSMOLOGICAL MODELS OF DARK ENERGY:\\ THEORY AND OBSERVATIONS }\\[2cm]

{\bf\normalsize OLGA AVSAJANISHVILI}\\[2cm]

{\it A dissertation submitted to the graduate division of \\
the Faculty of Natural Sciences and Medicine of \\
 Ilia State University in partial fulfillment of\\
 the requirements for the academic degree of \\
 Doctor Philosophy in Physics
}\\[2cm]
  Doctoral Program in Physics and Astronomy \\[3cm]

Supervisors: Prof. Tina Kahniashvili \& Prof. Lado Samushia

\vfill
\bf
ILIA STATE UNIVERSITY\\
Tbilisi, 2019
\end{center}
\end{titlepage}
\thispagestyle{empty}
\newpage

\thispagestyle{empty}
\newpage

\begin{center}
\vspace*{2cm}
 \foreignlanguage{russian}{\bf\large Космологические модели темной энергии:\\ теория и наблюдения}\\[2cm]

\foreignlanguage{russian}{
\normalsize Ольга Авсаджанишвили}\\[2cm]

 \foreignlanguage{russian}{\it Диссертация представлена
  на факультет естественных наук и медицины Государственного Университета Илии\\
  В соответствии с академическими требованиями доктора физических наук\\[2cm]
         Программа по физике и астрономии} \\[3cm]

\foreignlanguage{russian}{Научные руководители: профессор Тина Кахниашвили и профессор Ладо Самушиа}

\vfill
\foreignlanguage{russian}{
\bf
Государственный университет Илии\\
Тбилиси, 2019}
\end{center}
\thispagestyle{empty}

\newpage
%%%%%%%%%%%%%%%%%%%%%%%%%%%%%%%end of the declaration%%%%%%%%%%%%%%%%%%%%%%%%
%%%%%%%%%%%%%%%%%%%%%%%%%%%%% Abstact %%%%%%%%%%%%%%%%%%%%%%%%%%%%%%%%%%%%%%%
\addcontentsline{toc}{chapter}{Абстракт}

\begin{center}
{\bf \large{Абстракт}\\}
\end{center}
Стандартная Lambda Cold Dark Matter ($\Lambda$CDM) космологическая модель предполагает, что общая теория относительности является правильной теорией гравитации на пространственных и временных космологических масштабах, а ускорение вселенной обусловлено темной энергией или космологической постоянной $\Lambda$. Темная энергия оказывает отрицательное давление на пространство, т. е. она обладает свойством 'антигравитации', тем самым вызывая ускоренное расширение вселенной. Плотность энергии $\Lambda$ не зависит от времени и в последнее время становится доминирующей (в частности, плотность энергии, связанная с космологической постоянной, составляет около $69\%$ от полной плотности энергии вселенной в современную эпоху).
Около $26\%$ от полной плотности энергии во вселенной присутствует в виде холодной темной материи. Таким образом, в рамках стандартной $\Lambda$CDM модели $95\%$ от полной плотности энергии в современной вселенной присутствует в темной (невидимой) форме неизвестной природы (темная энергия и темная материя), и только $5\%$ является обычным веществом (барионы, лептоны). Теоретические предсказания $\Lambda$CDM модели хорошо соответствуют современным наблюдательным данным, но существуют нерешенные проблемы, связанные с этой моделью, в частности, проблема космологической постоянной (крайне маленькая величина космологической постоянной по сравнению с теоретической оценкой величины плотности энергии вакуума) и так называемая проблема совпадения величины плотности темной энергии и величины плотности  энергии темной материи. Для преодоления этих проблем были предложены динамические модели темной энергии. В этих моделях темная энергия присутствует в виде динамического скалярного поля, т. е. в этих моделях плотность темной энергии изменяется со временем.
В данной диссертации мы исследовали различные модели скалярного поля.
В частности, мы изучали эволюцию
фонового расширения и функции темпа роста флуктуаций плотности материи в модели скалярного поля с потенциалом Ратра-Пиблса. Мы ограничили модельный параметр этого потенциала $\alpha$ и параметр плотности материи, $\Omega_{\rm m}$, используя последние измерения функции темпа роста флуктуаций плотности материи и барионных акустических колебаний. Кроме того, мы  изучали ряд моделей скалярного поля с целью возможности определить более предпочтительные модели в сравнении со стандартной $\Lambda$CDM моделью, используя прогнозируемые данные будущих наблюдений DESI. С этой целью мы провели статистический анализ Байеса. Мы обнаружили, что результаты этого анализа служат убедительным доказательством в пользу $\Lambda$CDM модели.
Мы также применили MCMC анализ и получили ограничения на параметры моделей скалярного поля, сравнивая наблюдательные данные для: темпа расширения вселенной, углового расстояния и функции темпа роста флуктуаций плотности материи с соответственными данными сгенерированными для $\Lambda$CDM модели. Мы исследовали, насколько хорошо CPL параметризация аппроксимирует различные модели скалярного поля. Мы определили местоположение каждой модели скалярного поля в пространстве CPL параметров.
\\
\textit {\textbf {Ключевые слова:}} темная энергия, космологическая постоянная, скалярное поле, крупномасштабная структура, функция темпа роста флуктуаций плотности материи, статистика Байеса, метод Монте-Карло для цепей Маркова.
\newpage

%%%%%%%%%%%%%%%%%%%%% Acknowledgments %%%%%%%%%%%%%%%%%%%%%%%%%%%%%%%%%%%%%%
\addcontentsline{toc}{chapter}{Благодарность}
\begin{center}
\section*{Благодарность}
\end{center}
Я хочу выразить благодарность моему руководителю Тине Кахниашвили. Я благодарна ей за помощь и поддержку. Я высоко ценю ее лидерские качества и очень серьезное отношение к написанию статей. Тина стимулировала меня к саморазвитию, к самостоятельным исследованиям и к углублению моих знаний в области космологии.
Я очень благодарна моему руководителю Ладо Самушиа. Он познакомил меня с миром статистики Байеса и с методом Монте-Карло для цепей Маркова. Он научил меня писать компьютерные программы в области анализа данных. Мы обсуждали много проблем, связанных с созданием этих программ и с пониманием многочисленных явлений в космологии.
Мне бы хотелось поблагодарить моего соавтора Наталью Архипову. Мои первые шаги в космологии были связаны с этой доброй и умной женщиной. Мы независимо друг от друга выводили уравнения, писали компьютерные программы, а затем сравнивали и обсуждали полученные результаты.
Я очень признательна Василу Кухианидзе и Александру Тевзадзе за помощь и поддержку. В любой момент я всегда могла получить от них исчерпывающий ответ на любой мой вопрос, связанный с физикой изучаемых явлений и с написанием компьютерных программ в области космологии. Я глубоко благодарна Луке Понятовскому за помощь в оформлении диссертации в среде WinEdt.
Я благодарна Бхарат Ратра за полезные обсуждения, помощь и руководство.
Я также хочу поблагодарить моего соавтора Иен Хуан. Я благодарна Богдану Новосядлому за полезные обсуждения. Я бы хотела поблагодарить Геннадия Читова, Саяна Мандала и Биджита Бсингу за совместные обсуждения. Я хочу поблагодарить директора Абастуманской Астрофизической Обсерватории Майю Тодуа за помощь и поддержку. Мне бы хотелось выразить свою благодарность и уважение научному секретарю Абастуманской Астрофизической Обсерватории Мзии Баратели. Эта удивительная и добрая женщина поддерживала и помогала мне в трудные моменты. Я очень признательна Бидзине Чаргеишвили за полезные комментарии. Я от всей души благодарна моей дочери Марии за полезные комментарии и создание рисунков на странице 40. С большим удовольствием хочу поблагодарить Эрвина за моральную поддержку и вдохновение в моей работе.
Я благодарю Государственный Университет Ильи, Грузия, где эта диссертация была написана. Я благодарна за гостеприимство Католическому Университету в Лювене, Бельгия и физическому факультету Канзасского Государственного Университета, США. Во время посещения этих университетов мною частично была написана данная диссертации.

Эта работа была выполнена при частичной поддержке следующими грантами: Shota Rustaveli Georgian NSF (FR/339/6-350/14 and PhD\_F\_17\_196 ), the CRDF-SRNSF-GRDF Georgia Women's Research Fellowship Program (WRF-14-22), the Swiss NSF SCOPES (IZ 7370-152581), the SOLSPANET
FP7-PEOPLE-2010-IRSES, 269299 EC FP7-PEOPLE-2010-IRSES project 269299.

%%%%%%%%%%%%%%%%%%%%%%%%%%%%% list of figures %%%%%%%%%%%%%%%%%%%%%%%%%%%
\addcontentsline{toc}{chapter}{Список Рисунков}
\listoffigures
\newpage

%%%%%%%%%%%%%%%%%%%%%%%%%%%% list of tabels %%%%%%%%%%%%%%%%%%%%%%%%%%%%
\addcontentsline{toc}{chapter}{Список Таблиц}
\listoftables
\newpage

%%%%%%%%%%%%%%%%%%%%%%%% tabel of contatns %%%%%%%%%%%%%%%%%%%%%%%%%%%%%%
\tableofcontents
\newpage

%%%%%%%%%%%% set new page counter to arrabic n = 1 %%%%%%%%%%%%%%%%%%%%%%%
\pagenumbering{arabic}
\setcounter{page}{1}

%%%%%%%%%%%%%%%%%%%%%%%%%%%%%Notations%%%%%%%%%%%%%%%%%%%%%%%%%%%%%%%%%%%%

\addcontentsline{toc}{chapter}{Обозначения}
\section*{Обозначения}

\begin{table}[h!]
\begin{tabular}{C{6.5cm}| C{9.5cm}}
\hline
\multicolumn{2}{C{16cm}}{Естественная Система Единиц}\\
\hline
Наименование & Величина\\
\hline
\end{tabular}\\
\begin{tabular}{L{6.5cm} L{9.5cm}}
$\text{Время}$                       &$1~\text{Гэв}^{-1}=6.6\cdot10^{-25}~\text{с}$\\
$\text{Длина}$                       &$1~\text{Гэв}^{-1}=2\cdot10^{-14}~\text{см}$\\
$\text{Масса}$                       &$1~\text{Гэв}=1.8\cdot10^{-24}~\text{г}$\\
$\text{Плотность массы}$                       &$1~\text{Гэв}^4=2.3\cdot10^{17}~\text{эрг}~ \text{см}^{-3}$\\
$\text{Плотность числа частиц}$                       &$1~\text{Гэв}^3=1.3\cdot10^{41}~\text{см}^{-3}$\\
$\text{Плотность энергии}$                       &$1~\text{Гэв}^4=2.1\cdot10^{38}~\text{эрг}~\text{см}^{-3}$\\
$\text{Температура}$                       & $1~\text{Гэв}=1.16\cdot10^{13}~\text{К}$\\
$\text{Энергия}$                      &$1~\text{Гэв}=1.6\cdot10^{-3}~\text{эрг}$ \\
\hline\hline
\end{tabular}
\end{table}
\noindent
\\

\begin{table}[h!]
\begin{tabular}{C{5cm}| C{2.5cm}| C{4.5cm}|  C{4cm}}
\hline
\multicolumn{4}{C{16cm}}{Параметры}\\
\hline
\multirow{2}{*}\text{Наименование} & \multirow{2}{*}{Обозначение} & \multicolumn{2}{C{9cm}}{Единицы} \\ \cline{3-4}
                      &                           & \multicolumn{1}{C{4.5cm}|}{СГС} & \multicolumn{1}{C{4cm}}{Естественная Система} \\
\hline
\end{tabular}\\
\begin{tabular}{L{5cm}| C{2.5cm}| L{4.5cm}|  L{4cm}}
\text{Астрономическая единица}  & $\text{а.~е.}$                   & $1.4960\cdot10^{13}~\text{см}$                  & $7.5812\cdot10^{26}~\text{Гэв}^{-1}$\\
\text{Критическая плотность}     & $\rho_{\rm crit}$         & $1.8791h^2\cdot10^{-29}~\text{г~см}^{-3}$       & $8.0992h^2\cdot10^{-47}~\text{Гэв}^4$\\
\text{Масса Планка}         & $M_{\rm pl}$              & $2.1768\cdot10^{-5}~\text{г}$                   & $1.2211\cdot10^{19}~\text{Гэв}$\\
\text{Масса Солнца}         & $M_{\odot}$             & $1.989\cdot10^{33}~\text{г}$                     & $1.116\cdot10^{57}~\text{Гэв}$\\
\text{Мегапарсек}         & $\text{Мпс}$                 & $3.0856\cdot10^{24}~ \text{см}$                 & $1.5637\cdot10^{38}~\text{Гэв}^{-1}$ \\
\text{Парсек}               & $\text{пс}$                  & $3.0856\cdot10^{18}~\text{см}$                  & $1.5637\cdot10^{32}~\text{Гэв}^{-1}$ \\
\text{Постоянная Ньютона}  & $G$                   & $6.672\cdot10^{-8}~\text{см}^3~\text{г}^{-1}~\text{с}^{-2}$ & $6.707\cdot10^{-39}~\text{Гэв}^{-2}$\\
\text{Постоянная Хаббла}      & $H_0$                 & $3.241h\cdot10^{-18}~\text{с}^{-1}$            & $2.1332h\cdot10^{-42}~\text{Гэв}$ \\
\text{Скорость света}       & $\text{c}$                   & $2.9979\cdot10^{10}~\text{м~с}^{-1}$         & $1$ \\
\hline\hline
\end{tabular}
\end{table}
\newpage

\begin{table}[h!]
\begin{tabular}{C{4cm}| C{11cm}}
\hline
\multicolumn{2}{C{15cm}}{Переменные}\\
\hline
Символ & Значение\\
\hline
\end{tabular}\\
\begin{tabular}{L{4cm} L{11cm}}
$a$                       & Скалярный фактор\\
$a_{\rm l,m}$             & Мультипольные коэффициенты разложения по сферическим гармоникам\\
$c_s$                     & Скорость звука\\
$ds^2$                    & Метрика\\
$d_A$                     & Угловое расстояние\\
$d_L$                     & Фотометрическое расстояние\\
$e$                       & Математическое ожидание\\
$f(x)$                    & Плотность распределения\\
$f(a)$                    & Функция темпа роста флуктуаций плотности материи\\
$\Omega_{\rm m}(a)$       & Фракционная плотность материи\\
$h$                       & Нормированный параметр Хаббла\\
$\hbar$                   & Редуцированная постоянная Планка\\
$k$                       & Конформный импульс\\
$k_{\rm phys}$            & Физический импульс\\
$p$                       & Давление идеальной жидкости\\
$p_{\phi}$                & Давление скалярного поля\\
$p_{ij}$                  & Переходные вероятности\\
$\tilde{p}$               & Параллакс\\
$q$                       & Параметр ускорения вселенной\\
$r_s$                     & Сопутствующий размер звукового горизонта\\
$q_0$                     & Параметр ускорения вселенной для современной эпохи\\
$d$                       & Физическое расстояние\\
$m$                       & Видимая звездная величина\\
$m_{\rm ch}$              & Масса Чандрасекара\\
$r_{\bigoplus}$           & Радиус орбиты Земли\\
$t$                       & Физическое время\\
$v$                       & Радиальная скорость\\
$v_{\rm f}$               & Трехмерная скорость идеальной жидкости\\
$w$                       & Параметр уравнения состояния\\
$w_0$                     & Параметр уравнения состояния в современную эпоху\\
$w_a$                     & Производная параметра уравнения состояния по скалярному фактору в CPL параметризации\\
$w_{\rm vac}$             & Параметр уравнения состояния для вакуума\\
$w_\Lambda$               & Параметр уравнения состояния для космологической постоянной\\
$w_{\phi}$                & Параметр уравнения состояния для скалярного поля\\
$z$                       & Красное смещение\\
$z_{\rm dec}$             & Красное смещение при отсоединении фотонов от материи\\
$z_{\rm rec}$             & Красное смещение при рекомбинации\\
$C$                       & Ковариационная матрица\\
$D$                       & Линейный фактор роста\\
$D_V$                     & Масштаб расстояний\\
$E$                       & Энергия\\
$E(a)$                    & Нормированный параметр Хаббла\\
$\mathcal{E}$             & Интеграл доказательств Байеса\\
$F$                       & Поток\\
\hline\hline
\end{tabular}
\end{table}
\newpage

\begin{table}[h!]

\begin{tabular}{C{4cm}| C{11cm}}
\hline
\multicolumn{2}{C{15cm}}{Переменные}\\
\hline
Символ & Значение\\
\hline
\end{tabular}\\
\begin{tabular}{L{4cm} L{11cm}}
$G$                       & Постоянная Ньютона\\
$H$                       & Параметр Хаббла\\
$K$                       & Параметр кривизны\\
$L$                       & Светимость\\
$\mathcal{L}$             & Функция вероятности\\
$\mathcal{L}^{\rm f}$     & Функция вероятности для наблюдательных данных функции темпа роста флуктуаций плотности материи\\
$\mathcal{L^{\rm bao}}$   & Функция вероятности для BAO наблюдательных данных\\
$\mathcal{L}_{\phi}$      & Плотность Лагранжиана скалярного поля\\
$M$                       & Абсолютная звездная величина\\
$M_{\phi}$                & Масштаб массы частиц скалярного поля \\
$P_1$                     & Матрица перехода\\
$P_l$                     & Полиномы Лежандра\\
$P(k)$                    & Спектр мощности\\
$R$                       & Скаляр Риччи\\
${\bf R}$                 & Радиус\\
$S_M$                     & Действие для материи\\
$S$                       & Действие\\
$T_0$                     & Средняя температура CMBR в современную эпоху\\
$V(\phi)$                 & Потенциал скалярного поля\\
$V_0$                     & Модельный параметр в $\phi$CDM модели скалярного поля\\
$Y_{\rm lm}$              & Сферические гармоники\\
$\alpha$                  & Модельный параметр в $\phi$CDM Ратра-Пиблс модели скалярного поля\\
$\delta$                  & Флуктуация плотности материи\\
$\gamma$                  & Линдер $\gamma$-параметризация\\
$\gamma(a)$               & Эффективный индекс роста\\
$\delta\rho_{\rm b}$      & Флуктуация барионной плотности\\
$\delta^{\mu}_{\nu}$      & Дельта функция Кронекера\\
$\delta T_{\rm dipol}$    & Температурная анизотропия дипольной составляющей CMBR\\
$\eta$                    & Конформное время\\
$\lambda$                 & Длина волны\\
$\mu$                     & Модуль расстояния\\
$\xi(s)$                     & Двухточечная корреляционная функция\\
$\rho_0$                  & Средняя плотность энергии CMBR в современную эпоху\\
$\rho_{\rm b}$            & Плотность энергии барионов\\
$\rho_{\rm b0}$           & Плотность энергии барионов в современную эпоху\\
$\rho_{\rm m}$            & Плотность энергии материи\\
$\rho_{\rm ph}$           & Плотность энергии фотонов\\
$\rho_{\phi}$             & Плотность энергии скалярного поля\\
 $\rho_{\rm r}$           & Плотность энергии излучения\\
$\rho_{\rm K}$            & Плотность энергии кривизны\\
 $\rho_\Lambda$           & Плотность энергии вакуумной энергии\\
 $\rho_{\rm m0}$          & Плотность энергии материи в современную эпоху\\
 $\rho_{\rm r0}$          & Плотность энергии излучения в современную эпоху\\
\hline\hline
\end{tabular}
\end{table}

\begin{table}[t!]
\begin{tabular}{C{4cm}| C{11cm}}
\hline
\multicolumn{2}{C{15cm}}{Переменные}\\
\hline
Символ & Значение\\
\hline
\end{tabular}\\
\begin{tabular}{L{4cm} L{11cm}}
$\rho_{\rm K0}$           & Плотность энергии кривизны в современную эпоху\\
$\sigma$                 & Стандартное отклонение\\
 $\sigma_8$              & Среднеквадратичные линейные флуктуации в распределении плотности материи на масштабах $8h^{-1}$~\text{\rm Мпк}\\
 $\sigma^2$               & Дисперсия\\
 $\Delta\rho$             & Флуктуация плотности материи\\
 $\phi$                   & Амплитуда скалярного поля\\
 $\chi^2$                 & $\chi^2$ функция\\
$\omega_0$                & Частота плоской монохроматической волны\\
$\Gamma^\lambda_{\mu\nu}$ & Символы Кристоффеля\\
$\Lambda$                 & Космологическая постоянная\\
$\Omega_{\rm m}$          & Параметр плотности материи \\
$\Omega_{\rm m0}$         & Параметр плотности материи в современную эпоху\\
$\Omega_{\rm r}$          & Параметр плотности излучения\\
$\Omega_{\rm r0}$         & Параметр плотности излучения в современную эпоху\\
$\Omega_{\rm K}$          & Параметр плотности кривизны\\
$\Omega_{\rm K0}$         & Параметр плотности кривизны в современную эпоху\\
$\Omega_\Lambda$          & Параметр плотности вакуума\\
$\Omega_{\phi}$           & Параметр плотности скалярного поля\\
\hline\hline
\end{tabular}
\end{table}

\newpage
\begin{table}[h!]
\begin{tabular}{C{6cm}| C{9cm}}
\hline
\multicolumn{2}{C{15cm}}{Специальные Обозначения}\\
\hline
Обозначение & Значение \\
\hline
\end{tabular}\\
\begin{tabular}{p{6.5cm} p{9cm}}
\hline%\hline
$(+,-,-,-)$ & Пространственно-временная сигнатура\\
{\bf Соглашения для индексов}:\\
$(\alpha, \beta, \gamma, \mu, \nu)$ пробегают от 0 до 3 & Греческие буквы \\
$(i, j, k, l, m, n)$ пробегают от 1 до 3 & Латинские буквы\\
\\
 $(t,x,y,z)\equiv(x^0, x^1, x^2, x^3)=x^{\mu}$ & Четырехмерные координаты\\
 $(x,y,z)\equiv( x^1, x^2, x^3,)=x^{i}$ & Трехмерные декартовые координаты\\
 $ (r, \varphi)$  & Полярные координаты\\
$(r, \varphi, z)$ & Цилиндрические координаты\\
$(r, \theta, \varphi)$   & Сферические координаты\\
$ (\varrho, \varsigma, \varphi)$ & Псевдосферические координаты\\

{\bf Векторы}:\\
$A_i$                     & Ковариантный вектор\\
$A^i$                     & Контравариантный вектор\\

{\bf Тензоры}:\\
$A_{ij}$                  & Ковариантный тензор второго ранга\\
$A^{ij}$                  & Контравариантный тензор второго ранга\\
$A^i_j$                   & Смешанный тензор второго ранга\\
$g_{\mu\nu}$              & Метрический тензор пространства-времени\\
$u_\mu$                   & Четырехмерная скорость\\
$G_{\mu\nu}$              & Тензор Эйнштейна\\
$R_{ik}$                  & Тензор Риччи\\
$R_{iklm}$                & Тензор Римана\\
$T_{\mu\nu}$              & Тензор энергии-импульса\\
\hline\hline
\end{tabular}
\end{table}

%\newpage
\begin{table}[h!]
\begin{tabular}{C{5cm}| C{10cm}}
\hline
\multicolumn{2}{C{15cm}}{Аббревиатура}\\
\hline
Символы & Полная форма\\
\hline
\end{tabular}\\
\begin{tabular}{L{5cm} L{10cm}}
\bf{AIC}          &  Akaike Information Criterion\\
\bf{BIC}          &  Bayesian Information Criterion\\
\bf{BAO}          &  Baryon Acoustic Oscillations\\
\bf{CDM}          &  Cold Dark Matter \\
\bf{CMBR}         &  Cosmic Microwave Background Radiation\\
\bf{DESI}         &  Dark Energy Spectroscopic Instrument \\
\bf{HDM}          &  Hot Dark Matter\\
\bf{ISW}          &  Integrated-Sachs-Wolfe\\
\bf{CPL}          &  Chevallier-Polarsky-Linder  \\
\bf{FRII}         &  Fanaroff-Riley Type II\\
\bf{FLRW}         &  Friedmann-Lema\^\i tre-Robertson-Walker \\
\bf{MACHOs}       &  Massive Compact Halo Objects\\
\bf{MCMC}         &  Markov Chain Monte Carlo \\
\bf{SDSS}         &  Sloan Digital Sky Survey \\
\bf{SZ}           &  Sunyaev-Zel'dovich\\
\bf{WDM}          &  Warm Dark Matter\\
\bf{WFIRST}       &  Wide-Field Infrared Survey Telescope\\
\bf{WIMPs}         & Weakly Interacting Massive Particles\\
\bf{WMAP}         &  Wilkinson Microwave Anisotropy Probe\\
\bf{$\Lambda$CDM} &  Lambda Cold Dark Matter\\
\bf{$\phi$CDM}    &  Phi Cold Dark Matter\\
\bf{2dFGRS}       &  2dF Galaxy Redshift Survey\\
\bf{ОТО}          &  Общая Теория Относительности\\
\hline
\end{tabular}
\end{table}
%%%%%%%%%%%%%%%%%%%%%%%%%%%%%%%%%%%%%%%%%%%%%%%%%%%%%%%%%%%%%%%%%%%%%%%%%%
%%%%%%%%%%%%%%%%%%%       Beginng of chapter 1     %%%%%%%%%%%%%%%%%%%%%%%
%%%%%%%%%%%%%%%%%%%%%%%%%%%%%%%%%%%%%%%%%%%%%%%%%%%%%%%%%%%%%%%%%%%%%%%%%%
\chapter{Введение}\label{chapter:1}
В 1998 году было обнаружено ускоренное расширение нашей вселенной на основе измерений звездных величин сверхновых типа Ia, (\cite{Riess:1998cb}, \cite{Perlmutter:1998np}, \cite{riess07}). За это открытие  Солу Перлмуттеру, Брайану Шмидту и Адаму Риссу была присуждена в 2011 году Нобелевская премия по физике.
Ускоренное расширение вселенной подтверждается другими космологическими наблюдениями, в частности: измерениями анизотропии температуры и поляризации космического микроволнового фонового излучения, (\cite{Hinshaw:2008kr}, \cite{Nolta:2008ih}, \cite{Komatsu:2010fb},                                                                                                                                                                                                 \cite{Ade:2013zuv}, \cite{Ade:2015xua}); исследованиями крупномасштабной структуры во вселенной, (\cite{2dFGRS}, \cite{Eisenstein:2005su}, \cite{Percival:2007yw}, \cite{SDSS}).

Существуют многочисленные модели, объясняющие ускореннное расширение вселенной в современную эпоху, (\cite{Frieman:2008sn},  \cite{Caldwell2009}, \cite{Yoo:2012ug}). Наиболее популярная модель предполагает, что значительная часть
вселенной находится в форме {\it темной энергии} или {\it темной жидкости}, (обзоры: \cite{Peebles:2002gy}, \cite{Copeland:2006wr}, \cite{Tsujikawa:2010sc}, \cite{Tsujikawa:2010zza}). Необычным свойством
темной энергии является тот факт, что она оказывает отрицательное давление на пространство, т. е. темная энергия обладает свойством 'антигравитации'. Понимание природы и происхождения темной энергии является одной из наиболее важных и до сих пор нерешенных проблем современной космологии.

Простейшим описанием темной энергии является концепция энергии вакуума или космологическая постоянная $\Lambda$, впервые предложенная Альбертом Эйнштейном, (\cite{Einstein:1915by}, \cite{Einstein:1915ca}). Космологическая модель, основанная на таком описании темной энергии, называется Lambda Cold Dark Matter ($\Lambda$CDM) модель, которая является 'стандартной' моделью вселенной с 2003 года, (\cite{Zeldovich:1968gd}, \cite{Blumenthal:1984bp}); (монографии: \cite{Peebles:1994xt}, \cite{Dodelson2003}, \cite{Weinberg:2008zzc}); (обзоры: \cite{CPT}, \cite{Carroll:2000fy}, \cite{Peebles:2002gy}, \cite{Copeland2006}, \cite{Martin:2012bt}, \cite{Padilla:2015aaa}). Эта модель базируется на общей теории относительности (ОТО), разработанной Альбертом Эйнштейном для описания гравитации
во вселенной на пространственных и временных космологических масштабах.

Кроме того, в современной космологии существует до сих пор нерешенная проблема {\it темной материи} или скрытой массы во вселенной, которая, в частности, проявляется в аномально высокой скорости вращения  внешних областей галактик, (\cite{Rubin:1980zd}). Темная материя находится в галактиках и в скоплениях галактик. Частицы, формирующие темную материю, не взаимодействуют с электромагнитным излучением и слабо гравитационно взаимодействуют с обычной барионной материей.

Основываясь на ОТО, около $95\%$ энергии в современной вселенной находится в 'темной' форме, т. е. в виде темной энергии и темной материи. Последние результаты наблюдений космического телескопа Planck показывают, что вселенная состоит на $4,8\%$ из обычного вещества, на $26\%$ из темной материи и на $69,2\%$ из темной энергии, (\cite{Ade:2015xua}).

$\Lambda$CDM модель является {\it согласованной} моделью вселенной, т. к. эта модель находится в хорошем согласии с доступными на сегодняшний день космологическими наблюдениями.
 Однако $\Lambda$CDM модель имеет до сих пор нерешенные проблемы: {\it проблему космологической постоянной} или {\it проблему тонкой настройки} и {\it проблему совпадения}, (\cite{Weinberg:1988cp}, \cite{Padmanabhan:2002ji}, \cite{Padilla:2015aaa}). Проблема космологической постоянной заключается в том, что наблюдаемая величина космологической постоянной на 120 величин меньше ее теоретически предсказанной величины, (\cite{Weinberg:2000yb}). Проблема совпадения состоит в том, что основываясь на точных космологических наблюдениях, плотность темной энергии сопоставима с плотностью энергии темной материи в современную эпоху: $\rho_{\rm DM}/\rho_{\rm DE} \simeq 1/3$, $\rho_{\rm DE}$ и $\rho_{\rm DM}$ - плотность темной энергии и плотность темной материи, соответственно. Этот факт является загадкой, т. к. согласно стандартной $\Lambda$CDM модели, энергия космологической постоянной не зависит от времени, $\rho_{\rm DE}=\rho_\Lambda={\rm const}$, в то время как энергия темной материи изменяется со временем как, $\rho_{\rm DM}\sim a^{-3}(t)$, Рис.~(\ref{fig:f21}). Поэтому отношение этих величин должно быть зависящим от времени: $\rho_{\rm DM}/\rho_{\rm DE} \propto 1/a^3(t)$, $a(t)$ - скалярный фактор и $t$ - физическое время.

С целью решения проблем $\Lambda$CDM модели были разработаны альтернативные модели. Эти модели разделяются на два типа: на модели, основанные на гравитации ОТО и на модели с отличающейся гравитацией от ОТО на пространственных и временных космологических масштабах во вселенной (т. е. на масштабах, сопоставимых с современным размером вселенной). К первому типу моделей относятся динамические модели скалярного поля: модели квинтэссенции, (\cite{Ratra:1987aj}, \cite{Ratra:1987rm}, \cite{Wetterich:1987fk}), модели к-эссенции, (\cite{ArmendarizPicon:1999rj}, \cite{ArmendarizPicon:2000dh}, \cite{ArmendarizPicon:2000ah}), модели фантомного скалярного поля, (\cite{Caldwell1999}); модели взаимодействия темной энергии и материи, (\cite{Amendola:1999er}, \cite{Zimdahl:2001ar}); модель меняющейся массы нейтрино, (\cite{Farrar:2003uw}, \cite{Fardon:2003eh}); единая модель темной энергии и материи: газ Чаплыгина, (\cite{Kamenshchik:2001cp}, \cite{Bento:2002ps}) и модель к-эссенции, как единая модель темной энергии и материи, (\cite{Scherrer:2004au}); неоднородная модель Леметр-Толмэн-Бонди, (\cite{Lemaitre:1933gd}, \cite{Tolman:1934za}, \cite{Bondi:1947fta}, \cite{Tomita:2000jj}), и др. Ко второму типу моделей относятся: модели $f(R)$ гравитации, (\cite{Capozziello:2003tk}, \cite{Carroll:2003wy}, \cite{Mukhanov2005}, \cite{Nojiri:2006ri}); модель Двали-Габададзе-Поратти (brane world model), (\cite{dvali2001}); модели массивной гравитации, (\cite{Fierz:1939ix}, \cite{deRham:2010ik}, \cite{deRham:2010kj}, \cite{Hassan:2011zd}); квантовая гравитация и модификации гравитации, основанные на теории струн, (\cite{Polchinski:1998rq}, \cite{Polchinski:1998rr}); модели гравитации Галилея (Galileon), (\cite{Nicolis:2008in}); модели скалярно-тензорной гравитации, (\cite{Brans:1961sx}) и др.

Главной альтернативой $\Lambda$CDM модели являются модели динамического скалярного поля или $\phi$CDM модели, (\cite{Ratra:1987rm}, \cite{Ratra:1987aj}, \cite{Wetterich:1987fk}, \cite{Brax:2002vf}, \cite{Linder2007}, \cite{Cai:2009zp}, \cite{Bahamonde:2017ize}, \cite{Ryan:2019uor}). Модели скалярного поля позволяют избежать проблему космологической постоянной. В этих моделях параметр уравнения состояния, $w$, зависит от времени, $w\equiv p_{\rm DE}/\rho_{\rm DE}$, $p_{\rm DE}$ - давление темной энергии, тогда как в $\Lambda$CDM модели параметр уравнения состояния является константой, $w=-1$.

 В зависимости от величины параметра  уравнения состояния $\phi$CDM модели скалярного поля разделяются на: квинтэссенциальные, $-1<w<-1/3$, (\cite{Peebles:2002gy}, \cite{Caldwell:2005tm}, \cite{schimd06}) и фантомные, $w<-1$, (\cite{Caldwell1999}, \cite{Elizalde2004}, \cite{Scherrer:2008be}, \cite{Dutta:2009dr}, \cite{Frampton:2011sp},  \cite{Frampton:2011aa}, \cite{Ludwick2017}). Модели квинтэссенции подразделяются на два класса: на трекерные модели (или модели замерзания), в которых эволюция скалярного поля происходит медленее в сравнении с темпом расширения Хаббла, и на модели таяния, в которых эволюция скалярного поля происходит быстрее в сравнении с темпом расширения Хаббла, (\cite{Steinhardt:1999nw}, \cite{Caldwell:2005tm}, \cite{Dutta:2009dr}, \cite{Chiba:2012cb}, \cite{lima15}).

 В трекерных моделях квинтэссенции плотность энергии скалярного поля является субдоминантой на стадиях доминирования излучения и материи в эволюции вселенной, (\cite{Zlatev:1998tr}). Только в позднее время скалярное поле становится доминирующим и начинает вести себя как компонента с отрицательным давлением, что приводит к ускоренному расширению вселенной, (\cite{schimd06}, \cite{Linder:2015zxa}, \cite{Bag:2017vjp}). При определенной форме потенциала трекерные модели квинтэссенции  имеют  аттракторное решение, которое нечувствительно к начальным условиям, (\cite{Zlatev:1998tr}).
Простейшим примером трекерных моделей скалярного поля с аттракторным решением является модель скалярного поля с обратно-степенным потенциалом Ратра-Пиблса.  Эта модель была впервые предложена Бхаратом Ратра и Джимом Пиблсом в 1988 году, (\cite{Ratra:1987rm}, \cite{Ratra:1987aj}).

  Исследование скалярного поля квинтэссенции с потенциалом Ратра-Пиблса является одной из основных целей этой диссертации. В частности, мы исследовали динамику скалярного поля с этим потенциалом, влияние скалярного поля с потенциалом Ратра-Пиблса на динамику вселенной и на энергетические составляющие вселенной. Мы также изучали влияние модели скалярного поля с потенциалом Ратра-Пиблса на эволюцию крупномасштабной структуры во вселенной.

В последнее время среди космологов возрос интерес к фантомным моделям темной энергии в связи с тем, что некоторые современные наблюдательные данные согласуются с этими моделями, (\cite{Hinshaw:2012aka}, \cite{Ade:2015xua} и др.). Фантомные модели тёмной энергии имеют отрицательную неканоническую кинетическую составляющую в действии, в результате чего  плотность энергии в этих моделях увеличивается с течением времени, (\cite{Caldwell1999}, \cite{Scherrer:2008be}, \cite{scherrer07}, \cite{Ludwick2017}).
Во время ускоренного расширения вселенной, вызванного фантомным скалярным полем, может возникнуть разрыв между всеми
гравитационно связанным структурами во вселенной: начиная с распадом сверхскоплений галактик и заканчивая распадом атомных ядер.

С целью изучения истории расширения вселенной, крупномасштабной структуры во вселенной, природы темной энергии и темной материи, будут запущены в действие: инфракрасный телескоп с широким полем (WFIRST),
спектроскопический прибор для исследования темной энергии (DESI) и космический телескоп Эвклид (Euclid), (\cite{Amendola:2012ys}, \cite{Levi:2013gra}, \cite{Font-Ribera:2013rwa}, \cite{Spergel:2015sza}, \cite{Aghamousa:2016zmz}). По завершении этих миссий будут получены очень точные измерения темпа расширения вселенной, угловых расстояний и функции темпа роста флуктуаций плотности материи во вселенной до величины красного смещения, $z\approx2.0$. Эти точные измерения могут наложить ограничения на многочисленные модели темной энергии, и  некоторые из них могут быть отвергнуты.
Мы изучали 10 квинтэссенциальных и 7 фантомных $\phi$СDM моделей скалярного поля, которые впервые были представлены в статьях, (\cite{Frieman:1995pm}, \cite{Ferreira:1997hj}, \cite{Zlatev:1998tr}, \cite{Brax:1999gp}, \cite{Sahni:1999qe}, \cite{Barreiro:1999zs}, \cite{Albrecht:1999rm}, \cite{UrenaLopez:2000aj}, \cite{Caldwell:2005tm}, \cite{Scherrer:2008be}, \cite{Dutta:2009dr}, \cite{Rakhi:2009qf}, \cite{Chang:2016aex}, \cite{Bag:2017vjp}).
Мы предложили феноменологический метод исследования этих потенциалов.
В результате применения этого метода для каждого потенциала были найдены диапазоны величин начальных условий, модельных параметров и параметра уравнения состояния, при которых скалярное поле с данным потенциалом может проявить себя в ходе эволюции вселенной. Мы также исследовали, насколько хорошо различные модели скалярного поля могут быть аппроксимируемы Chevallier-Polarsky-Linder (CPL) параметризацией. Мы определили местоположение каждой модели в пространстве CPL параметров.
Одной из целей этого исследования является получение ответа на вопрос:
'Можно ли определить более предпочтительные модели в сравнении со стандартной $\Lambda$CDM моделью в современную эпоху, используя прогнозируемые данные будущих наблюдений DESI?'  Для этой цели были рассчитаны темп расширения, угловое расстояние и функция темпа роста флуктуаций плотности материи как для каждой исследуемой $\phi$СDM модели, так и для $\Lambda$CDM модели. Мы применяли критерии сравнения статистики Байеса, такие как коэффициенты Байеса, а также Акайке (Akaike) ($AIC$) и Байесовский (Bayesian) ($BIC$) информационные критерии.

Данная диссертация организована следующим образом: в Главе II рассматриваются теоретические основы космологии; в Главе III описываются типы расстояний, применяемых в космологии; в Главе IV представлены космологические наблюдения;  Глава V посвящена основам статистического анализа; в Главе VI рассматриваются модели темной энергии; в Главе VII описываются исследования Ратра-Пиблс $\phi$CDM модели скалярного поля; ограничения на модельные параметры Ратра-Пиблс $\phi$CDM модели рассмотрены в Главе VIII; наблюдательные ограничения на плоские квинтэссенции и фантомные модели скалярного поля обсуждаются в Главе IX; в Главе X содержится заключение; в Главе XI представлены планы будущих исследований.

Мы применяли естественную систему единиц: $c=\hbar=k_B=1$.

%%%%%%%%%%%%%%%%%%%%%%%%%%%%%%%%%%%%%%%%%%%%%%%%%%%%%%%%%%%%%%%%%%%%%%%%%%%
%%%%%%%%%%%%%%%%%%%%       Beginng of chapter 2     %%%%%%%%%%%%%%%%%%%%%%%
%%%%%%%%%%%%%%%%%%%%%%%%%%%%%%%%%%%%%%%%%%%%%%%%%%%%%%%%%%%%%%%%%%%%%%%%%%%

\chapter{Космология как наука}\label{chapter:2}

 Людей с древних времен всегда интересовало устройство мира в котором они живут. Завороженно вглядываясь в ночное небо, они задавались вопросами: 'Как возникла вселенная и как она устроена? Будет ли вселенная существовать вечно, а если нет, то как она завершит свое существование? Конечна ли вселенная и каковы ее размеры или же она бесконечна?' Именно любопытство людей узнать больше о вселенной побудило возникновение и развитие науки {\it космологии}.

Космология изучает вселенную как единое целое (как единую систему), исследует происхождение, эволюцию, динамику, структуру и окончательную судьбу вселенной. Особенностью этой науки является то, что объект ее исследования эксклюзивен и, по-видимому, существует в единственном экземпляре. Также составляет существенную трудность в изучении вселенной тот факт, что исследователю очень трудно делать объективные выводы о вселенной (о системе), частью которой он сам является. Эмпирическим фундаментом космологии является  внегалактическая астрономия, теоретическим фундаментом - основные физические теории, такие как общая теория относительности, теория поля и др.
Космология основана на результатах исследования наиболее общих свойств (таких как однородность, изотропность\footnote{Понятие однородности подразумевает, что
вселенная выглядит одинаково в каждой точке пространства; понятие изотропности означает, что вселенная выглядит
одинаково во всех направлениях. Из выполнения условия однородности автоматически не вытекает выполнение условия изотропности, как и наоборот. Только из требования выполнения условия изотропности относительно каждой точки пространства, вытекает выполнение условия однородности.}, расширение) той части вселенной, которая доступна для астрономических наблюдений.

В связи с тем, что скорость света имеет конечную величину, мы можем наблюдать только определенную часть расширяющейся вселенной, радиус которой  составляет приблизительно $14.25$ Гпс. На пространственных космологических масштабах, усредненная величина которых больше чем $100$ Mпс, крупномасштабная структура, которая включает в себя: галактики, скопления и сверхскопления галактик, не наблюдаема во вселенной. На этих масштабах применим принцип относительности или, так называемый {\it принцип Коперника}. Согласно этому принципу, во вселенной нет выделенных точек, и человеческие существа не являются привилегированными наблюдателями в ней. Таким образом, нашу вселенную можно считать {\it изотропной и однородной} на пространственных космологических масштабах.

\begin{figure}[h!]
\begin{center}
\psfig{file=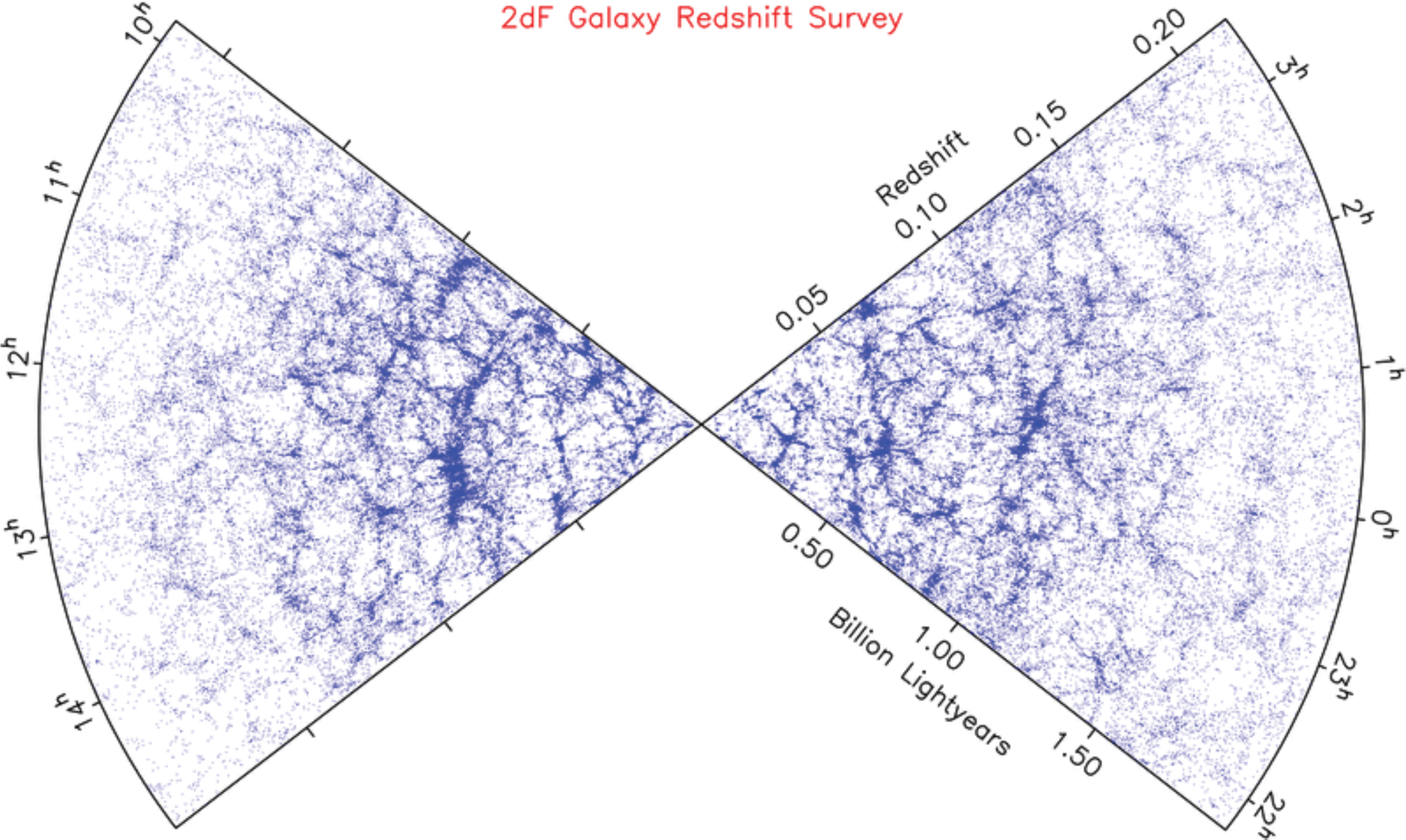,width=0.9\columnwidth}
\end{center}
 \caption {Пространственное распределение галактик по данным
 Two-degree-Field (2dF) Galaxy Redshift Survey. Радиальное направление соответствует скорости удаления или красному смещению галактики, полярный угол соответствует прямому восхождению галактики. Это распределение получено для 200 000 галактик с использованием 350 000 спектров, (\cite{Colless:2003wz}).}
 \label{fig:f1}
 \end{figure}

На  Рис.~(\ref{fig:f1}) представлено пространственное распределение близлежащих к нам галактик по данным Two-degree-Field (2dF) Galaxy Redshift Survey. В центре расположена наша галактика Млечный Путь (Milky Way). При увеличении расстояния (величины красного смещения) от нашей галактики, структурность в распределении галактик  становится менее четкой. На больших расстояниях (при больших величинах красного смещения) галактики расположены случайным образом, т. е. на этих масштабах наблюдается  изотропное и однородное распределение галактик.
На основе теоретических и экспериментальных результатов, Весто Слайфер, Джорж Леметр, Эдвин Хаббл обнаружили, что вселенная расширяется. Расширение является неотъемлемым свойством нашей вселенной.
Согласно модели горячей вселенной, наиболее распространенной в современной космологии,  вселенная начала свою эволюцию (расширение) приблизительно $13.7$ млрд лет назад в результате Большого Взрыва (Big Bang). В расширяющейся вселенной на ранней стадии развития вещество и излучение имели очень высокую температуру и плотность. Расширение вселенной привело к ее постепенному охлаждению, образованию атомов, а затем звезд, протогалактик, галактик, скоплений и сверхскоплений галактик, а также других космических тел, существующих сегодня.

\section{Расширение вселенной}

В 1917 году американский астроном Весто Слайфер, изучая спектры галактик, обнаружил смещение спектральных линий этих галактик к красному концу спектра\footnote{Красное смещение происходит благодаря эффекту Доплера. Этот эффект связан с  изменением частоты и, соответственно, длины волны излучения, воспринимаемого наблюдателем, вследствие движения источника излучения. При удалении источника излучения длина волны увеличивается, а при приближении источника излучения длина волны уменьшается.}.
На основе этих данных, Весто Слайфер сделал вывод, что галактики удаляются от нас.
В 1929 году американский ученый Эдвин Хаббл обнаружил, что радиальные скорости галактик, $v$, измеренные посредством доплеровского смещения спектральных линий, увеличиваются с увеличением физических расстояний до них, $d=|\vec{d}|$, (\cite{Hubble:1929ig}). Хаббл идентифицировал линейное соотношение между радиальными скоростями и физическими расстояниями\footnote{Определение понятия физического расстояния дано ниже.} между галактиками, $v\propto d$, которое называется законом Хаббла.
Математическая форма этого закона имеет вид:
\begin{equation}
\vec{v}=H_0\vec{d},
\label{eq:HL}
\end{equation}
где $H_0$ - коэффициент пропорциональности, который называется константой Хаббла\footnote{Коэффициент пропорциональности в законе Хаббла, $H_0$, является константой в современную эпоху. В общем случае этот коэффициент является функцией зависящей от времени (более подробное описание этой функции представлено ниже).}.
\begin{figure}[h!]
\begin{center}
\psfig{file=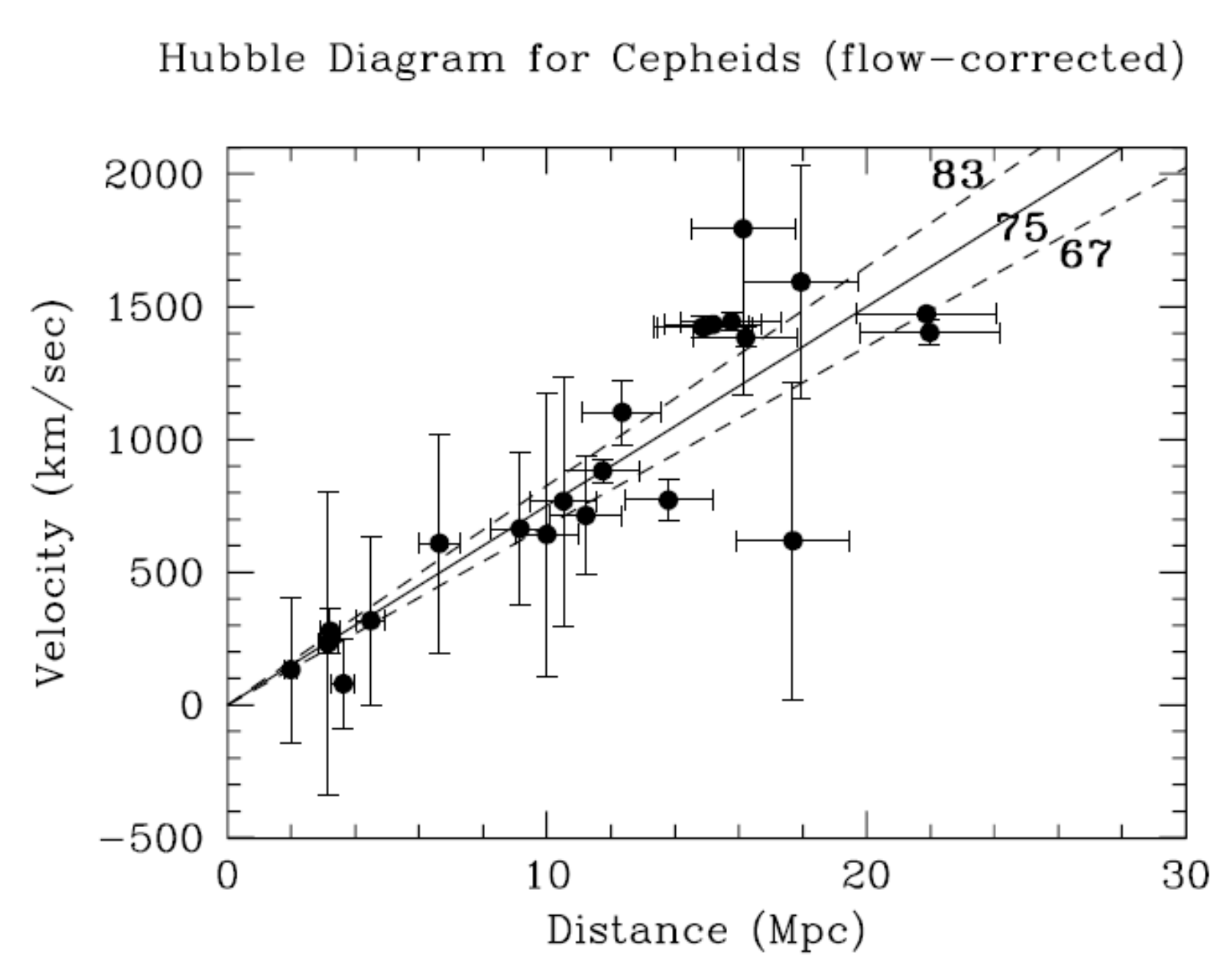,width= 0.6\columnwidth}
\end{center}
 \caption {Диаграмма Хаббла, построенная по наблюдательным данным проекта Hubble Space Telescope отдаленных цефеид. Сплошная линия соответствует закону Хаббла с $H_0=75~\text{км}~\text{с}^{-1}$~$\text{Mпс}^{-1}$, (\cite{Freedman:2000cf}).}
 \label{fig:f2}
 \end{figure}
На диаграмме Хаббла, Рис.~(\ref{fig:f2}), представлены величины радиальных скоростей в зависимости от величины физических расстояний, $d$. На этом рисунке точки аппроксимируются прямой, наклон которой определяется величиной константы Хаббла, $H_0$. Линейное увеличение радиальных скоростей галактик с увеличением физических расстояний до них можно интерпретировать удалением галактик друг от друга в результате расширения вселенной. При такой интерпритации радиальные скорости являются скоростями удаления галактик друг от друга (объяснение этого логического вывода приводится ниже). Расширение вселенной, названное расширением Хаббла, является одним из основных свойств нашей вселенной. \\
\\
Введем следующую терминологию\footnote{Подробная информация о разных видах расстояний, применяемых в космологии, содержится в Главе~III.}:\\
{\bf Собственное или физическое расстояние}

Физическое расстояние, $d$, является реальным, измеряемым расстоянием между двумя объектами в пространстве, где $t$ является космологическим или физическим временем.\\
\\
{\bf Сопутствующее или координатное расстояние}

Рассмотрим радиально расширяющуюся (или сжимающуюся) однородную сферу\footnote{Такое расширение (или сжатие) однородной сферы может служить моделью расширяющейся (или сжимающейся) вселенной.}. Выберем момент времени, $t=t_0$, который соответствует настоящему моменту времени и введем систему координат, $\vec{r}$, с началом координат, совпадающим с центром этой сферы. В результате расширения (или сжатия) сферы\footnote{В связи с тем, что  расширение (или сжатие) является радиальным, направление, $\vec{d}(t)$, будет оставаться постоянным.}, частица в настоящий момент времени будет находиться в положении, $\vec{d}(t_0)=\vec{r}$. В произвольный момент времени, $t$, частица будет находиться в положении, $\vec{d}(t)$:
\begin{equation}
\vec{d}(t)=a(t)\vec{r},
 \label{eq:Hubble2}
\end{equation}
 где функция $a(t)$ называется скалярным фактором. Скалярный фактор является функцией зависящей только от времени. Скалярный фактор описывает изменение пространственного разделения между объектами с течением времени и  характеризует расширение (или сжатие) вселенной. Для настоящего момента времени величина скалярного фактора обычно представлена в нормированном виде. В данной работе мы выбрали нормировку, при  которой величина скалярного фактора равна единице, $a(t_0)\equiv a_0=1$.

Наблюдатели, движущиеся согласно уравнению, Ур.~(\ref{eq:Hubble2}), называются сопутствующими наблюдателями.
В расширяющейся (сжимающейся) вселенной физическое расстояние между двумя сопутствующими объектами с течением времени увеличивается (уменьшается), в то время как сопутствующее расстояние между объектами, $r=|\vec{r}|$, не меняется с течением времени.\\
\\
{\bf Конформное время или сопутствующее время}

Конформное время - это время, прошедшее со времени Большого Взрыва по часам сопутствующего наблюдателя.
Дифференциал физического времени, $t$, и дифференциал конформного времени, $\eta$, взаимосвязаны следующим образом:
\begin{equation}
dt=a(t)d\eta.
 \label{eq:Conform}
\end{equation}
Величина конформного времени, $\eta$, может быть получить из уравнения, Ур.~(\ref{eq:Conform}):
\begin{equation}
\eta=\int_0^t \dfrac{dt'}{a(t')}.
 \label{eq:Conform1}
\end{equation}
Уравнение, Ур.~(\ref{eq:Conform1}), можно переписать как:
\begin{equation}
\eta=\int_0^a \dfrac{1}{a'H(a')}\dfrac{da'}{a'}.
 \label{eq:Conform2}
\end{equation}
Сопутствующее расстояние, $r$, и сопутствующе время, $\eta$, представленные в уравнениях, Ур.~(\ref{eq:Hubble2}) и Ур.~(\ref{eq:Conform1}), соответственно, образуют сопутствующую систему координат.
\section{Закон Хаббла}
Скорость сопутствующего наблюдателя может быть найдена как производная по времени от сопутствующего расстояния:
\begin{equation}
\vec{v}(d,t)=\dfrac{d}{dt}\vec{d}(t)=\dfrac{da}{dt}\vec{r}\equiv\dfrac{\dot{a}}{a}\vec{d}(t)\equiv H\vec{d}(t),
 \label{eq:ER}
\end{equation}
функция $H$ называется {\it темпом расширения вселенной}\footnote{Джорж Леметр (Georges Lemaître), основываясь на результатах исследований Весто Слайфера, предположил, что вселенная расширяется и впервые ввел понятие темпа расширения вселенной, $H$. Результаты его теоретических исследований были представлены в статье, (\cite{Lemaitre:1927}). Эта статья  была опубликована в 1927 году, за два года до публикации Эдвина Хаббла.}:
\begin{equation}
H=\dfrac{\dot{a}}{a}.
 \label{eq:Hubble}
\end{equation}
Закон Хаббла может быть записан в общем виде для  произвольного момента времени. Рассмотрим относительную скорость двух сопутствующих объектов, которые находятся в положении, $\vec{d}$ и $\vec{d}+d\vec{d}$, соответственно:
\begin{equation}
d\vec{v}(t)=\vec{v}(\vec{d}+d\vec{d}(t))-\vec{v}(\vec{d},t)=Hd\vec{d}(t).
 \label{eq:Hubble1}
\end{equation}
Таким образом, относительная скорость пропорциональна пространственному разделению сопутствующих объектов. Коэффициент пропорциональности, $H$, не зависит от положения наблюдателей, а зависит только от времени.
Параметр Хаббла для настоящего момента времени, $t=t_0$, называется постоянной Хаббла, $H(t_0)\equiv H_0$. Постоянная Хаббла обычно  представлена в параметризированном виде,
 $H_0=h\cdot~100~\text{км}~\text{с}^{-1}~\text{Mпс}^{-1}$, где $h$ - безразмерный параметр.
 В современную эпоху вселенная расширяется с ускорением, и гравитационно несвязанные астрономические объекты удаляются друг от друга,
поэтому $\dot{a}(t_0)>0$, т. е. масштабный фактор является возрастающей функцией в зависимости от времени.

Величина постоянной Хаббла, $H_0$, очень важна в космологии, так как она определяет возраст и темп расширения вселенной. Постоянная Хаббла определяет так называемое расстояние Хаббла или радиус сферы Хаббла, $r_{HS}$.  Радиус сферы Хаббла - это расстояние до объектов, которые удаляются от наблюдателя со скоростью света. Этот радиус определяет границу между объектами, движущимися медленнее или быстрее, чем движение объектов со скоростью света относительно наблюдателя в данный момент времени. В вообщем случае радиус сферы Хаббла, $r_{HS}$, вычисляется как\footnote{В нижеприведенных формулах сохраняется обозначение скорости света, $c$, для ясности.}: $r_{HS}(t)=c/H$. Соответственно, для настоящего времени радиус сферы Хаббла определяется как: $r_{HS}(t_0)=c/H_0$, и его величина равна $4.1$ Гпс.

\begin{figure}[h!]
\begin{center}
\includegraphics[width=0.6\columnwidth]{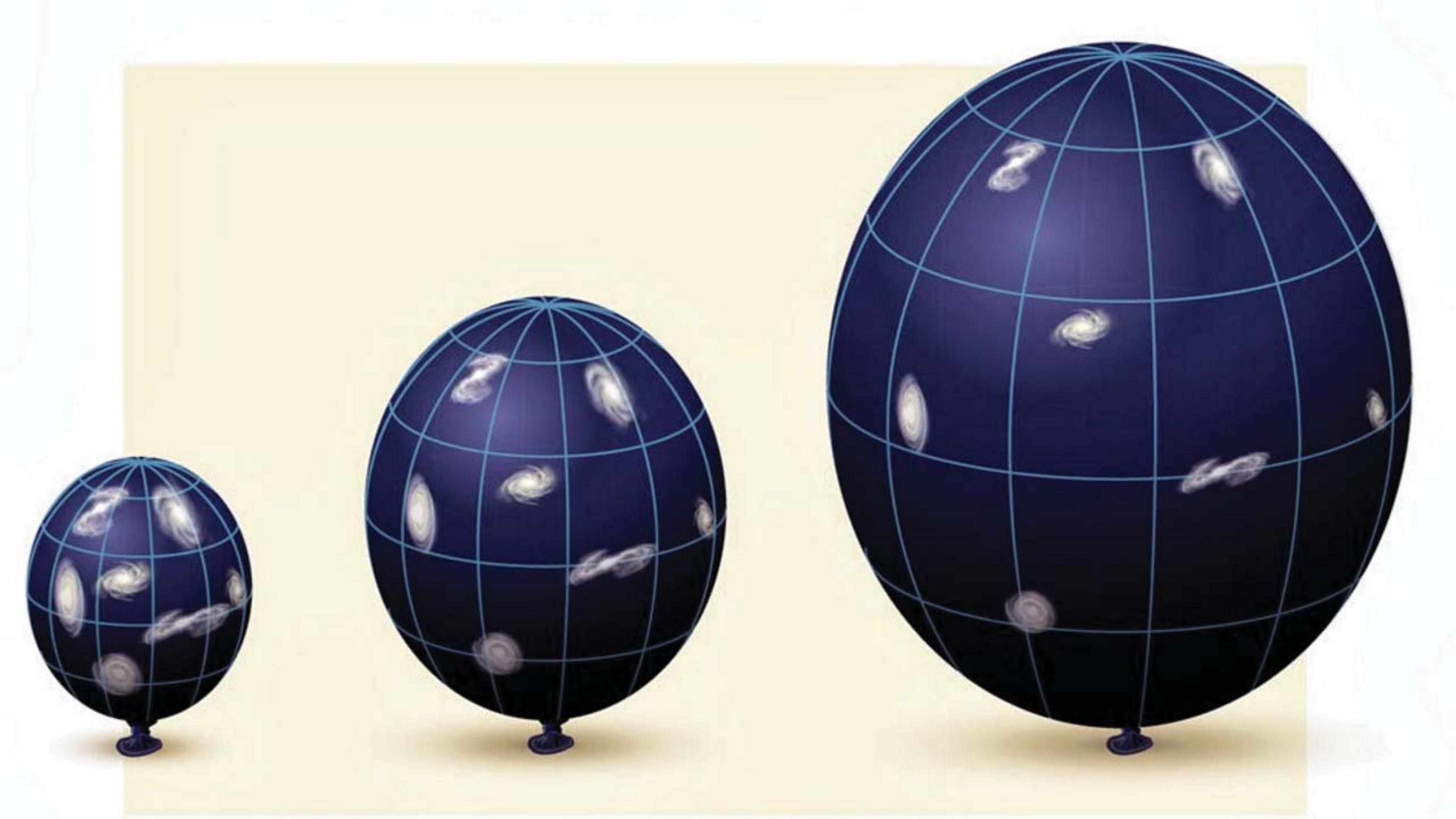}
\end{center}
 \caption {Расширение Хаббла, (https://www.nature.com).}
 \label{fig:f3}
 \end{figure}

Согласно закону Хаббла, Ур.~(\ref{eq:Hubble1}), в однородной и изотропной вселенной нет привилегированных точек,
и расширение будет одинаковым в любой точке пространства, Рис.~(\ref{fig:f3}). Это предположение согласуется с принципом Коперника.
Поэтому, являясь обобщенной характеристикой вселенной, величина постоянной Хаббла, $H_0$, одинакова для всех галактик и не зависит ни от напраления на галактику на небе, ни от расстояния до нее.

 Найдем производную от физического расстояния до некоторой галактики, $\vec{d}$, представленное в уравнении, Ур.~(\ref{eq:Hubble2}):
\begin{equation}
\vec{\dot{d}}(t)=\frac{\dot{a}}{a}\vec{d}(t)+\vec{u_p}(\vec{r},t),
 \label{eq:genvel}
\end{equation}
где $\vec{u_p}(\vec{r},t)$ - пекулярная скорость, которая определяет случайные движения галактики в пространстве. Пекулярная скорость характеризует отклонение движения близлежащей галактики от  однородного расширения Хаббла. На пространственных масштабах, меньших космологических, величина пекулярной скорости, $\vec{u_p}(\vec{r},t)$, в уравнении, Ур.~(\ref{eq:genvel}), превышает величину скорости движения галактики под воздействием расширения Хаббла, $\vec{v}=\dfrac{\dot{a}}{a}\vec{d}$. На этих масштабах движение галактик в большей мере определяется их случайным движением, чем влиянием расширения Хаббла, поэтому закон Хаббла не является точным на этих масштабах. С другой стороны, на пространственных космологических масштабах движение удаленных галактик полностью определяется расширением Хаббла, т. к. величины пекулярных скоростей галактик пренебрежимы в сравнении со скоростью расширения Хаббла. Движение астрономических объектов, обусловленное исключительно этим расширением, называется движением в соответствии с {\it потоком Хаббла}.

Открытия Весто Слайфера, Джоржа Леметра и Эдвина Хаббла являются фундаментом, на котором построена современная физическая космология. Эти открытия отмечены началом перехода космологии из описательной философской науки в точную науку, в которой каждая предложенная теория проверяется результатами наблюдательных экспериментов.

\section{Краткий обзор общей теории относительности}
\subsection{Метрика пространства-времени для криволинейных координат}
Общая теория относительности (ОТО) является теоретической основой современной космологии, (\cite{Einstein:1915by}, \cite{Einstein:1915ca}; (монографии:  \cite{Landau1971}, \cite{Weinberg:1972kfs}, \cite{Misner:1974qy}, \cite{Carroll:2004st}). В ОТО рассматривается пространство-время с четырехмерными криволинейными координатами, $x^{\mu}=(x^0, x^1, x^2, x^3)$. Пространственная часть пространства-времени обозначается как, $x^1, x^2,\\ x^3$, в то время как временная часть - $x^0=t$, где $t$ - физическое время. Расстояние между двумя соседними точками с координатами, $x^{\mu}$ и $x^{\mu} + dx^{\mu}$ задается линейным элементом, квадрат которого в криволинейных координатах является квадратичной формой дифференциалов, $dx^{\mu}$ или так называемой метрикой:
\begin{equation}
ds^2 \equiv g_{\mu\nu}dx^{\mu}dx^{\nu},
 \label{eq:FLRW0}
\end{equation}
где $g_{\mu\nu}$ - ковариантный метрический тензор пространства-времени, который является функцией координат. Величина метрики является инвариантом при переходе от одной системы координат к другой.
Ковариантный метрический тензор, $g_{\mu\nu}$, симметричен по индексам $\mu$ и $\nu$: $g_{\mu\nu}=g_{\nu\mu}$. Ковариантный метрический тензор
является обратным к контравариантному метрическому тензору, $g^{\mu\nu}$:
\begin{equation}
g_{m\mu}g^{\mu\nu}=\delta^\nu_m,
 \label{eq:MT1}
\end{equation}
где $\delta^\nu_m$  дельта-функция Кронекера.\\
\\
{\bf Дельта-функция Кронекера}

Дельта-функция Кронекера - это единичный четырехмерный тензор,
 который определен как:
\begin{equation}
\delta^\nu_m x^m=x^\nu,
\label{eq:DFK}
\end{equation}
в матричной форме это выражение имеет вид:
\begin{equation}\label{eq:SK}
\delta^\nu_m =
\begin{cases}
1, \quad m=\nu \\
0, \quad m\neq \nu
\end{cases}
\end{equation}
След\footnote{След (или Spur-нем.) матрицы — это сумма элементов главной диагонали матрицы. Если $b_{ij}$ - элементы матрицы $B$, то след этой матрицы определяется как, ${\rm tr}(B)=\sum_i b_{ii}$.} дельта-функции Кронекера равен, $\sum_i\delta_i^i=4$. Дельта-функция Кронекера обладает следующим свойством - компоненты этой функции одинаковы в любой системе координат.

\subsection{Преобразование криволинейных координат}
Рассмотрим преобразование скаляра, вектора и тензора от одной криволинейной системы координат, ($x^0, x^1, x^2, x^3$), в другую, ($x^{'0}, x^{'1}, x^{'2}, x^{'3}$).\\
\\
{\bf Скаляр (тензор нулевого ранга)}

Скаляр - это величина, которая в любой координатной системе полностью определяется одним числом (или функцией). Величина скаляра не меняется при переходе от одной системы координат в другую.
Если $\varphi$ - величина скаляра в одной системе координат, ($x^0, x^1, x^2, x^3$), и $\varphi'$ - величина скаляра в другой системе координат, ($x^{'0}, x^{'1}, x^{'2}, x^{'3}$), то
\begin{equation}
\varphi'(x^{'0}, x^{'1}, x^{'2}, x^{'3})=\varphi(x^0, x^1, x^2, x^3).
 \label{eq:Scalar}
\end{equation}
Обычно скаляр имеет одну компоненту. Примеры скаляров: давление, плотность, температура, объем, длина, площадь и т. д.\\
\\
{\bf Векторы (тензоры первого ранга)}

Четырехмерный вектор - это величина, определяемая в четырехмерной криволинейной системе координат четырьмя числами  в случае  контравариантного вектора как: $A^i=A^0, A^1, A^2, A^3$;  в случае ковариантного вектора как: $A_i=A_0, A_1, A_2, A_3$.

Например, в преобразованиях Лоренца при переходе от одной четырехмерной системы координат в другую, компоненты четырехмерного контравариантного вектора, $A^i$, преобразуются следующим образом\footnote{В этой формуле сохраняется обозначение скорости света, $c$, для ясности.}:
\begin{equation}
A^0=\frac{A^{'0}+(V/c)A^{'1}}{\sqrt{1-V^2/c^2}},~~~A^1=\frac{A^{'1}+(V/c)A^{'0}}{\sqrt{1-V^2/c^2}},~~~A^2=A^{'2},~~~A^3=A^{'3'},
  \label{eq:Lorens}
\end{equation}
где $V$ - скорость движения одной инерциальной системы координат  относительно  другой инерциальной системы координат.

Ковариантный вектор, $A_i$, является ковектором контравариантному вектору, $A^i$. Элементы ковариантных, $A_i$, и контравариантных векторов, $A^i$, взаимосвязаны следующим образом:
\begin{equation}
 A_0=A^0,~~~A_1=-A^1,~~~A_2=-A^2,~~~A_3=-A^3.
  \label{eq:Vector}
\end{equation}
Компоненты четырехмерного вектора могут быть записаны как:
\begin{equation}
A^i=(A^0, \vec{A}),~~~A_i=(A^0, -\vec{A}),
\label{eq:Vect_comp}
\end{equation}
где $A^0$ - временная координата, являющаяся скалярной величиной; $\vec{A}$ - трехмерный вектор, содержащий пространственные координаты.
Квадрат величины четырехмерного вектора определен как\footnote{В тензорном анализе применяют правило Эйнштейна, согласно которому: повторение индексов в выражении  дважды (при этом один из них стоит наверху, а другой внизу) означает суммирование, при этом знак суммы опускается.}:
\begin{equation}
\sum_{i=0}^3A^iA_i=A^0A_0+A^1A_1+A^2A_2+A^3A_3.
  \label{eq:Square}
\end{equation}
Связь между ковариантными и контравариантными векторами осуществляется через метрический тензор, $g_{\mu\nu}$, который применяют для повышения или понижения индексов векторов и тензоров\footnote{В частном случае, при рассмотрении пространства Минковского, символ Кронекера, $\delta^\nu_m$, используется для повышения или понижения индексов.}:
\begin{equation}
g^{ik}A_k=A^i,~~g_{ik}A^k=A_i.
\end{equation}
В общем случае, в криволинейных координатах контравариантные, $A^i$, и ковариантные, $A_i$, четырехмерные вектора преобразуются следующим образом:
\begin{equation}
A^i=\dfrac{\partial x^i}{\partial x^{'k}}A^{'k},~~~~~A_i=\dfrac{\partial x^{'k}}{\partial x^i}A_k^{'}.
 \label{eq:TV}
\end{equation}
\\
{\bf Тензоры (тензоры второго и выше рангов)}

Четырехмерным тензором второго ранга называется совокупность $4^2=16$ компонент этого тензора, которые при переходе от одной системы координат в другую преобразуются как произведение компонент двух четырехмерных векторов. Аналогичным образом можно дать определение четырехмерным тензорам третьего ранга (с $4^3=64$ компонентами) и тензорам высшего N-го ранга, которые включают в себя $4^N$ компонент.

Компоненты четырехмерного тензора второго ранга могут быть представлены как: контравариантные, $A^{ik}$, ковариантные, $A_{ik}$, и смешанные, $A^i_k$.

Контравариантный тензор второго ранга, $A^{ik}$, образуется в результате произведения двух четырехмерных контравариантных векторов $A^i=\dfrac{\partial x^i}{\partial x^{'l}}A^{'l}$ и $A^k=\dfrac{\partial x^k}{\partial x^{'m}}A^{'m}$. При переходе от одной системы координат в другую, компоненты контравариантного тензора второго ранга преобразуются как:
\begin{equation}
A^{ik}=A^i\cdot A^k=\dfrac{\partial x^i}{\partial x^{'l}} \dfrac{\partial x^k}{\partial x^{'m}} A^{'lm}.
\label{eq:Contr_tenz}
\end{equation}
Ковариантный тензор второго ранга, $A_{ik}$, образуется в результате произведения двух ковариантных четырехмерных векторов, $A_i=\dfrac{\partial x^{'l}}{\partial x^i} A^{'}_{l}$ и $A_k=\dfrac{\partial x^{'m}}{\partial x^k} A^{'}_{m}$. При переходе от одной системы координат в другую, компоненты ковариантного тензора второго ранга преобразуются как:
\begin{equation}
A_{ik}=A_i\cdot A_k=\dfrac{\partial x^{'l}}{\partial x^i} \dfrac{\partial x^{'m}}{\partial x^{k}} A^{'}_{lm}.
\label{eq:Covar_tenz}
\end{equation}
Смешанный тензор второго ранга, $A^i_k$, образуется в результате произведения четырехмерного контравариантного вектора, $A^i=\dfrac{\partial x^{i}}{\partial x^{'l}}A^{'l}$, и ковариантного четырехмерного вектора, $A_k= \dfrac{\partial x^{'m}}{\partial x^{k}}A^{'}_{m}$.  При переходе от одной системы координат в другую, компоненты смешанного тензора второго ранга преобразуются как\footnote{Здесь и выше использовались следующие обозначения: $A^{'lm}=A^{'l}\cdot A^{'m}$, $A^{'}_{lm}=A^{'}_{l}\cdot A^{'}_{m}$, $A^{'l}_{m}=A^{'l}\cdot A^{'}_{m}$}:
\begin{equation}
A^i_k=A^i\cdot A_k=\dfrac{\partial x^{i}}{\partial x^{'l}}\dfrac{\partial x^{'m}}{\partial x^{k}} A^{'l}_m.
 \label{eq:TT}
\end{equation}
Четырехмерные тензоры (контравариантные, ковариантные, смешанные) $N$-го ранга преобразуются в результате произведения  $N$ четырехмерных (контравариантных, ковариантных, смешанных) векторов, соответственно.  При переходе от одной системы координат в другую, компоненты  тензоров (контравариантных, ковариантных, смешанных)  $N$-го ранга преобразуются, соответственно, как:
\begin{equation}
A^{\beta_1...\beta_N}=\dfrac{\partial x^{\beta_1}}{\partial x^{'\gamma_1}}...\dfrac{\partial x^{\beta_N}}{\partial x^{'\gamma_N}}A^{'\gamma_1...\gamma_N},
 \label{eq:CTN}
\end{equation}
\begin{equation}
A_{\beta_1...\beta_N}=\dfrac{\partial x^{'\gamma_1}}{\partial x^{\beta_1}}...\dfrac{\partial x^{'\gamma_N}}{\partial x^{'\beta_N}}A_{'\gamma_1...'\gamma_N},
 \label{eq:CovTN}
\end{equation}
\begin{equation}
A^{\beta_1...\beta_l}_{\beta_{1+1}...\beta_N}=\dfrac{\partial x^{\beta_1}}{\partial x^{'\gamma_1}}...\dfrac{\partial x^{\beta_l}}{\partial x^{'\gamma_l}}\dfrac{\partial x^{'\gamma_{l+1}}}{\partial x^{\beta_{l+1}}}...\dfrac{\partial x^{'\gamma_N}}{\partial x^{'\beta_N}}A^{'\gamma_1...'\gamma_l}_{'\gamma_{l+1}...'\gamma_N}.
 \label{eq:MTN}
\end{equation}
\\
\\
{\bf Математические операции, производимые над тензорами}
\begin{itemize}
\item[$\bullet$] Сложение тензоров: $A^{\alpha\beta}_{\gamma\delta}+B^{\alpha\beta}_{\gamma\delta}=C^{\alpha\beta}_{\gamma\delta}$
\item[$\bullet$] Вычитание тензоров: $A^{\alpha\beta}_{\gamma\delta}- B^{\alpha\beta}_{\gamma\delta}=F^{\alpha\beta}_{\gamma\delta}$
\item[$\bullet$]  Произведение тензоров: $A^{\alpha\beta}_{\gamma\delta}B^{\eta\nu}_{\gamma\delta}=C^{\alpha\beta\eta\nu}_{\gamma\delta\gamma\delta}$
\item[$\bullet$] Сворачивание тензоров в результате суммирования по одинаковым индексам: $B^{\lambda\chi}_{\chi\xi}=H^{\lambda}_{\xi}$
\item[$\bullet$] Внутреннее произведение тензоров: $F^{\alpha\beta}_{\phi\sigma}K^{\sigma\psi}_{\gamma\omega}=M^{\alpha\beta\sigma\psi}_{\phi\sigma\gamma\omega}=N^{\alpha\beta\psi}_{\phi\gamma\omega}$
\end{itemize}
\subsection{Ковариантные производные}
Рассмотрим вектор, $A_i$, в криволинейных координатах. Дифференциал, $dA_i$, этого вектора не является вектором, а производная, $\partial A_i/\partial x^k$, не является тензором. Это происходит из-за того, что дифференциал, $dA_i$, является разностью векторов, расположенных в различных точках искривленного пространства. Векторы в искривленном пространстве в различных точках преобразуются по-разным законам, поэтому для криволинейных координат применяют специальный тип производных - ковариантные или контравариантные производные.

Ковариантные производные для контравариантных и ковариантных векторов определяются как:
\begin{equation}
A^i_{;j}=\dfrac{\partial A^i}{\partial x^j} + \Gamma^i_{kj}A^k,~~~~A_{i;j}=\dfrac{\partial A_i}{\partial x^j} - \Gamma^k_{ij}A_k,
 \label{eq:CDV}
\end{equation}
где функции,  $\Gamma^\lambda_{\mu\nu}$, называются символами Кристоффеля или афинной связностью. Символы Кристоффеля выражаются через производные метрического тензора следующим образом:
\begin{equation}
\Gamma^{\lambda}_{\mu\nu}=\frac{1}{2} g^{\lambda\kappa}\left( \frac{\partial g_{\kappa\mu}}{\partial x^\nu} + \frac{\partial g_{\kappa\nu}}{\partial x^\mu} - \frac{\partial g_{\mu\nu}}{\partial x^\kappa}\right).
\label{eq:CS}
\end{equation}
Ковариантные производные для тензоров второго ранга: контравариантного, $A^{ik}$, ковариантного, $A_{ik}$, и смешанного типа, $A^i_k$, определяются как:
\begin{equation}
A^{ik}_{;j}=\dfrac{\partial A^{ik}}{\partial x^j} + \Gamma^i_{mj}A^{mk}+ \Gamma^k_{mj}A^{im},
 \label{eq:CDT}
\end{equation}
\begin{equation}
A_{ik;j}=\dfrac{\partial A_{ik}}{\partial x^j} - \Gamma^m_{ij}A_{km} - \Gamma^m_{kj}A_{im},
 \label{eq:CDT1}
\end{equation}
\begin{equation}
A^i_{k;j}=\dfrac{\partial A^i_k}{\partial x^j} - \Gamma^m_{kj}A^i_{m}+ \Gamma^i_{mj}A^m_k.
 \label{eq:CDT2}
\end{equation}
Контравариантные производные могут быть образованы из ковариантных с помощью повышения индекса, который означает дифференцирование. Это можно сделать с помощью контравариантного метрического тензора:
\begin{equation}
A_i^{;k}=g^{kj}A_{i;j},~~~~A^{i;k}=g^{kj}A^i_{;j}.
 \label{eq:CDV5}
\end{equation}
\section{Тензоры Римана-Кристоффеля, Риччи, и Эйнштейна. Скаляр Риччи. Их свойства.}
{\bf Тензор Римана-Кристоффеля}

Комбинация символов Кристоффеля и их производных образуют тензор кривизны или тензор Римана-Кристоффеля четвертого ранга, $R^i_{klm}$:
\begin{equation}
 R^i_{klm}=\dfrac{\partial \Gamma^i_{km}}{\partial x^l}-\dfrac{\partial \Gamma^i_{kl}}{\partial x^m} + \Gamma^i_{nl}\Gamma^n_{km}-\Gamma^i_{nm}\Gamma^n_{kl}.
 \label{eq:RT}
\end{equation}
Тензор Римана-Кристоффеля обладает следующими свойствами:\\
$\bullet$ Свойство цикличности: $R^i_{klm}+R^i_{mkl}+R^i_{lmk}=0$\\
$\bullet$ Свойство антисимметрии $l$, $m$ индексов: $R^i_{klm}=-R^i_{kml}$\\
$\bullet$ Свойство симметрии: $R_{iklm}=R_{lmik}$\\
$\bullet$ Свойство асимметрии: $R_{iklm}=-R_{kilm}=-R_{ikml}$\\
$\bullet$ Первое тождество Бьянки: $R_{iklm}+R_{imkl}+R_{ilmk}=0$\\
$\bullet$ Второе тождество Бьянки: $R^n_{ikl;m}+R^n_{imk;l}+R_{ilm;k}=0$

Равенство или неравенство нулю тензора Римана-Кристоффеля, $R^i_{klm}$, является критерием для определения является ли четырехмерное пространство-время плоским или искривленным. При этом верна как прямая теорема: если четырехмерное пространство-время является плоским (искривленным), то тензор кривизны равен нулю (не равен нулю), так и обратная: если тензор кривизны равен нулю (не равен нулю), то  четырехмерное пространство-время является плоским (искривленным).\\
\\
{\bf Тензор Риччи}

Тензор Риччи второго ранга, $R_{ik}$, получается в результате свертки тензора Римана-Кристоффеля:
\begin{equation}
g^{lm}R_{limk}=R^l_{ilk}=R_{ik}.
 \label{eq:RicciT}
\end{equation}
Тензор Риччи определен как:
\begin{equation}
R_{ik}=\frac{\partial\Gamma^l_{ik}}{\partial x^l} - \frac{\partial\Gamma^l_{il}}{\partial x^k} + \Gamma^l_{ik}\Gamma^m_{lm} - \Gamma^m_{il}\Gamma^l_{km}.
\label{eq:RicciT1}
\end{equation}
Из уравнения, Ур.~(\ref{eq:RicciT1}), очевидна симметричность тензора Риччи, $R_{ik}=R_{ki}$.\\
\\
{\bf Cкаляр Риччи}

Сворачивая тензор Риччи, $R_{ik}$, мы получим скалярную величину, $R$, которая называется скаляром Риччи или скалярной кривизной:
\begin{equation}
R=g^{ik}R_{ik}=g^{il}g^{km}R_{iklm}.
 \label{eq:RicciS}
\end{equation}
Скаляр Риччи является следом тензора Риччи, $R_{ik}$: $R=\sum_iR_{ii}$.

В ОТО  действие для гравитационного поля, $S_G$, выражается через  интеграл по четырёхмерному объему, $d\Omega$, от скалярной плотности кривизны, $R\sqrt{-g}$, следующим образом:
\begin{equation}
S_G=8\pi G\int_M R\sqrt{-g}d\Omega,
 \label{eq:Action_Grav}
 \end{equation}
 где $g$ - определитель, составленный из элементов матрицы метрического тензора, $g_{\mu\nu}$.\\
\\
 {\bf Тензор Эйнштейна}

Комбинация тензора Риччи, $R_{\mu\nu}$, скаляра Риччи, $R$, и метрического тензора, $g_{\mu\nu}$, определяют тензор Эйнштейна:
\begin{equation}
 G_{\mu\nu}=R_{\mu\nu}-\dfrac{1}{2}g_{\mu\nu}R.
 \label{eq:ET}
 \end{equation}
Тензор Эйнштейна, $G_{\mu\nu}$, является тензором второго ранга  в N-мерном пространстве-времени. Он содержит $N(N+1)/2$ независимых компонентов.
Этот тензор может быть построен только из квадратичных по первым производным от метрики или линейным по второй производной от метрики членов.

Тензор Эйнштейна симметричен ввиду симметричности составляющих его компонент: тензора Риччи, $R_{\mu\nu}$, и метрического тензора, $g_{\mu\nu}$:
\begin{equation}
 G_{\mu\nu}=G_{\nu\mu}.
 \label{eq:ET1}
 \end{equation}
Тензор Эйнштейна инвариантен при ковариантном дифференцировании, т. е. ковариантная производная тензора Эйнштейна тождественно равна нулю:
\begin{equation}
 G_{\mu\nu;\lambda}=0.
  \label{eq:TE}
 \end{equation}
\subsection{Тензор энергии-импульса}
В ОТО понятие тензора энергии-импульса или тензора напряжения, $T_{\mu\nu}$, включает в себя все возможные формы материи и энергии\footnote{В соответствии с принципом эквивалентности массы и энергии в ОТО.}, которые могут искривлять пространство-время. Тензор энергии-импульса характеризует все, что может содержаться в определенной области пространства-времени: поток энергии и поток импульса, плотность  энергии и плотность импульса, а также энергию и массу.
Тензор энергии-импульса определяется как поток четырехмерного импульса, проходящего через трехмерную поверхность постоянных координат.

Тензор энергии-импульса, $T_{\mu\nu}$, является тензором второго ранга. Его свойства идентичны со свойствами тензора Эйнштейна, $G_{\nu\mu}$, такими как: симметричность тензора энергии-импульса, $T_{\mu\nu}=T_{\nu\mu}$, и равенство нулю ковариантной производной тензора энергии-импульса или выполнение закона сохранения для тензора энергии-импульса:
\begin{equation}
 T_{\mu\nu ;\nu}=0.
\label{eq:EC}
 \end{equation}
В предельном случае метрики Минковского (которая  описывается ниже в уравнении, Ур.~(\ref{eq:Minkovski})), ковариантная производная трансформируется в обычную производную:
\begin{equation}
 \dfrac{T_{\mu\nu}}{\partial x^{\nu}}=0.
\label{eq:EC1}
 \end{equation}
В присутствии гравитационного поля, закон сохранения тензора энергии-импульса принимает следующий вид:
\begin{equation}
 T_{\mu\nu;\nu}=\dfrac{\partial T_{\mu\nu}}{\partial x^{\nu}}+ \Gamma^{k}_{\mu\nu}T_{k \nu}+\Gamma^{k}_{k\nu} T_{\mu \nu}=0.
 \label{eq:EC2}
 \end{equation}
Рассмотрим различные формы тензора энергии-импульса, $T_{\mu\nu}$, для случая идеальной жидкости, вакуума и пыли.

\subsubsection{Идеальная жидкость}
Идеальная жидкость изотропна относительно системы координат, в которой она покоится. Идеальную жидкость можно полностью охарактеризовать ее плотностью массы покоя, $\rho$, и изотропным давлением, $p$, которые связаны между собой уравнением состояния, $p=f(\rho)$. Эта жидкость не имеет вязкости или теплопроводности. В космологии модель идеальной жидкости применяется для описания ранней вселенной на стадии доминирования энергии излучения.

Для любой системы координат тензор энергии-импульса для идеальной жидкости  имеет вид:
\begin{equation}
T_{\mu\nu}=(\rho + p)u_{\mu}u_{\nu} - pg_{\mu\nu},
 \label{eq:EE}
 \end{equation}
здесь $u_{\mu}$ - это четырехмерная скорость.

Четырехмерная скорость определяется формулой:
\begin{equation}
u_{\mu}\equiv\dfrac{dx_{\mu}}{ds}.
\label{eq:Velocity}
 \end{equation}

Четырехмерная скорость нормирована как, $u^{\mu}u_{\mu}\equiv1$\footnote{В геометрическом представлении, $u_{\mu}$ - это единичный четырехмерный вектор, касательный к мировой линии частицы.}. Поэтому для наблюдателя в сопутствующей системе координат, относительно которой идеальная жидкость покоится, четырехмерная скорость, $u_{\mu}$,  принимает вид, $u_{\mu}=(1,0,0,0)$.

В сопутствующей системе координат тензор энергии-импульса для идеальной жидкости принимает вид:
\begin{equation} \label{eq:PFM}
T_{\mu\nu}=\left(\begin{array}{cccc}
           \rho & 0 & 0 & 0 \\
           0 & p & 0  & 0\\
           0 & 0 & p & 0 \\
           0 & 0 & 0 & p \\
           \end{array}
           \right).
\end{equation}

Из уравнения сохранение энергии-импульса, Ур.~(\ref{eq:EC}), следует уравнение непрерывности:
\begin{equation}
\dfrac{\partial \rho}{\partial t} +\nabla(\rho \vec{v_f})=0,
\label{eq:CE4}
\end{equation}
где $\vec v_{f}$ - вектор трехмерной скорости идеальной жидкости, $v_{f}=|\vec v_{f}|$.

Это уравнение описывает динамику идеальной жидкости и отражает факт сохранения материи. В самом деле,
сходящееся поле скоростей приводит к увеличению плотности жидкости. И наоборот, расходящееся поле скоростей приводит к уменьшению плотности жидкости.

\subsubsection{Вакуум}
В этом случае не существует ни полей, ни энергии, ни материи в выбранной области пространства-времени.
Компоненты тензора энергии-импульса, $T_{\mu\nu}$, для вакуума равны нулю:
\begin{equation}
T_{\mu\nu}=0.
 \label{eq:STV}
 \end{equation}

 \subsubsection{Пылевидная материя}
В космологии материя во вселенной аппроксимируется моделью пылевой жидкости (или пылевидной материей)\footnote{Правомерность такого приближения связана с тем, что в гравитационных задачах астрофизики и космологии материя испытывает очень большие напряжения, поэтому становится текучей.}, состоящей из одинаковых, электрически нейтральных массивных частиц, движущихся с одинаковыми скоростями намного меньшими, чем скорость света, $u\ll c$. Пылевая жидкость характеризуется нулевым давлением, плотностью покоя, $\rho$, и четырехмерной скоростью, $u(\vec{r},t)$\footnote{Реальная вселенная содержит в себе многокомпонентные потоки пылевидной материи.}.

В этом случае тензор энергии-импульса для любой системы координат определется как:
\begin{equation}
T_{\mu\nu}=\rho u_{\mu}u_{\nu}.
 \label{eq:SED}
 \end{equation}
В сопутствующей системе координат тензор энергии-импульса принимает вид:
\begin{equation}
T_{\mu\nu}=\left(\begin{array}{cccc}
           \rho & 0 & 0 & 0 \\
           0 & 0 & 0  & 0\\
           0 & 0 & 0 & 0 \\
           0 & 0 & 0 & 0 \\
           \end{array}
           \right).
\end{equation}
В предельном случае малых скоростей и  нулевом  давлении, модель идеальной жидкости сводится к модели пылевидной жидкости. Модель пылевидной жидкости применяется при описании вселенной в более позднюю эпоху чем  доминирование энергии излучения - в эпоху доминирования энергии материи.

\subsection{Материя во вселенной}
Материю во вселенной образуют нерелятивистские частицы, составленные из барионов, массивного нейтрино и темной материи. Общим свойством для всех частиц, образующих материю, является то, что эти частицы скапливаются под действием сил гравитации. Плотность числа частиц, $n(t)$, и плотность энергии, $\rho(t)$, материи изменяются со временем одинаковым образом\footnote{Этот результат справедлив только для холодной темной материи.}, $\rho(t)\sim n(t)\propto a^{-3}$(t).

Наблюдаемая вселенная состоит на $26 \%$ из темной материи; на $4.8\%$ из обычной, барионной материи; на $0.1 \%$ из нейтрино, согласно наблюдениям космической обсерватории Planck 2015, (\cite{Ade:2015xua}).

\subsubsection{Барионная материя}
Барионная материя состоит из барионов. Согласно Стандартной модели физики элементарных частиц, барионы принадлежат к семейству андронов. Барионы сформированы из трех кварков. В то же время, обладая полуцелым спином, барионы являются фермионами. Самыми легкими барионами являются нуклоны: протоны и нейтроны. Протоны состоят из одного $d$-кварка и двух $u$-кварков, $p=uud$, а нейтроны состоят из одного $u$-кварка и двух $d$-кварков, $n=ddu$, (\cite{Okun:1988}).

Барионы являются компонентами ядер атомов обычного вещества. Барионы составляют большую часть массы видимого вещества во вселенной, также барионы могут составлять невидимую  барионную темную материю. Современная величина плотности барионов составляет, $\rho_{\rm b0}\approx2.4\cdot10^{-7}$~ГэВ/см$^3$. Для поздней вселенной, характеризуемой величиной средней температуры, $\langle T\rangle \leq100$~КэВ, сохраняется постоянным отношение плотности числа барионов к плотности числа фотонов, $\eta_{\rm b}\equiv n_{\rm b}/n_{\gamma}\approx6.1\cdot10^{-10}$, (\cite{Rubakov:2015yza}).
\subsubsection{Массивное нейтрино}
Нейтрино принадлежат к семейству лептонов. Нейтрино, являясь лептоном, может участвовать только в слабых гравитационных взаимодействиях. Лептоны являются фермионами, их спин равен 1/2. Лептоны не имеют структуры, поэтому являются истинно элементарными частицами. Будучи нейтральной элементарной частицей, нейтрино имеет три разновидностей: электронное, $\nu_e$, мюонное, $\nu_{\mu}$, и тау-нейтрино, $\nu_{\tau}$. Если нейтрино является фермионом Дирака, тогда существуют антинейтрино, соответственно: $\tilde{\nu}_e, \tilde{\nu}_{\mu}, \tilde{\nu}_{\tau}$. Если же нейтрино является фермионом Майорана, тогда оно не имееет своей античастицы и является, как и фотон, истинно нейтральной частицей.

В современную эпоху величина плотности числа частиц для каждого типа нейтрино: $n_{\nu_{\alpha}0}=110$~см$^{-3}$, где $n_{\nu_{\alpha}}=\nu_e, \nu_{\mu}, \nu_{\tau}$. Плотность энергии для всех типов нейтрино: $\rho_{\nu}\sim6\cdot10^{-7}$~ГэВ/см$^3$.
Суммарная масса всех типов нейтрино: $\sum m_{\nu}<0.23$~эВ, (\cite{Ade:2015xua}).

\subsubsection{Темная материя}
Предположительно, темная материя состоит из стабильных массивных частиц, природа которых пока что не известна.
 Частицы темной материи не взаимодействуют с наблюдаемым электромагнитным излучением и слабо  гравитационно взаимодействуют с обычной барионной материей.

 Темная материя находится в галактиках и в скоплениях галактик.
 Термин 'темная материя' впервые был введен Фрицем Цвикки в 1933 году. Цвикки измерил  угловые скорости для восьми галактик в созвездии Кома, $v(R)$, в зависимости от расстояния от центра галактики, $R$. Он пришел к выводу, что для устойчивости скопления общая масса галактики должна быть в десятки раз больше, чем масса входящих в нее звезд.

\begin{figure}[h!]
\begin{center}
\includegraphics[width=\columnwidth]{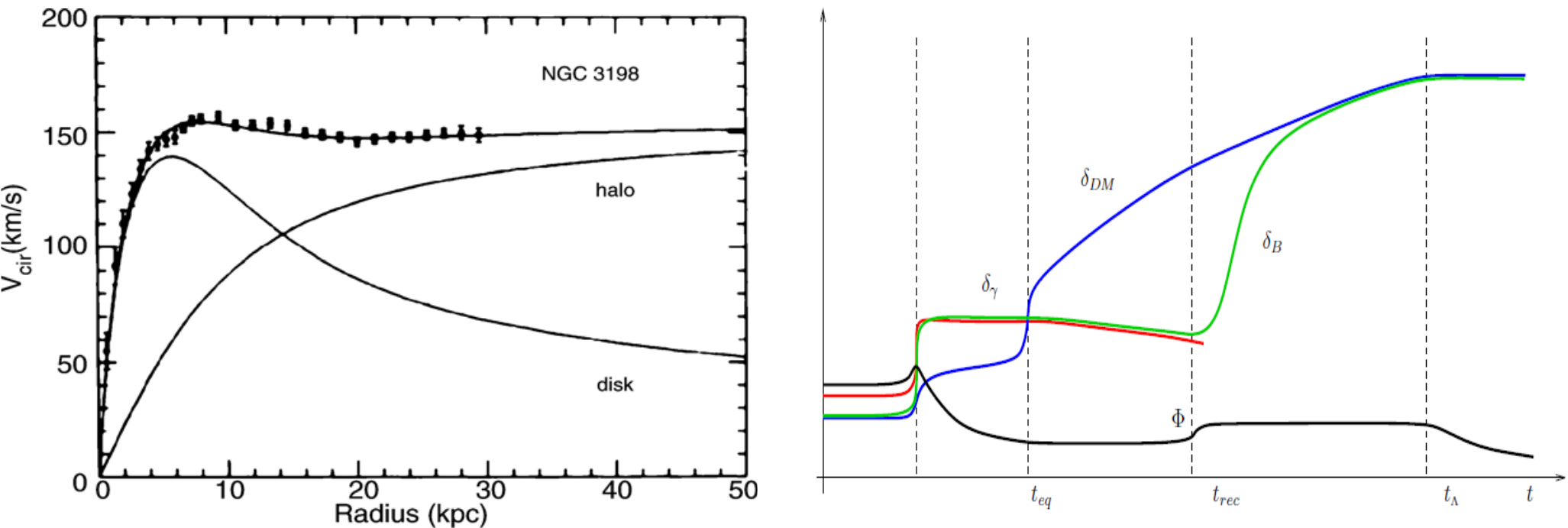}
\end{center}
 \caption {Левая панель: плоская кривая вращения спиральной галактики NGC 3198 (верхняя кривая), являющаяся совмещением вращения видимой материи (кривая 'disk') и темной материи (кривая 'halo'), (\cite{Begeman:1991iy}). Правая панель: эволюция потенциала Ньютона, $\Phi$, и контраста относительной плотности для: темной материи, $\delta_{\rm DM}$, барионов,  $\delta_{\rm B}$, и фотонов,  $\delta_{\gamma}$. $t_{\rm eq}$ - момент времени перехода от эпохи доминирования излучения к эпохе доминирования материи; $t_{\rm rec}$ - начало эпохи рекомбинации; $t_{\Lambda}$ - момент времени перехода от замедленного к ускоренному расширению вселенной, (\cite{Rubakov:2015yza}).}
 \label{fig:f4}
 \end{figure}

Вера Рубин и Кент Форд были первыми, кто представил точные вычисления, указывающие на существование темной материи в галактиках, (\cite{Rubin:1980zd}).  Они обнаружили, что в спиральных галактиках большинство звёзд, не слишком близко расположенных от центра галактик, движутся по орбитам с одинаковой угловой скоростью, $v(R)={\rm const}$, Рис.~(\ref{fig:f4})~(левая панель). Для областей, заполненных видимым веществом (учитывая только видимое вещество), $v(R)\propto\sqrt{R}$, Рис.~(\ref{fig:f4})~(левая панель). При больших расстояниях от центра галактик, т. е. для периферийных областей галактик, $v(R)\propto1/\sqrt{R}$, Рис.~(\ref{fig:f4})~(левая панель). Такое расхождение в угловых скоростях звезд можно объяснить, если предположить, что видимое вещество галактик находится в облаке намного большего размера - в галактическом гало. Галактическое гало содержит значительную массу невидимого вещества, частицы которого не взаимодействуют с фотонами.

На ранних стадиях эволюции вселенной частицы темной материи находились в термодинамическом равновесии с частицами первичной плазмы. При расширении вселенной в определенный момент времени температура первичной плазмы снизилась настолько, что взаимодействие частиц темной материи с барионным веществом прекратилось, и произошло отсоединение частиц темной материи от первичной плазмой, Рис.~(\ref{fig:f4})~(правая панель).
В зависимости от температуры, при которой  произошло это отсоединение (или же в зависимости от массы частиц темной материи в этот момент), темная материя подразделяется  на {\it холодную темную материю} (Cold Dark Matter (CDM)), на {\it теплую темную материю} (Warm Dark Matter (WDM)) и на {\it горячую темную материю} (Hot Dark Matter (HDM)). Холодную темную материю составляют тяжелые частицы, масса которых, $m_{\rm CDM}\geq100$~КэВ. Кандидатами для холодной темной материи являются медленно движущиеся гипотетические частицы, так называемые {\it слабо взаимодействующие массивные частицы} (WIMPs). Теплую темную материю составляют частицы, c массой, $m_{\rm WDM}\approx 3-30$~КэВ. В момент выхода из равновесия с первичной плазмой эти частицы были релятивистскими. Энергия частиц горячей темной материи при их отсоединении от первичной плазмой намного превосходила их массу, т. е. частицы были ультрарелятивистскими.
Легкие частицы, такие как нейтрино, могли составлять горячую темную материю.
Темная материя играет очень важную роль в формировании крупномасштабной структуры во вселенной. Галактики сформировались в регионах с повышенной плотностью темной материи. Отсоединение частиц темной материи  от первичной плазмы происходило намного раньше, чем отсоединение барионов. В связи с этим флуктуации плотности темной материи возникли намного раньше, чем флуктуации плотности барионной материи, Рис.~(\ref{fig:f4})~(правая панель). Барионы попадали в потенциальную яму, образованную темной материей, поэтому после рекомбинации флуктуации плотности темной материи и флуктуации плотности барионов развивались совместно, неотделимо друг от друга, Рис.~(\ref{fig:f4})~(правая панель).

Существует множество возможных кандидатов на роль темной материи. Темная материя может быть барионного и небарионного происхождения. Барионная темная материя, или так называемые {\it массивные компактные объекты гало} (MACHOs), обладают низкой светимостью. Ими могут быть коричневые карлики, темные галактические гало, массивные планеты, компактные объекты на конечных стадиях эволюции: нейтронные звезды, белые и черные карлики, черные дыры. Небарионной темной материей могут быть легкие или тяжелые нейтрино, аксионы, суперсимметричные частицы. Кроме того, темной материей могут быть первичные черные дыры и топологические дефекты пространства-времени.

\subsection{Уравнения Эйнштейна}
Основными уравнениями ОТО являются уравнения гравитационного поля или так называемые {\it уравнения Эйнштейна}:
\begin{equation}
 G_{\mu\nu}\equiv R_{\mu\nu}-\dfrac{1}{2}g_{\mu\nu}R=8\pi G T_{\mu\nu}.
 \label{eq:EQ0}
\end{equation}
Уравнения Эйнштейна связывают метрику искривленного пространства-времени, $g_{\mu\nu}$, тензор кривизны Риччи, $R_{\mu\nu}$, скаляр Риччи, $R$, со свойствами заполняющей это пространство материи, которую характеризует тензор энергии-импульса, $T_{\mu\nu}$. Эти уравнения устанавливают взаимосвязь между кривизной (геометрией) пространства-времени (левая часть уравнений) и материей, а также с её движением (правая часть). Таким образом, уравнения Эйнштейна описывают как кривизна пространства-времени воздействует на материю во вселенной, и наоборот, как  материя во вселенной влияет на кривизну (геометрию) пространства-времени.

Уравнения гравитационного поля являются нелинейными дифференциальными уравнениями второго порядка в частных производных. Эта нелинейность связана с воздействием гравитации самой на себя, т. к. гравитационное поле переносит энергию и импульс. Ввиду того, что уравнения Эйнштейна нелинейны, принцип суперпозиции несправедлив для гравитационных полей. Линеаризация уравнений Эйнштейна возможна в случае рассмотрения  гравитационных волн с низкой амплитудой или для слабых гравитационных полей (например, для гравитационных полей в Ньютоновском пределе). Для таких полей отклонения метрических составляющих в уравнениях от их величин для плоского пространства-времени незначительны и, соответственно, так же мала порождаемая ими кривизна пространства-времени. В этом случае может быть применим принцип суперпозиции полей.

В случае слабых гравитационных полей, создаваемых нерелятивистским движущимся веществом, нулевая компонента тензора Эйнштейна, $G_{00}$, определяется как:
\begin{equation}
G_{00}\approx\nabla^2g_{00},
 \label{eq:G00}
\end{equation}
для этого Ньютоновского предела уравнения Эйнштейна принимают вид:
\begin{equation}
G_{00}=-8 \pi G T_{00}.
 \label{eq:G001}
\end{equation}
Получим альтернативный вид уравнений Эйнштейна, Ур.~(\ref{eq:EQ0}), произведя свертку с контравариантным метрическим тензором, $g^{\mu\nu}$:
\begin{equation}
 R=-8\pi G T.
 \label{eq:EQ1}
\end{equation}
Подставляя уравнение, Ур.~(\ref{eq:EQ0}), в уравнение, Ур.~(\ref{eq:EQ1}), получим еще одну форму записи уравнений Эйнштейна:
\begin{equation}
 R_{\mu\nu}=8\pi G (T_{\mu\nu}-\dfrac{1}{2}g_{\mu\nu}T).
 \label{eq:EQ2}
\end{equation}
В связи с тем, что для вакуума величина тензора энергии-импульса равна нулю, $T_{\mu\nu}=0$, Ур.~(\ref{eq:STV}), из уравнения, Ур.~(\ref{eq:EQ2}), следует, что для вакуума выполняется условие:
\begin{equation}
 R_{\mu\nu}=0.
 \label{eq:EQ3}
\end{equation}
{\bf Результат, полученный в уравнении, Ур.~(\ref{eq:EQ3}), не означает, что пустое пространство является плоским, и в нем отсутствуют гравитационные поля. Для такого утверждения требуется выполнение дополнительного условия - равенство нулю тензора Римана-Кристоффеля, $R^i_{klm}=0$.}
В пространстве-времени двух или трех измерений условие, $R_{\mu\nu}=0$, означает равенство нулю полного тензора Римана-Кристоффеля, и, сответственно, отсутствие гравитационных полей.
Для вакуума в пространстве-времени для четырех и выше измерений  при выполнении условия, $R_{\mu\nu}=0$, полный тензор Римана-Кристоффеля может быть не равным  нулю, поэтому гравитационные поля могут существовать.

\section{Пространственные метрики}
\subsection{Плоское пространство Эвклида}
Геометрия Эвклида основана на пяти аксиомах:
\begin{enumerate}[1.]
 \item Аксиома принадлежности
  \item Аксиома порядка
   \item Аксиома равенства отрезков и углов
   \item Аксиома параллельных прямых
   \item Аксиома непрерывности (или аксиома Архимеда)
   \end{enumerate}
\begin{figure}[h!]
\begin{center}
\includegraphics[width=\columnwidth]{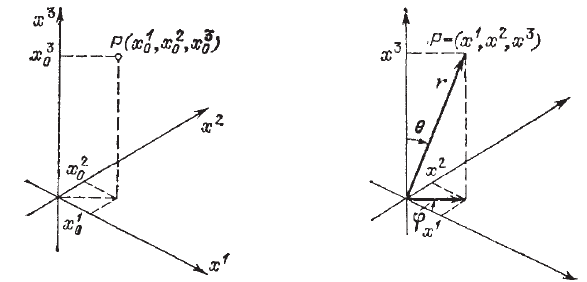}
\end{center}
 \caption {Левая панель: трехмерные декартовые координаты. Правая панель: сферические координаты, (\cite{Dubrovin:1979}).}
 \label{fig:f5}
 \end{figure}
Из 'Аксиомы параллельных прямых' следует утверждение 'Сумма внутренних углов треугольника равна $180^{\circ}$', которое является важной особенностью пространства Евклида.
Пространство Евклида представляет собой трехмерное плоское пространство. Каждая точка этого пространства определяется ортогональными декартовыми координатами, $x^1, x^2, x^3=x, y, z$, Рис.~(\ref{fig:f5})~(левая панель).

Инвариантная метрика в декартовых координатах определена как:
\begin{equation}
ds^2=\sum_{i=1}^3dx^i=(x^1)^2 + (x^2)^2 + (x^3)^2.
 \label{eq:LCart}
\end{equation}
Компактная форма этой метрики имеет вид:
 \begin{equation}
ds^2=g_{\mu\nu}dx^\mu dx^\nu,
 \label{eq:LCart1}
\end{equation}
где $g_{\mu\nu}=\delta_{\mu\nu}$. \\
Метрический тензор для пространства Евклида в декартовых координат имеет вид:\footnote{Изотропия и однородность пространства выражается в диагональной форме метрического тензора, и наоборот, метрический тензор для изотропного и однородного пространства должен быть диагональным.}
\begin{equation}
g_{\mu\nu}=\delta_{\mu\nu}=\left(\begin{array}{cccc}
           1 & 0 & 0\\
           0 & 1 & 0\\
           0 & 0 & 1\\
          \end{array}
           \right).
\end{equation}
Инвариантная метрика в декартовых координатах, ($dx^\mu, dx^\nu$), может быть выражена через произвольные координаты, ($dx^{m'}, dx^{n'}$), следующим образом:
 \begin{equation}
ds^2=\delta_{\mu\nu}dx^\mu dx^\nu=\delta_{\mu\nu}\Bigl(\dfrac{\partial x^i }{\partial x^{m'}}dx^{m'}\Bigr)\Bigl(\dfrac{\partial x^j }{\partial x^{k'}}dx^{k'}\Bigr)=g_{m'k'}dx^{m'}dx^{k'},
\end{equation}
$g_{m'k'}$ - пространственный метрический тензор в произвольной системе координат.

Рассмотрим метрику Евклида в полярных, цилиндрических и сферических координатах.\\
\\
{\bf Полярные координаты}

На плоскости декартовые координаты, $(x^1, x^2)$, выражаются через полярные координаты, $(x^1=r,~ x^2=\varphi)$, следующим образом:
\begin{equation}
x^1=r\cos \varphi,~~x^2=r\sin \varphi
 \label{eq:PolCart}
\end{equation}
и
\begin{equation}
g_{m'k'}=\delta_{\mu\nu}=\left(\begin{array}{cccc}
           1 & 0 \\
           0 & r^2 \\
         \end{array}
           \right).
\end{equation}
Метрика, представленная в полярных координатах:
\begin{equation}
ds^2=(dr)^2+r^2(d\varphi)^2.
 \label{eq:MetPol}
\end{equation}
\\
{\bf Цилиндрические координаты}

Декартовые координаты, $(x^1, x^2, x^3)$, выражаются через цилиндрические координаты, $(y^1=r,~ y^2=\varphi, ~y^3=z)$, как:
\begin{equation}
x^1=r\cos \varphi,~~x^2=r\sin \varphi,~~x^3=z
 \label{eq:PolCart1}
\end{equation}
и
\begin{equation}
g_{m'k'}=\left(\begin{array}{cccc}
           1 & 0 & 0\\
           0 & r^2 & 0\\
           0 & 0 & 1\\
         \end{array}
           \right).
\end{equation}
Метрика, представленная  в цилиндрических координатах:
\begin{equation}
ds^2=(dr)^2+r^2(d\varphi)^2+\sin^2 (d\varphi)^2.
 \label{eq:MetPol1}
\end{equation}\\
{\bf Сферические координаты}

Декартовые координаты, $(x^1, x^2, x^3)$, выражаются через сферические координаты, $(y^1=r, y^2=\theta, y^3=\varphi)$, Fig.~(\ref{fig:f5})~(правая панель), следующим образом:
\begin{equation}
x^1=r\cos \varphi\sin\theta,~~x^2=r\sin \varphi\sin\theta,~~x^3=r\cos\theta
 \label{eq:PolCart4}
\end{equation}
и
\begin{equation}
g_{m'k'}=\left(\begin{array}{cccc}
           1 & 0 & 0\\
           0 & r^2 & 0\\
           0 & 0 & r^2\sin^2\varphi\\
         \end{array}
           \right).
\end{equation}
Метрика, представленная  в сферических координатах:
\begin{equation}
ds^2=dr^2+r^2[(d\theta)^2+r^2\sin^2 \theta (d\varphi)^2].
 \label{eq:MetPol6}
\end{equation}
\subsection{Пространство-время Минковского}
В 1908 году Герман Минковский впервые представил четыре координаты для описания четырехмерного векторного пространства или пространственно-временного континуума. Точки этого пространства-времени называются событиями или {\it мировыми точками}. Каждое событие соответствует набору из четырех чисел, $(x^0, x^1, x^2, x^3)$, где $x^0=t$ - момент времени, когда произошло событие и $(x^1, x^2, x^3)$ - местоположение события. В четырехмерном пространстве-времени процесс жизни для каждого объекта идентифицируется линией, $x^{i}(t)~(i=1, 2, 3)$, которая называется {\it мировой линией}.
  Координаты, $(x^0, x^1, x^2, x^3)$, представляют собой декартовые координаты в пространственно-временном континууме. Таким образом, пространственно-временной континуум можно рассматривать как четырехмерное декартовое пространство. С другой стороны, трехмерное пространство, в котором разворачивается классическая геометрия, будет поверхностью постоянного уровня, где $t={\rm const}$.

Метрический тензор пространства-времени Минковского определен как\footnote{Здесь и далее применяется сигнатура, $(1,-1,-1,-1)$ ; наряду с данной сигнатурой, также может быть применен другой, эквивалентный вариант, $(-1,1,1,1)$.}:
\begin{equation}
\label{eq:Minkovski}
\eta_{\mu\nu}=\delta_{\mu\nu}=\left(\begin{array}{cccc}
           1 & 0 & 0 & 0 \\
           0 & -1 & 0 & 0\\
           0 & 0 & -1 & 0 \\
           0 & 0 & 0 & -1 \\
           \end{array}
           \right).
\end{equation}
Этот метрический тензор описывает плоское четырехмерное изотропное и однородное пространство-время.
Метрика для метрического тензора Минковского представлена как:
\begin{equation}
ds^2 = \eta_{\mu\nu}dx^{\mu}dx^{\nu}.
 \label{eq:IMin}
\end{equation}
\begin{figure}[t!]
\begin{center}
\includegraphics[width=\columnwidth]{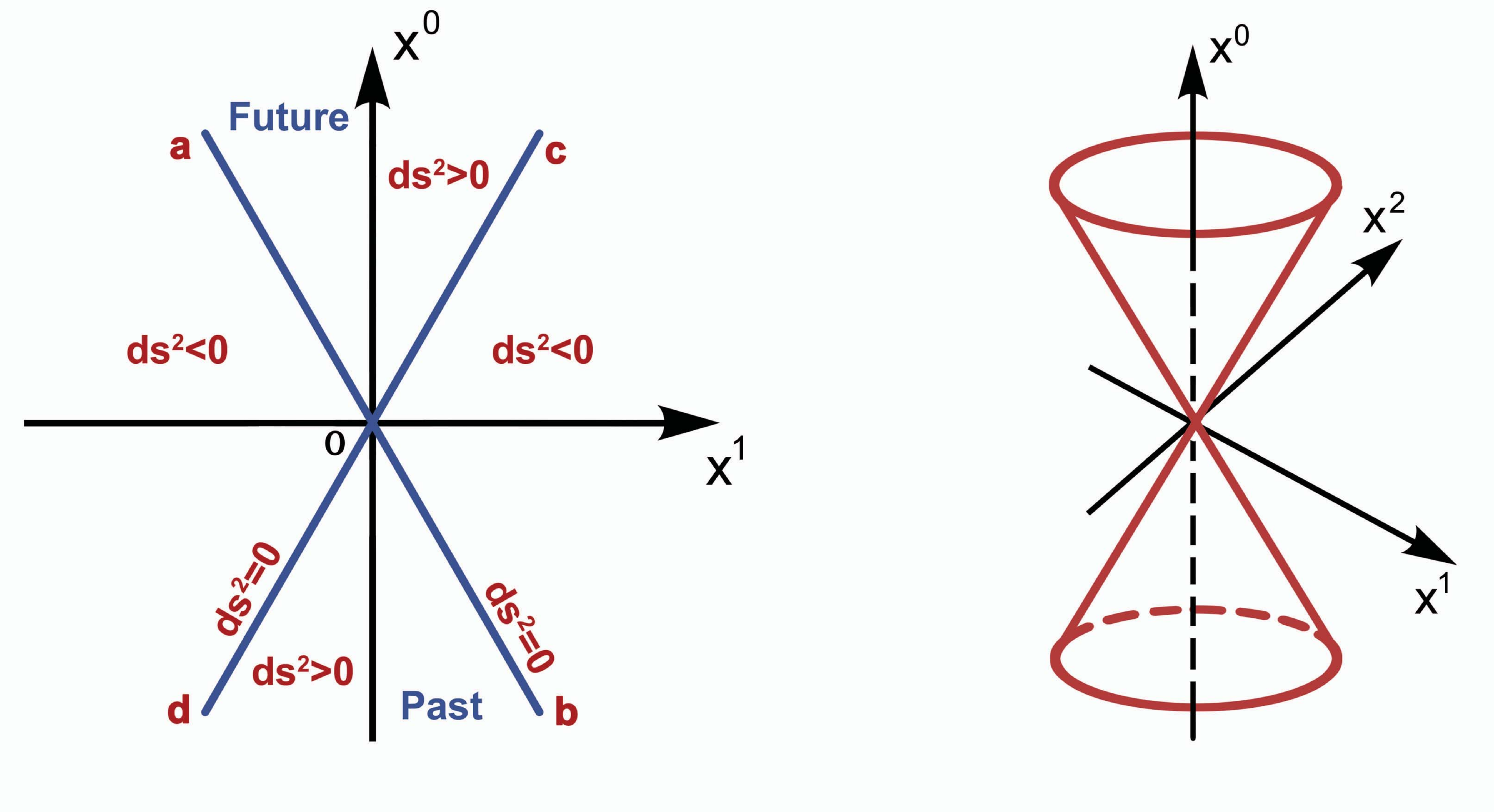}
\end{center}
 \caption {Левая панель: двухмерная диаграмма Минковского. Правая панель: трехмерный световой конус, (\cite{Dubrovin:1979}).}
 \label{fig:f6}
 \end{figure}
 В четырехмерном пространстве-времени метрика, $ds^2$, может принимать следующие величины: быть равной нулю, быть положительной или отрицательной. Метрика, $ds^2=0$, соответствует распространению сигнала со скоростью света или движению безмассовых частиц в четырехмерном пространстве-времени. Нулевая метрика, $ds^2=0$, описывает светоподобные события. Положительная метрика, $ds^2>0$, описывает времениподобные события. Для времениподобных событий существует система координат, в которой эти события могут происходить в одном и том же месте. В этом случае линейный интервал между этими событиями, $ds$, является вещественным числом. Отрицательная метрика, $ds^2<0$, описывает пространственноподобные события. Для пространственноподобных событий существует система координат, в которой эти события могут происходить одновременно. В этом случае линейный интервал между двумя событиями, $ds$, является мнимым числом.

На двумерной диаграмме Минковского, $(x^0, x^1)$,  Рис.~(\ref{fig:f6})~(левая панель), представлены вышеупомянутые типы событий. Начало координат, $O$, соответствует настоящему моменту времени. Линии $ab$ и $cd$ согласуются с двумя разными сигналами, распространяющимися со скоростью света, поэтому для них $ds^2=0$. Пространственноподобные события содержатся в областях  $dOa$ и $cOb$ с $ds^2<0$, в то время как области $aOc$ и $dOb$ соответствуют времениподобным событиям с $ds^2>0$.
Так как для области $aOc$ время имеет положительную величину, $t>0$, то все события из этой области будут происходить в будущем по отношению к настоящему моменту времени, $O$. Для области  $dOb$, для которой время имеет отрицательную величину, $t<0$, и события из этой области происходили в прошлом по отношении к моменту времени $O$. Другими словами, события из области $aOc$ можно назвать 'абсолютно будущими', а события из области  $dOb$ - 'абсолютно прошедшими' относительно настоящего момента времени, $O$.
 Так как для событий с времениподобным интервалом можно однозначно определить какое из них произошло раньше, а какое позже, то для таких событий имеет смысл понятия причины и следствия. Эти события  могут быть причинно-связанными друг с другом.

Метрика для пространства-времени Минковского, Ур.~(\ref{eq:IMin}), является времениподобной, поэтому она может быть расположена в областях $aOc$ и $dOb$ на диаграмме Минковского.
Эта метрика может быть записана в более расширенной форме как:
\begin{equation}
ds^2 = (x^0)^2 - (x^1)^2 - ( x^2)^2 - (x^3)^2.
 \label{eq:IMin1}
\end{equation}
Уравнение, Ур.~(\ref{eq:IMin1}), описывает так называемый световой конус или конус причинно-следственных событий.
Трехмерные координаты Минковского, $(x^0, x^1, x^2)$, могут быть выражены через псевдосферические координаты, $(\varrho, \varsigma, \varphi)$, следующим образом:
\begin{equation}
\chi(r)= \left\{
     \begin{array}{lll}
        x^0=\varrho\cosh \varsigma  \\
        x^1=\varrho \sinh \varsigma \\
        x^2=\varrho \sinh \varsigma \sin \varphi \\
     \end{array}.
   \right.
\label{eq:Psevd}
\end{equation}
Из уравнения, Ур.~(\ref{eq:Psevd}), следует:
\begin{equation}
(x^0)^2 - (x^1)^2 - ( x^2)^2 = \varrho^2>0.
 \label{eq:IMin2}
\end{equation}
Поэтому координаты, $(\varrho, \varsigma, \varphi)$, определяются только в области,  $(x^0)^2 - (x^1)^2 - ( x^2)^2>0$. В трехмерном пространстве-времени эта область расположена внутри светового конуса, $(x^0)^2 = (x^1)^2 + ( x^2)^2$, Рис.~(\ref{fig:f6})~(правая панель).
Метрика для этой области имеет вид:
\begin{equation}
ds^2=d\varrho^2-\varrho^2[(d\chi)^2+\sinh^2\chi(d\varphi)^2].
 \label{eq:IMin3}
\end{equation}
\subsection{Уравнение геодезической линии}
Предположим, что точка с координатами $x^i$ движется по некоторой траектории с четырехмерной скоростью, $u^i=x^i/ds$. Согласно ОТО свободная материальная точка движется в гравитационном поле в четырехмерном пространстве-времени так, что её мировая линия является экстремальной. Эта экстремальная мировая линия называется {\it геодезической} линией между двумя заданными мировыми точками.

Движение частицы в гравитационном поле определяется  {\bf принципом наименьшего действия}, согласно которому функционал действия принимает минимальную величину:
\begin{equation}
\delta S=\delta\int ds=0,
 \label{eq:PND}
\end{equation}
где $ds^2=g_{ik}dx^idx^k$ - метрика в четырехмерном искривленном пространстве-времени.

Применяя принцип наименьшего действия, получим уравнение движения частицы в гравитационном поле.

Проварьируем выражение для метрики:
\begin{equation}
\delta ds^2=2ds\delta ds=\delta(g_{ik}dx^idx^k)=dx^idx^k\frac{\partial g_{ik}}{dx^l}\delta x^l+2g_{ik}dx^id\delta x^k.
 \label{eq:PND1}
\end{equation}
Подставляя полученный результат в уравнение, Ур.~(\ref{eq:PND}), получим:
\begin{equation}
S=\int\Bigl(\frac{dx^i}{ds}\frac{dx^k}{ds}\frac{dg_{ik}}{dx^l}\delta x^l+g_{ik}\frac{dx^i}{ds}\frac{d\delta x^k}{ds}\Bigl)ds=0.
 \label{eq:PND2}
\end{equation}
Проинтегрируем уравнение, Ур.~(\ref{eq:PND2}), по частям, учитывая, что во втором слагаемом на границах интегрирования $\delta x^k=0$:
\begin{equation}
S=\int\Bigl(\frac{1}{2}\frac{dx^i}{ds}\frac{dx^k}{ds}\frac{dg_{ik}}{dx^l}\delta x^l-\frac{d}{ds}\Bigl(g_{ik}\frac{dx^i}{ds}\Bigl)\delta x^k\Bigl)ds=0.
 \label{eq:PND3}
\end{equation}
Во втором слагаемом уравнения, Ур.~(\ref{eq:PND3}), заменим индекс $k$ на индекс $l$, в результате получим:
\begin{equation}
\frac{1}{2}u^iu^k\frac{dg_{ik}}{dx^l}-\frac{d}{ds}(dg_{il}u^i)=\frac{1}{2}u^iu^k\frac{dg_{ik}}{dx^l}-g_{il}\frac{du^i}{ds}-u^iu^k\frac{dg_{il}}{dx^k}=0.
 \label{eq:PND4}
\end{equation}
Представим третье слагаемое в уравнении, Ур.~(\ref{eq:PND4}), в виде:
\begin{equation}
u^iu^k\frac{dg_{il}}{dx^k}=\frac{1}{2}u^iu^k\Bigl(\frac{dg_{il}}{dx^k}+\frac{dg_{kl}}{dx^i}\Bigl).
 \label{eq:PND5}
\end{equation}
Умножим левую и правую части уравнения, Ур.~(\ref{eq:PND4}) на $g^{im}$:
\begin{equation}
g^{im}g_{il}\frac{du^i}{ds}+\frac{1}{2}g^{im}u^iu^k\Bigl(\frac{dg_{il}}{dx^k}+\frac{dg_{kl}}{dx^i}-\frac{dg_{ik}}{dx^l}\Bigl)=0.
 \label{eq:PND6}
\end{equation}
Учитывая, что $g^{im}g_{il}=\delta^m_l$, заменим индекс $l$ на индекс $m$ в выражении в скобках в уравнении, Ур.~(\ref{eq:PND6}):
\begin{equation}
\frac{du^i}{ds}+\frac{1}{2}g^{im}u^iu^k\Bigl(\frac{dg_{il}}{dx^k}+\frac{dg_{km}}{dx^i}-\frac{dg_{ik}}{dx^m}\Bigl)=0.
 \label{eq:PND7}
\end{equation}
Заменим индекс $i$ на индекс $l$ в выражении в скобках в уравнении, Ур.~(\ref{eq:PND7}) и введём символы Кристоффеля, $\Gamma^{i}_{kl}=\frac{1}{2} g^{im}\left( \frac{\partial g_{mk}}{\partial x^l} + \frac{\partial g_{ml}}{\partial x^k} - \frac{\partial g_{kl}}{\partial x^m}\right)$. В результате получаем уравнение движения материальной точки в гравитационном поле по геодезической линии:
\begin{equation}
\frac{d^2x^i}{ds^2}+\Gamma^{i}_{kl}\frac{dx^k}{ds}\frac{dx^l}{ds}=0.
 \label{eq:PND8}
\end{equation}
В четырехмерном пространстве-времени, $(x^0, x^1, x^2, x^3)$, геодезическая линия имеет искривленную форму, в то время как движение частицы не является равномерным и прямолинейным.
\subsection{Изотропная четырехмерная метрика пространства-времени}
Метрический тензор для четырехмерного однородного и изотропного пространства-времени, который пространственно расширяется или сжимается в зависимости от  масштабного фактора, $a(t)$\footnote{Данный метрический тензор описывает расширяющееся пространство-время, т. к. масштабный фактор является возрастающей функцией в зависимости от физического времени, $\dot{a}(t)>0$.}, определяется следующим образом:
\begin{equation}
\label{eq:emdiagonal}
g_{\mu\nu}=\left(\begin{array}{cccc}
           1 & 0 & 0 & 0 \\
           0 & -a^2(t) & 0  & 0\\
           0 & 0 & -a^2(t) & 0 \\
           0 & 0 & 0 & -a^2(t) \\
           \end{array}
           \right).
\end{equation}

Метрика для такого пространства-времени принимает вид:
\begin{equation}
ds^2 \equiv g_{\mu\nu}dx^{\mu}dx^{\nu} = dt^2 - a^2(t)\gamma_{ij}dx^idx^j,
 \label{eq:FLRW1}
\end{equation}
где $\gamma_{ij}$ - метрика трехмерного пространства.

В сферических координатах, ($r,\theta,\varphi$), функция $\gamma_{ij}$  в Ур.~(\ref{eq:FLRW1}) представлена как:
\begin{equation}
\gamma_{ij}=dr^2+\chi(r)^2(d\theta^2 + \sin^2{\theta}d\varphi^2),
 \label{eq:Mgamma}
\end{equation}
здесь $\chi(r)$ является функцией пространственной кривизны:
\begin{equation}
\chi(r)= \left\{
     \begin{array}{cc}
       \frac{1}{\sqrt{\rm K}}\ \mathrm{sin}\big(\sqrt{\rm K}\ r\big) & \ \ \ \ \text{для}\ {\rm K}>0 \\
       r & \ \ \ \ \text{для}\ {\rm K}=0 \\
       \frac{1}{\sqrt{-\rm K}}\ \mathrm{sinh}\big(\sqrt{-{\rm K}}\ r\big) & \ \ \ \ \text{для}\ {\rm K}<0 \\
     \end{array},
   \right.
\label{eq:chi}
\end{equation}
где ${\rm K}$ - параметр кривизны.

Заменяя переменную $x=\chi(r)$ в уравнении, Ур.~(\ref{eq:chi}), и выражая переменную $r$ через $x$, найдем квадрат дифференциала, $dr^2$:
\begin{equation}
dr^2= \left\{
     \begin{array}{cc}

       \frac{1}{1-{\rm K}x^2}dx^2& \ \ \ \ \text{для}\ {\rm K}>0 \\

       dx^2 & \ \ \ \ \text{для}\ {\rm K}=0 \\
       \vspace{0.2cm}
       \frac{1}{1-{\rm K}x^2}dx^2 & \ \ \ \ \text{для}\ {\rm K}<0 \\
     \end{array}.
   \right.
\label{eq:chi1}
\end{equation}
Подставляя уравнение, Ур.~(\ref{eq:chi}), и уравнение, Ур.~(\ref{eq:chi1}), в
 уравнение, Ур.~(\ref{eq:FLRW1}), мы получим  метрику пространства-времени Фридмана-Леметра-Робертсона-Уокера (Friedmann-Lema\^\i tre-Robertson-Walker) (FLRW):
\begin{equation}
ds^2 =  dt^2 - a^2(t)\left[\dfrac{dr^2}{1-{\rm K}r^2}+r^2(d\theta^2 + \sin^2{\theta}d\varphi^2)\right].
\label{eq:FLRW}
\end{equation}
Эта метрика описывает однородное и изотропное расширяющееся пространство. Координаты, ($r$, $\theta$, $\varphi$), являются сопутствующими, т. е. движущийся объект находится в покое относительно сопутствующей системы координат, связанной с этими  координатами.

 Метрика FLRW в декартовых координатах может быть записана как:
\begin{equation}
ds^2 =  dt^2 - a^2(t)\dfrac{1}{(1+\frac{\rm K}{4}r^2)^2}\delta_{ij}dx^i dx^j.
\label{eq:FLRW2}
\end{equation}
\begin{figure}[h!]
\begin{center}
\includegraphics[width=0.6\columnwidth]{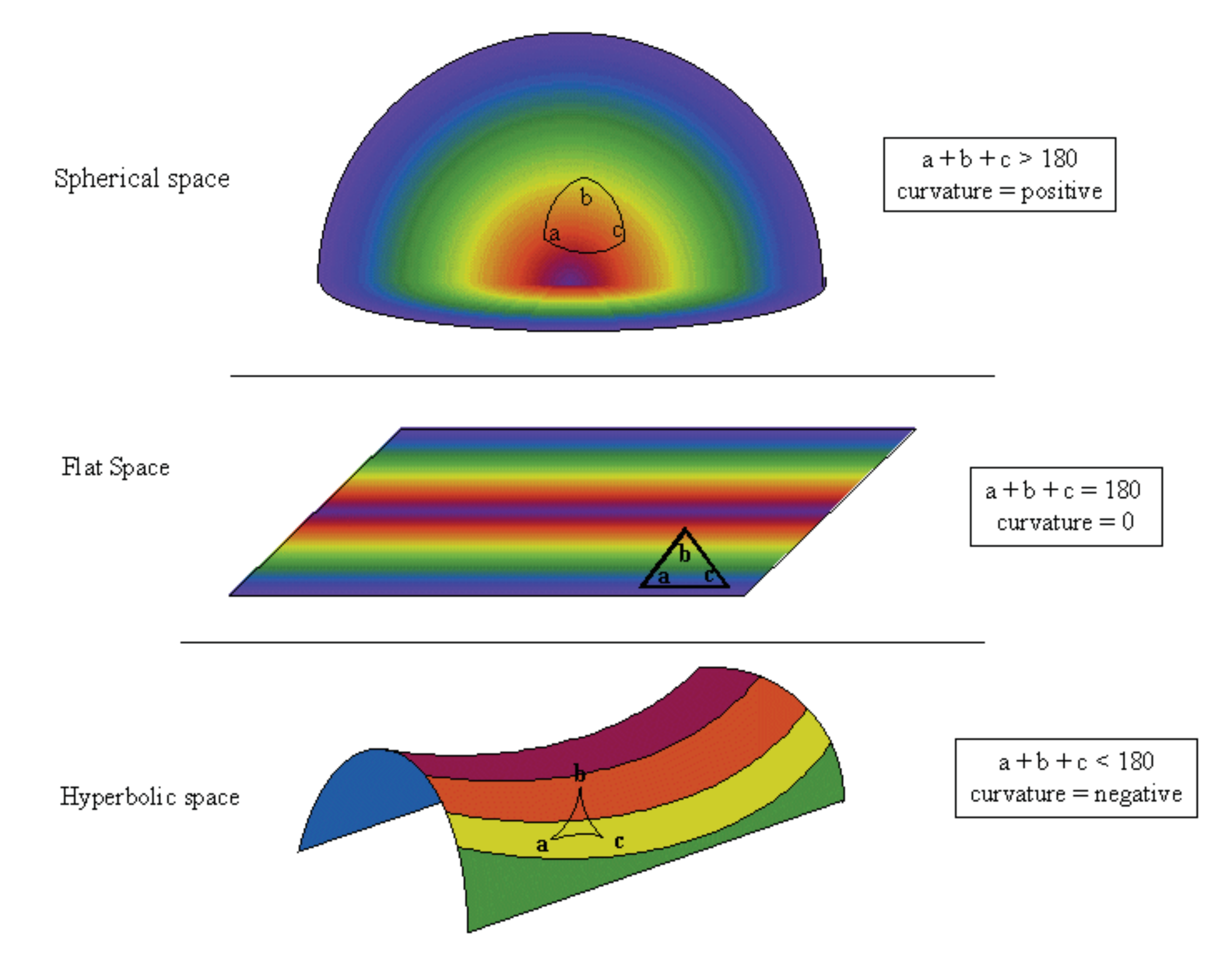}
\end{center}
 \caption {Примеры замкнутого, плоского и открытого двумерных пространств, (http://www.astro.cornell.edu/academics/courses/astro201/).}
 \label{fig:f7}
 \end{figure}
 В зависимости от знака параметра кривизны, ${\rm K}$, уравнение, Ур.~(\ref{eq:FLRW}), описывает геометрически различные типы вселенной. Случай ${\rm K} > 0$ соответствует {\it закрытой вселенной} или сферическому трехмерному пространству. Двумерный аналог такой вселенной - поверхность сферы, Рис.~(\ref{fig:f7}), а функция $1/\sqrt{\rm K}$ может быть интерпретирована как ее радиус кривизны. Случай ${\rm K} = 0$ соответствует {\it плоской вселенной} или евклидовому трехмерному пространству, Рис.~(\ref{fig:f7}). Случай ${\rm K} <0$ соответствует {\it открытой вселенной} или трехмерному гиперболическому пространству. Двумерным аналогом такой вселенной является поверхность седла, Рис.~(\ref{fig:f7}).

При изучении определенных процессов кривизной вселенной можно пренебречь. Например, при свободном движении фотона в однородной и изотропной вселенной, длина волны фотона намного меньше радиуса пространственной кривизны вселенной (в случае открытой или замкнутой вселенной). В этом случае вселенную можно рассматривать как пространственно-плоскую, и можно использовать метрику, представленную в уравнении, Ур.~(\ref{eq:FLRW1}).
В терминах конформного времени, которое определено в уравнении, Ур.~(\ref{eq:Conform1}), уравнение, Ур.~(\ref{eq:FLRW1}), принимает вид:
\begin{equation}
ds^2 = a^2(\eta)d\eta^2 - a^2(\eta)\gamma_{ij}dx^idx^j=a^2(\eta)[\eta^2-\gamma_{ij}dx^idx^j].
 \label{eq:ConfM}
\end{equation}
Из уравнения, Ур.~(\ref{eq:ConfM}), следует соотношение между метрическим тензором Минковского, $\eta_{\mu\nu}$, и метрическим тензором, $g_{\mu\nu}$:
\begin{equation}
g_{\mu\nu}=a^2(\eta)\eta_{\mu\nu}.
 \label{eq:MinM}
\end{equation}
Таким образом, метрический тензор, $g_{\mu\nu}$, имеет конформно-плоскую форму в координатах, $(\eta, x^{\mu})$.

Для разных типов кривизны, уравнение,  Ур.~(\ref{eq:ConfM}), имеет вид:
\begin{equation}
ds^2=a^2(\eta)(d\eta^2-d\xi^2-\varpi^2(d\theta^2 + \sin^2{\theta}d\varphi^2)),
\label{eq:ConfM1}
\end{equation}
где переменная $\varpi$ определена как:
\begin{equation}
\varpi= \left\{
     \begin{array}{cc}

      \sin\xi & \ \ \text{для}\ \ \ {\rm K}>0,\ r=a(\eta)\sin\xi, \ \xi\in[0, \pi]\\

      \xi & \ \ \text{для}\ \  \ {\rm K}=0,\ r=a(\eta)\xi, \ \xi\in[0, \infty]\\

      \sinh\xi & \ \ \text{для}\ \  \ {\rm K}<0,\ r=a(\eta)\sinh\xi, \ \xi\in[0, \infty] \\
     \end{array}.
   \right.
\label{eq:ConfM2}
\end{equation}

\subsection{Уравнения Фридмана}
На пространственных космологических масштабах все компоненты массы-энергии во вселенной можно аппроксимировать моделью баротропной жидкости\footnote{Баротропная жидкость представляет собой жидкость, плотность которой зависит только от давления.}. Соотношение между плотностью энергии, $\rho$, и давлением, $p$, для баротропной жидкости определяется уравнением состояния:
\begin{equation}
p=w\rho,
\label{eq:EOS1}
\end{equation}
где $w$ - параметр уравнения состояния, величина которого различна для каждой компоненты массы-энергии во вселенной.

\begin{figure}[h!]
\begin{center}
\includegraphics[width=0.7\columnwidth]{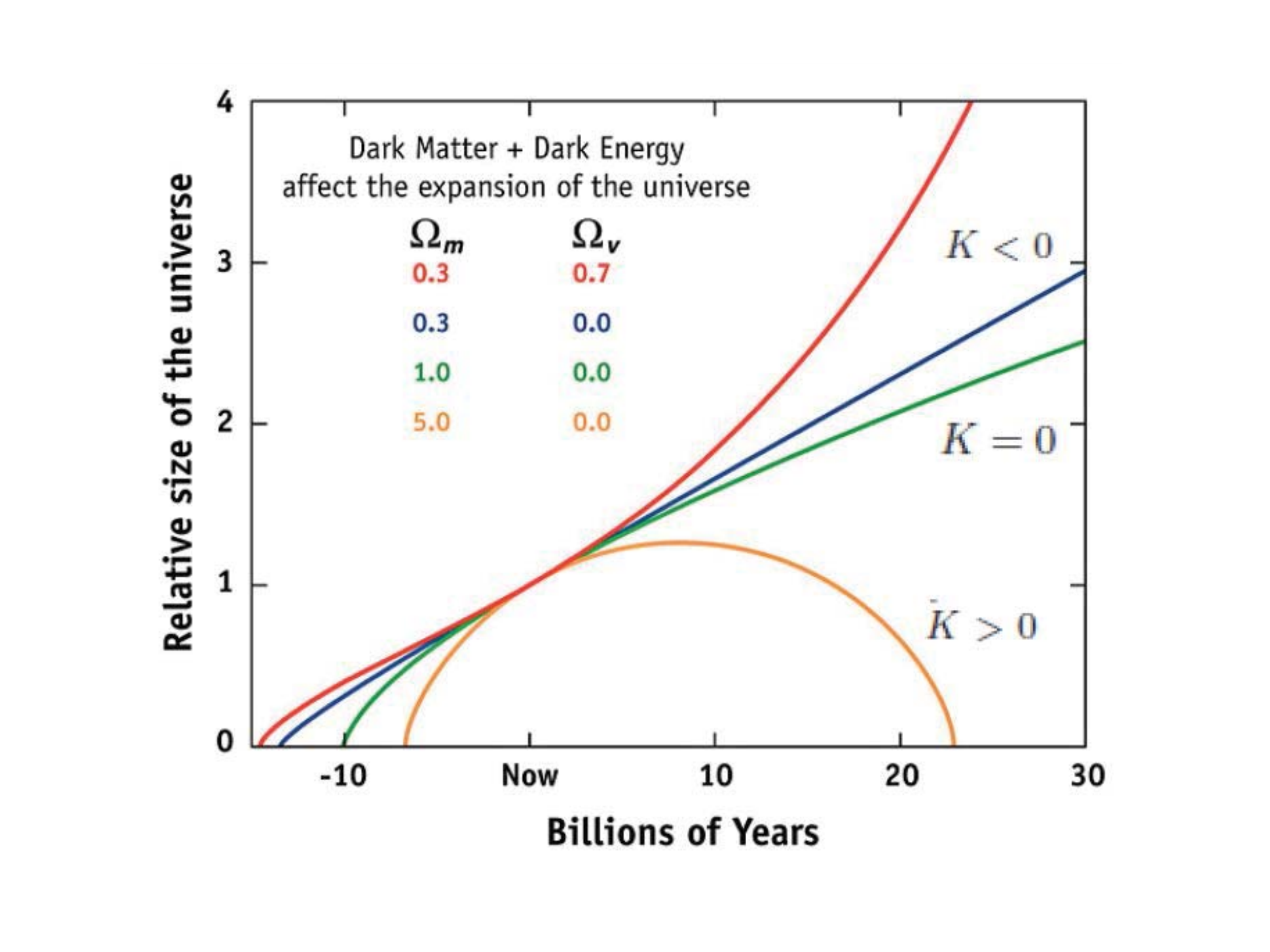}
\end{center}
 \caption{Эволюция скалярного фактора, $a(t)$, при разных величинах параметра кривизны, $K$, (https://wmap.gsfc.nasa.gov/universe/).}
 \label{fig:f8}
 \end{figure}

Подставляя FLRW метрику, Ур.~(\ref{eq:FLRW}), и тензор энергии-импульса, Ур.~(\ref{eq:EE}), в уравнения Эйнштейна,
Ур.~(\ref{eq:EQ0}), можно получить первое и второе уравнения Фридмана:
\begin{equation}
\frac{\dot{a}^2}{a^2}=\frac{8\pi G}{3}\rho-\frac{\rm K}{a^2}
\label{eq:FR1}
\end{equation}
и
\begin{equation}
\frac{\ddot{a}}{a}=-\frac{4\pi G}{3}(\rho+3p).
\label{eq:FR2}
\end{equation}
Если мы знаем эволюцию масштабного фактора, $a(t)$, который характеризует историю расширения вселенной, то мы можем определить величину параметра кривизны и величины компонент массы-энергии во вселенной, используя уравнения Фридмана. И наоборот, если мы знаем величину параметра кривизны и величины компонент массы-энергии во вселенной, то мы можем вычислить эволюцию масштабного фактора, $a(t)$. Например: история расширения вселенной зависит от величины параметра кривизны, ${\rm K}$: для ${\rm K<0}$ ({\it открытая вселенная}) вселенная будет расширяться вечно, Рис.~(\ref{fig:f8}); для ${\rm K=0}$ ({\it плоская вселенная}) вселенная будет расширяться вечно, но при $t\rightarrow\infty$ расширение будет происходить с постоянной скоростью, т. е. $\dot{a}(t)\rightarrow0$, Рис.~(\ref{fig:f8}); для ${\rm K>0}$ ({\it закрытая вселенная}) вселенная будет расширяться до определенного момента времени, после чего расширение сменится сжатием, в результате которого вселенная сколлапсирует, Рис.~(\ref{fig:f8}).

Решая уравнение непрерывности, Ур. (\ref{eq:CE4}), и уравнение Фридмана, Ур.~(\ref{eq:FR1}), для плоской вселенной, $\rm K=0$, мы получим следующие уравнения:
\begin{equation}
\rho \propto a^{-3(1+w)},~~~a(t)\propto t^{\frac{2}{3(1+w)}}~~\Rightarrow~~H=\dfrac{2}{3(1+w)t},
\label{eq:DW}
\end{equation}
где величина параметра уравнения состояния, $w$, не зависит от времени и $w\neq1$.

 Уравнение, Ур.~(\ref{eq:DW}), описывает эволюцию плотности энергии, $\rho$, скалярного фактора, $a$, и параметра Хаббла, $H$, в зависимости от параметра уравнения состояния, $w$, и физического времени, $t$. Мы можем проанализировать уравнение, Ур.~(\ref{eq:DW}), для различных величин параметра уравнения состояния, $w$, в предположении, что только одна компонента массы-энергии, содержится во вселенной, которая описывается данным параметром уравнения состояния, $w$.

Величина параметра уравнения состояния, $w=1/3$, соответствует жидкости релятивистских частиц (фотонов и нейтрино), которая называется излучением. Для этого случая уравнения, Ур.~(\ref{eq:DW}), принимают вид:
\begin{equation}
\rho_{\rm r} \propto a^{-4},~~~a(t)\propto t^{\frac{1}{2}}~~\Rightarrow~~H=\dfrac{1}{2t}.
\label{eq:DWR}
\end{equation}
Величина параметра уравнения состояния, $w=0$, соответствует жидкости нерелятивистских частиц или пылевидной материи, которая состоит из холодной темной материи и барионов. Соответственно, для этого случая уравнения, Ур.~(\ref{eq:DW}), принимают вид:
\begin{equation}
\rho_{\rm m} \propto a^{-3},~~~a(t)\propto t^{\frac{2}{3}}~~\Rightarrow~~H=\dfrac{2}{3t}.
\label{eq:DWM}
\end{equation}
Величина параметра уравнения состояния, $w=-1/3$, соответствует вселенной с ненулевой пространственной кривизной, т. е. закрытой или открытой вселенной. В этом случае уравнения, Ур.~(\ref{eq:DW}), принимают вид:
\begin{equation}
\rho_{\rm K} \propto a^{-2},~~~a(t)\propto t~~\Rightarrow~~H=\dfrac{1}{t}.
\label{eq:DWM9}
\end{equation}
Подставляя параметр уравнения состояния, которое определено в уравнении, Ур.~(\ref{eq:EOS1}), во второе уравнение Фридмана, Ур.~(\ref{eq:FR2}), мы получим:
\begin{equation}
\frac{\ddot{a}}{a}=-\frac{4\pi G \rho}{3}(1+3w).
\label{eq:CE8}
\end{equation}
Если величина параметра уравнения состояния, $w$, удовлетворяет условию, $-1\leq w < -1/3$, тогда $\ddot{a}(t)\leq-1$ и из уравнения, Ур.~(\ref{eq:Decsel}), следует, что вселенная расширяется с ускорением, Рис.~(\ref{fig:f7}).

Ускоренное расширение вселенной объясняется наличием в ней темной энергии. Случай $w=-1$ соответствует простейшей модели темной энергии, так называемой  энергии вакуума или космологической постоянной $\Lambda$. В этом случае вселенная ускоряется с постоянной плотностью энергии вакуума, $\rho_\Lambda$, и с постоянной величиной параметра Хаббла, тогда как скалярный фактор экспоненциально изменяется со временем:
\begin{equation}
\rho_\Lambda = {\rm const},~~~a(t)\propto e^{Ht}~~\Rightarrow~~H={\rm const}.
\label{eq:DELambda}
\end{equation}
Полная плотность энергии вселенной включает в себя все разновидности плотностей энергий, рассмотренных выше: плотности излучения, материи, кривизны и темной энергии:
\begin{equation}
\rho=\rho_{\rm r} + \rho_{\rm m} + \rho_{\rm K} + \rho_\Lambda.
\label{eq:rho}
\end{equation}
Учитывая зависимость компонент плотности энергии от скалярного фактора, представленные в уравнениях, Ур.~(\ref{eq:DWR}) - Ур.~(\ref{eq:DELambda}):
\begin{equation}
\rho=\rho_{\rm r0}a^{-4} + \rho_{\rm m0}a^{-3} + \rho_{\rm K0}a^{-2} + \rho_\Lambda,
\label{eq:rho1}
\end{equation}
где $\rho_{\rm r0}$, $\rho_{\rm m0}$, $\rho_{\rm K0}=- {\rm K}/H_0^2$, и $\rho_\Lambda$ - плотности энергии излучения, материи, кривизны и темной энергии  сегодня, соответственно.
Уравнение для полной плотности энергии, $\rho_0$, для настоящего момента времени, $a=a_0=1$, имеет вид:
\begin{equation}
\rho_0=\rho_{\rm r0} + \rho_{\rm m0} + \rho_{\rm K0} + \rho_\Lambda.
\label{eq:rho2}
\end{equation}
Уравнение, Ур.~(\ref{eq:rho1}), можно представить в более удобной форме через безразмерные параметры плотности энергии, которые применяются для описания массы и энергии во вселенной:
\begin{equation}
\Omega=\rho/\rho_{\rm cr}=\Omega_{\rm r0}a^{-4} + \Omega_{\rm m0}a^{-3} + \Omega_{\rm K0}a^{-2} + \Omega_\Lambda,
\label{eq:Omega}
\end{equation}
где  $\Omega$ - параметр полной плотности энергии для произвольного момента времени; $\Omega_{ i0}$ - параметр плотности энергии для '$ i$'-ой компоненты для настоящего момента времени,
характеризующийся соответствующей плотностью энергии, $\rho_{i0}$; $\rho_{\rm cr}$ - критическая плотность во вселенной для настоящего момента времени\footnote{Критическая плотность представляет собой полную плотность энергии во вселенной, которая необходима для того, чтобы вселенная была пространственно-плоской.}, величина которой равна, $\rho_{\rm cr}=1.8791h^2\cdot10^{-29}~\text{г~см}^{-3}$.

Для настоящего момента времени уравнение, Ур.~(\ref{eq:Omega}), имеет вид:
\begin{equation}
\Omega_{0}=\rho_{i0}/\rho_{\rm cr}=\sum_i\Omega_{i0}=\Omega_{\rm r0} + \Omega_{\rm m0} + \Omega_{\rm K0} + \Omega_\Lambda,
\label{eq:Omega1}
\end{equation}
где $\Omega_{0}$ - параметр полной плотности энергии для настоящего момента времени. Этот параметр является одним из важнейших космологических параметров.

Первое уравнение Фридмана, определенное в уравнении, Ур.~(\ref{eq:FR1}), может быть выражено через параметры плотности энергии, $\Omega_{i0}$, следующим образом:
\begin{equation}
H(a)=H_0(\Omega_{\rm r0}a^{-4} + \Omega_{\rm m0}a^{-3} + \Omega_{\rm K0}a^{-2} + \Omega_\Lambda)^{1/2}.
\label{eq:Omega3}
\end{equation}
Уравнение, Ур.~(\ref{eq:Omega3}), можно представить в виде:
\begin{equation}
E(a)=(\Omega_{\rm r0}a^{-4} + \Omega_{\rm m0}a^{-3} + \Omega_{\rm K0}a^{-2} + \Omega_\Lambda)^{1/2},
\label{eq:OmegaE}
\end{equation}
где $E(a)=H(a)/H_0$ - безразмерный параметр Хаббла.

Для настоящего момента времени уравнение, Ур.~(\ref{eq:Omega3}), можно переписать как:
\begin{equation}
\Omega_0-1=\frac{\rm K}{H_0^2}.
\label{eq:OmegaK}
\end{equation}
Из уравнения, Ур.~(\ref{eq:OmegaK}), следует, что величина полного параметра плотности, $\Omega_{0}>1$, соответствуют замкнутой вселенной с положительной кривизной, $\rm K>0$, Рис.~(\ref{fig:f8}). Величина полного параметра плотности, $\Omega_{0}<1$, соответствуют открытой вселенной, где величина  кривизны отрицательна, $\rm K<0$, Рис.~(\ref{fig:f8}). Величина полного параметра плотности, $\Omega_{0}=1$, соответствует плоской вселенной с нулевой кривизной, $\rm K=0$, Рис.~(\ref{fig:f8}).
Согласно данным Planck 2015, (\cite{Ade:2015xua}), параметр плотности кривизны для настоящего момента времени, $\Omega_{\rm K0}=0.006$ (на уровне  достоверности $68 \%$). Таким образом, величина параметра плотности кривизны для настоящего момента времени приблизительно равна нулю, $\Omega_{\rm K0}\simeq0$, т. е. наша вселенная является плоской с точностью до $1\%$. Таким образом, величина критической плотности во вселенной соответствует величине средней плотности во вселенной, $\rho_{\rm cr}=\langle \rho\rangle$, с точностью порядка $1\%$.

\subsection{Параметр ускорения вселенной}
Найдем производную от параметра Хаббла, представленного в уравнении, Ур.~(\ref{eq:Hubble}):
\begin{equation}
\dot{H}=\dfrac{a\ddot{a}-\dot{a}^2}{a^2}=-H^2+\dfrac{\ddot{a}}{a}=-H^2\Bigl(1-\dfrac{\ddot{a}}{H^2a}\Bigr)=-H^2(1-q),
 \label{eq:Decsel}
\end{equation}
где $q$ - безразмерный {\it параметр ускорения}\footnote{Обычно в литературе используют так называемый параметр замедления, определенный как, $q\equiv -\ddot{a}/aH^2$. В данной работе мы используем название 'параметр ускорения', т. к. такое название лучше описывает современное состояние вселенной.}, который определяется как:
\begin{equation}
q\equiv \dfrac{\ddot{a}}{aH^2}.
 \label{eq:Decsel1}
\end{equation}
Параметр ускорения характеризует состояние ускорения или замедления вселенной. Положительная величина этого параметра, $q>0$, соответствует ускоренному расширению вселенной, при котором $\ddot{a}(t)>0$, а отрицательная величина, $q<0$, соответствует замедленному расширению вселенной, при котором $\ddot{a}(t)<0$.

Параметр ускорения можно выразить через величины параметра уравнения состояния, $w_{ i}$, и параметра плотности энергии, $\Omega_{i}$:
\begin{equation}
q(t)=-\dfrac{1}{2}\sum_i(1+3w_{i})\Omega_{ i}(t),
 \label{eq:Decsel2}
\end{equation}
здесь индекс $i$ характеризует определенную компоненту плотности энергии во вселенной и соответственный параметр уравнения состояния.
Используя величины для параметра состояния материи, излучения и вакуума, соответственно: $w_{\rm m}=0, w_{\rm r}=1/3, w_\Lambda=-1$, получим:
\begin{equation}
q(t)=-(\Omega_{\rm m}/2+\Omega_{\rm r}-\Omega_\Lambda).
 \label{eq:Decsel3}
\end{equation}

Параметр ускорения в настоящий момент времени, $q_0$, определяется как:
\begin{equation}
q_0\equiv \dfrac{1}{H_0^2}\Bigl(\dfrac{\ddot{a}}{a}\Bigr)_0.
 \label{eq:Decsel1}
\end{equation}

 Используя данные Planck 2015, (\cite{Ade:2015xua}), можно вычислить  величину параметра ускорения вселенной в современную эпоху:
\begin{equation}
[q_0]_{\rm Planck}\approx0.54.
 \label{eq:Decsel4}
\end{equation}
Положительный знак параметра ускорения вселенной в современную эпоху указывает на то, что в  настоящий момент времени вселенная находится в состоянии ускоренного расширения, которое началось при величине скалярного фактора, $a \approx0.60$ (или при величине красного смещения, $z\approx0.65$), согласно данным Planck 2015, (\cite{Ade:2015xua}).

%%%%%%%%%%%%%%%%%%%%%%%%%%%%%%%%%%%%%%%%%%%%%%%%%%%%%%%%%%%%%%%%%%%%%%%%%%%
%%%%%%%%%%%%%%%%%%%%       Beginng of chapter 3     %%%%%%%%%%%%%%%%%%%%%%%
%%%%%%%%%%%%%%%%%%%%%%%%%%%%%%%%%%%%%%%%%%%%%%%%%%%%%%%%%%%%%%%%%%%%%%%%%%%

\chapter{Расстояния в космологии}\label{chapter:3}

\section{Понятие расстояния в космологии}
Определение величины расстояния между астрономическими объектами в расширяющейся вселенной является одной из основных и наисложнейших задач в космологии.

{\it В космологии не существует понятия единого расстояния.}
Применяются  различные типы космологических расстояний, такие как: физическое расстояние, сопутствующее расстояние, фотометрическое расстояние, угловое расстояние и т. д. Эти расстояния отличаются друг от друга методами их определения и измерения. Общее между вышеперечисленными типами космологических расстояний то, что эти расстояния являются мерой  разделения двух объектов, находящихся на одной радиальной траектории друг от друга.

{\it В космологии понятие 'точного расстояния' до удаленного объекта является раcплывчатым.} Величина космологического расстояния зависит от выбранной космологической модели и, следовательно, является функцией параметров модели. Таким образом, точность в определении расстояний зависит как от правильности рассматриваемой космологической модели так и от точности определения модельных параметров\footnote{В космологии все величины, полученные из наблюдательных данных (расстояния, модельные параметры и т. д.), вычисляются с применением методов статистики или теории вероятности (более подробная информация об этом содержится в Главе~V). Поэтому при упоминании вычисленной величины всегда указывается точность, с которой была получена данная величина. Обычно указываются  интервалы уровня доверия, $1\sigma$, $2\sigma$, $3\sigma$, где  $\sigma$ - среднеквадратичное отклонение в распределении Гаусса, или соответствующие им уровни достоверности, $68.27\%$, $95.45\%$, $99.73\%$.}.

Знание точных расстояний до удаленных астрономических объектов необходимо для получения трехмерных данных, характеризующих эти объекты, а также  для определения физических параметров вселенной.
Ярким примером важности определения точных космологических расстояний является доказательство существования темной энергии во вселенной, которое во многой степени основано на измеренных космологических фотометрических расстояний до суперновых типа Ia.
\section{Тригонометрический параллакс}
\begin{figure}[h!]
\begin{center}
\includegraphics[width=0.8 \columnwidth]{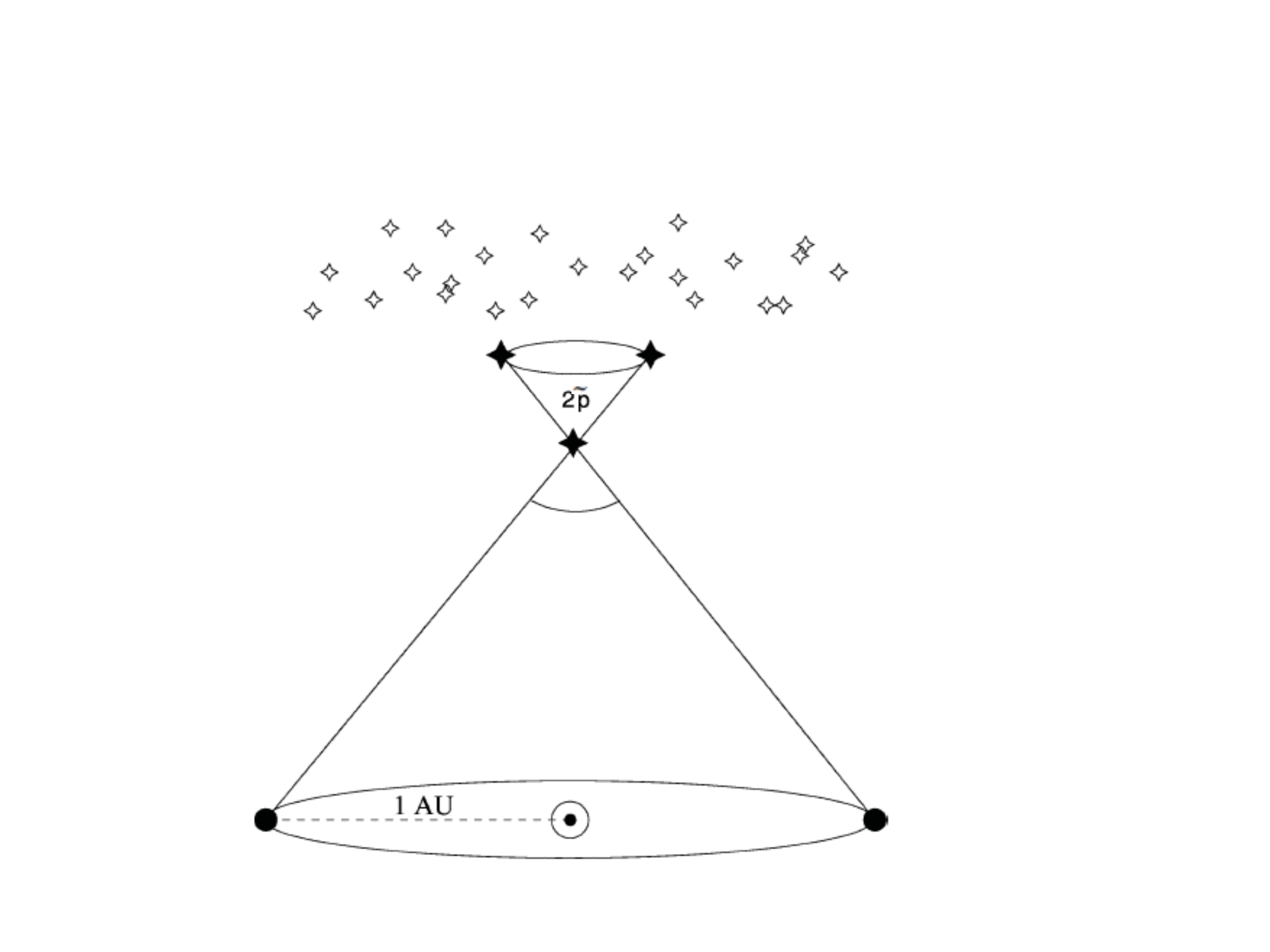}
\end{center}
 \caption{Иллюстрация эффекта тригонометрического параллакса, (\cite{Schneider:2006}).}
 \label{fig:f9}
 \end{figure}

Тригонометрический параллакс является одним из наиболее важных методов определения расстояний, применяемых в астрономии.
Этот метод основан на следующем геометрическом эффекте: из-за вращения Земли вокруг Солнца для наблюдателя, расположенного на поверхности Земли, положения соседних звезд меняются на фоне отдаленных объектов, Рис.~(\ref{fig:f9}).

 В течение года видимое положение ближайшей звезды следует за эллипсом на сфере, Рис.~(\ref{fig:f9}). Большая полуось этого эллипса называется {\it параллаксом}, $\tilde{p}$. Величина параллакса, $\tilde{p}$, зависит от физического расстояния до звезды, $d$, и радиуса орбиты Земли, $r_{\bigoplus}$, который равен одной астрономической единице, $1~\text{а.~е.}$\footnote{Точнее, $1~\text{а.~е.} = 1.496\cdot10^{13}$~см - большая полуось эллипсоидальной орбиты Земли.}, Рис.~(\ref{fig:f9}). Величина тригонометрического параллакса определяется как:
\begin{equation}
\dfrac{r_{\bigoplus}}{d}=\dfrac{1~\text{а.~е.}}{d}=\tan{\tilde{p}}\approx \tilde{p},
\label{eq:Par1}
\end{equation}
где $\tilde{p}\ll1$, при этом параллакс, $\tilde{p}$, выражен в радианах.

Величина тригонометрического параллакса, выраженная в секундах дуги связана с величиной тригонометрического параллакса, выраженной в радианах, следующим образом:
\begin{equation}
\tilde{p}''=\dfrac{180\cdot3600}{\pi}\tilde{p}\approx 206 265 \tilde{p}.
\label{eq:Par_sec}
\end{equation}
Подставляя уравнение Ур.~(\ref{eq:Par1}) в уравнение Ур.~(\ref{eq:Par_sec}):
\begin{equation}
\tilde{p}''\approx\dfrac{206 265}{d}~\text{а.~е.}
\label{eq:Par_sec1}
\end{equation}
\noindent
Тригонометрический параллакс также применяется  для определения одной из основных единиц расстояния в астрономии - {\it парсека}. Парсек (пс)\footnote{В космологии оперируют с более большими масштабами, поэтому используют величину $1~\text{Мпс} =10^6~\text{пс}$ как единицу измерения.} - это расстояние до объекта, для которого параллакс равен одной секунде. Полагая в  уравнении Ур.~(\ref{eq:Par_sec1}) $\tilde{p}''=1''$, получим:
\begin{equation}
1~\text{пс}=206265~\text{а.~е.} = 3.086\cdot10^{18}~\text{см}.
\label{eq:Par3}
\end{equation}
Если тригонометрический параллакс выражен в секундах дуги, то физическое расстояние до объекта, выраженное в парсеках, определяется простой формулой:
\begin{equation}
d(\text{пс})=1/\tilde{p}''.
\label{eq:Par_dist}
\end{equation}
Метод тригонометрического параллакса для определения расстояний очень точен, но его можно использовать только для близко расположенных звезд, находящихся на расстоянии в пределах $\sim5$~кпс, (\cite{Gaia}, \cite{Brown:2018dum}).

\section{Космологическое красное смещение}
{\bf Релятивистское смещение Доплера}\footnote{В этом разделе сохраняется обозначение скорости света, $c$, для ясности.}

Рассмотрим некоторый удаленный источник излучения, который  последовательно излучает световые сигналы в моменты физического времени, $t_{\rm em}$ и $t_{\rm em}+\Delta t_{\rm em}$, соответственно. Отсчеты времени ведутся по часам, которые неподвижны относительно данного источника излучения.
Этот источник излучения движется относительно наблюдателя со скоростью, $\vec{u}$, Рис.~(\ref{fig:f8}).
\begin{figure}[h!]
\begin{center}
\includegraphics[width=0.7 \columnwidth]{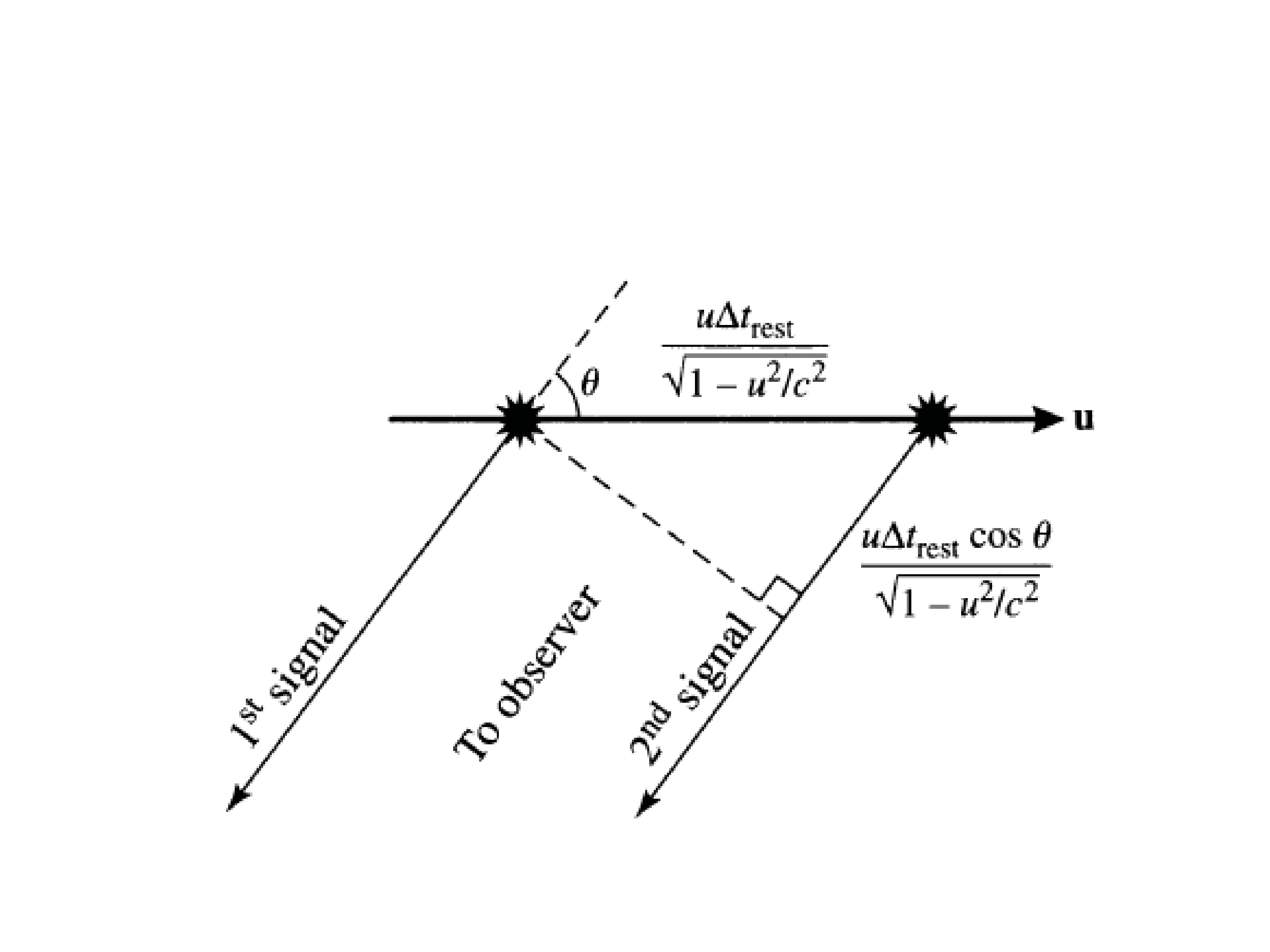}
\end{center}
 \caption{Иллюстрация эффекта релятивистского смещения Доплера, (\cite{Carroll:2009}).}
 \label{fig:f10}
 \end{figure}
На величину интервала физического времени между двумя световыми сигналами, регистрируемыми наблюдателем, $\Delta t_{\rm obs}$, будут влиять: релятивистский эффект замедления времени, связанный с движением источника излучения, $\Delta t_{\rm em}/\sqrt{1-u^2/c^2}$, и эффект, связанный с разницей расстояний, пройденными двумя сигналами от движущегося удаленного источника излучения до наблюдателя, $\Delta d=u \cos {\theta} \Delta t_{\rm em}/\sqrt{1-u^2/c^2}$, Рис.~(\ref{fig:f10}).

Таким образом, временной интервал между двумя сигналами, регистрируемые наблюдателем, равен:
\begin{equation}
\Delta t_{\rm obs}=\dfrac{\Delta t_{\rm em}}{\sqrt{1-u^2/c^2}}+\dfrac{u/c\Delta t_{\rm em}\cos {\theta}}{\sqrt{1-u^2/c^2}}
=\dfrac{\Delta t_{\rm em}}{\sqrt{1-u^2/c^2}}(1+u/c\cos {\theta}).
\label{eq:Relshift}
\end{equation}
Предположим, что излучение фотона с длиной волны $\lambda_{\rm em}$ (или с частотой $\nu_{\rm em}$\footnote{Длина волны и частота электромагнитного излучения взаимосвязаны как, $\lambda\nu=c$.}) произошло в момент физического времени, $t_{\rm em}$. Этот фотон будет наблюдаться в момент времени, $t_{\rm obs}$, с длиной волны $\lambda_{\rm obs}$  (или с частотой $\nu_{\rm obs}$). Величины интервала физического времени между двумя последовательными световыми сигналами излучаемые источником, $\Delta t_{\rm em}$, и зарегистрированные наблюдателем, $\Delta t_{\rm obs}$, связаны с частотой излученных фотонов, $\nu_{\rm em}$, и частотой регистрируемых фотонов, $\nu_{\rm obs}$, как $\nu_{\rm em}=1/\Delta t_{\rm em}$ и $\nu_{\rm obs}=1/\Delta t_{\rm obs}$. Используя эти соотношения, можно переписать уравнение, Ур.~(\ref{eq:Relshift}), как:
 \begin{equation}
\nu_{\rm obs}=\nu_{\rm em}\dfrac{\sqrt{1-u^2/c^2}}{1+u/c\cos{\theta}}=\nu_{\rm em}\dfrac{\sqrt{1-u^2/c^2}}{1+{v/c}},
\label{eq:Relshift1}
\end{equation}
где $v=u\cos{\theta}$ - радиальная скорость источника излучения относительно наблюдателя.
Уравнение, Ур.~(\ref{eq:Relshift1}), описывает {\bf релятивистское смещение Доплера}.

Рассмотрим проекции скорости движения объекта в двух перепендикулярных направлениях: поперечном и радиальном (продольном) к лучу зрения. В уравнении, Ур.~(\ref{eq:Relshift1}), полагая $\theta=90^{\circ}$, получим уравнение для  {\bf поперечного релятивистского смещения Доплера}:
\begin{equation}
\nu_{\rm obs}=\nu_{\rm em}\sqrt{1-u^2/c^2}.
\label{eq:Relshift2}
\end{equation}
Поперечное релятивистское смещение Доплера происходит благодаря эффекту сокращения времени, связанного с движением источника излучения относительно наблюдателя.

В уравнении, Ур.~(\ref{eq:Relshift1}), при движении источника излучения от наблюдателя, полагаем $\theta=0^{\circ}$ и $v=u$, а при движении источника излучения к наблюдателю, полагаем $\theta=180^{\circ}$ и $v=-u$. В результате получим уравнение {\bf радиального релятивистского смещения Доплера}:
\begin{equation}
\nu_{\rm obs}=\nu_{\rm em}\sqrt{\dfrac{1-{v/c}}{1+{v/c}}}.
\label{eq:Relshift3}
\end{equation}
Соответственно, уравнение, Ур.~(\ref{eq:Relshift3}), для длин волн, $\lambda_{\rm obs}$ и $\lambda_{\rm em}$, будет иметь вид:
 \begin{equation}
\lambda_{\rm obs}=\lambda_{\rm em}\sqrt{\dfrac{1+{v/c}}{1-{v/c}}}.
\label{eq:Relshift4}
\end{equation}
 \\
{\bf Определение красного смещения}

Красное смещение (или синее смещение), $z$, определяется относительной разницей между наблюдаемой и излучаемой длинами волн (или, соответственно, частотами):
\begin{equation}
z=\dfrac{\lambda_{\rm obs}-\lambda_{\rm em}}{\lambda_{\rm em}}=\dfrac{\nu_{\rm em}-\nu_{\rm obs}}{\nu_{\rm obs}}.
\label{eq:redshift}
\end{equation}
При красном смещении, при котором $z> 0$, источник излучения удаляется от наблюдателя, при этом регистрируемая наблюдателем энергия излучения смещается  к более низким величинам.
 При синем смещении, при котором $z<0$, источник излучения приближается к наблюдателю, при этом регистрируемая наблюдателем энергия излучения смещается к более высоким величинам.

Из уравнения, Ур.~(\ref{eq:redshift}), получим:
\begin{equation}
1+z=\dfrac{\lambda_{\rm obs}}{\lambda_{\rm em}}=\dfrac{\nu_{\rm em}}{\nu_{\rm obs}}.
\label{eq:redshift1}
\end{equation}
\\
{\bf Релятивистское красное смещение}

Подставим полученные результаты из уравнения, Ур.~(\ref{eq:Relshift3}) (или из уравнения, Ур.~(\ref{eq:Relshift4})) в уравнение, Ур.~(\ref{eq:redshift}), получим уравнение для {\bf релятивистского красного смещения}:
 \begin{equation}
z=\sqrt{\dfrac{1+{v/c}}{1-{v/c}}}-1.
\label{eq:Relshift5}
\end{equation}
\\
{\bf Красное смещение при маленьких скоростях движения источника излучения}

Рассмотрим предельный случай маленькой величины радиальной скорости движения источника, $v\ll c$, в уравнении, Ур.~(\ref{eq:Relshift5}):
\begin{eqnarray}
z&=&\lim_{{v/c}\rightarrow0}\Bigl(\sqrt{\dfrac{1+{v/c}}{1-{v/c}}}-1\Bigr), \nonumber \\
&=&\lim_{{v/c}\rightarrow0}\Bigl(\sqrt{1+\dfrac{2{v/c}}{1-{v/c}}}-1\Bigr)\approx \dfrac{v/c}{1-{v/c}}\approx{v/c}.
\label{eq:Relshift6}
\end{eqnarray}
\\
{\bf Связь космологического красного смещения со скалярным фактором.}

Рассмотрим  систему координат, которая описывается FLRW метрикой. В центре этой системы координат находится наблюдатель. Световой луч движется к наблюдателю в радиальном направлении по нулевой геодезической линии, которую описывает метрика, $ds^2=0$, при $d\theta=d\phi=0$.\\
Из уравнения, Ур.~(\ref{eq:FLRW}) можно получить:
\begin{equation}
dt = \pm a(t)\dfrac{dr}{\sqrt{1-{\rm K}r^2}}.
\label{eq:FLRW4}
\end{equation}
В уравнении, Ур.~(\ref{eq:FLRW4}), мы выберем отрицательный знак, из-за того что световой луч исходит от источника излучения, находящегося на расстоянии, $r=r_{\rm em}$, двигаясь в направлении центра координат, $r=r_{\rm obs}=0$, поэтому $dr<0$ и $dt>0$:
\begin{equation}
\int_{t_{\rm em}}^{t_{\rm obs}} \dfrac{dt}{a(t)}=\int_0^{r_{\rm em}}\dfrac{dr}{\sqrt{1-{\rm K}r^2}}.
\label{eq:FLRW5}
\end{equation}
Дифференцируя уравнение, Ур.~(\ref{eq:FLRW5}), и учитывая, что радиальная координата, $r_{\rm em}$, сопутствующего источников не зависит от времени:
\begin{equation}
\dfrac{\Delta t_{\rm em}}{a(t_{\rm em})}=\dfrac{\Delta t_{\rm obs}}{a(t_{\rm obs})}.
\label{eq:LS}
\end{equation}
Предполагая, что световые сигналы являются последовательными гребнями волн, испускаемая и наблюдаемая частоты определяются, соответственно, как,
$\nu_{\rm em}=1/\Delta t_{\rm em}$ и $\nu_{\rm obs}=1/\Delta t_{\rm obs}$.
Мы можем переписать уравнение, Ур.~(\ref{eq:LS}), как:
\begin{equation}
\nu_{\rm obs}/\nu_{\rm em}= a(t_{\rm em})/a(t_{\rm obs}).
\label{eq:LS1}
\end{equation}
Скалярный фактор, $a(t)$, является возрастающей функцией, тогда как частота, $\nu(t)$ - убывающая функция в соответствии с величиной множителя $(1+z)$, Ур.~(\ref{eq:redshift1}). Объединяя уравнение, Ур.~(\ref{eq:redshift1}) и уравнение, Ур.~(\ref{eq:LS1}), мы получим:
\begin{equation}
1+z= a(t_{\rm obs})/a(t_{\rm em})=a_0/a(t_{\rm em}).
\label{eq:LS2}
\end{equation}
 Соотношение между красным смещением и масштабным фактором, Ур.~(\ref{eq:LS2}), является очень важным в космологии. Красное смещение может быть измерено, и оно является подчас единственной информацией о расстояниях для большинства удаленных объектов.

\section{Сопутствующее расстояние}
Сопутствующее расстояние - это расстояние между двумя астрономическими объектами, измеренное вдоль геодезической линии (по радиальному направлению) в настоящий момент космологического времени. Сопутствующее расстояние и конформное время образуют {\it сопутствующую систему координат}. Сопутствующее расстояние между двумя объектами в сопутствующей системе координат остается постоянной величиной при условии, что эти объекты движутся только с потоком Хаббла\footnote{Солнце (вместе с солнечной системой) движется с пекулярной скоростью $370.6\pm0.4$~км с$^{-1}$ относительно потока Хаббла в направлении созвездия Льва, определяемое экваториальными координатами, $(\alpha,\delta)=(11.2^h, -7^{\circ}$).}.

Исходя из симметрии задачи, используем четырехмерную метрику Минковского, представленную в сферических координатах:
\begin{equation}
ds^2=g_{\mu\nu}dx^{\mu}dx^{\nu}=dt^2-a^2(t)[dr^2+r^2(d\theta)^2+r^2\sin^2 \theta (d\varphi)^2].
\label{eq:MM}
\end{equation}
В уравнении, Ур.~(\ref{eq:MM}), полагаем $ds^2=0$ и $d\theta=d\phi=0$.
Сопутствующее расстояние от наблюдателя до удаленного объекта определяется следующим образом:
\begin{equation}
r=\int_{t_{\rm em}}^{t_0}\dfrac{dt'}{a(t')}=\int_{a_{\rm em}}^{a_0}\dfrac{da}{a\dot{a}}=\dfrac{1}{a_0H_0}\int_{0}^{z}\dfrac{dz'}{E(z')},
\label{eq:com_dist}
\end{equation}
где $t_{\rm em}$, $a_{\rm em}$ и $z_{\rm em}$ - величины физического времени, скалярного фактора и красного смещения, на котором находится источник излучения, соответственно; $t_0$ и $a_0$ - величины физического времени и скалярного фактора в момент регистрирации излучения  наблюдателем, соответственно.

Рассмотрим чему равно сопутствующее расстояние для разных величин параметра кривизны, $K$, для FLRW метрики, определенной в уравнении, Ур.~(\ref{eq:FLRW}). В уравнении, Ур.~(\ref{eq:FLRW}) полагаем $ds^2=0$ и $d\theta=d\phi=0$. В результате получим:
\begin{equation}
r= \left\{
     \begin{array}{cc}
       \vspace{0.2cm}
       \frac{1}{\sqrt{\rm K}}\sin \Bigl(\frac{\sqrt{\rm K}}{a_0H_0}\int_{0}^{z}\frac{dz'}{E(z')}\Bigr)& \ \ \ \text{для}\ {\rm K}>0 \\
       \vspace{0.2cm}
       \frac{1}{a_0H_0}\int_{0}^{z}\frac{dz'}{E(z')} & \ \ \ \text{для}\ {\rm K}=0 \\
       \vspace{0.2cm}
       \frac{1}{\sqrt{-\rm K}}\sinh \Bigl(\frac{\sqrt{-\rm K}}{a_0H_0}\int_{0}^{z}\frac{dz'}{E(z')}\Bigr) & \ \ \  \text{для}\ {\rm K}<0 \\
     \end{array}.
   \right.
\label{eq:FLRW_com}
\end{equation}
В уравнении, Ур.~(\ref{eq:FLRW_com}), выразим параметр кривизны, ${\rm K}$, через параметр плотности кривизны для современной эпохи, $\Omega_{\rm K0}$, в результате получим:
\begin{equation}
r= \left\{
     \begin{array}{cc}
       \vspace{0.2cm}
       \frac{1}{H_0\sqrt{\Omega_{\rm K0}}}\sin \Bigl(\frac{\sqrt{\Omega_{\rm K0}}}{H_0}\int_{0}^{z}\frac{dz'}{E(z')}\Bigr)& \ \ \   \text{для}\ \Omega_{\rm K0}>0 \\
       \vspace{0.2cm}
       \frac{1}{a_0H_0}\int_{0}^{z}\frac{dz'}{E(z')} & \ \ \  \text{для}\ \Omega_{\rm K0}=0 \\
       \vspace{0.2cm}
       \frac{1}{H_0\sqrt{-\Omega_{\rm K0}}}\sinh \Bigl(\frac{\sqrt{-\Omega_{\rm K0}}}{H_0}\int_{0}^{z}\frac{dz'}{E(z')}\Bigr) & \ \ \  \text{для}\ \Omega_{\rm K0}<0 \\
     \end{array}.
   \right.
\label{eq:FLRW_com1}
\end{equation}
\section{Физическое расстояние}
Физическое расстояние - это расстояние до удаленного объекта, которое можно измерить в некоторый момент космологического времени, $t$, физической линейкой. Величина физического расстояния меняется вследствие расширения вселенной.

Для определения расстояний до астрономических объектов, находящихся на красном смещении с маленькой величиной, $z\ll1$, можно применять следующий метод.
При маленьких величинах красного смещения выполняется соотношение между радиальной скоростью и красным смещением объекта\footnote{Для маленьких величин красного смещения, $v\approx cz$, но в нашем соглашении, $c=1$.}, $v\approx z$, согласно уравнению, Ур.~(\ref{eq:Relshift6}). В этом случае закон Хаббла, представленный в Ур.~(\ref{eq:HL}), преобразуется в {\it локальный закон Хаббла}:
\begin{equation}
z\approx H_0d~~\Rightarrow~~d\approx\dfrac{z}{H_0}~~\text{для}~z\ll1.
\label{eq:HL1}
\end{equation}
Физическое расстояние, полученное этим методом называется расстоянием, определенным по красному смещению.

Взаимосвязь между физическим, $d(t)$, и сопутствующим, $r$, расстояниями устанавливает следующее выражение:
 \begin{equation}
d(t)=a(t)r.
\label{eq:phys_dist}
\end{equation}
Согласно выражению, Ур.~(\ref{eq:phys_dist}), величины физического и сопутствующего расстояний равны между собой только в настоящий момент времени:
 \begin{equation}
d(t_0)=a(t_0)r~~\Rightarrow~~d(t_0)=r=\dfrac{1}{H_0}\int_{0}^{z}\dfrac{dz'}{E(z')}.
\label{eq:phys_dist1}
\end{equation}

Разложим интеграл в уравнении, Ур.~(\ref{eq:com_dist}) в ряд Тейлора вблизи $z=0$. Также применим соотношение, $(\dot{H})_0=-H_0^2(1-q_0)$, из уравнения, Ур.~(\ref{eq:Decsel}), где $q_0$ параметр ускорения, который был определен в уравнении, Ур.~(\ref{eq:Decsel1}):
 \begin{equation}
d(t_0)=\dfrac{z}{H_0}\int_0^z\Bigl[1-(1-q_0)z'+\Bigl(\frac{1}{2}+2q_0-\frac{3}{2}q_0^2+\frac{1}{2}\Omega_{K0}\Bigr)z'^2\Bigr]dz'+....
\label{eq:phys_dist2}
\end{equation}
В результате интегрирования уравнения, Ур.~(\ref{eq:phys_dist2}), получим:
 \begin{equation}
d(t_0)=\dfrac{1}{H_0}\Bigl[z-(1-q_0)z^2+\Bigl(\frac{1}{6}-\frac{2}{3}q_0-\frac{1}{2}q_0^2+\frac{1}{6}\Omega_{K0}\Bigr)z^3\Bigr]+....
\label{eq:phys_dist3}
\end{equation}
Ограничимся двумя первыми слагаемыми разложения Тейлора в уравнении, Ур.~(\ref{eq:phys_dist3}):
 \begin{equation}
d(t_0)\simeq\dfrac{z}{H_0}\Bigl[1-(1-q_0)z\Bigr]~~\text{для}~z\ll1.
\label{eq:phys_dist4}
\end{equation}
Уравнение, Ур.~(\ref{eq:phys_dist4}), является приближенным выражением для определения физического расстояния до объекта с учетом ускорения вселенной. Второе слагаемое в этом уравнении является отклонением от классического определения физического расстояния в законе Хаббла, Ур.~(\ref{eq:HL1}).
 С увеличением величины параметра плотности материи, $\Omega_{\rm m0}$, величина параметра ускорения, $q_0=-(\Omega_{\rm m0}/2+\Omega_{\rm r0}-\Omega_\Lambda)$, уменьшается, т. е. увеличение массы (энергии) во вселенной препятствует ускоренному расширению вселенной. Это приводит к уменьшению величины физического расстояния до объекта, Ур.~(\ref{eq:phys_dist4}).
\section{Интервал космологического времени между двумя событиями}
Фотон был испущен источником излучения, который находится на красном смещении, $z$, а затем зарегистрирован приемником излучения, который находится на красном смещении, $z=0$. Фотон находился в пути в течение физического времени $\Delta t=d/c$, где $d$ - физическое расстояние.

Рассмотрим  FLRW систему координат, центре которой находится наблюдатель. Свет распространяется по нулевой геодезической линии, которую описывает нулевая метрика, $ds^2=0$, Рис.~(\ref{fig:f5})~(левая панель). В уравнении, Ур.~(\ref{eq:FLRW}), полагаем $ds^2=0$ и $d\theta=d\phi=0$. Из уравнения, Ур.~(\ref{eq:FLRW}), найдем интервал физического времени, прошедший между двумя моментами времени $t(z)$ и $t(0)$:
\begin{equation}
\Delta t=t(z)-t(0)=\int_{t(0)}^{t(z)}dt=\int_{a_0}^{a(z)}d (d(t))=\int_{a_0}^{a(z)}a d r.
\label{eq:interval}
\end{equation}
Используя уравнения, Ур.~(\ref{eq:Omega3}) и Ур.~(\ref{eq:com_dist}), перейдем к дифференциалу $da=-dza_0/(1+z)^2$ при $a=a_0/(1+z)$. Перепишем уравнение,  Ур.~(\ref{eq:interval}), следующим образом:
\begin{eqnarray}
\Delta t&=&\dfrac{1}{a_0H_0}\int_{0}^{z}\dfrac{dz'}{(1+z)E(z')},\nonumber \\
&=&\dfrac{1}{a_0H_0}\int_{0}^{z}\dfrac{dz'}{(1+z)\sqrt{\Omega_{\rm r0}(1+z')^4 + \Omega_{\rm m0}(1+z')^3 + \Omega_{\rm K0}(1+z')^2 + \Omega_\Lambda}}.\nonumber \\
\label{eq:interval1}
\end{eqnarray}
Из уравнения, Ур.~(\ref{eq:interval1}), следует, что интервал физического времени между двумя событиями однозначно связан с величиной красного смещения. Величина интервала физического времени зависит от выбранной космологической модели и, соответственно, от параметров этой модели.

Уравнение, Ур.~(\ref{eq:interval1}), можно использовать для определения возраста вселенной, устремив верхний предел интегрирования к бесконечности, $z\rightarrow\infty$:
\begin{equation}
\Delta t=\dfrac{1}{a_0H_0}\int_{0}^{\infty}\dfrac{dz'}{(1+z)\sqrt{\Omega_{\rm r0}(1+z')^4 + \Omega_{\rm m0}(1+z')^3 + \Omega_{\rm K0}(1+z')^2 + \Omega_\Lambda}}.
\label{eq:interval2}
\end{equation}
Согласно данным Planck 2015, (\cite{Ade:2015xua}), в предположении правильности  $\Lambda$CDM модели, возраст нашей вселенной составляет $t_0=13.799 \pm 0.038$ млрд. лет (на уровне  достоверности в $68 \%$).
\section{Фотометрическое расстояние}
Фотометрическим расстоянием, $d_L$, называется расстояние, с которого астрономический объект, характеризуемый болометрической светимостью\footnote{Болометрическая светимость - полная мощность излучения, измеренная в Ваттах.}, $L$, и находящийся на красном смещении, $z$, создает болометрический (т. е. интегрированный по всем частотам) поток, $F$,  в предположении выполнения следующего соотношения между светимостью и потоком:
\begin{equation}
F=\dfrac{L}{4 \pi d_L^2}.
\label{eq:LD}
\end{equation}
 Таким образом, фотометрическое расстояние до объекта определяется как:
\begin{equation}
d_L=\sqrt{\dfrac{L}{4 \pi F}}.
\label{eq:LD1}
\end{equation}
Фотометрическое расстояние, $d_L$, является мерой величины потока, $F$, создаваемого объектом с известной светимостью, $L$.

В связи с расширением вселенной, абсолютная болометрическая светимость, $L$, создаваемая источником излучения, который находится на красном смещении, $z$, отличается от светимости, $L_{\rm obs}$, регистрируемой приемником излучения, который находится на красном смещении,  $z=0$. Абсолютная болометрическая светимость, $L$, определяется как энергия, $E_{\rm em}$, излучаемая источником излучения на красном смещении, $z$, за некоторый интервал физического времени, $\Delta t_{\rm em}$:
\begin{equation}
L=\dfrac{E_{\rm em}}{\Delta t_{\rm em}}.
\label{eq:Lum}
\end{equation}
Соответственно, наблюдаемая  болометрическая светимость, $L_{\rm obs}$, определяется как регистрируемая приемником излучения энергия, $E_{\rm obs}$, за интервал физического времени, $\Delta t_{\rm obs}$:
\begin{equation}
L_{\rm obs}=\dfrac{E_{\rm obs}}{\Delta t_{\rm obs}}.
\label{eq:Lum1}
\end{equation}
Рассмотрим отношение абсолютной болометрической светимости, $L$, к наблюдаемой болометрической светимости, $L_{\rm obs}$:
\begin{equation}
\dfrac{L}{L_{\rm obs}}=\dfrac{E_{\rm em}}{\Delta t_{\rm em}}\cdot\dfrac{\Delta t_{\rm obs}}{E_{\rm obs}}=\dfrac{E_{\rm em}}{E_{\rm obs}}\cdot\dfrac{\Delta t_{\rm obs}}{\Delta t_{\rm em}}.
\label{eq:Lum2}
\end{equation}
Ввиду того, что энергия фотона прямо пропорциональна его частоте и учитывая результаты, полученные в уравнениях, Ур.~(\ref{eq:LS1}) и Ур.~(\ref{eq:LS2}), получим:
\begin{equation}
\dfrac{E_{\rm em}}{E_{\rm obs}}=\dfrac{\nu_{\rm em}}{\nu_{\rm obs}}=1+z.
\label{eq:Lum3}
\end{equation}
 Этот результат отражает факт уменьшения энергии фотонов из-за красного смещения, возникшего вследствие расширения вселенной.

С другой стороны, учитывая полученные в уравнениях, Ур.~(\ref{eq:LS}) и Ур.~(\ref{eq:LS2}),  соотношения:
\begin{equation}
\dfrac{\Delta t_{\rm obs}}{\Delta t_{\rm em}}=1+z.
\label{eq:Lum4}
\end{equation}
Этот результат иллюстрирует тот факт, что из-за расширения вселенной происходит увеличение времени распространения фотонов, что приводит к уменьшению интенсивности поступления фотонов на приемник излучения.

Таким образом, основываясь на результатах, полученных в выражениях, Ур.~(\ref{eq:Lum3}) и Ур.~(\ref{eq:Lum4}), перепишем уравнение, Ур.~(\ref{eq:Lum2}), как:
\begin{equation}
\dfrac{L}{L_{\rm obs}}=(1+z)^2.
\label{eq:Lum5}
\end{equation}
Поток энергии определяется как энергия, $E_{\rm em}$, переносимая в единицу времени и через единицу площади некоторой поверхности, $S$. Согласно этому определению, можно написать, $F=L_{\rm obs}/S$. Энергия, $E_{\rm em}$, излучаемая источником, была распределена по сферической поверхности с радиусом, $R=a_0r$, в момент регистрации приемником излучения, который находится на красном смещении, $z=0$. Таким образом, регистрируемый поток энергии, определяется как:
\begin{equation}
F=\dfrac{L}{4 \pi d_L^2}=\dfrac{L_{\rm obs}}{4 \pi (a_0r)^2}.
\label{eq:flux}
\end{equation}
Из этого соотношения следует:
\begin{equation}
d_L^2=(a_0r)^2\dfrac{L}{L_{\rm obs}}.
\label{eq:lum_dist1}
\end{equation}
Подставляя уравнение, Ур.~(\ref{eq:Lum5}) в уравнение, Ур.~(\ref{eq:lum_dist1}), получим:
\begin{eqnarray}
d_L^2&=&(a_0r)^2(1+z)^2,\nonumber \\
\Rightarrow~~d_L&=&a_0r(1+z).
\label{eq:lum_dist2}
\end{eqnarray}
Подставим выражения, определяющие сопутствующее расстояние, $r$,  Ур.~(\ref{eq:FLRW_com1}), в уравнение, Ур.~(\ref{eq:lum_dist2}). В результате получим выражения для фотометрического расстояния в зависимости от космологических параметров\footnote{Предполагая, что темная энергия представлена космологической постоянной, $\Lambda$.}:
\begin{equation}
d_L(z)=  \left\{
     \begin{array}{cc}
       \vspace{0.2cm}
       \frac{(1+z)}{H_0\sqrt{\Omega_{\rm K0}}}\sin \Bigl(\frac{\sqrt{\Omega_{\rm K0}}}{H_0}\int_{0}^{z}\frac{dz'}{E(z')}\Bigr)& \ \ \ \ \text{для}\ \Omega_{\rm K0}>0 \\
       \vspace{0.2cm}
       \frac{(1+z)}{a_0H_0}\int_{0}^{z}\frac{dz'}{E(z')} & \ \ \ \ \text{для}\ \Omega_{\rm K0}=0 \\
       \vspace{0.2cm}
       \frac{(1+z)}{H_0\sqrt{-\Omega_{\rm K0}}}\sinh \Bigl(\frac{\sqrt{-\Omega_{\rm K0}}}{H_0}\int_{0}^{z}\frac{dz'}{E(z')}\Bigr) & \ \ \ \ \text{для}\ \Omega_{\rm K0}<0 \\
     \end{array}.
   \right.
\label{eq:lum_dist3}
\end{equation}
Коэффициент $(1 + z)$ характеризует потерю  потока  энергии из-за эффектов, связанных с расширением вселенной:
(i) происходит уменьшение интенсивности поступления фотонов из-за продления времени распространения фотонов; (ii) происходит уменьшение энергии фотонов из-за красного смещения, возникшего вследствие расширения вселенной. Следовательно, объект со светимостью, $L_{\rm obs}$, будет казаться более отдаленным, чем он есть на самом деле.

При маленьких величинах красного смещения, $z$, фотометрическое расстояние можно определить как:
\begin{equation}
d_L\simeq\dfrac{z}{H_0}\Bigl[1+\dfrac{1}{2}(1+q_0)z\Bigr]~~~\text{для}~z\ll1.
\label{eq:aprLD}
\end{equation}
Сравнивая уравнения, Ур.~(\ref{eq:phys_dist3}) и Ур.~(\ref{eq:aprLD}), можно заключить, что физическое и фотометрическое расстояния до объекта, определенные в настоящий момент времени, равны между собой только при очень маленьких величинах красного смещения, когда первое слагаемое доминирует в этих уравнениях. При более больших величинах красного смещения, величина фотометрического расстояния больше величины физического расстояния, $d_L>d(t_0)$.

 Сверхновые типа Ia имеют небольшой разброс величин светимости. В космологии эти объекты являются {\it стандартными свечами} при определении расстояний до удаленных объектов. Измеряя поток энергии, полученный от сверхновых типа Ia на разных красных смещениях, $z$,  можно определить фотометрические расстояния до этих объектов другим способом и, применяя уравнение, Ур.~(\ref{eq:lum_dist3}), найти более точные величины параметров изучаемой космологической модели.

\section{Угловое расстояние}
Рассмотрим астрономический объект, находящийся на красном смещении, $z$, c линейным поперечный диаметром, ${\bf R}$, и с видимым угловым диаметром, $\theta$, измеренным в радианах.
Угловое расстояние до этого объекта, обозначенное, $d_A$,  определяется как отношение его линейного поперечного диаметра, ${\bf R}$, к видимому угловому диаметру, $\theta$:
\begin{equation}
d_A=\dfrac{\bf R}{\theta}.
\label{eq:ADD}
\end{equation}
Введем  FLRW систему координат, в центре которой находится наблюдатель.
В FLRW системе координат астрономический объект с красным смещением, $z$, имеет сопутствующую координату, $r$. Линейный поперечный диаметр этого объекта  является физическим  расстоянием между двумя событиями, находящимися на одном и том же красном смещении, $z$, и  разделенными в пространстве на маленький угол, $d\theta$. В FLRW метрике, описываемой в уравнении, Ур.~(\ref{eq:FLRW}), полагаем, $dt=dr=d\phi=0$. В результате получим:
\begin{eqnarray}
ds^2=a(t)^2r(t)^2d\theta^2,\nonumber \\
\Rightarrow~~ds=d{\bf R}=a(t)r(t)d\theta.
\label{eq:ADM}
\end{eqnarray}
В уравнении, Ур.~(\ref{eq:ADM}), проинтегрируем FLRW метрику в поперечном направлении к лучу зрения:
\begin{equation}
{\bf R}=a(t)r(t)\theta.
\label{eq:TrD}
\end{equation}
Подставим этот результат в уравнение, Ур.~(\ref{eq:ADD}):
\begin{equation}
d_A(z)=\dfrac{ar(z)\theta}{\theta}=\dfrac{r(z)}{(1+z)}.
\label{eq:ADD1}
\end{equation}
Подставляя значения сопутствующего расстояния, $r$, из уравнения, Ур.~(\ref{eq:FLRW_com1}), в уравнение, Ур.~(\ref{eq:ADD1}), получим значения углового расстояния, в зависимости от модельных параметров:
\begin{equation}
d_A(z)= \left\{
     \begin{array}{cc}
       \vspace{0.2cm}
       \frac{1}{(1+z)H_0\sqrt{\Omega_{\rm K0}}}\sin \Bigl(\frac{\sqrt{\Omega_{\rm K0}}}{H_0}\int_{0}^{z}\frac{dz'}{E(z')}\Bigr)& \ \ \ \ \text{для}\ \Omega_{\rm K0}>0 \\
       \vspace{0.2cm}
       \frac{1}{(1+z)H_0}\int_{0}^{z}\frac{dz'}{E(z')} & \ \ \ \ \text{для}\ \Omega_{\rm K0}=0 \\
       \vspace{0.2cm}
       \frac{1}{(1+z)H_0\sqrt{-\Omega_{\rm K0}}}\sinh \Bigl(\frac{\sqrt{-\Omega_{\rm K0}}}{H_0}\int_{0}^{z}\frac{dz'}{E(z')}\Bigr) & \ \ \ \ \text{для}\ \Omega_{\rm K0}<0 \\
     \end{array}.
   \right.
\label{eq:ADD2}
\end{equation}
Связь между фотометрическим  и угловым расстояниями выражается через уравнение:
\begin{equation}
d_L(z)=(1+z)^2d_A(z).
\label{eq:DLDA}
\end{equation}
Фотометрические и угловые расстояния, определяемые в уравнениях,  Ур.~(\ref{eq:lum_dist3}) и Ур.~(\ref{eq:ADD2}), зависят от выбранной космологической модели. Эти расстояния совпадают при маленьких величинах красного смещения, $z\ll1$, при которых можно пренебречь пространственной кривизной. При больших величинах красного смещения (соответственно при больших расстояниях) уже сказываются специфические космологические эффекты, такие как нестационарность и пространственная кривизна, поэтому понятие однозначного расстояния до объекта становится  неприменимым.

Радиогалактики Fanaroff-Riley Type II (FRII), имеют небольшой разброс в величинах их линейного поперечного диаметра, поэтому эти объекты могут служить в качестве {\it стандартной линейки} для определения расстояний до удаленных объектов в космологии, (\cite{Buchalter:1997vz}). Зная угловой размер, $\theta$, и величину красного смещения, на котором находятся эти объекты, $z$, можно определить угловое расстояние до этих объектов другим способом и, применяя уравнение, Ур.~(\ref{eq:ADD2}), можно уточнить параметры рассматриваемой космологической модели.

%%%%%%%%%%%%%%%%%%%%%%%%%%%%%%%%%%%%%%%%%%%%%%%%%%%%%%%%%%%%%%%%%%%%%%%%%%
%%%%%%%%%%%%%%%%%%%       Beginning of chapter 4     %%%%%%%%%%%%%%%%%%%%%%%
%%%%%%%%%%%%%%%%%%%%%%%%%%%%%%%%%%%%%%%%%%%%%%%%%%%%%%%%%%%%%%%%%%%%%%%%%%
%%%%%%%%%%%%%%%%%%%%%%%%%%%%%%%%%%%%%%%%%%%%%%%%%%%%%%%%%%%%%%%%%%%%%%%%%%
\chapter{Космологические наблюдения}\label{chapter:4}

\section{Сверхновые Типа Ia}
Взрыв сверхновой наблюдается как внезапное увеличение яркости звезды примерно на 10 порядков. В результате этого взрыва сверхновые в максимуме кривой блеска светят как все звезды галактики вместе взятые. Сверхновые регистрируются из далеких галактик вплоть до величины красного смещения, $z\approx1.7$. В зависимости от спектральных свойств сверхновые делятся на два основных типа: I - в спектрах которых отсутствуют линии водорода и II - в спектрах которых присутствуют линии водорода. Тип I сверхновых подразделяется в свою очередь на: Ia - кривые блеска имеют универсальную форму, Ib - кривые блеска аналогичны кривым блеска сверхновых типа II и Ic - в спектрах которых отсутствуют линии $He$ и их кривые блеска подобны кривым блеска сверхновых типа II.

Наиболее правдоподобной моделью сверхновых типа Ia считается модель термоядерного взрыва белого карлика с радиусом, ${\bf R}\sim10^3$~км, масса которого достигла массы Чандрасекара, $m_{\rm ch}\approx 1.44~M_{\odot}$, в результате аккреции массы со спутника звезды с высвобождением энергии,  $E\approx2\cdot10^{52}$~эрг. Этот взрыв вызван термоядерным
углеродным слиянием и радиоактивным распадом никеля, $\,^{56}{\rm Ni}$, $\,^{56}{\rm Ni}\rightarrow\,^{56}{\rm Co}\rightarrow\,^{56}{\rm Fe}$.  Радиоактивный распад $\,^{56}{\rm Ni}$ является основным источником наблюдаемых кривых блеска сверхновых типа Ia и определяет форму этих кривых.
Светимость в максимуме кривых блеска зависит только от массы выброшенного никеля, $\,^{56}{\rm Ni}$, $L_{\rm max}\approx1.4\cdot10^{43}$~эрг/сек, для массы никеля $m_{\rm Ni}=0.5~M_{\odot}$. Эта светимость соответствует абсолютной звездной величине, $M_{\rm max}=-19^m.2$\footnote{Определение понятия абсолютной звездной величины дано ниже.}.  Можно ожидать, что все сверхновые типа Ia излучают одинаковое количество света, в предположении, что белый карлик полностью выгорает.

Поскольку механизм взрыва универсален, все сверхновые типа Ia, находящиеся на одном расстоянии от нас, должны иметь примерно одинаковую светимость в максимуме кривых блеска, поэтому эти объекты используются как {\it стандартные свечи} для определения расстояний до далеких галактик. Самая далекая галактика, в которой зарегистрирована сверхновая типа Ia (1997ff), находится на красном смещении, $z=1.7$.

\begin{figure}[h!]
\begin{center}
\includegraphics[width=\columnwidth]{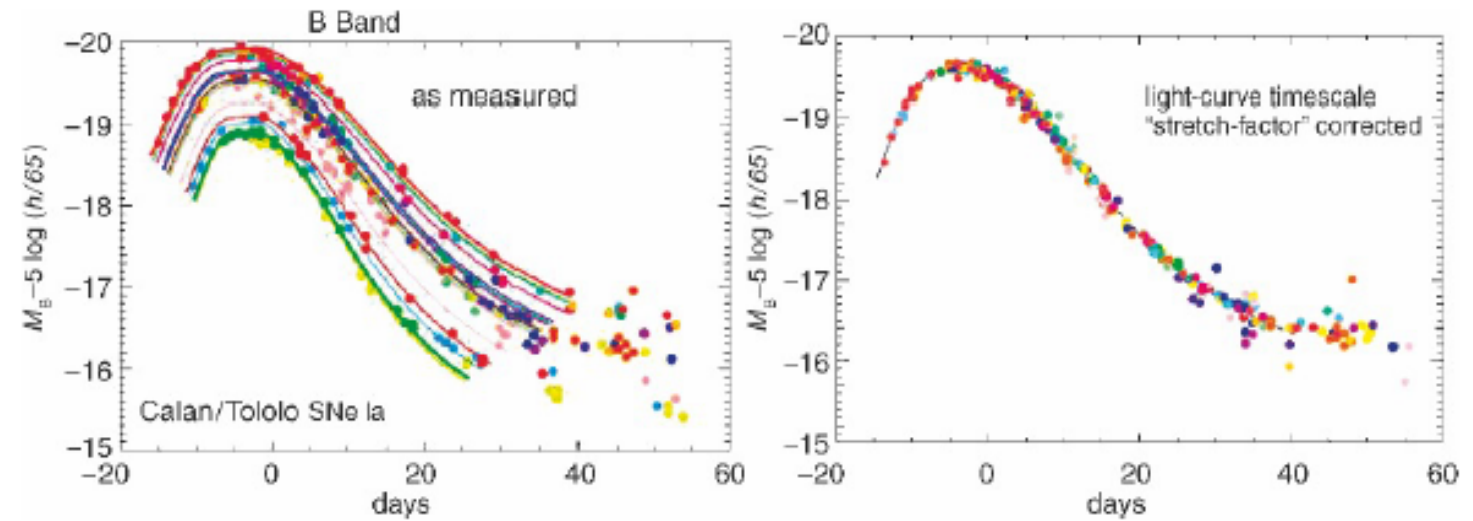}
\end{center}
 \caption {Левая панель: кривые блеска в B-диапазоне для разных сверхновых типа Ia, взятых из обзора Калан-Толоно, (\cite{Heitmann:2006hr}). Правая панель: те же самые кривые блеска после коррекции одного из параметров, (\cite{Kim:2003mq}).}
  \label{fig:f11}
\end{figure}
Среди различных образцов кривых блеска существует дисперсия в формах этих кривых, а также в максимальных величинах светимости (дисперсия достигает 0.4 звездной величины в B-диапазоне), Рис.~(\ref{fig:f11})~(левая панель).
Этот эффект вызван влиянием красного смещения на наблюдаемые спектры объектов в расширяющейся вселенной, т. к. эти наблюдения были проведены в определенном диапазоне длин волн. Эти кривые можно нормировать, применяя эмпирически найденную корреляцию, так называемую K-коррекцию между максимальной светимостью и шириной кривой блеска, Рис.~(\ref{fig:f11})~(правая панель). После выполнения этой коррекции, кривые блеска сверхновых типа Ia можно использовать как {\it стандартные свечи} для определения расстояний до объектов.\\
\\
{\bf Модуль расстояния}

Модуль расстояния - это способ определения расстояний до удаленных объектов, основанный на логарифмической шкале сравнения звездных величин.

Модуль расстояния, $\mu$, определяется как разность между видимой звездной величиной,
 $m$, и абсолютной звездной величиной, $M$, удаленного объекта с соответствующими болометрическими потоками энергии, $F_{m}$ и $F_{M}$.
Видимая звездная величина, $m$ - это звездная величина объекта, расположенного на фотометрическом расстоянии, $d_L$, а абсолютная звездной величина, $M$, определяется как видимая звездная величина, которую объект имел бы, если бы он находился на расстоянии $d_L=10$ пс. Из закона Погсона, (\cite{Pogson:1857}), связывающего звездную величину астрономического объекта и болометрический поток энергии, зарегистрированный от него, следующим образом, $10^m\propto F^{-2.5}$, мы получим:
\begin{multline}
\mu=m-M=-2.5 \log_{10}{\Bigl(\dfrac{F_m}{F_M}\Bigr)}=5\log_{10}{\Bigl(\dfrac{d_L}{\rm 10~pc}\Bigr)},\\
     =5\log_{10}(H_0d_L)- 5\log_{10}H_0+25.
\label{eq:DM}
\end{multline}
Из уравнения, Ур.~(\ref{eq:DM}), следует, что модуль расстояния, $\mu$, определяется фотометрическим расстоянием, $d_L$, до объекта. Современный параметр Хаббла, $H_0$, в этом уравнении считается неудобным параметром, и он является причиной неточности в определении абсолютных звездных величин сверхновых типа Ia.

\begin{figure}[h!]
\begin{center}
\includegraphics[width=0.8\columnwidth]{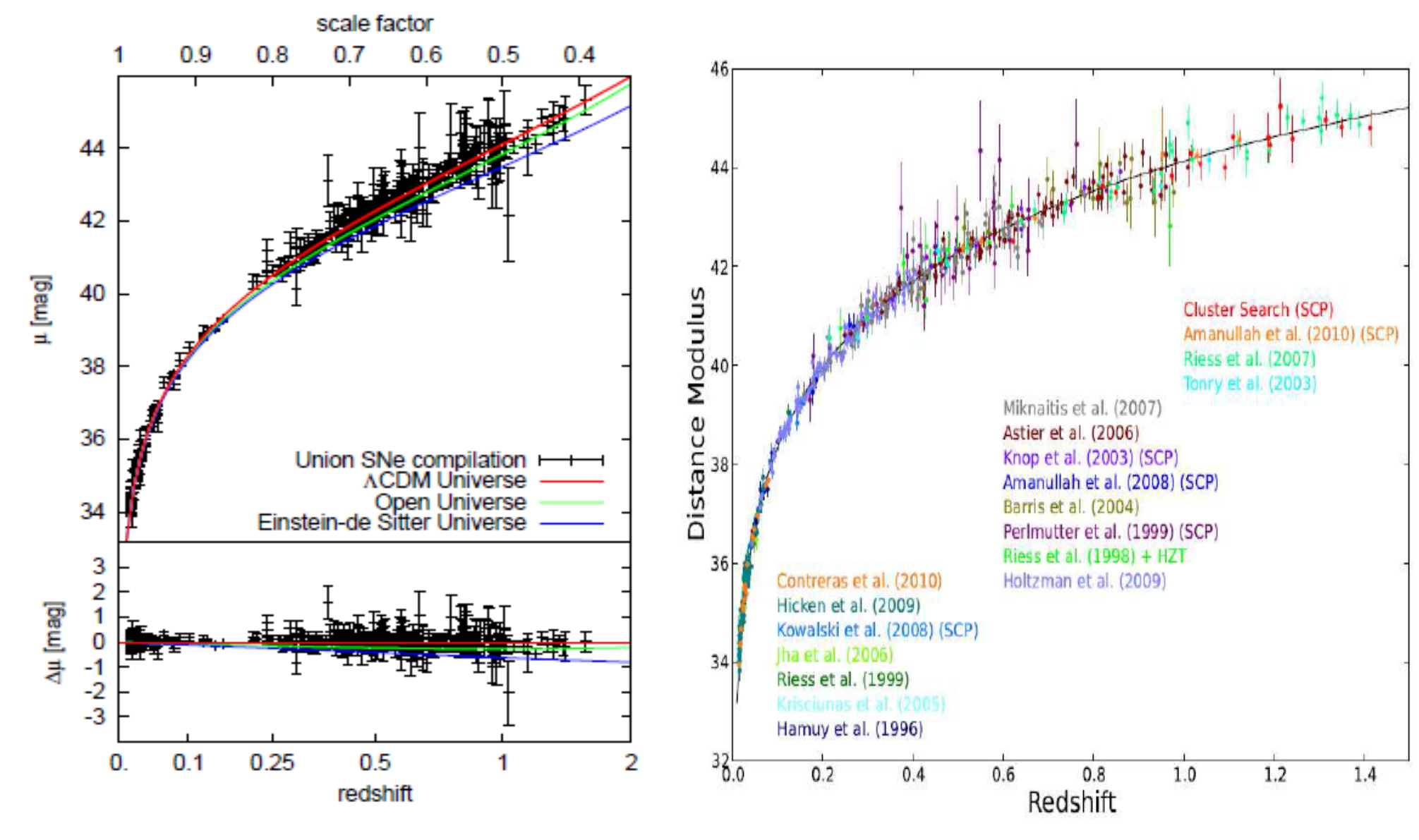}
\end{center}
 \caption {Левая панель: диаграмма Хаббла для 307 сверхновых типа Ia из  данных Union, (\cite{Kowalski:2008ez}). Левая панель, верхняя часть: красная линия соответствует $\Lambda$CDM модели ($\Omega_{\rm m}=0.28$, $\Omega_\Lambda=0.72$), зеленая линия  соответствует открытой вселенной ($\Omega_{\rm m}=0.28$, $\Omega_\Lambda=0$), синяя линия соответствует Эйнштейн - де Ситтер вселенной ($\Omega_{\rm m}=1$, $\Omega_\Lambda=0$). Левая панель, нижняя часть: разница величин модуля расстояния для рассматриваемых космологических моделей  и для $\Lambda$CDM модели. Правая панель: диаграмма Хаббла для данных Union2.1, (\cite{Suzuki:2011hu}).  $\Lambda$CDM модель лучшим образом аппроксимирует данные Union2.1, она представлена в виде черной сплошной линии.}
 \label{fig:f12}
 \end{figure}

Обычно при вычислении реальной величины модуля расстояния учитывают величину скорости света, $c=3\cdot10^5$~\text{км с$^{-1}$}. Таким образом, при вычислении фотометрического расстояния, необходимо применять выражение $c\cdot d_L$, где величина $d_L$ получена из уравнения, Ур.~(\ref{eq:lum_dist3}). В этом случае, применяя уравнение, Ур.~(\ref{eq:DM}), получим выражение для реальной величины модуля расстояния, $\mu$, в зависимости от красного смещения и модельных параметров:
\begin{multline}
\mu=42.3856 - 5\log_{10}(h) + 5\log_{10}(1+z)
+ 5\log_{10} \left\{
     \begin{array}{cc}
       \vspace{0.2cm}
       \frac{1}{\sqrt{\Omega_{\rm K0}}}\sin \Bigl(\frac{\sqrt{\Omega_{\rm K0}}}{H_0}\int_{0}^{z}\frac{dz'}{E(z')}\Bigr)& \ \ \text{для} \ \Omega_{\rm K0}>0 \\
       \vspace{0.2cm}
       \int_{0}^{z}\frac{dz'}{E(z')} & \ \ \text{для} \ \Omega_{\rm K0}=0 \\
       \vspace{0.2cm}
       \frac{1}{\sqrt{-\Omega_{\rm K0}}}\sinh \Bigl(\frac{\sqrt{-\Omega_{\rm K0}}}{H_0}\int_{0}^{z}\frac{dz'}{E(z')}\Bigr) & \ \ \text{для} \ \Omega_{\rm K0}<0 \\
     \end{array}.
   \right.
 \label{eq:DM1}
\end{multline}
Модуль расстояния является функцией космологических параметров, Ур.~(\ref{eq:DM1}), поэтому величина модуля расстояния очень чувствительна к изменению величин космологических параметров, Рис.~(\ref{fig:f12})~(левая панель). Данные для сверхновых типа Ia лучше всего соответствуют величинам модуля расстояния для $\Lambda$CDM модели, как показано на Рис.~(\ref{fig:f12})~(правая панель).

В середине девяностых годов двадцатого века две независимые астрономические группы: Supernova Cosmology Project (SCP), которой руководил Сол  Перлмуттер,   (\cite{Riess:1998cb}, \cite{Perlmutter:1998np}) и High-Z Supernova Cosmology Team (HZSNS Team), которую возглавлял Брайан Шмидт, (\cite{Schmidt:1998ys}), наблюдали сверхновые типа Ia с целью определения расстояний до этих объектов.

 При обработке собранной информации учёные надеялись получить подтверждение замедляющегося расширения вселенной.
 Обе группы исследователей независимо друг от друга обнаружили, что суперновые типа Ia, находящиеся на красном смещении, $z=0.5$, были тусклее на $0.25$ звездной величины в сравнении со звездной величиной, предсказанной моделью с  космологическими параметрами: $\Omega_{\rm m0}=0.3$ и $\Omega_\Lambda=0$, которая описывает  открытую замедляющуюся вселенную. Так называемая модель Эйнштейн - де Ситтер с  космологическими параметрами: $\Omega_{\rm m0}=1$ и $\Omega_\Lambda=0$, которая описывает плоскую замедляющуюся вселенную, так же не смогла верно аппроксимировать полученные результаты. Таким образом, суперновые типа Ia оказались на более далеком расстоянии, чем предсказывали  космологические модели, описывающие открытую и плоскую замедляющуюся вселенную.

Космологическая модель, которая описывает плоскую ускоряющуюся вселенноую, с космологическими параметрами, $\Omega_{\rm m0}=0.3$ и $\Omega_\Lambda=0.7$, хорошо аппроксимировала результаты, полученные этими наблюдателями. Таким образом было сделано открытие ускоренного расширения нашей вселенной по данным сверхновых типа Ia. За это открытие Солу Перлмуттеру, Брайану Шмидту и Адаму Риссу была присуждена Нобелевская премия в 2011 году.

\section{Космическое микроволновое фоновое излучение}

\subsection{Описание микроволнового фонового излучения}
\subsubsection{Возникновение микроволнового фонового излучения}
Эпоха {\it рекомбинации} во вселенной началась приблизительно $t_{\rm rec}=350 000$ лет после Большого Взрыва\footnote{В 1946 году Георгий Гамов разработал теорию 'горячей вселенной', которую также называют теорией Большого Взрыва, (\cite{Gamov:1946}). На основе этой теории, Георгий Гамов, Ральф Альфер и Роберт Херман предсказали существование реликтового излучения или микроволнового фонового излучения (CMBR), (\cite{Alpher:1948a}, \cite{Alpher:1948b}).
В 1965 году американские радиоастрономы  Арно Пензиас и Роберт Вилсон абсолютно случайно зарегистрировали это изотропное реликтовое излучение, (\cite{Penzias:1965}).  Обнаружение реликтового излучения, возникшего в эпоху первичной рекомбинации водорода, является одним из главных подтверждений верности теории Большого Взрыва. В 1978 году Арно Пензиасу и Роберт Вилсону была присуждена Нобелевская премия за открытие реликтового излучения.}, при красном смещении $z_{\rm rec}\approx1400$, при средней температуре во вселенной, $\langle T \rangle_{\rm rec}\approx 3800$ К, (\cite{Gorbunov:2011zz}).
Вследствие расширения, и, соответственно, охлаждения вселенной в эпоху рекомбинации, заряженные электроны и протоны становятся связаными, образуя электрически нейтральные атомы водорода\footnote{До рекомбинации барионная материя состояла на $75\%$ из протонов и на $25\%$ из $\alpha$-частиц, ядер гелия, $\,^4{\rm He}$. Энергия ионизации гелия больше чем энергия ионизации водорода, следовательно рекомбинация гелия произошла значительно раньше, (\cite{Peebles:1966zz}). Первая рекомбинация гелия, ${\rm He}^{++} +e^-\rightarrow {\rm He}^+ +\gamma$, произошла при красном смещении, $z\approx6000$. Вторая рекомбинация гелия, ${\rm He}^+ +e^-\rightarrow {\rm He} +\gamma$, произошла при красном смещении, $z\approx2500$, (\cite{Hu:1995em}). Несмотря на то что после рекомбинации гелия вселенная остается все еще оптически непрозрачной, рекомбинация гелия влияет на температурный спектр мощности микроволнового фона, увеличивая амплитуду 2-го, 3-го и 4-ого пиков на $0.2\%$, $0.4\%$ и на $1\%$, соответственно, (\cite{Hu:1995fqa}, \cite{Hu:1995em}).}, (\cite{Peebles:1968ja}). При этом материя  из состояния плазмы, непрозрачной для большей части электромагнитного излучения, переходит в газообразное, электрически нейтральное состояние.

Реликтовое излучение возникло в конце эпохи рекомбинации, в период последнего рассеяния фотонов на электронах, в так называемый период {\it отсоединения фотонов} от атомов водорода. Последнее рассеяние фотонов произошло $t_{\rm dec}\approx379 000$ лет после Большого Взрыва, при красном смещении $z_{\rm dec}\approx1100$,  при средней температуре во вселенной, $\langle T \rangle_{\rm dec}\approx 3100~K$. Как следствие отсоединения излучения от материи, реликтовые фотоны больше не взаимодействовали с нейтральными атомами водорода. Длина свободного пробега реликтовых фотонов становится больше размера горизонта Хаббла, и эти фотоны начинают свободно распространяться во вселенной. Таким образом, современный наблюдатель регистрирует реликтовые фотоны, которые в последний раз взаимодействовали с материей при красном смещении, $z_{\rm dec}$.

 Согласно модели Большого Взрыва, реликтовое излучение начало свое распространение во вселенной от поверхности сферы, называемой {\it поверхностью последнего рассеяния}, радиус которой составляет\footnote{В реальности, из-за того, что рекомбинация не является мгновенным процессом и протекает в течение конечного диапазона величин красного смещения, реликтовые фотоны в последний раз рассеиваются внутри поверхности конечной толщины. Толщина этой поверхности при рекомбинации приблизительно равна длине {\it диффузии фотонов}, поэтому этот эффект значим на тех же масштабах длин, что и затухание Силка (эффект затухания Силка описан ниже), (\cite{Schneider:2006}).}:
\begin{equation}
r_{\rm dec}=\dfrac{1}{a_0H_0}\int_{0}^{z_{\rm dec}}\dfrac{dz'}{E(z')}.
 \label{eq:PR}
 \end{equation}

\subsubsection{Свойства микроволнового фонового излучения}
В 1989 году был запущен спутник Cosmic Background Explorer (COBE) с целью исследования реликтового излучения. Результатами измерений, проведенных  с этого спутника являются: открытие планковского спектра CMBR (проект Differential Microwave Radiometer) (DMR), (\cite{Mather:1993ij}, \cite{Mather:1998gm}) и открытие анизотропии температуры CMBR\footnote{В 1983 году в СССР был проведен с борта космического спутника 'Прогноз-9' эксперимент 'Реликт-1' по изучению анизотропии микроволнового излучения. Советским ученым не удалось зарегистрировать анизотропию микроволнового излучения.} (проект Far-InfraRed Absolute Spectrophotometer) (FIRAS), (\cite{Bennett:1996ce}). Руководители этих проектов Джордж Смут (проект DMR) и Джон Мазер (проект FIRAS)  получили  Нобелевскую премию в 2006 году.

Реликтовое излучение является тепловым излучением, его спектр соответствует спектру абсолютно чёрного тела с температурой в современную эпоху,
$T_{0}\simeq 2,72548\pm0,00057~{\rm K}$, Рис.~(\ref{fig:f12})~(левая панель).
Эта температура соответствует средней температуре реликтового излучения в современную эпоху, $\langle T_{\gamma} \rangle=T_0$.
Максимум планковского спектра приходится на частоту $160,4~\text{ГГц}$, что соответствует длине волны $1,9~\text{мм}$, Рис.~(\ref{fig:f12})~(левая панель).
Плотность энергии микроволнового излучения приблизительно равна,  $\rho_{\gamma}=(\pi^2/15)T_{0}^4\simeq4.64\cdot10^{-34}$~г~\text{см}$^{-3}\simeq 0.26~{\rm eV}$~\text{см}$^{-3}$. Массовая плотность реликтового излучения составляет, $n_{\gamma}=(2\zeta(3)/\pi^2)T_{0}^3\simeq411$~см$^{-3}$, где $\zeta$-функция Римана, $\zeta(3)=1.202$, (\cite{Scott:2010yx}).
\subsubsection{Анизотропия микроволнового фонового излучения}
\begin{figure}[h!]
\begin{center}
\includegraphics[width=\columnwidth]{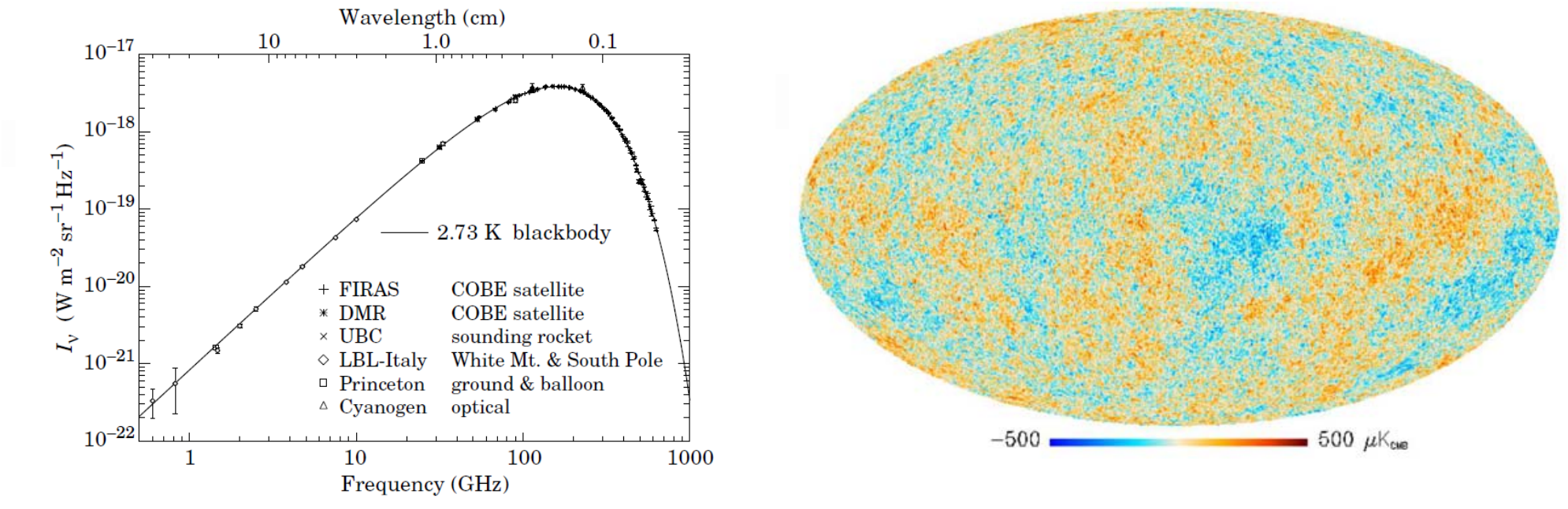}
\end{center}
 \caption {Левая панель: планковский спектр микроволнового излучения, полученный различными измерениями: FIRAS, DMR, UBC, LBL-Italy, Princeton, Cyanogen, (\cite{Smoot:1997cv}). Правая панель: температурные флуктуации космического микроволнового фонового излучения относительно средней температуры на основе результатов исследования проекта Planck 2013, (\cite{Ade:2013sjv}). Данные представлены с вычетом непланковского излучения, исходящего от диска Галактики и дипольной анизотропии, связанной с движением солнечной системы относительно космического микроволнового фонового излучения. Амплитуда температурных колебаний относительно фона составляет $\delta T/T_0 \sim 10^{-5}$.}
 \label{fig:f13}
 \end{figure}

Температура микроволнового фона, регистрируемая в направлении, $(\theta,\varphi)$, на небе, $T(\theta,\varphi)$, является  основным измерением в исследованиях микроволнового фона. Величина $\theta$ определяет полярный угол на сфере, а величина $\varphi$ - азимутальный угол. Безразмерная величина анизотропии температуры СМВR определена как:
\begin{equation}
\dfrac{\delta T(\theta,\varphi)}{T_0}=\dfrac{T(\theta,\varphi)-T_0}{T_0}.
\label{eq:TA}
\end{equation}
Микроволновое излучение изотропно и однородно на уровне величины флуктуации температуры, $\delta T(\theta,\varphi)/T_0\simeq10^{-4}$, Рис.~(\ref{fig:f13})~(правая панель).

На Рис.~(\ref{fig:f13})~(правая панель) представлена карта температурных анизотропий микроволнового фонового излучения, полученная проектом Planck 2013, (\cite{Ade:2013sjv}). В современную эпоху анизотропия температуры реликтовых фотонов составляет, $\delta T(\theta,\varphi)/T_0\simeq10^{-5}$.

\subsection{Угловой спектр мощности микроволнового фонового излучения}
Ввиду того, что величина анизотропии температуры микроволнового фона зависит от направления наблюдения, величину температурной анизотропии можно представить в виде разложения по сферическим ортонормированным гармоникам, $Y_l^m(\theta,\varphi)$. Это разложение является аналогом  разложения Фурье на сферической поверхности:
\begin{equation}
\dfrac{\delta T(\theta,\varphi)}{T_0}=\sum_{l=1}^{\infty}\sum_{m=-l}^la_{l,m}Y_l^m(\theta,\varphi),
\label{eq:TA1}
\end{equation}
где $a_{l,m}$ - мультипольные коэффициенты разложения по сферическим гармоникам,  $Y_l^m(\theta,\varphi)$. Коэффициенты $a_{l,m}$ характеризуют амплитуду температурных флуктуаций на разных угловых масштабах и обладают следующим свойством, $a^\ast_{l,m}=(-1)^m a_{l,-m}$.

Для анализа распределения анизотропии температуры микроволнового фона важны исследования статистических свойств коэффициентов $a_{l,m}$. Коэффициенты $a_{l,m}$ могут иметь как положительную так и отрицательную величину. Квадрат величины\footnote{По установленной договоренности, знак модуля применяется для того, чтобы комплексные величины сферических гармоник переводить в действительные величины.}, $|a_{l,m}|^2$, определяет отклонение коэффициента $a_{l,m}$ от нуля и, таким образом, определяет амплитуду температурной анизотропии. Согласно наблюдательным данным, распределение температурных флуктуаций микроволнового фона образует случайное гауссово поле.
В предположении изотропной и однородной вселенной, коэффициенты $a_{l,m}$ для разных величин  индексов $l$ и $m$ статистически независимы друг от друга, (\cite{Mukhanov2005}):
\begin{equation}
\langle a_{l,m}a^{\ast}_{l',m'}\rangle=C_{lm}\delta_{ll'}\delta_{mm'}.
\label{eq:coeff}
\end{equation}
Величина коэффициентов $C_{lm}$ определяет {\bf температурный угловой спектр мощности} анизотропии CMBR.

Требование независимости статистических свойств коэффициентов $a_{l,m}$ от выбора начала координат для любого направления наблюдения, или так называемое требование выполнения вращательной инвариантности, приводит к тому, что величина углового спектра мощности, $C_{l,m}$, не зависит от величины индекса, $m$, а зависит только от индекса $l$, т. е. $C_{l,m}=C_l$, (\cite{Durrer:1998ya}). Поэтому уравнение, Ур.~(\ref{eq:coeff}), при совпадении индексов, $l=l'$,  можно переписать как, (\cite{Mukhanov2005}):
\begin{equation}
\langle |a_{l,m}|^2\rangle=C_l.
\label{eq:coeff6}
\end{equation}
Угловые скобки, $\langle \rangle$, в уравнениях, Ур.~(\ref{eq:coeff}) и Ур.~(\ref{eq:coeff6}), обозначают усреднение по гипотетическому ансамблю вселенных, подобной нашей.  В предположении, что наша вселенная является эргодической динамической системой\footnote{Эргодические системы характеризуются совпадением  математического ожидания по временным рядам с математическим ожиданием по пространственным рядам.}, эти угловые  скобки можно интерпретировать как усреднение по всем возможным наблюдателям в нашей вселенной.  Дело в том, что каждый наблюдатель во вселенной может наблюдать только одну реализацию из всех возможных наблюдаемых вселенных. Например, наблюдатели с Земли могут изучать микроволновой фон, видимый только с Земли. Каждый наблюдатель во вселенной регистрирует реликтовые фотоны со своим собственным распределением температурных флуктуаций, которые отличаются друг от друга.  Разница между нашей областью наблюдаемой вселенной в сравнении со средней областью наблюдаемой вселенной называется {\it космической дисперсией} (cosmic variance).
Величина космической дисперсии для каждого измерения, $C_l$, определяется как, (\cite{Scott:2010yx}):
\begin{equation}
(\Delta C_l)^2=\dfrac{2}{2l+1}C_l^2.
\label{eq:coeff10}
\end{equation}
Величина космической дисперсии пренебрежительно мала на маленых угловых масштах, она становится значимой для угловых масштабов, $\vartheta\geq10^{\circ}$.

 Величина углового спектра мощности, $C_l$, характеризует размер температурных флуктуаций на угловом масштабе, $\vartheta=180^{\circ}/l$. Индекс $l$ определяет  величину углового масштаба. Маленькая величина индекса $l$ соответствует большому угловому масштабу и наоборот, большая величина $l$ соответствует маленькому угловому масштабу. С увеличением величины индекса $l$, сферические гармоники имеют вариации на меньших угловых масштабах. В современных наблюдениях применяют  величину индекса $l$ в пределах от единицы до нескольких тысяч.

Величину индекса $l=1$ определеляет свойства микроволнового излучения, названное {\bf диполем}. В 1969 году в реликтовом излучении была обнаружена дипольная составляющая, проявляющая себя в том, что в направлении созвездия Льва температура этого излучения на $0,1$~K выше средней температуры микроволнового излучения, соответственно, в противоположном направлении - на такую же величину ниже.
 Эта температурная анизотропия объясняется эффектом Доплера вследствие  движением
солнечной системы относительно реликтового излучения в направлении созвездия Льва со скоростью, $370.6\pm0.4$~км с$^{-1}$. Скорость этого движения определяет величину дипольной составляющей температурной анизотропии,
$\delta T_{\rm dipol}=3.355\pm0.008$~мK, (\cite{Hinshaw:2008kr}). Максимальная величина флуктуаций температуры для дипольной составляющей,  усредненной  за год, составляет: $\delta T/ T_0 \simeq1.23\cdot10^{-3}$. Наблюдательные данные о дипольной составляющей не несут в себе информацию о свойствах, присущих микроволновому излучению. В связи с этим диполь рассматривают отдельно, а изучение реликтового излучения начинают с минимальной величиной индекса $l=2$, с так называемой квадрупольной анизотропии.

Рассмотрим анализ анизотропии реликтового излучения без учета диполя:
\begin{equation}
\dfrac{\delta T(\theta,\varphi)}{T_0}\equiv \dfrac{T(\theta,\varphi)-T_0-\delta T_{\rm dipol}}{T_0}=\sum_{l=2}^{\infty}\sum_{m=-l}^la_{l,m}Y_l^m(\theta,\varphi).
\label{eq:FT}
\end{equation}
Сферические гармоники, $Y_l^m(\theta,\varphi)$, выражаются через функции Лежандра, $P_l^m(\cos\vartheta)$, следующим образом, (\cite{Arfken:1985vv}):
\begin{equation}
Y_l^m(\theta,\varphi)=(-1)^m \sqrt{\dfrac{2l+1}{2}\dfrac{(l-m)!}{(l+m)!}} P_l^m(\cos\vartheta)e^{im\varphi}.
\label{eq:SG}
\end{equation}
Требование выполнения вращательной инвариантности или выполнения условия изотропности относительно величины азимутального угла, $\varphi$, эквивалентно равенству нулю величины $m$, $m=0$. В этом случае уравнение, Ур.~(\ref{eq:SG}), принимает вид:
\begin{equation}
Y_l(\theta,\varphi)=\sqrt{\dfrac{2l+1}{2}} P_l(\cos\vartheta).
\label{eq:SG1}
\end{equation}
Таким образом, в уравнении, Ур.~(\ref{eq:SG1}), сферические гармоники сводятся к обычным многочленам Лежандра, $P_l(\cos\vartheta)$.

В этом случае температурная корреляционная функция между двумя направлениями имеет вид:
\begin{equation}
\Big\langle \dfrac{\delta T(\theta_1,\varphi_1)}{T_0}\cdot \dfrac{\delta T(\theta_2,\varphi_2)}{T_0}\Big\rangle =\sum_l\dfrac{2l+1}{4\pi}C_lP_l(\cos \vartheta),
\label{eq:coeff1}
\end{equation}
величина $\vartheta$ является полярным углом между направлениями ($\theta_1,\varphi_1$) и ($\theta_2,\varphi_2$).
Коэффициенты $C_l$ устанавливают корреляцию между флуктуациями температуры в разных направлениях.

Выражение для среднеквадратичных флуктуаций температуры является частным случаем уравнения, Ур.~(\ref{eq:coeff1}):
\begin{equation}
\Big\langle \dfrac{\delta T(\theta_1,\varphi_1)}{T_0}\cdot \dfrac{\delta T(\theta_2,\varphi_2)}{T_0}\Big\rangle=\sum_l\dfrac{2l+1}{4\pi}C_l\approx\int\dfrac{l(l+1)}{2\pi}C_l d\ln l,
\label{eq:rmsT}
\end{equation}
при выводе этой формулы учитывалось, что полярный угол между двумя коллинеарными сонаправленными векторами равен нулю, $\vartheta=0$, и $P_l(\cos 0)=1$.
Величина $\frac{l(l+1)C_l}{2\pi}$ определяет общий вклад угловых моментов одинакого порядка.

\begin{figure}[h!]
\begin{center}
\includegraphics[width=0.8\columnwidth]{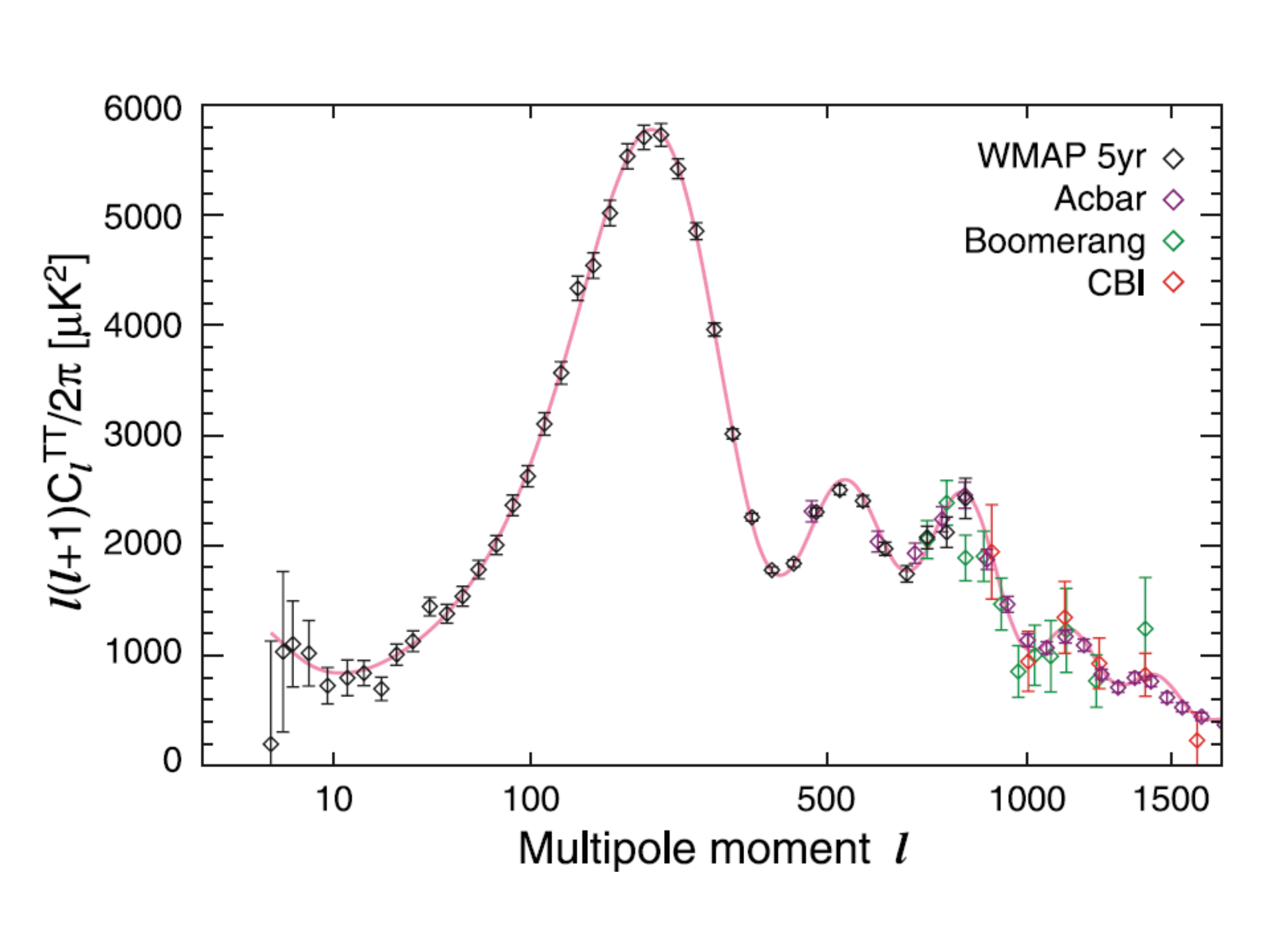}
\end{center}
 \caption {Температурный угловой спектр мощности анизотропии CMBR, полученный экспериментами: WMAP 5 year, Acbar, Boomerang, CBI, (\cite{Nolta:2008ih}).}
 \label{fig:f14}
 \end{figure}

Зависимость углового спектра мощности анизотропии температуры СМВR, $\frac{l(l+1)C_l}{2\pi}T^2_0$, от углового момента, $l$, представлен на Рис.~(\ref{fig:f14}).

\subsection{Первичные температурные анизотропии микроволнового фонового излучения}
Температурные флуктуации, образованные при отсоединении фотонов в эпоху рекомбинации, называются {\it первичной анизотропией}.

\begin{figure}[h!]
\begin{center}
\includegraphics[width=0.8\columnwidth]{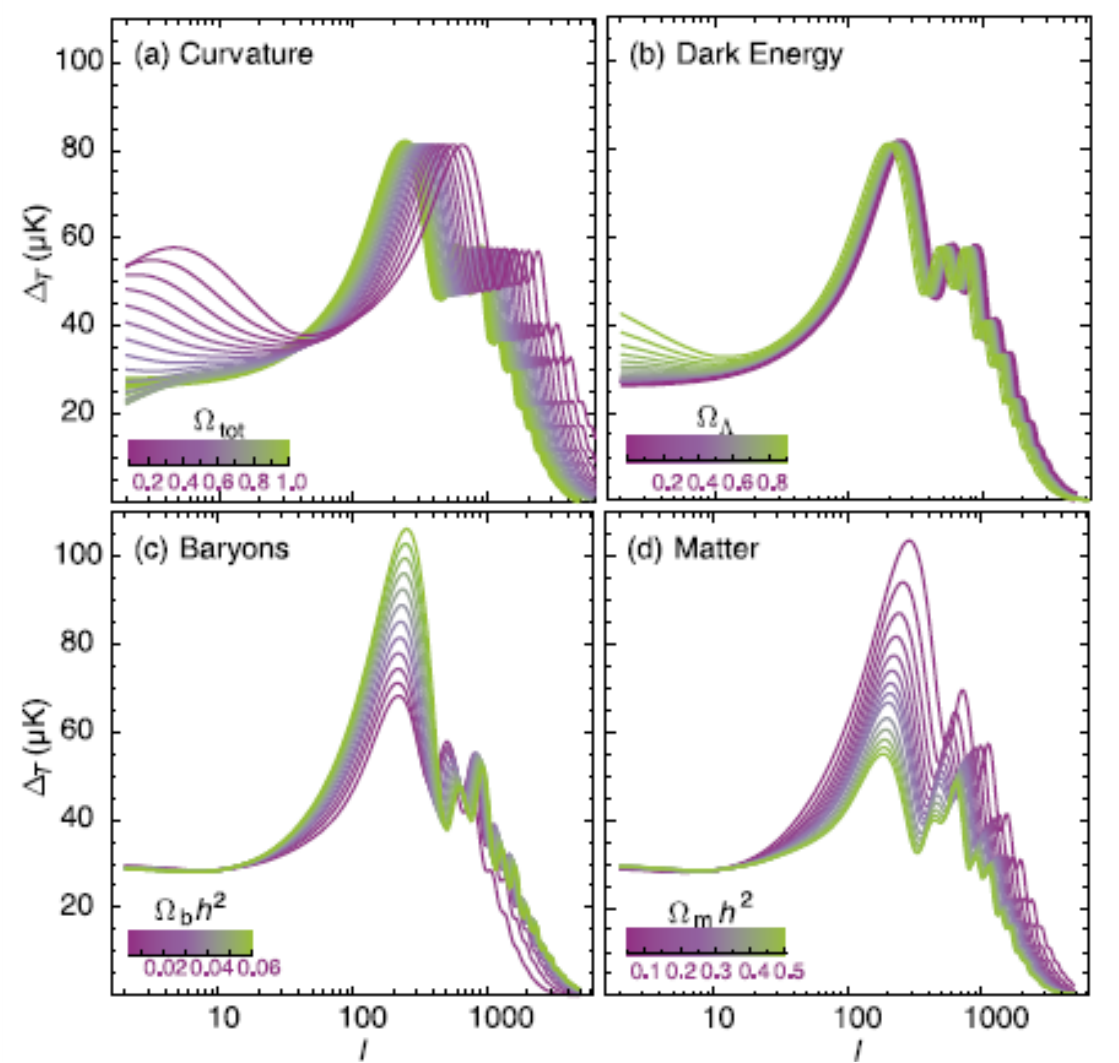}
\end{center}
 \caption {Влияние космологических параметров на угловой спектр мощности анизотропии температуры СМВR. Величины квадратного корня от спектра мощности, $\delta T=\sqrt{l(l+1)C_l/2\pi T_0}$, построены в зависимости от  логарифмического масштаба углового момента, $l$, (\cite{Hu:2001kj}).}
 \label{fig:f15}
 \end{figure}

 Рассмотрим угловой спектр мощности анизотропии температуры CMBR, представленный на Рис.~(\ref{fig:f15}).
Угловой спектр мощности анизотропии температуры CMBR в основном характеризуется тремя областями величин углового момента, $l$: $l\leq100$, $l\geq100$ and $l\geq1000$, Рис.~(\ref{fig:f15}), (\cite{Hu:2001kj}, \cite{Scott:2010yx}).

 Для первой области c $l\leq100$, функция $(2l+1)/4\pi$ является почти плоской\footnote{Этот результат получен вследствие применения спектра мощности Харрисона-Зельдовича в расчетах для получения данного графика. Спектр мощности $P(k)=k^{n_s}$ с $n_s=1$ называется спектром Харрисона-Зельдовича, где $k$ - конформный импульс.}. Вторая область c $l\geq100$, содержит пики с разными амплитудами. Причиной этих пиков являются акустические колебания, возникшие в барион-фотонной плазме до отсоединения фотонов от барионов в эпоху рекомбинации. После окончания рекомбинации их положения были сдвинуты в результате расширения вселенной. Поэтому положения и амплитуды акустических пиков содержат важную информацию об эволюции вселенной. Первый акустический пик в анизотропии температуры CMBR определяет размер звукового горизонта барионов, величина которого служит {\it стандартной линейкой} для определения расстояний в космологии. С другой стороны, размер звукового горизонта можно определить путем измерения угловой шкалы первого звукового пика. В третьей области c $l\geq1000$, амплитуда спектра анизотропии мощности резко уменьшается из-за затухания Силка (описание этого эффекта дано ниже).

\subsection{Основные механизмы, вызывающие первичные анизотропии температуры микроволнового фонового излучения}
\begin{itemize}
\item[$\bullet$]Флуктуации плотности материи в первичной плазме, (\cite{Hu:2001kj}, \cite{Kosowsky:2001ue})

Плотность барионов напрямую связана с плотностью темной материи. На масштабах, больших чем горизонт событий во время рекомбинации, распределение барионов следует за распределением темной материи. На меньших масштабах давление барион-фотонной плазмы является эфффективным, так как до рекомбинации эти компоненты были тесно связаны рассеянием Томпсона.
В областях с повышенной плотностью темной материи плотность  барионов так же повышена. В таких областях  темпертура барионов увеличивается из-за их адиабатического сжатия, что  приводит к увеличению температуры фотонов.

\item[$\bullet$]Эффект Доплера, (\cite{Schneider:2006})

 Электроны, рассеивающие фотоны СМВR в последний раз во время рекомбинации, обладают дополнительными пекулярными скоростями относительно потока Хаббла. Эти скорости связаны с флуктуациями плотности материи и, соответственно, с флуктуациями температуры. Как следствие эффекта Доплера, реликтовые фотоны, удаляющиеся от нас со скоростями, большими чем скорость расширения Хаббла, испытывают дополнительное красное смещение. Это приводит к уменьшению температуры, измеренной в этом направлении.

\item[$\bullet$]Затухание Силка, (\cite{Hu:2001kj}, \cite{Kosowsky:2001ue}, \cite{Scott:2010yx})

   Затухание Силка или затухание диффузии фотонов - это физический процесс, который уменьшает анизотропию плотности энергии реликтовых фотонов, (\cite{Silk:1967kq}). Поскольку средняя длина свободного пробега фотонов конечна, барионы и фотоны становятся разделенными друг от друга на малых пространственных масштабах. Следовательно, температурные флуктуации могут быть размыты
диффузией фотонов на малых масштабах длин (для $l\geq1000$), Рис.~(\ref{fig:f15})~(d).

\item[$\bullet$] Интегральный эффект Сакса-Вольфа, (\cite{Sachs:1967er}, \cite{White:1997vi}, \cite{Hu:2001kj}, \cite{Scott:2010yx})

Пространственное распределение потенциала во вселенной меняется в эпоху
доминирования излучения или темной энергии. При прохождении реликтовых фотонов через этот эволюционирующий потенциал, происходит изменение энергии этих фотонов, т. е. происходит дифференциальное гравитационное красное смещение фотонов.
Это так называемый {\it интегральный эффект Сакса-Вольфа} (Integrated Sachs–Wolfe (ISW) effect), (\cite{Sachs:1967er}). Эффект ISW в основном влияет на низкие величины мультиполей CMBR, Рис.~(\ref{fig:f15})~(a). На больших масштабах температурная анизотропия CMBR связана с флуктуациями плотности из-за ISW эффекта, (\cite{White:1997vi}).

\item[$\bullet$]Первичные тензорные возмущения метрики, (\cite{Hu:1997hp}, \cite{Scott:2010yx})

Причиной возникновения первичной температурной анизотропии CMBR являются возмущения метрики. Эти возмущения могут генерировать скалярные, векторные и тензорные моды. Тензорные моды (поперечные с нулевым следом возмущения метрики) или гравитационные волны генерируют первичные температурные анизотропии CMBR благодаря суммарному эффекту анизотропного расширения пространства, (\cite{Scott:2010yx}). Вклад тензорной моды в температурный угловой спектр мощности анизотропии CMBR может происходить при $\vartheta>1$, соответственно при $l<180$.
Тензорную моду можно выделить из температурного углового спектра мощности анизотропии CMBR при использовании данных о поляризации реликтового излучения (информация об этом представлена ниже).

 \end{itemize}
\noindent

\subsection{Вторичные анизотропии температуры микроволнового фонового излучения}

 При распространении во вселенной реликтовое излучение может испытывать на себе ряд искажений, которые могут изменить  температурное распределение фотонов CMBR на небе. В угловом спектре мощности анизотропии температуры микроволнового излучения эти эффекты рассматриваются как {\it вторичные анизотропии}. Рассмотрим эффекты, вызывающие вторичные анизотропии температуры:

\begin{itemize}
\item [$\bullet $]Повторное рассеяние Томсона фотонов CMBR на свободных электронах, (\cite{Hu:2001bc}, \cite{Schneider:2006})

Повторное рассеяние Томсона фотонов CMBR на свободных электронах происходило в диапазоне величин красного смещения, $z \in(6; 20)$. Эти свободные электроны появились в результате реионизации атомов водорода  карликовыми галактиками, и/или звездами первого поколения,  и/или первыми квазарами.
Рассеяние Томсона изотропно, поэтому направления движения фотонов после этого рассеяния становятся почти независимыми к их первоначальным направлениям движения.
Вторично рассеянные реликтовые фотоны образуют изотропную составляющую излучения с микроволновым фоновым излучением. В результате этого эффекта происходит подавление первичной температурной анизотропии, т. е. измеренные величины флуктуаций температуры CMBR будут уменьшаться за счет доли фотонов, которые испытывали рассеяние Томпсона. Кроме подавления первичной температурной анизотропии CMBR, перерассеяние реликтовых фотонов вызывает дополнительную поляризацию CMBR на большим углах и дополнительный эффект Доплера на больших углах, (\cite{Hu:2001bc}).

\item [$\bullet $] Гравитационное линзирование реликтовых фотонов, (\cite{Hu:2001bc}, \cite{Schneider:2006})

Гравитационное поле флуктуаций плотности материи во вселенной вызывает гравитационное линзирование (гравитационное отклонение) реликтовых фотонов, что приводит к
 изменению первоначального направления движения фотонов. Это означает, что если мы сегодня наблюдаем два фотона, разделенных  углом, $\theta$, то в период отсоединения фотонов физическое разделение между ними будет отличаться от величины, $d_A(z_{\rm dec})\theta$, по причине гравитационного отклонения фотонов. Это означает, что в результате этого эффекта корреляционная функция температурных флуктуаций становится немного размазанной.
 Влияние этого эффекта становится значимым на маленьких угловых масштабах.
\item [$\bullet $] Эффект Сюняева-Зельдовича, (\cite{Scott:2010yx}, \cite{Yoo:2012ug})

  Скопления галактик оставляют отпечаток на реликтовые фотоны так называемым эффектом {\it Сюняева-Зельдовича} (Sunyaev–Zel'dovich (SZ) effect)\footnote{SZ эффект является рассеивающим, а его величина не зависит от красного смещения, поэтому скопления галактик могут быть найдены на любых расстояниях. Измерения SZ эффекта применяются для поиска скоплений галактик, для оценки их масс, для уточнения величины постоянной Хаббла, (\cite{Scott:2010yx}). Кроме того, в сочетании с точными величинами красного смещения для скопления галактик (например, с рентгеновскими наблюдениями) и оценкам их массы, SZ эффект может быть применен как  {\it стандартная линейка} в космологии,  (\cite{Cooray:2001av}).}, (\cite{Sunyaev:1970eu}).
Если реликтовые фотоны движутся сквозь скопление галактики, то они испытывают обратное комптоновское рассеяние на высокоэнергетических электронах, находящихся в этом скоплении. В результате этого рассеяния энергия и температура реликтовых фотонов увеличивается. Таким образом, спектр микроволнового излучения становится искаженным.
\end{itemize}

\subsubsection{Влияние космологических параметров на угловой спектр мощности анизотропии температуры микроволнового фонового излучения}

Влияние величин космологических параметров на угловой спектр мощности анизотропии температуры СМВR показано на Рис.~(\ref{fig:f15}).
Зависимость углового спектра мощности от величины параметра плотности кривизны показана на Рис.~(\ref{fig:f15})~(a). Существует два эффекта, связанных с влиянием пространственной кривизны на угловой спектр мощности CMBR: смещение минимумов и максимумов доплеровских пиков и сильная зависимость спектра в области с $l\leq100$ от величины полного параметра плотности энергии, $\Omega_{\rm tot}$, (\cite{Hu:2001bc}, \cite{Schneider:2006}). Последний эффект является следствием эффекта ISW, поскольку увеличение величины параметра плотности кривизны приводит к более сильному изменению от времени гравитационного потенциала. Сдвиг акустических пиков происходит из-за того, что величина углового расстояния, $d_A(z_{\rm rec})$,  чувствительна к изменению величины  параметра плотности кривизны, поэтому шкала угловых расстояний, соответствующая звуковому горизонту, также изменяется.
Влияние темной энергии (космологической постоянной $\Lambda$) на угловой спектр мощности CMBR в случае пространственно-плоской вселенной представлено на Рис.~(\ref{fig:f15})~(b). Можно заметить, что расположение акустических пиков почти не зависит от величины параметра плотности темной энергии, $\Omega_\Lambda$. Зависимость от величины параметра плотности энергии барионов показана на Рис.~(\ref{fig:f15})~(c). Увеличение величины параметра плотности энергии барионов,  $\Omega_{\rm b} h^2$, приводит к увеличению амплитуды первого акустического пика и к уменьшению амплитуды второго акустического пика.
Влияние величины параметра плотности энергии материи, $\Omega_{\rm m} h^2$, на спектр мощности СМВR представлено на Рис.~(\ref{fig:f15})~(d). Изменение величины этого параметра вызывает изменения амплитуд акустических пиков и расположения этих пиков, (\cite{Hu:2001bc}, \cite{Schneider:2006}).

\subsection{Поляризация микроволнового фонового излучения}
\begin{figure}[h!]
\begin{center}
\includegraphics[width=\columnwidth]{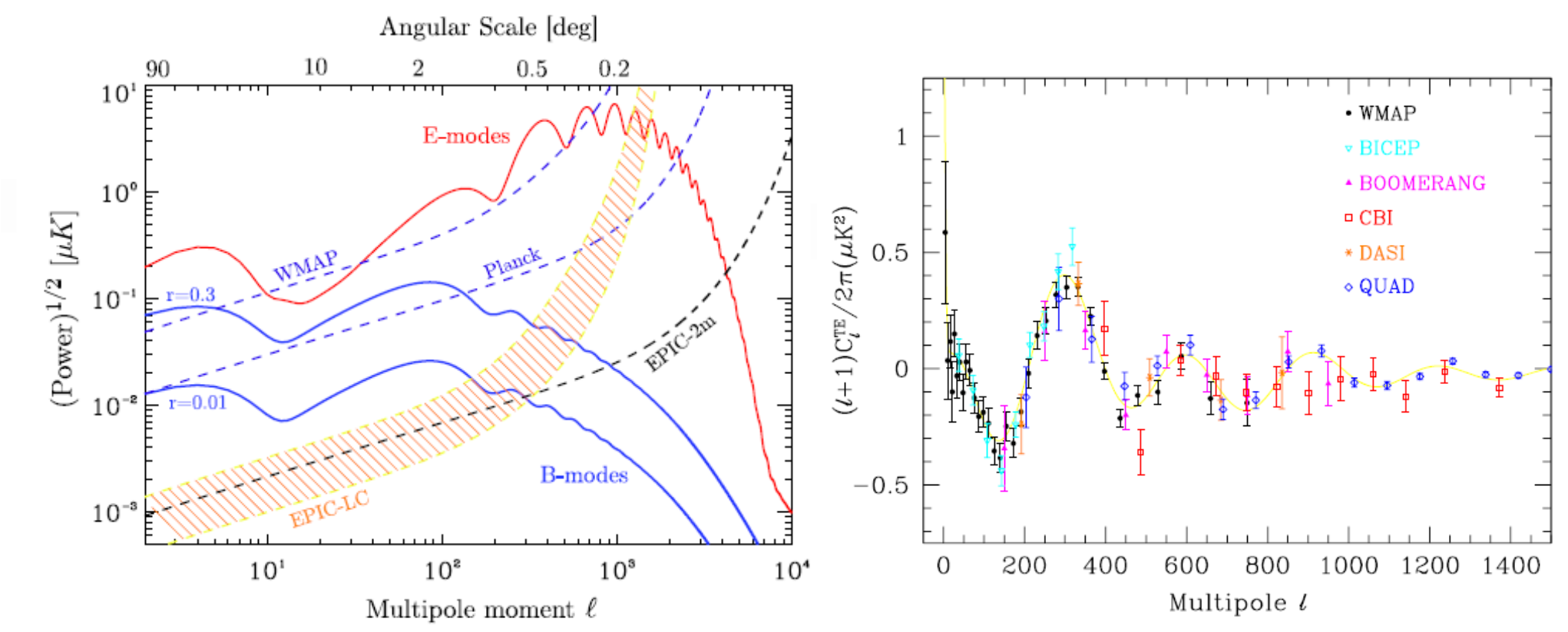}
\end{center}
 \caption {Левая панель: предсказанные спектры поляризации E-моды (красная кривая) и B-моды (синие кривые) совмещенные с результатами экспериментов WMAP, Planck и EPIC, (\cite{Dodelson:2009kq}). Правая панель: совмещение углового спектра мощности анизотропии  температуры и сигнала Е-моды поляризации, по результатам экспериментов: BICEP,
BOOMERANG, CBI, DASI, и QUAD, (\cite{Scott:2010yx}).}
 \label{fig:f16}
 \end{figure}

Микроволновое излучение поляризовано на уровне нескольких мкК, (\cite{Hu:1997hp}).
Причиной возникновения как температурной анизотропии микроволнового излучения, так  и ее поляризации являются скалярные и тензорные гравитационные возмущения метрики\footnote{Обычно векторные возмущения не рассматриваются ввиду их отсутствия в стандартном космологическом сценарии.}. Так как источники возникновения температурной анизотропии и  поляризации микроволнового излучения одни и те же, их спектры мощности должны быть скоррелированы, (\cite{Kosowsky:2001ue}, \cite{Scott:2010yx}. На  Рис.~(\ref{fig:f16})~(правая панель) представлены совмещение углового спектра мощности анизотропии  температуры CMBR и сигнала Е-моды поляризации CMBR, согласно результатам экспериментов: BICEP,
BOOMERANG, CBI, DASI, и QUAD, (\cite{Scott:2010yx}).

\subsubsection{Параметры Стокса}
Математически вектор поляризации электромагнитных волн описывается  {\it параметрами Стокса}, (\cite{Kosowsky:1994cy}).

Предположим, что плоская монохроматическая волна характеризуется частотой $\omega_0$.  Эта волна распространяется вдоль направления $z$. Проекции вектора напряженности, $\vec{E}$, на оси $x$ и $y$ представлены, соответственно, как, (\cite{Kosowsky:1994cy}, \cite{Kosowsky:2001ue}):
\begin{equation}
E_x=a_x(t)\cos(\omega_0t-\beta_x(t))~~~E_y=a_y(t)\cos(\omega_0t-\beta_y(t)),
\label{eq:proect}
 \end{equation}
где амплитуды проекций вектора напряженности, $a_x$ и $a_y$, а также фазовые углы, $\beta_x$ и $\beta_y$, являются медленно изменяющимися функциями в зависимости от времени по сравнению с обратной величиной частоты электромагнитной волны.

Параметры Стокса определены через усредненные по времени величины проекций  амлитуд и фаз вектора напряженности следующим образом:
\begin{eqnarray}
I\equiv \langle a_x^2\rangle+\langle a_y^2\rangle, \label{eqn:Ipar}\\
Q\equiv \langle a_x^2\rangle-\langle a_y^2\rangle, \label{eqn:Qpar}\\
U\equiv \langle 2a_xa_y\cos(a_x-a_y)\rangle, \label{eqn:Upar}\\
V\equiv \langle 2a_xa_y\sin(a_x-a_y)\rangle. \label{eqn:Vpar}
\end{eqnarray}
Величина параметра $I$ является интенсивностью электромагнитного излучения, поэтому величина этого параметра всегда положительна. Величины и знак параметров, $Q$, $U$ и $V$,  характеризуют состояние поляризации электромагнитной волны. Для естественного неполяризованного света величины этих параметров равны нулю, $Q=U=V=0$. Величина параметра $V$ определяет разницу между интенсивностями право- и левосторонней круговой (роторной) поляризаций. Параметр $V$ зависит от поворота осей системы координат, тогда как параметры $Q$ и $U$ инвариантны по отношению к повороту осей системы координат.

Линейная поляризация электромагнитной волны определяется параметрами $Q$ и $U$. Эти параметры формируют матрицу линейной поляризации следующим образом:
\begin{equation}
A=\left(\begin{array}{cc}
          Q& U\\
          U&-Q  \\
         \end{array}
           \right).
\end{equation}
Определить этой матрицы является алгебраическим инвариантом матрицы $A$:
\begin{equation}
{\rm det}(A)=-(Q^2+U^2).
\end{equation}
Линейная поляризация отсутствует, если определитель матрицы $A$ равен нулю. Предположим, что электромагнитное излучение линейно, т. е. $Q^2+U^2\neq0$. Тогда можно определить степень линейной поляризации, $p$, и величину позиционного угла, $\psi$, по отношению к оси $x$ как:
\begin{eqnarray}
p=\dfrac{\sqrt{Q^2+U^2}}{I},~~~
\psi=\dfrac{1}{2}\arctan\dfrac{U}{Q},
\end{eqnarray}
 величина параметра, $I$, определяет интенсивность электромагнитного излучения, (\cite{Kosowsky:1994cy}).

 \subsubsection{Градиентная и роторная составляющие поляризации микроволнового излучения}
\begin{figure}[h!]
\begin{center}
\includegraphics[width=0.8\columnwidth]{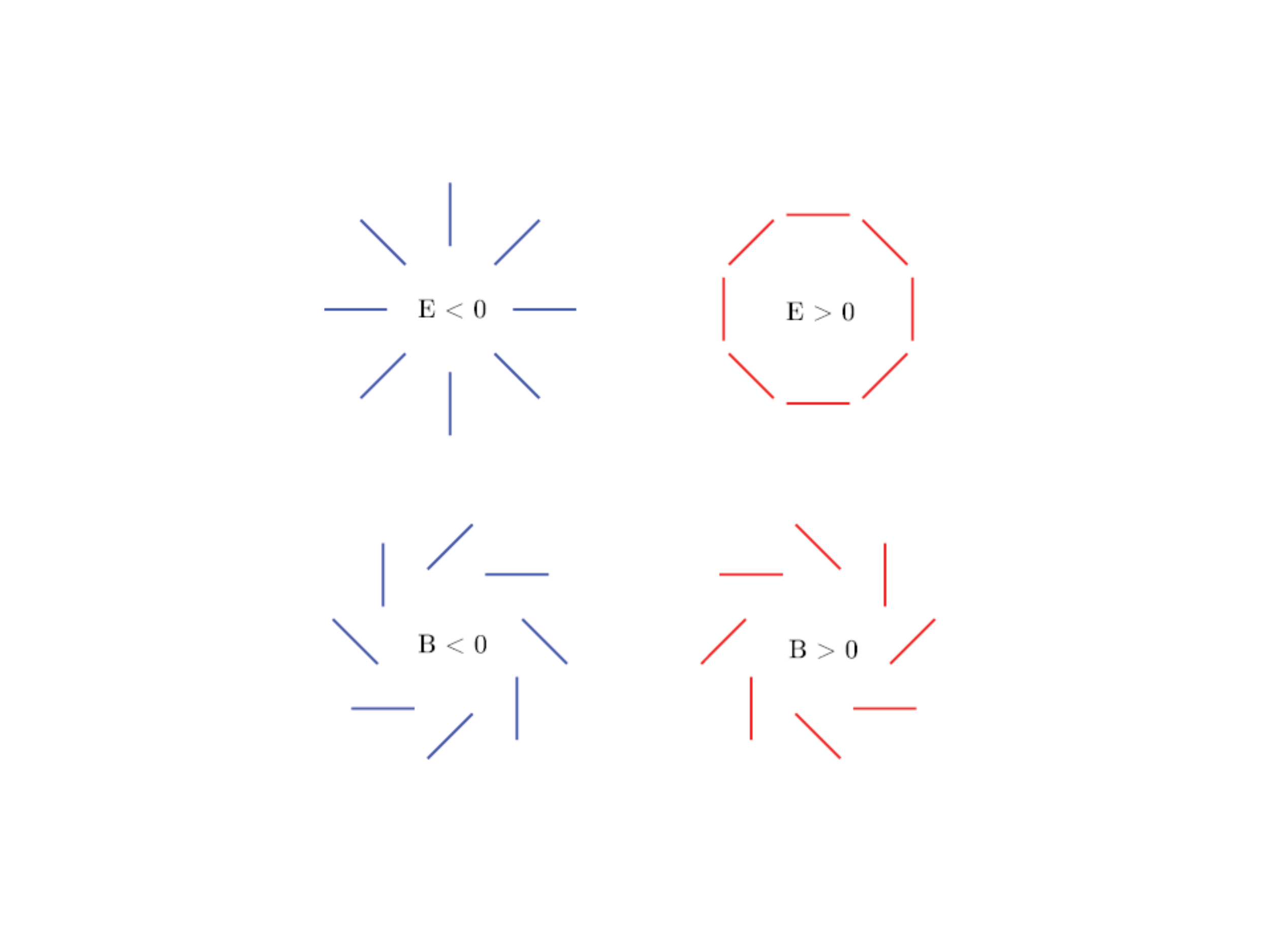}
\end{center}
 \caption {Градиентная  E-мода и роторная B-мода поляризационного поля, (\cite{Dodelson:2009kq}).}
 \label{fig:f17}
 \end{figure}

 В поляризации микроволнового излучения выделяют  градиентную составляющую, так называемую градиентную E-моду и роторную составляющую, B-моду, (\cite{Kosowsky:1994cy}).
 Направление поляризации В-моды повернуто на $45^{\circ}$ относительно направления поляризации Е-моды, Рис.~(\ref{fig:f17}).
 Е-мода поляризации микроволнового фона имеет четность, $(-1)^l$, подобную сферическим гармоникам, Рис.~(\ref{fig:f17}), тогда как B-мода имеет четность, $(-1)^{l+1}$.
  Скалярные возмущения не могут порождать В-моду поляризации. Вклад векторных возмущений в образование В-моды в $6$ раз выше чем в образование Е-моды, в то время как вклад тензорных возмущений в формирование В-моды меньше чем в  формирование Е-моды фактором $8/13$, (\cite{Hu:1997hp}).
Возникновение Е-моды связано с рассеянием Томпсона на электронах реликтовыми фотонами, распространяющимися  в неоднородной плазме, (\cite{Kosowsky:1998mb}, \cite{Kosowsky:2001ue}). В 2002 году E-мода была зарегистрирована экспериментом, Degree Angular Scale Interferometer (DASI), (\cite{Leitch:2002nzn}), Рис.~(\ref{fig:f16})~(правая панель).

Максимальная амплитуда В-моды СМВR поляризации составляет порядка $0.1~$ мкК, (\cite{Hu:1997hp}). Космологи предсказывают существование двух типов В-моды СМВR поляризации. Возникновение первого типа В-моды связано с взаимодействием реликтового излучения с реликтовыми гравитационными волнами (тензорной модой) или с вращательными, вихревыми возмущениями (векторной модой\footnote{В стандартном космологическом сценарии векторная мода затухает уже на инфляционной стадии. Присутствие нейтрино, (\cite{Lewis:2004ef}), или/и первичных магнитных полей, (\cite{Kahniashvili:2005xe}), могут противодействовать затуханию векторной моды. Принимая во внимание эти эффекты, необходимо учитывать вклад векторной моды.}), образовавшимися во время инфляции. Реликтовые гравитационные волны порождены тензорными возмущениями метрики.

Второй тип В-моды связан с гравитационным линзированием Е-моды или с так называемым эффектом космологического двулучепреломления (cosmological birefrin-\\gence), основанного на взаимодействии электромагнитного поля и скалярного поля, (\cite{Lepora:1998ix}, \cite{Galaverni:2014gca}). Второй тип В-моды возник в более позднее время, чем первый тип В-моды. Кроме того, В-моду  поляризации СМВR может также вызвать взаимодействие реликтового излучения с частицами фоновой галактической пыли. Второй тип В-моды был зарегистрирован в 2013 году совместными исследованиями South Pole Telescope и Herschel, (\cite{Hanson:2013hsb}).

Для космологов представляет огромный интерес обнаружение и изучение первого типа В-моды. Амплитуда первого типа В-моды соответствует амплитуде реликтовых гравитационных волн и, соответственно, определяет энергетическую шкалу инфляции, (\cite{Gawiser:2000az}). Поэтому регистрация этого типа В-моды, т. е. регистрация реликтовых гравитационных волн являлась бы прямым доказательством правильности теории инфляции.  В марте 2014 года было объявлено о регистрации первого типа В-моды экспериментом BICEP2, (\cite{Ade:2014gua}). Однако, более поздний анализ, проведённый другой группой исследователей с использованием данных обсерватории Planck показал, что результат, полученный экспериментом BICEP2, вызван рассеянием реликтового излучения на частицах галактической пыли, (\cite{Adam:2014bub}). До сих пор первый тип В-моды не обнаружен.

Сложность обнаружения первого типа В-моды связана с маленькой величиной амплитуды В-моды поляризации СМВR, а также с влиянием эффекта двулучепреломления на В-моду, (\cite{Zhao:2014rya}), и с влиянием межгалактической среды (в частности, галактической пыли). Эффект двулучепреломления влияет на векторные и тензорные флуктуации. В результате этого эффекта происходит преобразование В-моды в Е-моду, также происходит смешивание тензорных возмущений, порождающих В-моду и Е-моду, (\cite{Lepora:1998ix}).

В данной работе мы получили ограничения на модельные параметры $\alpha$ и $\Omega_{\rm m}$ для $\phi$CDM Ратра-Пиблс модели скалярного поля применяя BAO/CMBR анализ. В BAO/CMBR анализе мы сравнивали отношения величины сопутствующего углового расстояния в период отсоединения фотонов от барионов к величине масштаба расширения, полученные из наблюдений и из теоретических расчетов. (Более подробное описание BAO/CMBR анализа и результаты его проведения представлены в Главе~VIII.)

 \section{Барионные акустические осцилляции}
До эпохи рекомбинации фотоны, барионы и электроны были тесно взаимосвязаны между собой. В первичной плазме случайным образом могли образовываться области  повышенной плотности материи, состоящие из темной материей и барионов. С одной стороны такие области притягивают к себе другую материю, а с другой стороны как результат взаимодействия барионов и фотонов, создается сильное радиационное давление.  Противоположно направленные гравитационное  и радиационное давления индуцируют совместные колебания барионов и фотонов.
 Эти колебания называются {\it Барионными Акустическими Осцилляциями} (BAO), которые являются звуковыми волнами, и они характеризуются возмущениями, $\delta_{\rm b}$, в барион-фотонной среде.

Радиальное давление приводит к возникновению сферической звуковой волны как барионов, так и фотонов, движущейся наружу от области повышенной  плотности материи. Барион-фотонная среда до рекомбинации почти релятивистская, т. е. величина плотности энергии фотонов, $\rho_{\gamma}$, больше величины плотности энергии барионов, $\rho_{\rm b}$: $\rho_{\rm b}<\rho_{\gamma}$. Давление фотонов, $P_{\gamma}$,  связано с плотностью энергии фотонов, $\rho_{\gamma}$ как: $P_{\gamma}=1/3\rho_{\gamma}$. Величина скорости звуковой волны в первичной плазме определяется как, (\cite{Rubakov:2015yza}):
\begin{equation}\label{eq:bph0}
 v_{\rm s}=\sqrt{\partial P_{\gamma}/\partial \rho_{\gamma}}=\sqrt{1/3}\approx0.58.
\end{equation}
Таким образом, величина скорости звука или скорости звуковой волны ненамного  больше половины  скорости света\footnote{При учитывании величины скорости света, эта формула имеет вид: $v_{\rm s}\approx0.58c$.}.
 Темная материя взаимодействует только гравитационно, и поэтому она остается в центре звуковой волны, являясь первопричиной возникновения области  повышенной плотности материи.

 В конце эпохи первичной рекомбинации водорода происходит отсоединение фотонов от барионов при величине красного смещения, $z_{\rm dec}$. Если до отсоединения барионы и фотоны двигались от центра области повышенной плотности материи вместе, то после отсоединения фотоны перестают взаимодействовать с барионами и рассеиваются. В результате чего радиационное давление на область повышенной плотности материи уменьшается, и формируется область повышенной плотности материи с фиксированным радиусом, который называется {\it звуковым горизонтом}, $r_{\rm s}$.
Сопутствующий размер звукового горизонта, $r_{\rm s}$, в период отсоединения фотонов определяется уравнением\footnote{Физический размер звукового горизонта во время отсоединения равен $a(t_{\rm dec})r_{\rm s}$.}:
\begin{equation}\label{eq:bph}
 r_{\rm s}=\int_0^{t_{\rm dec}} v_s\frac{dt'}{a(t')}.
\end{equation}
Распределение  энергии BАО в пределах звукового горизонта определяется как, (\cite{Rubakov:2015yza}):
\begin{equation}\label{eq:bph1}
\delta_{\rm b}\sim\cos{(k r_{\rm s})}=\cos{\left(\int_0^{t_{\rm dec}}v_{\rm s}\frac{k}{a(t')}dt'\right)},
\end{equation}
здесь $k$ - конформный импульс\footnote{Физический импульс описывается уравнением, $k_{\rm phys}=k/a(t)$.}.

Распределение  энергии BАО за пределами звукового горизонта определяется как, $\delta_{\rm b}=\rm const$, т. е. барионные флуктуации плотности являются вмороженными.
Согласно уравнению, Ур.~(\ref{eq:bph1}), до рекомбинации барион-фотонные флуктуации плотности являются осциллирующей функцией в зависимости от конформного импульса, $k$:
\begin{equation}
\delta\rho_{\rm b}(k)\approx \rho_{\rm b}\delta\rho_{\gamma}(k)\sim\rho_{\rm b}\cos(k r_{\rm s}).
\label{eq:bden}
\end{equation}

\begin{figure}[h!]
\begin{center}
\includegraphics[width=0.6\columnwidth]{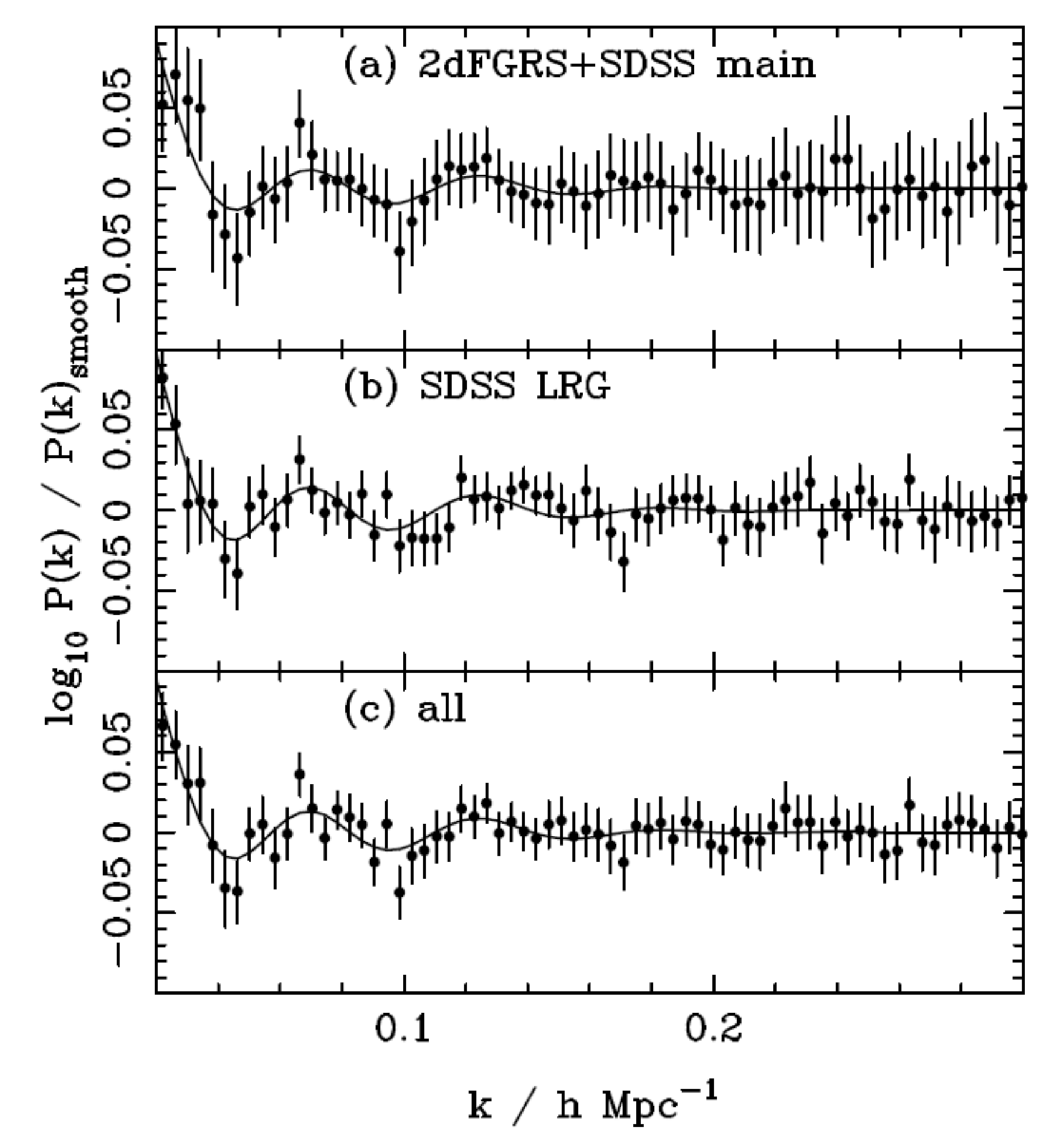}
\end{center}
 \caption {Барионные акустические колебания в спектре мощности материи, обнаруженные: (a) в обзорах галактик 2dFGRS и SDSS , (b) в обзорах галактик SDSS LRG, (c) Объединенные данные из обзоров 2dFGRS,  SDSS и SDSS LRG. Сплошные линии наилучшим образом аппроксимируют  наблюдательные данные, (\cite{Percival:2007yw}).}
 \label{fig:f18}
 \end{figure}
Осцилляции барионной плотности, $\delta\rho_{\rm b}$, сохраняются до настоящего времени. На  Рис.~(\ref{fig:f18}) представлены осцилляции барионной плотности  в спектре мощности материи, $P(k)$, как крошечные  флуктуации.

После рекомбинации барионы остаются  друг от друга на расстоянии равному величине звукового горизонта, $r_{\rm s}$, а темная материя расположена
в центре повышенной плотности. Темная материя и барионная материя притягивают друг друга\footnote{Вследствие доминирования темной материи, гравитационный потенциал, образованный ею, является также доминирующим. Барионная материя следуя за этим гравитационным потенциалом, скатывается в его потенциальную яму.}, что в конечном итоге приводит к формированию галактик во вселенной. Таким образом, галактики разделены друг от друга звуковым горизонтом или BАО сигналом, величина которого увеличивается со временем вследствие расширения Хаббла, (\cite{Rubakov:2015yza}). Теоретические предсказания величины современного сопутствующего звукового горизонта BAO дают следующие результаты, (\cite{Yoo:2012ug}):
\begin{equation}
r_{\rm s} = \int^{\infty}_{t_{\rm dec}}\dfrac{c_{\rm s} dt}{a}= \int^{\infty}_{t_{\rm dec}}\dfrac{c_{\rm s}}{H(z)}dz\sim150~\text{Мпс}\sim100h^{-1}\text{Мпс},
\label{eq:SH}
\end{equation}
предполагая, что $h=0.678$, согласно данным Plank 2015,  (\cite{Ade:2015xua}).

Используя данные о  крупномасштабной структуре галактик, можно измерить шкалу звукового горизонта и сравнить полученные  результаты с теоретическими предсказаниями размера этой шкалы.
\begin{figure}[h!]
\begin{center}
\includegraphics[width=0.6\columnwidth]{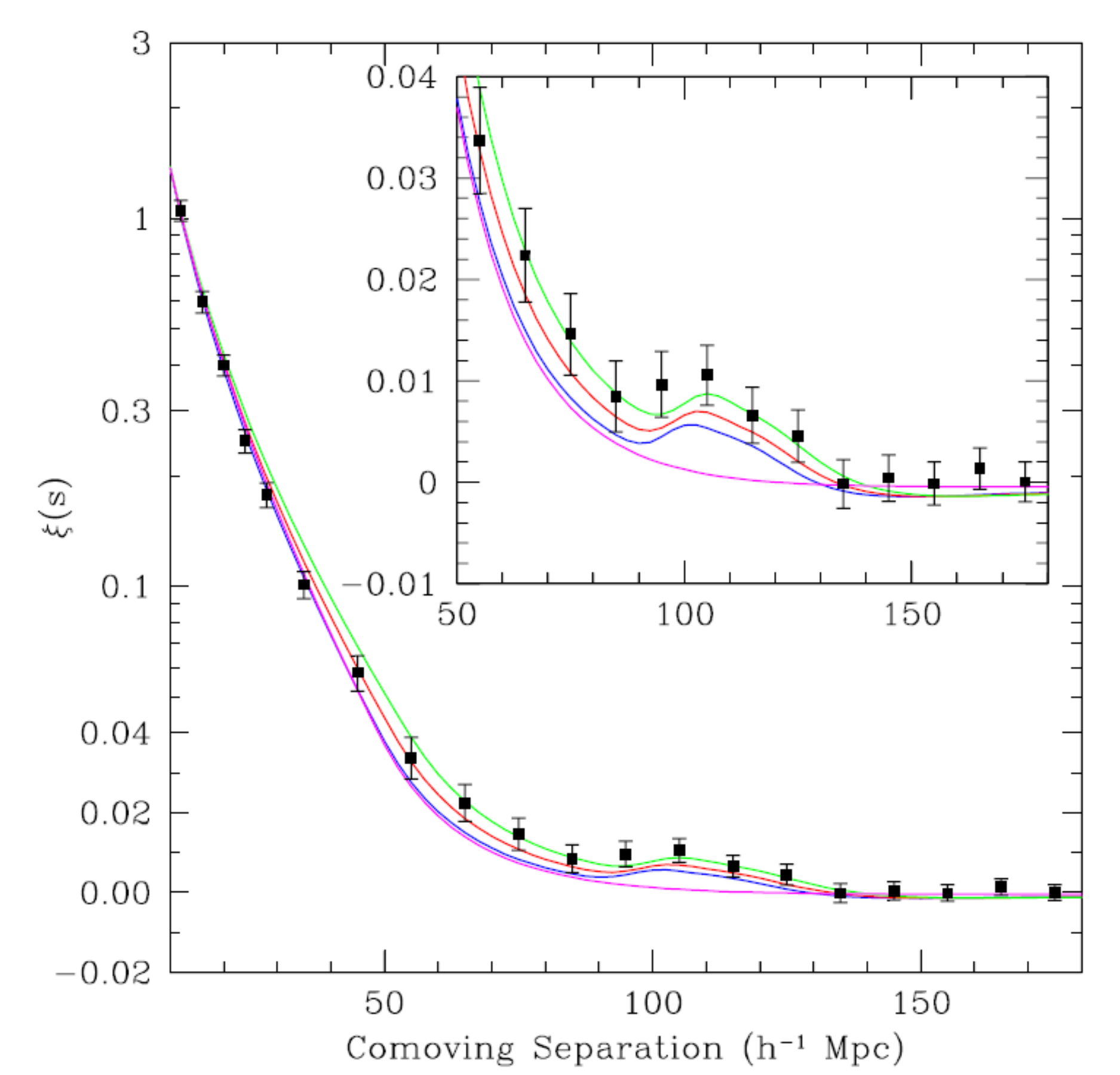}
\end{center}
 \caption {Крупномасштабная двухточечная корреляционная функция, $\xi(s)$, построенная по данных обзора SDSS, (\cite{Eisenstein:2005su}).}
 \label{fig:f19}
 \end{figure}
Двухточечная корреляционная функция, $\xi(s)$, является функцией сопутствующего расстояния галактики, $s$, и описывает вероятность того, что одна галактика будет найдена на заданном расстоянии от другой, (\cite{Rubakov:2015yza}). Каталог Sloan Digital Sky Survey (SDSS) предоставляет распределение
галактик до величины красного смещения, $z=0.47$, (\cite{Eisenstein:2005su}). Эта информация может быть применена для оценки величины BAO сигнала. Двухточечная корреляционная функция
фиксирует BAO сигнал на расстоянии $100h^{-1}$ Mпс, в диапазоне величин красного смещения,  $z\in(0.16;0.47)$, Рис.~(\ref{fig:f19}). Величина BAO сигнала применяется как {\it стандартная линейка} для определения масштаба расстояний в космологии, (\cite{Yoo:2012ug}).

Сравнивая Рис.~(\ref{fig:f14}) и Рис.~(\ref{fig:f19}), можно сделать вывод, что результаты измерения температурного углового спектра мощности анизотропии CMBR
и измерения сигнала BAO показывают, что современный радиус звукового горизонта составляет приблизительно $150$ Мпк. Этот результат совпадает с теоретически рассчитанной величиной сигнала BAO в уравнении, Ур.~(\ref{eq:SH}).

\section{Статистика крупномасштабной структуры во вселенной}
Наблюдаемая в современную эпоху крупномасштабная структура во вселенной, которая включает в себя: галактики, скопления галактик и сверхскопления галактик, сформировалась в результате эволюции первоначальных крошечных флуктуаций плотности материи в расширяющейся вселенной, (\cite{Lifshitz:1946}).
\subsection{Влияние гравитационной неустойчивости  во вселенной на формирование крупномасштабной структуры во вселенной}
Температурные  флуктуации реликтового излучения, зарегистрированные  спутником COBE, вызваны флуктуациями плотности материи, возникшими в ранней вселенной, (\cite{Kosowsky:2001ue}). Причиной возникновения флуктуаций плотности материи могли послужить квантовые флуктуации скалярного поля или топологические дефекты, возникшие в результате фазовых переходов во время инфляции, (\cite{Kamionkowski:1997av}). Теория, описывающая образование и рост этих флуктуаций, основана на нестабильности Джинса или гравитационной неустойчивости флуктуаций плотности материи, (\cite{Jeans:1902}). Флуктуации плотности материи, являясь источником дополнительного гравитационного поля, притягивают к себе окружающее вещество. В результате этого процесса происходит увеличение размеров этих флуктуаций, т. к. сила радиационного давления, действующая на эти флуктуации, преобладает над силой гравитации. Рост флуктуаций плотности материи продолжается до достижения равновесия между силой гравитации и силой радиационного давления. Это равновесие происходит при критическом размере флуктуаций, или при так называемой {\it длине волны Джинса}, $\lambda_J$. Величина длины волны Джинса определяется скоростью звуковой волны, $v_s$, и величиной средней плотности среды, $\langle\rho\rangle$, в которой развиваются флуктуации плотности материи, (\cite{Gorbunov:2011zzc}):
\begin{equation}
\lambda_J=v_s \sqrt{\dfrac{\pi}{G \langle\rho\rangle}}.
\label{eq:Jeans}
\end{equation}
  После того как размер флуктуаций плотности материи  достигает величины длины волны Джинса, сила гравитации преобладает над силой радиационного давления. При этом, процесс увеличения размеров флуктуаций плотности материи сменяется процессом адиабатического сжатия. В результате чего происходит релаксация или коллапс флуктуаций плотности материи. Частицы стремятся к общему гравитационному центру, в конце концов большинство частиц концентрируются в центре и образуется новый объект - протогалактика. Возникновение протогалактик во вселенной происходит при красном смещении, $z\approx10$.

 В результате гравитационного коллапса первичных флуктуаций плотности материи, последующего за ним образования протогалактик и их эволюции на стадии доминирования материи во вселенной, сформировалась наблюдаемая сегодня крупномасштабная структура во вселенной.

\subsection{Теория линейных возмущений}
\subsubsection{Относительный контраст плотности}
Величину флуктуаций плотности материи во вселенной определяют через относительный контраст плотности материи, $\delta$:
\begin{equation}
\delta=\dfrac{\delta\rho(\vec{r},t)}{\langle \rho \rangle}=\dfrac{\rho(\vec{r},t)-\langle \rho\rangle}{\langle \rho \rangle},
\label{eq:DensContr}
\end{equation}
где $\rho(\vec{r},t)$ - величина плотности во вселенной в направлении, $\vec{r}$, и в момент времени, $t$.

Из уравнения, Ур.~(\ref{eq:DensContr}), следует, что $\delta\geq-1$ так как $\rho>0$. Маленькая величина анизотропии температуры микроволнового фона, $\delta T/T_0 = 1/3\delta \sim 10^{-5}$, предполагает, что $|\delta|\ll1$ в момент, $z_{\rm dec}$.  Протогалактики характеризуются большим контрастом плотности c $\delta >1$.

Динамику расширения Хаббла вселенной определяет гравитационное поле, образованное средней плотностью материи, $\langle \rho \rangle$. Флуктуации плотности материи от средней величины, $\delta\rho(\vec{r},t)=\rho(\vec{r},t)-\langle \rho \rangle$, порождают дополнительное гравитационное поле во вселенной.

\subsubsection{Линейное уравнение флуктуаций плотности материи}
Рассмотрим рост флуктуаций плотности материи на пространственных масштабах, значительно меньших величины радиуса Хаббла\footnote{На этих масштабах рост структур во вселенной описывается теорией гравитации Ньютона. Рассматривая
рост флуктуаций плотности материи на пространственных масштабах, сопоставимых или
больше чем радиус Хаббла, необходимо учитывать влияние кривизны пространства-времени и, следовательно, необходимо применять
ОТО.}.  Для простоты предположим, что материя во вселенной является пылевидной. Применим аппроксимацию пылевидной жидкости, которая характеризуется: плотностью материи, $\rho(\vec{r},t)$, трехмерной скоростью, $v(\vec{r},t)$, и нулевым давлением, $p$.\\
Поведение пылевидной жидкости описывается:
\begin{enumerate}[1.]
\item
{\it Уравнением непрерывности}, представленным ранее, Ур.~(\ref{eq:EC}).
\item
{\it Уравнением Эйлера}\footnote{Уравнение Эйлера выражает закон сохранения импульса. Это уравнение также описывает поведение пылевидной материи при действии на него сил, которые представлены через градиент давления, $\nabla p$, и градиент гравитационного потенциала Ньютона, $\nabla\Phi$.}:
\begin{eqnarray}
\frac{\partial\vec{v}}{\partial
    t}+(\vec{v}\cdot\nabla)\vec{v}+
\nabla\Phi+\dfrac{\nabla p}{\rho}=0, \label{eq:enpert}
\end{eqnarray}
где  $\Phi$ - гравитационный потенциал Ньютона, соответствующий уравнению Пуассона.
\item
{\it Уравнением Пуассона}\footnote{Уравнение Пуассона задается как $0$-$0$ компонента уравнений Эйнштейна, Ур.~(\ref{eq:EQ0}). Поскольку  только пылевидная материя рассматривается для исследования роста флуктуаций плотности материи, в уравнении, Ур.~(\ref{eq:Poison}), давление равно нулю, $p=0$.}:
\begin{equation}
\nabla^2\Phi=4\pi G\left(\rho+3p\right).
\label{eq:Poison}
\end{equation}
\end{enumerate}

Решая систему уравнений: уравнение непрерывности, Ур.~(\ref{eq:EC}), уравнение Эйлера, Ур.~(\ref{eq:enpert}), и уравнение Пуассона, Ур.~(\ref{eq:Poison}), а затем  линеризируя найденное решение, при $|\delta|\ll1$, можно получить линейное уравнение флуктуаций плотности материи, (\cite{pwb10}):
\begin{equation}
\delta^{''}+\Bigl(\frac{3}{a}+\frac{E^{'}}{E}\Bigr)\delta^{'}-\frac{3\Omega_{\rm {m0}}}{2a^{5}E^{2}}\delta=0,
\label{eq:deltaeq}
\end{equation}
здесь штрих обозначает производную по масштабному фактору, $^\prime = d/da$.

Линейное уравнение флуктуаций плотности материи, Ур.~(\ref{eq:deltaeq}), полностью описывает эволюцию флуктуаций плотности материи во вселенной.

\subsubsection{Функция темпа роста флуктуаций плотности материи}
Эволюция флуктуаций плотности материи выражается через линейный фактор роста флуктуаций плотности материи, $D(a)$, который обычно нормируется произвольным способом. Мы выбрали нормировку, в которой величина линейного фактора роста флуктуаций плотности материи равно единице в современную эпоху, т. е. $D(a_{\rm 0})=1$.
Таким образом, линейный фактор роста флуктуаций плотности материи определяется как:
\begin{equation}
D(a)=\delta(a)/\delta(a_0),
\end{equation}
где $\delta(a_0)$ - величина контраста плотности материи в современную эпоху. Соотношение $D(a)=a$ для $a\ll1$ выполняется для стадии доминирования материи.

Фракционная плотность материи определяется как:
\begin{equation}
\Omega_{\rm m}(a)=\Omega_{\rm m0}a^{-3}/E^2(a).
\label{eq:FM}
\end{equation}
Скорость роста флуктуаций плотности материи описывается логарифмической производной от линейного фактора роста или функцией темпа роста флуктуаций плотности материи, (\cite{wang1998}):
\begin{equation}
f(a) ={d \ln D(a)}/{d \ln a}.
\label{eq:GR}
\end{equation}
Функция темпа роста флуктуаций плотности материи, $f(a)$, зависит от
фракционной плотности материи, $\Omega_{\rm m}(a)$, и эта
зависимость может быть параметризована степенным законом, (\cite{wang1998}):
\begin{equation}
f(a) \approx (\Omega_{\rm m}(a))^{\gamma(a)},
 \label{eq:f1f2}
\end{equation}
где $\gamma(a)$ - {\it эффективный индекс роста}, который является функцией зависящей от времени. Величина эффективного индекса роста зависит как от модели темной энергии, так и от теории гравитации. Зависимость эффективного индекса роста, $\gamma(a)$, от масштабного фактора может быть определена по формуле, представленной в уравнении, Ур.~(\ref{eq:f1f2}), (\cite{Wu:2009zy}):
\begin{equation}
\gamma(a)=\dfrac{\mathrm{ln}f(a)}{\mathrm{ln}(\Omega_{\rm m}(a))}.
\label{eq:effgamma}
\end{equation}

\subsection{Линдер $\gamma$-параметризация}
В предположении верности ОТО, эффективный индекса роста, $\gamma(a)$, может быть параметризован независимым способом, так называемой Линдер $\gamma$-параметризацией, (\cite{Linder:2007hg}):
\begin{eqnarray}
\gamma=\left\{\begin{array}{rl}
    0.55+0.05(1+w_0+0.5w_a),& \mbox {if } w_0 \ge-1;\\
    0.55+0.02(1+w_0+0.5w_a),& \mbox {if } w_0<-1,
        \end{array}\right.\
\label{eq:Lgamma}
\end{eqnarray}
где $w_0=w(z=0)$ и $w_a=(dw/dz)|_{z=1}$.

Мы определили, что эта параметризация точна вплоть до величины красного смещения, $z=5$ ($a=0.2$), Рис.~(\ref{fig:f28})~(правая панель).
Величина Линдер $\gamma$-параметризации зависит от характеристик модели темной энергии, например, от параметра уравнения состояния $w$.

В $\Lambda$CDM модели, основанной на ОТО, величина Линдер $\gamma$-параметризации, $\gamma\approx0.55$, (\cite{Linder:2007hg}). В космологической модели, которая базируется на другой теории гравитации чем ОТО, величина $\gamma$ отличается от величины $\gamma$ в модели, которая базируется на ОТО. Например, для модели Двали-Габададзе-Поратти, $\gamma\approx0.68$, (\cite{Dvali:2000hr,linder2005,Linder:2007hg}).
Величина Линдер $\gamma$-параметризации, полученная из наблюдательных данных в сочетании с ограничениями на другие космологические параметры, может быть использована для проверки достоверности ОТО на космологических масштабах, (\cite{Pouri2014}, \cite{Taddei2014}).

%%%%%%%%%%%%%%%%%%%%%%%%%%%%%%%%%%%%%%%%%%%%%%%%%%%%%%%%%%%%%%%%%%%%%%%%%%
%%%%%%%%%%%%%%%%%%%       Beginng of chapter 5    %%%%%%%%%%%%%%%%%%%%%%%
%%%%%%%%%%%%%%%%%%%%%%%%%%%%%%%%%%%%%%%%%%%%%%%%%%%%%%%%%%%%%%%%%%%%%%%%%%
\chapter{Элементы статистического анализа}\label{chapter:5}
\section{Распределение Гаусса}
\subsection{Определение распределения Гаусса}
Нормальное распределение или распределение Гаусса некоторой случайной величины, $x$, описывается плотностью вероятности:
\begin{equation}
f(x)=\dfrac{1}{\sigma\sqrt{2\pi}}e^{-(x-e)/2\sigma^2}.
\label{eq:Gauss}
\end{equation}
Распределение Гаусса определено параметрами: $e$ и $\sigma$. Параметр $e$ - это математическое ожидание, а параметр $\sigma$ - среднеквадратичное отклонение случайной величины, $x$. Величина $\sigma^2$ является  дисперсией случайной величины, $x$. Величины $1\sigma$, $2\sigma$ и $3\sigma$ определяют вероятности реализации события или уровни достоверности в: $68.27\%$, $95.45\%$, $99.73\%$, соответственно.
\subsection{Функция $\chi^2$ и функция вероятности}
\subsubsection{Функция $\chi^2$ и функция вероятности для независимых друг от друга измерений}
Предположим, что модельные параметры, ${\bf p}$, распределены согласно распределению Гаусса.
Для определения величин модельных параметров, ${\bf p}$, было произведено  $N$ независимых измерений, $ X^{\rm obs}(z_i)$, где $z_i$ величина красного смещения для каждого измерения. Известно среднеквадратичное отклонение для каждого измерения, $\sigma_i$.
Теоретическая модель предсказывает соответственные величины, $ X^{\rm th}({\bf p},z_i)$.

Функция $\chi^2({\bf p})$ является функцией модельных параметров, ${\bf p}$:
\begin{equation}
\chi^2({\bf p})=\sum_{i=1}^N\dfrac{[X^{\rm obs}(z_i)-X^{\rm th}({\bf p},z_i)]^2}{\sigma_i^2}.
\label{eq:HS0}
\end{equation}
Функция $\chi^2({\bf p})$ определяет отклонение теоретических предсказаний от наблюдательных величин  параметров, ${\bf p}$. Маленькая величина функции $\chi^2({\bf p})$ означает хорошую аппроксимацию выбранной теории наблюдательных данных, и, соответственно, большая величина функции $\chi^2({\bf p})$ означает плохую аппроксимацию данной теорией наблюдательных данных.

Функция вероятности, $\mathcal{L}({\bf p})$, для независимых переменных:
\begin{equation}
\mathcal{L}({\bf p})=\exp\Bigl\{-\dfrac{1}{2}\chi^2({\bf p})\Bigr\}.
\label{eq:HS}
\end{equation}
Функция вероятности, $\mathcal{L}({\bf p})$, определяет вероятность того, что теоретические предсказания величин параметров, ${\bf p}$, совпадают с наблюдательными величинами этих параметров. Большая величина функции вероятности, $\mathcal{L}({\bf p})$, означает хорошую аппроксимацию данной теорией наблюдательных данных, при этом теоретические величины параметров, ${\bf p}$, являются наиболее подходящими величинами\footnote{Необходимо провести различие между понятиями {\bf наиболее подходящими величинами} (best fit values) параметров, ${\bf p}$, и {\bf истинными величинами} (true values) параметров, ${\bf p}$. Функция вероятности, $\mathcal{L}({\bf p})$, показывает с какой вероятностью  произвольная величина параметров, ${\bf p}$, будет истинной величиной (величина которой нам не известна). Тогда наиболее подходящая величина - это такая величина параметров ${\bf p}$, которая, скорее всего, будет истинной величиной.}.
 И наоборот, маленькая величина функции вероятности, $\mathcal{L}({\bf p})$, означает плохую аппроксимацию данной теорией наблюдательных данных.

В случае совмещения $M$ типов независимых переменных, $ p_1, p_2,...,p_{\rm M}$, результирующая  величина функции $\chi^2({\bf p})$ вычисляется как сумма функций, $\chi^2( p_1)... \chi^2( p_{\rm M})$, характеризующих каждый тип независимых переменных:
\begin{equation}
\chi^2({\bf p})=\chi^2( p_1)+...+\chi^2( p_{\rm M-1})+\chi^2( p_{\rm M}).
\label{eq:HSM}
\end{equation}
 В этом случае результирующая функция вероятности вычисляется как произведение функций вероятностей, $\mathcal{L}( p_1), \mathcal{L}( p_2),... ,\mathcal{L}( p_{\rm M})$, определенных для каждого типа независимых переменных:
\begin{equation}
\mathcal{L}({\bf p})=\mathcal{L}({ p_1})\cdot \mathcal{L}({ p_2})... \mathcal{L}({ p_{M-1}})\cdot\mathcal{L}({p_{\rm M}}).
\label{eq:HSM1}
\end{equation}

\subsubsection{Функция $\chi^2$ и функция вероятности для зависимых друг от друга измерений}
В случае зависимых друг от друга измерений, функция $\chi^2({\bf p})$ определяется как:
\begin{equation}
\chi^2({\bf p})=[{\bf X^{\rm obs}}(z_i)-{\bf X^{th}(p},z_i)]^TC^{-1}[{\bf X^{\rm obs}}(z_i)-{\bf X^{th}(p},z_i)],
\label{eq:HS2}
\end{equation}
где $C={\rm cov}[X_i,X_j]$ - ковариационная матрица зависимых измерений; ${\bf X^{\rm obs}}(z_i)$ - вектор, составленный из величин зависимых друг от друга измерений; ${\bf X^{\rm th}(p},z_i)$ - вектор теоретически предсказанных величин; верхний индекс $T$ означает операцию транспонирования вектора.

Функция вероятности для зависимых друг от друга измерений имеет вид:
\begin{equation}
\mathcal{L}({\bf p})=\exp\Bigl\{-\dfrac{1}{2}\Bigl[{\bf X^{\rm obs}}(z_i)-{\bf X^{th}(p},z_i)\Bigr]^TC^{-1}\Bigl[{\bf X^{\rm obs}}(z_i)-{\bf X^{th}(p},z_i)\Bigr]\Bigr\}.
\label{eq:LF1}
\end{equation}

\subsection{Формализм Фишера}
{\bf Обратной матрицей Фишера}, $[F^{-1}]$, является матрица, обратная ковариационной  матрице, $[C]$:
\begin{equation}  %
  \label{eq:Finv}
  \left [ F \right ]^{-1} =
  \left [ C \right ] =
  \left[
    \begin{array}{cc}
     \sigma^2_{p_1} & \sigma_{p_1p_2}\vspace{0.07in}\\
      \sigma_{p_1p_2} & \sigma^2_{p_2}
    \end{array}
  \right],
\end{equation}
где среднеквадратичные отклонения, $\sigma^2_{p_1}$ и $\sigma^2_{p_2}$, являются $1\sigma$ неопределенностями параметров $p_1$ и $p_2$;  $\sigma_{p_1p_2}=\varrho\sigma_{p_1}\sigma_{p_2}$; $\varrho$ - коэффициент корреляции. Абсолютная величина коэффициента корреляции, $\varrho$, не превышает единицы: $\mid\varrho\mid\leq1$. Если $\varrho=0$, то параметры, $p_1$ и $p_2$, независимы друг от друга, т. е. являются некоррелированными между собой. Если $\mid\varrho\mid=1$, то параметры являются полностью коррелированными между собой. Если $\mid\varrho\mid<1$, то параметры являются частично коррелированными между собой.

Рассмотрим функцию $\chi^2(p_1, p_2)$, зависящую от двух параметров, $p_1$ и $p_2$.
 Элементы матрицы Фишера являются коэффициентами разложения второго порядка в ряд Тейлора функции $\chi^2(p_1, p_2)$ вблизи её минимума.

 Двумерная матрица Фишера, $[F]$, вычисляется как:
\begin{equation}  %
  \label{eq:Fish}
  \left [ F \right ] =
  \frac{1}{2}
  \left[
    \begin{array}{cc}
      \frac{\partial^2}{\partial p_1^2}& \frac{\partial^2}{\partial p_1 \partial p_2} \vspace{0.07in}\\
       \frac{\partial^2}{\partial p_1 \partial p_2}& \frac{\partial^2}{\partial p_2^2}
    \end{array}
  \right]
  \chi^2.
\end{equation}
Другими словами, элементы матрицы Фишера, $[F]$, вычисляются как вторые производные от функции $\chi^2$ по параметрам $p_1$ и $p_2$:
\begin{equation}
F_{p_1 p_2}= \frac{1}{2} \frac{\partial \chi^2 }{\partial p_1 \partial p_2}.
\label{eq:Fish1}
\end{equation}
Ковариационная матрица, $[C]$, определена через матрицу Фишера следующим образом: $[C]= [F]^{-1}$.

\subsubsection{Трансформация переменных}
{\bf Постановка задачи}: матрица Фишера, $[F]$, определена через переменные\footnote{Количество переменных может быть сколь угодно большим, ${\bf p}=p_1, p_2...p_{\rm N}$. В данном случае мы ограничились числом переменных $N=3$.}, ${\bf p}=p_1, p_2, p_3$. Эти переменные зависят от других переменных, ${\bf p'}=p'_1, p'_2, p'_3$. Необходимо вычислить матрицу Фишера, $[F']$, относительно переменных, ${\bf p'}=p'_1, p'_2, p'_3$, основываясь на информации об исходной матрице Фишера, $[F]$.

Элементы матрицы Фишера, $[F'_{\rm mn}]$, вычисляются по законам дифференцирования сложной функции:
\begin{equation}
F'_{mn}= \sum_{ij}\frac{\partial p_i}{\partial p'_m}\frac{\partial p_j}{\partial p'_n }F_{ij}.
\label{eq:Fish2}
\end{equation}
\\
Матрица Фишера, $[F']$, может быть получена как, (\cite{Coe:2009xf}):
\begin{equation}
[F^\prime] = [M]^T [F] [M].
\label{eq:Fish3}
\end{equation}
Матрица, $[M]$, определена как:
\begin{equation}
 \left [ M \right ] =
  \left[
    \begin{array}{ccc}
       \frac{\partial p_1}{\partial p'_1} & \frac{\partial p_1}{\partial p'_2} & \frac{\partial p_1}{\partial p'_3} \vspace{0.1in}\\
      \frac{\partial p_2}{\partial p'_1} & \frac{\partial p_2}{\partial p'_2} & \frac{\partial p_2}{\partial p'_3} \vspace{0.1in}\\
      \frac{\partial p_3}{\partial p'_1} & \frac{\partial p_3}{\partial p'_2} & \frac{\partial p_3}{\partial p'_3}
    \end{array}
  \right].
   \label{eq:Finv}
\end{equation}
Таким образом, элементы матрицы, $[M]$, вычисляются как: $M_{ij}=\partial p_i/\partial  p'_j$.

\subsection{Оптимальные модельные параметры}

Независимо от типа наблюдений, модельные параметры, ${\bf p_0}$, при которых функция $\chi^2({\bf p})$ принимает минимальную величину, называются {\bf оптимальными модельными параметрами} для данной теории. Минимальная величина функции $\chi^2({\bf p_0})$ определяет наименьшее квадратичное отклонение для оптимальных модельных параметров в данной теории. Для модели с двумя параметрами, границы интервалов доверия в
$1\sigma$, $2\sigma$ и $3\sigma$ определяюся, соответственно, как: $\chi^2({\bf p})=\chi^2({\bf p_0})+2.3$, $\chi^2({\bf p})=\chi^2({\bf p_0})+6.17$ и $\chi^2({\bf p})=\chi^2({\bf p_0})+11.8$.

Функция вероятности, $\mathcal{L}({\bf p})$, имеет максимальную величину при оптимальных модельных параметрах, ${\bf p_0}$. Величины параметров, ${\bf p_0}$, при которых функция вероятности максимальна, имеют максимальную вероятность быть истинными параметрами.

\section{Метод Монте-Карло для цепей Маркова}
Метод Монте-Карло для цепей Маркова (MCMC) применяется при построении векторов для многомерных функций распределения. В статистике этот метод используется для изучения апостериорных распределений параметров модели.

\subsection{Определение цепей Маркова. Переходные вероятности.}
В 1907 году А. А. Марков разработал новый тип случайных процессов.
 В этом процессе результат данного эксперимента влияет на результат последующего эксперимента. Такой тип процессов называется цепью Маркова.

Цепь Маркова можно описать следующим образом.
Рассмотрим набор состояний, $S = {s_1, s_2, ..., s_r}$.
Процесс начинается в одном из этих состояний и последовательно перемещается из одного состояния в
другое. Каждое перемещение называется шагом. Если цепь в настоящее время находится в состоянии $s_i$, то она переходит в состояние $s_j$ на следующем шаге с вероятностью, обозначенной $p_ {ij}$, и эта
вероятность не зависит от того, в каких состояниях цепь находилась до текущего состояния.
Вероятности $p_ {ij}$ называются {\bf вероятностями перехода}.
Начальная вероятность
распределения, $S$, определяет начальное состояние\footnote{Часто цепи Маркова сравнивают с
прыгающей лягушкой в пруду с одной лилии на другую. Лягушка начинает прыгать с одной из лилий, а затем
прыгает от лилии к лилии с соответствующими вероятностями перехода, $p_ {ij}$, (\cite{Howard:1971}).}.

\subsubsection{Матрица перехода. Однородная цепь Маркова.}
В обозначении $p_ {ij}$ первый индекс указывает номер предыдущего состояния $i$, а второй индекс указывает номер последующего состояния $j$. Процесс, при котором цепь остается в одном и том же состоянии, происходит с вероятностью, $p_ {ii}$.

Допустим, что число состояний конечное и равно $k$.

{\bf Матрица перехода} системы представляет собой матрицу, содержащую все вероятности перехода этой системы, (\cite{Gmurman:2003}):
\begin{equation}
\label{eq:transMarix}
P_{1}=\left(\begin{array}{cccc}
           p_{11} & p_{12} & ...& p_{1k} \\
          p_{21} & p_{22} & ... & p_{2k}\\
           ... & ... & ... & ... \\
          p_{k1} & p_{k2}& ... & p_{kk} \\
           \end{array}
           \right).
\end{equation}
Так как вероятности перехода событий из состояния $i$ в состояние $j$, помещенные в каждую строку матрицы,  образуют полную группу, то сумма вероятностей этих событий равна единице. Другими словами, сумма вероятностей перехода для каждой строки этой матрицы равна единице:
\begin{equation}
\sum_{j=1}^k{p_{ij}}=1,~(i=1,2...,k).
\label{eq:TrM}
\end{equation}
Цепь Маркова называется {\bf однородной}, если условная вероятность, $p_{ij}$, не зависит от номера испытания.

\subsubsection{Равенство Маркова}
Обозначим через $P_{ij}(n)$ вероятность того, что система, $S$, перейдет из состояния $i$ в состояние $j$ в результате $n$ шагов (или испытаний). Например, $P_{25}(10)$ - вероятность перехода от второго состояния к пятому состоянию в результате 10 шагов. Подчеркнем, что при $n = 1$ мы получим вероятности перехода, $ p_ {ij} $:
\begin{equation}
P_{ij}(1)=p_{ij}.
\end{equation}
{\bf Задача Маркова}: зная вероятности перехода, $p_ {ij}$, найти вероятности $P_ {ij}$ перехода системы из состояния $i$ в состояние $j$ за $n$ шагов.

Введем  между состояниями $i$ и $j$ промежуточное состояние $r$. Другими словами, мы предполагаем, что система будет переходить из начального состояния $i$ в промежуточное состояние $r$ с вероятностью $P_ {ir}(m)$ за $m$ шагов. После этого система перейдет из промежуточного состояния $r$ в конечное состояние $j$ с вероятностью $P_{rj}(n-m)$ за $(n-m)$ шагов.

Вероятность, $P_{ij}$, перехода системы из состояния $i$ в состояние $j$ за $n$ шагов можно найти с помощью равенства Маркова:
\begin{equation}
P_{ij}(n)=\sum_{r=1}^k{P_{ir}(m)P_{rj}(n-m)}.
\end{equation}

В наших расчетах применяется нормальное распределение случайной величины $x$, которое описывается уравнением, Ур.~(\ref{eq:Gauss}).

\subsection{Метод Монте-Карло}
В 1949 году Н. Метрополис и С. Улам опубликовали статью 'Метод Монте-Карло', в которой этот метод был представлен. Метод Монте-Карло является статистическим методом изучения проблем, основывающимся на использовании случайных чисел, аналогичных числам, генерируемых в азартных играх. При применении метода Монте-Карло требуется найти набор случайных чисел, соответствующих определенному распределению вероятности.

\subsubsection{Сущность метода Монте-Карло}
Требуется найти величину математического ожидания, $e$, некоторой случайной величины. Для этой цели выбирается такая случайная переменная, $X$, математическое ожидание которой равно $e$:
\begin{equation}
 M(X)=e.
\label{eq:MMC}
\end{equation}
В реальности производятся $n$ испытаний, в результате чего получают $n$ возможных величин $X = {x_1, x_2, ..., x_k}$, после этого вычисляют их среднеарифметическую величину, $\bar{x}$:
\begin{equation}
\bar{x}=\dfrac{\sum_{i=1}^k x_i}{n}.
\label{eq:AM}
\end{equation}
Величина $\bar{x}$ рассматривается как приблизительная величина $e^*$ числа $e$:
\begin{equation}
e\simeq e^*=\bar{x}
\label{eq:AM1}
\end{equation}
Поскольку метод Монте-Карло требует большого количества тестов, его часто называют методом {\bf статистических тестов}. Для применения метода Монте-Карло необходим достоверный набор случайных чисел. Такие числа трудно получить, поэтому обычно используются псевдослучайные числа. Они должны быть некоррелированными числами и равномерно распределенными по предварительно определенному диапазону чисел.

\subsubsection{Метод преобразования}
Метод преобразования используется для поиска случайных (или псевдослучайных) чисел из известных распределений вероятностей.
Требуется найти (или воспроизвести) непрерывную случайную переменную, $X$, т. е. получить последовательность ее возможных величин, $X = {x_1, x_2, ..., x_k}$, которая характеризуется функцией распределения, $F(x)$.

{\bf Теорема}: рассмотрим возможную случайное величину, $x_i$, с функцией распределения, $F(x)$. Величина случайного числа, $r_i$, соответствует величине $x_i$, если оно является решением следующего уравнения:
\begin{equation}
F(x_i)=r_i.
\label{eq:ther1}
\end{equation}
Другими словами, чтобы найти возможную величину, $x_i$, непрерывной случайной переменной, $X$, определяемой плотностью распределения, $f(x)$, мы должны выбрать случайное число,
$r_i$ и решить одно из уравнений относительно $x_i$:
\begin{equation}
\int_{-\infty}^{x_i}f(x)dx=r_i~~\text{или}~~\int_b^{x_i}f(x)dx=r_i,
\label{eq:ther2}
\end{equation}
где $b$ является конечной, наименьшей величиной случайной переменной $X$.

%%%%%%%%%%%%%%%%%%%%%%%%%%%%%%%%%%%%%%%%%%%%%%%%%%%%%%%%%%%%%%%%%%%%%%%%%%
%%%%%%%%%%%%%%%%%%%       Beginng of chapter 6    %%%%%%%%%%%%%%%%%%%%%%%
%%%%%%%%%%%%%%%%%%%%%%%%%%%%%%%%%%%%%%%%%%%%%%%%%%%%%%%%%%%%%%%%%%%%%%%%%%
\chapter{Темная энергия}\label{chapter:6}
Как было описано в Главе~I, наша вселенная находится в состоянии ускоренного расширения. Одним из возможных объяснений этого явления, является существование так называемой {\it темной энергии}. Отрицательное эффективное давление темной энергии вызывает ускоренное расширению вселенной. Темная энергия
характеризуется параметром уравнения состояния, ${w}$, определяемым как
соотношение между давлением, $p_{\rm DE}$, и плотностью энергии, $\rho_{\rm DE}$: $w \equiv p_{\rm DE}/\rho_{\rm DE}$. Ускоренное расширение протекает при $w <-1/3$. Темная энергия составляет приблизительно $69\%$ от общей плотности энергии в современной вселенной, (\cite{Ade:2015xua}). Пространственное распределение темной энергии очень однородно и изотропно. Природа темной энергии до сих пор остается неразгаданной для космологов.

\section{Космологическая постоянная $\Lambda$}
Самой простой моделью темной энергии является концепция энергии вакуума или независимая от времени космологическая постоянная, которая впервые была предложена Альбертом Эйнштейном, (\cite{Einstein:1917ce}), (обзоры: \cite{Carroll:2000fy}, \cite{Peebles:2002gy}, \cite{Martin:2012bt}).

С целью получить статическое решение, $\dot{a} = 0$, в 1917 году Альберт Эйнштейн ввел новое слагаемое, $\Lambda g_{\mu\nu}$, в тензор Эйнштейна, Ур.~(\ref{eq:ET}), (\cite{{Einstein:1917ce}}). В результате чего уравнения Эйнштейна, Ур.~(\ref{eq:EQ0}), приняли вид:
\begin{equation}
 R_{\mu\nu}-\dfrac{1}{2}g_{\mu\nu}R - \Lambda g_{\mu\nu}=8\pi G T_{\mu\nu},
 \label{eq:EQ4}
\end{equation}
где $\Lambda$ - космологическая постоянная. При добавлении этого слагаемого нарушается условие перехода сильных гравитационных полей в слабые гравитационные поля (переход к Ньютоновскому пределу), наложенное на тензор Эйнштена в уравнениях, Ур.~(\ref{eq:G00}) и Ур.~(\ref{eq:G001}).
Для соблюдения условия этого перехода, величина космологической постоянной должна быть пренебрежимо малой.

У Эйнштейна не было реальной физической интерпретации постоянной $\Lambda$.
 После того, как в 1929 году Эдвином Хабблом было открыто расширение вселенной, (\cite{Hubble:1929ig}), в 1931 году Эйнштейн удалил космологическую постоянную из своих уравнений, назвав введение постоянной $\Lambda$ в свои уравнения самой большой своей ошибкой (biggest blunder), (\cite{Gamov:1956}). С 1930-х до конца 1990-х годов космологи не учитывали космологическую постоянную, предполагая ее величину равной нулю. После открытия ускоренного расширенния вселенной в 1998 году, (\cite{Riess:1998cb}, \cite{Schmidt:1998ys}, \cite{Perlmutter:1998np}), космологи стали использовать космологическую постоянную с положительной ненулевой величиной для объяснения этого явления. Дело в том, что при учитывании космологической постоянной в уравнениях Фридмана, Eq.~(\ref{eq:FR1}) и  Eq.~(\ref{eq:FR2}), нестатическое решение, описывающее расширение вселенной,  может быть найдено.

 В настоящее время принято, что космологическая постоянная эквивалентна конечной плотности энергии вакуума, (\cite{Zeldovich:1968gd}). Таким образом, если космологическая постоянная определяется плотностью энергии вакуума, $\rho_{\rm vac}$, тогда плотность энергии космологической постоянной, $\rho_\Lambda$, не зависит от времени:
\begin{equation}
\rho_\Lambda=\rho_{\rm vac}=\mathrm{const}.
\end{equation}
Плотность энергии космологической постоянной определена как:
\begin{equation}
\rho_\Lambda=\frac{\Lambda}{8\pi G},
\end{equation}
где $\Lambda=4.33\cdot10^{-66}~{\rm eV}^2$.

Уравнение состояния для космологической постоянной:
 \begin{equation}
 p_\Lambda=-\rho_\Lambda=\mathrm{const}.
 \end{equation}
Таким образом, параметр уравнения состояния для космологической постоянной определяется как:
 \begin{equation}
w_\Lambda=-1.
\end{equation}
Действие для космологической постоянной:
\begin{equation}
\textsl{S}=-\frac{1}{16\pi G} \int d^4x\sqrt{-g}(R+2\Lambda)+\textsl{S}_{\rm M},
\end{equation}
где $\textsl{S}_{\rm M}$ - действие для материи.

Уравнения Фридмана с космологической постоянной имеют вид:
\begin{equation}
\frac{\dot{a}^2}{a^2}=\frac{8\pi G}{3}\rho-\frac{K}{a^2}+\frac{\Lambda}{3}
\end{equation}
и
\begin{equation}
\frac{\ddot{a}}{a}=-\frac{4\pi G}{3}(\rho+3p)+\frac{\Lambda}{3}.
\end{equation}
\section{Космологическая $\Lambda$CDM Модель}
Lambda Cold Dark Matter ($\Lambda$CDM) модель является {\it стандартной} моделью вселенной. Эта модель описывает пространственно-плоскую вселенную и является самой простой параметризацией космологической модели Большого Взрыва.
В $\Lambda$CDM модели темная энергия представлена в виде энергии вакуума или космологической постоянной $\Lambda$. Темной материей в $\Lambda$CDM модели является {\it холодная темная материя}. $\Lambda$CDM модель базируется на ОТО для описания гравитации во вселенной на космологических масштабах.

\begin{figure}[h!]
\begin{center}
\includegraphics[width=\columnwidth]{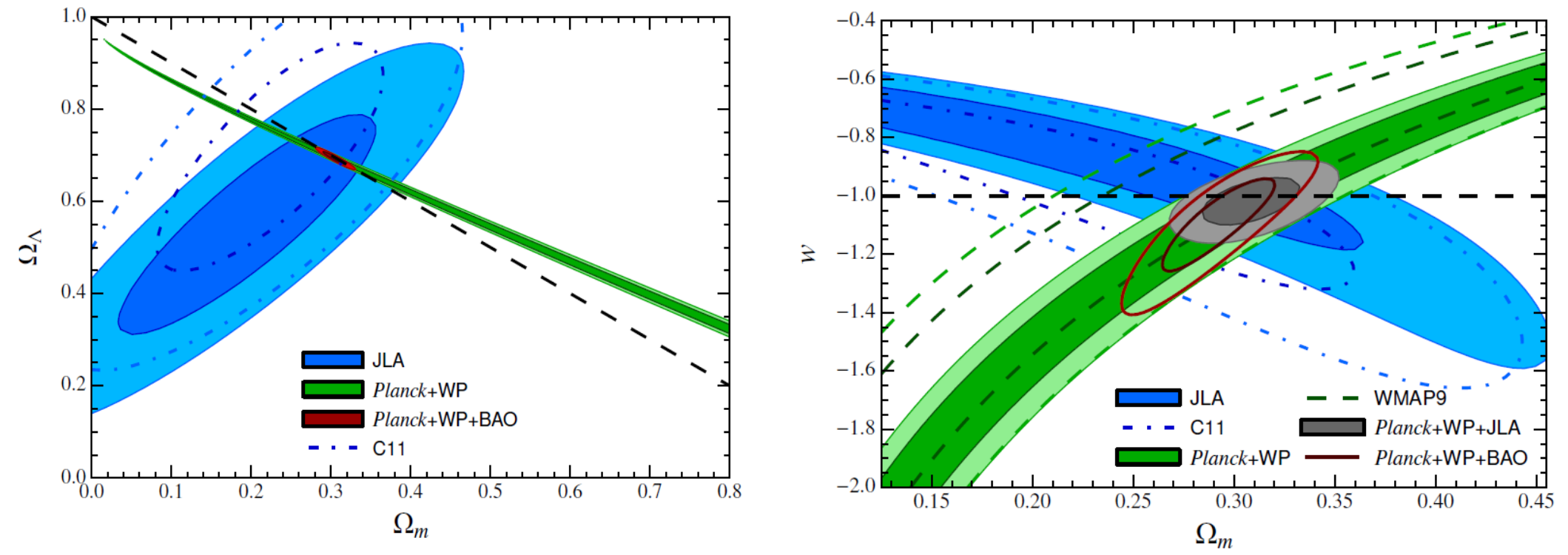}
\end{center}
 \caption {Контуры с $68\%$ и $95\%$ уровнями доверия как результат различных измерений: SNIa (JLA) и SNIa (C11) компиляции, комбинация Planck температуры, WMAP поляризации (Planck + WP) и компиляция  BAO масштаба. Левая панель: для $\Omega_{\rm m}$ и $ \Omega_\Lambda$ космологических параметров для $\Lambda$CDM модели. Черная пунктирная линия соответствует пространственно-плоской вселенной. Правая панель: для космологических параметров $\Omega_{\rm m}$ и $w$ для пространственно-плоской $w-\Lambda$CDM модели. Черная пунктирная линия соответствует $\Lambda$CDM модели, (\cite{Betoule:2014frx}).}
 \label{fig:f20}
\end{figure}

$\Lambda$CDM модель является {\it согласованной} (concordance) моделью вселенной, т. к. эта модель находится в хорошем согласии с доступными на сегодняшний день космологическими наблюдениями, Рис.~(\ref{fig:f20}).
Кроме того, $\Lambda$CDM модель объясняет: ускоренное расширение вселенной, крупномасштабную структуру в распределении галактик, анизотропию микроволнового фона, химический состав вселенной (содержание водорода, гелия и лития\footnote{Процесс образования этих химических элементов во вселенной начался в эпоху {\it первичного нуклеосинтеза} (primordial nucleosynthesis). Эта эпоха началась при температуре порядка 1~МэВ, когда возраст вселенной составлял приблизительно 1 сек. В это время прекращаются реакции: $e^-+p\leftrightarrow n + \nu_e$ и происходит 'вымерзание'  нейтронов из реакций. Приблизительно от 10 сек до 20 мин после Большого Взрыва происходят термоядерные реакции формирования более сложных элементов: $p+n\rightarrow^2H+\gamma,~^2H+p\rightarrow^3He+\gamma,~^3He+^2H\rightarrow^4He+p$, ..., вплоть до $^7Li$, (\cite{Rubakov:2015yza}).}), (\cite{Schneider:2006}).

$\Lambda$CDM модель характеризуется основными шестью {\it независимыми параметрами}: параметром физической барионной плотности, $\Omega_{\rm b}^2$; параметром физической плотности темной материи, $\Omega_{\rm c} h^2$; величиной возраста вселенной, $t_0$; скалярным спектральным индексом, $n_{\rm s}$; амплитудой флуктуаций кривизны, $\Delta^2_{\rm R}$; оптической глубиной в период повторной ионизации\footnote{Повторная ионизация - это процесс ионизации нейтральных атомов водорода, который происходил во вселенной на диапазоне величин красного смещения, $z\in(6;20)$.}, $\tau$. Кроме этих параметров, $\Lambda$CDM модель описывается шестью расширенными {\it фиксированными параметрами}: параметром общей плотности энергии, $\Omega_{\rm tot}$; параметром уравнения состояния, $w$; суммарной массой трех типов нейтрино, $\sum m_{\nu}$; эффективным числом релятивистских степеней свободы, $N_{\rm eff}$; тензор/скаляр отношением, $r$; бегущим скалярным индексом, $d n_{\rm s}/d \ln k$.

Согласно $\Lambda$CDM модели, современная вселенная состоит на $69,2 \%$  из темной энергии;  на $26 \%$ из темной материи; на $4.8\%$ из обычной, барионной материи; на $0.1 \%$ из нейтрино; на $0.01 \%$ из фотонов, в соответствии c наблюдениями космической обсерватории Planck 2015, (\cite{Ade:2015xua}).

Первое уравнение Фридмана, описывающее расширение вселенной для пространственно-плоской $\Lambda$CDM модели, имеет вид:
 \begin{equation}
 E(a)=(\Omega_{\rm r0}a^{-4} + \Omega_{\rm m0}a^{-3} + \Omega_\Lambda)^{1/2}, \label{eq:FLCDM}
 \end{equation}
 где $\Omega_{\rm r0}$, $\Omega_{\rm m0}$ и $\Omega_\Lambda$ параметры плотности энергии: излучения, материи и вакуума, соответственно, в современную эпоху. До момента нерелятивизации (neutrinos non-relativization)\footnote{Переход нейтрино из релятивистского в нерелятивистское состояние происходит в эпоху доминирования материи. Чем раньше происходит этот переход, тем более большую массу приобретает нейтрино.}, нейтрино является релятивистской частицей, поэтому параметр плотности энергии нейтрино, $\Omega_{\nu}$, изменяется в зависимости от скалярного фактора как, $a^{-4}$. Таким образом, до момента нерелятивизации нейтрино полная плотность энергии излучения состоит из плотностей энергий релятивистских частиц: фотонов и нейтринно. После момента нерелятивизации, нейтрино становится нерелятивистской частицей и параметр плотности энергии нейтрино, $\Omega_\nu$, эволюционирует как, $a^{-3}$. Поэтому параметр плотности энергии материи, $\Omega_{\rm m0}$, содержит в себе все нерелятивистские компоненты, включая нерелятивистское нейтрино.

\subsection{Проблемы $\Lambda$CDM модели}
Если в действительности происхождением космологической постоянной является энергия вакуума, то тогда существует проблема, связанная с энергетической шкалой космологической постоянной. Теоретически предсказанная плотность энергии космологической постоянной определяется как:
\begin{equation}
\rho_\Lambda\sim\hbar M_{\rm pl}^4\sim 10^{72}~\text{\rm Gev}^4\sim 2\cdot10^{110}~{\rm erg/cm^3},
\label{eq:ther}
\end{equation}
где $M_{\rm pl}\sim10^{18}$~{\rm ГэВ} - шкала массы Планка; $\hbar$ - приведенная постоянная Планка\footnote{В соответствии с принятым нами соглашением, $\hbar=1$.}.
Результат, полученный в уравнении, Ур.~(\ref{eq:ther}), подтверждается измерениями вакуумных флуктуаций эффектом Казимира, проведенными в лабораторных условиях, (\cite{Casimir:1948dh}).
Однако, космологические наблюдения космологической постоянной как темной энергии показывают абсолютно другой результат:
\begin{equation}
\rho^{\rm obs}_\Lambda\sim10^{-48}~{\rm Gev}^4\sim 2\cdot10^{-10}~{\rm erg/cm^3}.
\label{eq:obs}
\end{equation}
Таким образом, наблюдаемая величина энергетической шкалы космологической постоянной на 120 величин
меньше, чем величина космологической постоянной, полученная из теоретических предсказаний.
Это расхождение в 120 величинах энергетической шкалы
 называется проблемой космологической постоянной или проблемой {\it тонкой настройки}, (\cite{CPT}, \cite{Carroll:2000fy}).
Второй проблемой космологической постоянной является проблема {\it совпадения}. Суть этой проблемы состоит в том, что плотность энергии темной энергии сравнима с плотностью энергии темной материи в настоящий момент времени.
Плотности энергии излучения, материи и темной энергии изменяются в зависимости от скалярного фактора по разным законам, которые представлены в уравнениях, Ур.~(\ref{eq:DWR}), Ур.~(\ref{eq:DWM}), и Ур.~(\ref{eq:DELambda}): для излучения, $\rho_{\rm r}\sim a^{-4}(t)$, для материи, $\rho_{\rm m}\sim a^{-3}(t)$, а для космологической постоянной, $\rho_\Lambda=\rm const$. Точные космологические наблюдения показывают, что соотношение между плотностью материи и плотностью темной энергии в современную эпоху составляет порядка единицы, $\rho_{\rm m}/\rho_\Lambda \simeq 1/3$. Этот факт является загадкой, поскольку стандартная $\Lambda$CDM модель предсказывает, что это отношение должно быть зависящим от времени, $\rho_{\rm m}/\rho_\Lambda \propto a^{-3}(t)$.
\begin{figure}[h!]
\begin{center}
\includegraphics[width=0.6\columnwidth]{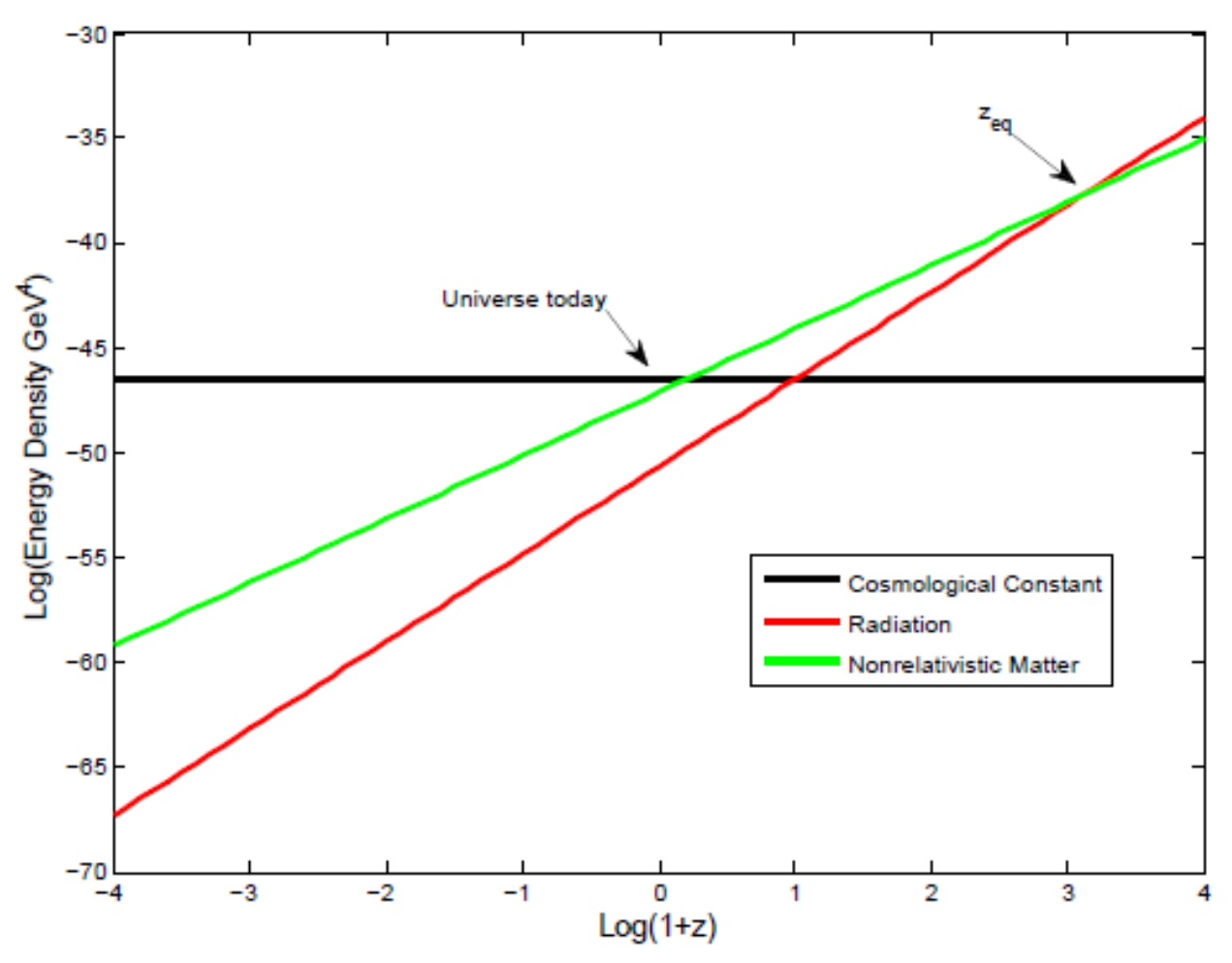}
\end{center}
 \caption {Эволюция плотности энергии излучения, материи и космологической постоянной $\Lambda$, (\cite{Samushia:2009dd}).}
 \label{fig:f21}
 \end{figure}
Поскольку энергия вакуума не изменяется со временем, она была незначительной в периоды доминирования излучения и материи. В то время как энергия вакуума стала доминирующей составляющей только недавно,  при величине скалярного фактора $a\approx0.76$ (или при величине красного смещения $z\approx0.31$), согласно данным Planck 2015, Ref.~(\cite{Ade:2015xua}), и она будет единственной составляющей во вселенной в будущем,  Рис.~(\ref{fig:f21}).
Величины плотности энергии материи и плотности энергии космологической
постоянной сравнимы в течение очень короткого интервала времени, Рис.~(\ref{fig:f21}),  поэтому возникает вопрос: 'Почему так совпало, что мы живем в этом коротком (по космологическим масштабам) временном промежутке?' Ведь этот факт находится в противоречии с принципом Коперника.

Так называемый {\it антропный} принцип, предложенный Стивеном Вайнбергом в 1987 году, (\cite{Weinberg:1987dv}),  может дать объяснение обеим проблемам
$\Lambda$CDM модели и ответить на вопросы: 'Почему плотность энергии космологической постоянной настолько мала?' и 'Почему ускоренное расширение вселенной началось недавно?' Согласно антропному принципу, величина плотности энергии космологической постоянной, наблюдаемая сегодня, $\rho_\Lambda$, должна быть подходящей для эволюции разумных существ во вселенной, (\cite{Barrow:1988yia}).

\section{Модели скалярного поля}
Cуществуют многочисленные модели темной энергии, альтернативные для $\Lambda$CDM модели, (\cite{Copeland2006}, \cite{Yoo:2012ug}). Несмотря на многообразие этих моделей,  $\Lambda$CDM модель до сих пор остается базовой моделью, моделью сравнения с остальными моделями темной энергии.

Главной альтернативой $\Lambda$CDM модели являются модели динамического  скалярного поля\footnote{Скалярное поле - это поле, которое характеризуется  скалярной величиной, определенной в любой точке этого поля. Такое поле инвариантно при трансформациях Лоренца.} или так называемые $\phi$CDM модели, (\cite{Wetterich1987}, \cite{Ratra:1987aj}, \cite{Peebles:2002gy}). В этих моделях темная энергия представлена в виде медленно изменяющегося космологического однородного скалярного поля, $\phi$. Взаимодействующее с самим собой пространственно-однородное скалярное поле, $\phi$, минимально связано с гравитацией на космологических масштабах. В $\phi$CDM моделях не существует проблемы тонкой настройки, свойственной для $\Lambda$CDM  модели. Эти модели имеют более естественное объяснение наблюдаемого низкоэнергетического масштаба темной энергии. Если для $\Lambda$CDM модели  параметр уравнения состояния является постоянной величиной, $w=-1$, то для $\phi$CDM модели величина параметра уравнения состояния является функцией зависящей от времени.  Когда плотность энергии скалярного поля начинает доминировать над плотностью энергии материи, вселенная начинает стадию ускоренного расширения.

 На ранних этапах развития вселенной, т. е. при больших величинах красного смещения, динамическое скалярное поле в $\phi$CDM моделях отличается от поведения $\Lambda$CDM модели. В более позднюю эпоху развития вселенной, т. е. при маленьких величинах красного смещения, динамическое скалярное поле почти не отличимо от поведения космологической постоянной.

Модели $\phi$CDM  подразделяются на два класса: на {\it модели квинтэссенции} (quintessence models), (\cite{Zlatev:1998tr}), и на {\it фантомные модели} (phantom models), (\cite{Caldwell1999}, \cite{Caldwell2003}). Эти модели различаются друг от друга:
\begin{itemize}
\item[$\bullet$]{\it По величине параметра уравнения состояния}\\
Для полей квинтэссенции $-1/3<w_\phi<-1$, для фантомных полей $w_\phi<-1$.
\item[$\bullet$] {\it В знаке кинетической составляющей в Лагранжиане}\\
Позитивный знак для полей квинтэссенции и отрицательный знак для фантомных полей.
\item[$\bullet$] {\it В динамике скалярного поля}\\
Поле квинтэссенции скатывается к минимуму своего потенциала, фантомное поле стремится к максимуму своего потенциала.
\item[$\bullet$] {\it В динамике темной энергии}\\
Для полей квинтэссенции темная энергия почти не изменяется со временем, а для фантомных полей она увеличивается со временем.
\item[$\bullet$] {\it В прогнозе будущего вселенной}\\
В моделях квинтэссенции предсказано либо вечное расширение вселенной, либо повторный коллапс в зависимости от величины параметра кривизны вселенной. Для фантомных моделей предсказано разрушение, разрыв любых гравитационно-связанных структур во вселенной. В зависимости от асимптотики параметра Хаббла, $H(t)$, сценарии будущего вселенной подразделяются: на большой разрыв, для которого  $H(t)\rightarrow\infty$ за конечное время, $t=\rm const$; на маленький разрыв, для которого  $H(t)\rightarrow\infty$ за бесконечное время, $t\rightarrow\infty$, и на псевдо разрыв, для которого  $H(t)\rightarrow \rm const$ за бесконечное время, $t\rightarrow\infty$.
\end{itemize}
Полное действие для скалярного поля определяется как:
\begin{equation}
S=\int{d^{4}x\sqrt{-g}\Bigr[-\frac{M_{\rm pl}^2}{16\pi}R+\mathcal{L}_{\phi}\Bigl]} + S_{\rm M},
\label{eq:S}
\end{equation}
где $\mathcal{L}_{\phi}$ - плотность Лагранжиана скалярного поля, форма которой зависит от выбранного типа модели.
\subsection{Скалярное поле квинтэссенции}
Скалярное поле квинтэссенции описывается плотностью Лагранжиана:
\begin{equation}
\mathcal{L}_{\phi}=\frac{1}{2}g^{\mu\nu}\partial_\mu\phi\partial_\nu\phi-V(\phi).
\label{eq:QLD}
\end{equation}
 Существует много различных потенциалов квинтэссенции, но до сих пор не отдано предпочтения ни одному из них.
\begin{table}[h!]
\begin{tabular}{|p{3cm}|p{8cm}|p{3.5cm}|}
\hline %\hline
\hspace{0.5 cm}\textrm{Название} & \hspace{3cm}\textrm{Форма}& \hspace{1cm}\textrm{Ссылка} \\
\hline\hline
\hline
Ratra-Peebles & $V(\phi)=V_0M_\mathrm{pl}^2\phi^{-\alpha}$; $\alpha ={\rm const} >0$  &  \cite{Ratra:1987aj} \\
\hline
Ferreira-Joyce & $V(\phi)=V_0\exp(-\lambda\phi/M_{\rm pl})$; $\lambda = {\rm const}>0$ & \cite{Ferreira:1997hj}\\
\hline
Zlatev-Wang-Steinhardt & $V(\phi)=V_0(\exp({M_{\rm pl}/\phi})-1)$  & \cite{Zlatev:1998tr}\\
\hline
Sugra & $V(\phi)=V_0\phi^{-\chi}\exp(\gamma\phi^2/M_{\rm pl}^2)$; \par $\chi, \gamma={\rm const}>0$ & \cite{Brax:1999gp}\\
\hline
Sahni-Wang & $V(\phi)=V_0(\cosh(\varsigma\phi)-1)^g$; \par $\varsigma={\rm const}>0$, $g={\rm const}<1/2$ & \cite{Sahni:1999qe} \\
\hline
Barreiro-Copeland-Nunes & $V(\phi)=V_0(\exp(\nu\phi) + \exp(\upsilon\phi))$;\par $\nu$, $\upsilon={\rm const}\geq0$ & \cite{Barreiro:1999zs}\\
\hline
Albrecht-Skordis  & $V(\phi)=V_0((\phi-B)^2 + A)\exp(-\mu\phi)$; \par $A$, $B={\rm const}\geq0$, $\mu={\rm const}>0$ & \cite{Albrecht:1999rm}\\
[0.2cm]
\hline
Ur\~{e}na-L\'{o}pez-Matos & $V(\phi)=V_0\sinh^m(\xi M_\mathrm{pl}\phi)$; \par $\xi={\rm const}>0$, $m={\rm const}<0$ & \cite{UrenaLopez:2000aj}\\
\hline
Inverse exponent potential & $V(\phi)=V_0\exp({M_{\rm pl}/\phi})$ & \cite{Caldwell:2005tm}\\
\hline
Chang-Scherrer & $V(\phi)=V_0(1+\exp(-\tau\phi))$; $\tau={\rm const}>0$ & \cite{Chang:2016aex}\\
[0.2cm]
\hline
\end{tabular}
\caption{\rm Список потенциалов квинтэссенции.}
\label{table:QP}
\end{table}
 Неполный список потенциалов квинтэссенции\footnote{Потенциал Феррейры-Джойса ранее исследовали Лучин и Матаррезе, (\cite{LM1985}), Ратра и Пиблс (\cite{Ratra:1987rm}). Хотя полное подробное описание этой модели было дано Феррейра и Джойс (\cite{Ferreira:1997hj}).} представлен в Таблице\footnote{В Таблице~\ref{table:QP} и  Tаблице~\ref{table:PP}  модельный параметр, $V_0$, имеет размерность $\text{\rm ГэВ}^4$. Этот модельный параметр связан с параметром плотности темной энергии для настоящего момента времени.} \ref{table:QP}.

Тензор энергии-импульса для скалярного поля квинтэссенции, $T_{\mu\nu}$, определяется как:
\begin{equation}
T_{\mu\nu}=2\frac{\partial{\mathcal{L}_{\phi}}}{\partial{g^{\mu\nu}}} - g_{\mu\nu}\partial{\mathcal{L}_{\phi}}.
\label{eq:SET1}
\end{equation}
Подставляя уравнение, Ур.~(\ref{eq:QLD}), в уравнение, Ур.~(\ref{eq:SET1}), мы получим:
\begin{equation}
T_{\mu\nu}=\partial_\mu \phi \partial_\nu\phi - g_{\mu\nu}\left[\frac{1}{2}g^{\alpha\beta}\partial_\alpha \phi \partial_\beta \phi - V(\phi)\right].
\label{eq:SET2}
\end{equation}
Компоненты тензора энергии-импульса скалярного поля квинтэссенции, $T_{\mu\nu}$, определяются как:
\begin{align}
\label{eq:enmomuni5}
T_{00}&\equiv\rho_{\phi}=\frac{1}{2}\dot{\phi}^2 + V(\phi),\\
\label{eq:enmomuni6}
T_{0i}&=0,\\
\label{eq:enmomuni7}
T_{ij}&=0 \ (i\neq j),\\
\label{eq:enmomuni8}
T_{ii}&\equiv p_{\phi}=\frac{1}{2}\dot{\phi}^2 - V(\phi),
\end{align}
где $\rho_{\phi}$ и $p_{\phi}$ - плотность энергии и давление скалярного поля, соответственно, в предположении, что данное скалярное поле описывается моделью идеальной баротропной жидкости.

Компоненты тензора энергии-импульса скалярного поля могут быть представлены в матричной форме, как в уравнении, Ур.~(\ref{eq:PFM}).
\noindent
Параметр уравнения состояния для скалярного поля квинтэссенции определен как:
\begin{equation}
w_{\phi}\equiv\dfrac {p_{\phi}}{\rho_{\phi}}=\dfrac{\dot{\phi}^2/2 - V(\phi)}{\dot{\phi}^2/2 + V(\phi)}.
\label{eq:SPEOP}
\end{equation}
Уравнение движения Клейна-Гордона для скалярного поля квинтэссенции может быть получено в результате вариации действия в уравнении, Ур.~(\ref{eq:S}), где плотность Лагранжиана определена  уравнением, Ур.~(\ref{eq:QLD}):
\begin{equation}
\ddot{\phi}+3H{\dot\phi}+\frac{\partial V(\phi)}{\partial \phi}=0,
\label{eq:KleinGordon}
\end{equation}
точка обозначает производную по физическому времени, $t$.

Влияние скалярного поля, $\phi$, на динамику вселенной отражено в первом уравнении Фридмана:
\begin{equation}
E(a)=(\Omega_{\rm r0}a^{-4} + \Omega_{\rm m0}a^{-3} +  \Omega_{\rm \phi}(a))^{1/2},
\label{eq:freqphi}
\end{equation}
где $\Omega_{\rm \phi}(a)$ - зависящий от времени параметр плотности энергии скалярного поля. Эволюция функции $\Omega_{\rm \phi}(a)$ определяется формой потенциала скалярного поля, $V(\phi)$.

\begin{figure}[h!]
\begin{center}
\includegraphics[width=\columnwidth]{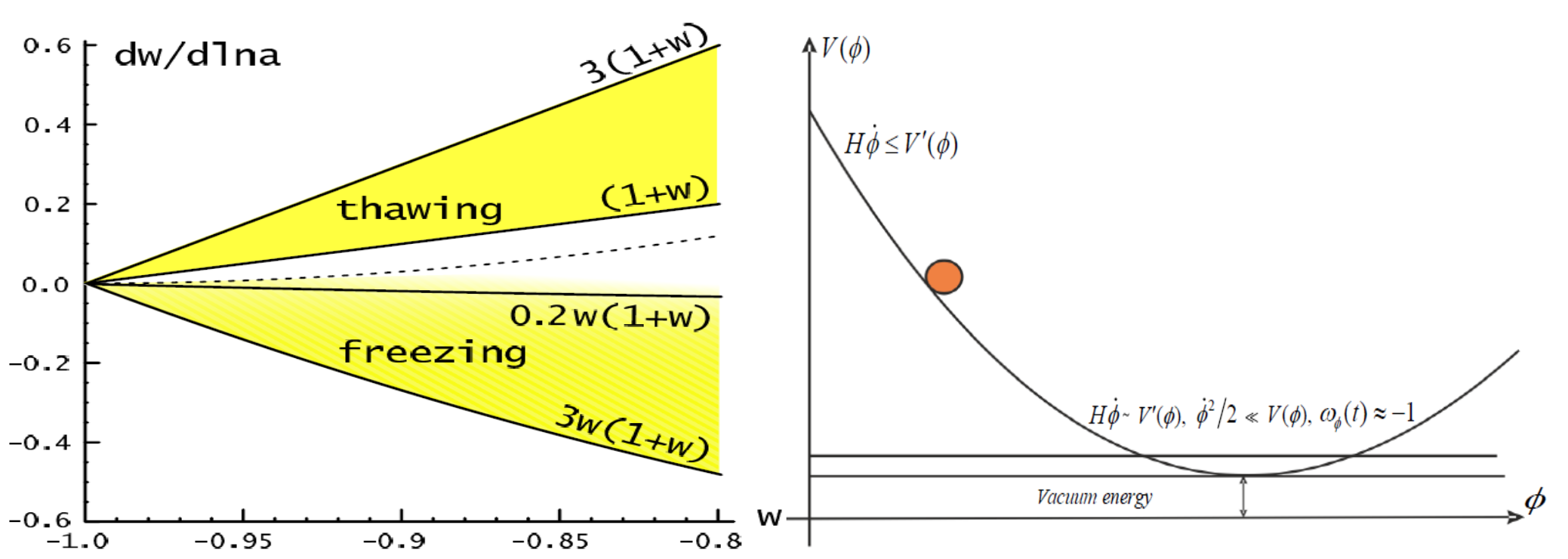}
\end{center}
 \caption {Левая панель: расположение моделей таяния и моделей замерзания на плоскости $w_{\phi}-dw_{\phi}/d\ln a$, (\cite{Caldwell:2005tm}). Правая панель: режимы быстрого  и медленного скатывания для модели замерзания скалярного поля, $\phi$, к минимуму своего потенциала.}
\label{fig:f22}
 \end{figure}

В зависимости от формы  потенциалов, модели квинтэссенции подразделяются на модели {\it таяния}~(thawing models) и на модели {\it замерзания}~(freezing models), (\cite{Caldwell:2005tm}).  На $w_{\phi} - dw_{\phi}/d\ln a$ плоскости модели таяния и модели замерзания могут быть расположены в строго обозначенных областях для каждой из них, Fig.~(\ref{fig:f22})~(левая панель).
 В ранние эпохи эволюции вселенной скалярное поле в моделях таяния было слишком подавлено  тормозящим влиянием расширения Хаббла, представленное слагаемым, $3H{\dot\phi}$, в уравнении,
Ур.~(\ref{eq:KleinGordon}). Таким образом, эволюция скалярного поля происходила  значительно медленнее в сравнении с темпом расширения Хаббла. Результатом подавляющего действия расширения Хаббла на скалярное поле в моделях таяния является вмороженность скалярного поля.  Это скалярное поле проявляет себя как энергия вакуума с параметром уравнения состояния, $w_{\phi}=-1$. Темп расширения Хаббла, $H(a)$, является убывающей со временем функцией. После того, как величина темпа расширения Хаббла достигает величины, $H<\sqrt{{\partial}^2 V(\phi)/\partial t^2}$, скалярное поле скатывается к минимуму своего потенциала. Это приводит к тому, что величина параметра уравнения состояния скалярного поля, $w_{\phi}$, увеличивается со временем и достигает величины $w_{\phi}>-1$.

В моделях замерзания скалярное поле всегда подавлено (оно является затухающим), т. е. $H > \sqrt{{\partial}^2 V(\phi)/\partial t^2}$.
В моделях замерзания существуют режимы быстрого и медленного скатывания. Уравнение движения скалярного поля, Ур.~(\ref{eq:KleinGordon}), в зависимости от соотношения слагаемого $3H{\dot\phi}$ и слагаемого $\partial V(\phi)/\partial t$, описывает: режим быстрого скатывания, который происходит при $3H{\dot\phi}< \partial V(\phi)/\partial t$, т. е. $\ddot{\phi}\gg V(\phi)$, или режим медленного скатывания, который происходит при $3H{\dot\phi}< \partial V(\phi)/\partial t$.
 При режиме медленного скатывания скалярное поле стремится к минимуму своего потенциала и почти не изменяется со временем, $\ddot{\phi}\ll V(\phi)$, поэтому из уравнения, Ур.~(\ref{eq:SPEOP}), следует, что $w_{\phi}\approx-1$, Рис.~(\ref{fig:f22})~(правая панель).

В моделях замерзания скалярное поле имеет так называемое следящее решение (tracker solution). Плотность энергии скалярного поля в моделях замерзания почти постоянна с течением времени. Ее вклад в полную
плотность энергии вселенной в эпоху доминирования излучения и в эпоху доминирования материи незначителен. Поэтому плотность энергии скалярного поля остается субдоминантной в эти эпохи и отслеживает сперва плотность энергии излучения, а затем плотность энергии материи.
С течением времени плотность энергии излучения и плотность энергии материи уменьшаются  из-за расширения вселенной. Плотность энергии скалярного поля увеличивается со временем, в конечном итоге она становится доминирующей составляющей и начинает вести себя как компонента с отрицательным эффективным давлением. На поздних этапах эволюции вселенной это проявляется в ускоренном расширении вселенной.
\subsection{Фантомное скалярное поле}
 Плотность Лагранжиана для фантомного скалярного поля описывается уравнением:
\begin{equation}
\mathcal{L}_{\phi}= - \frac{1}{2}g^{\mu\nu}\partial_\mu\phi\partial_\nu\phi-V(\phi).
\label{eq:QLD1}
\end{equation}
Тензор энергии-импульса для фантомного скалярного поля, $T_{\mu\nu}$, определен как:
\begin{equation}
T_{\mu\nu}= - 2\frac{\partial{\mathcal{L}_{\phi}}}{\partial{g^{\mu\nu}}} - g_{\mu\nu}\partial{\mathcal{L}_{\phi}}.
\label{eq:SET3}
\end{equation}
Подставляя уравнение, Ур.~(\ref{eq:QLD1}), в уравнение, Ур.~(\ref{eq:SET3}), мы получим:
\begin{equation}
T_{\mu\nu}= - \partial_\mu \phi \partial_\nu\phi - g_{\mu\nu}\left[\frac{1}{2}g^{\alpha\beta}\partial_\alpha \phi \partial_\beta \phi - V(\phi)\right].
\label{eq:SET4}
\end{equation}
Компоненты тензора энергии-импульса для фантомного скалярного поля, $T_{\mu\nu}$, представлены как:
\begin{align}
\label{eq:enmomuni1}
T_{00}&\equiv\rho_{\phi}= - \frac{1}{2}\dot{\phi}^2 + V(\phi),\\
\label{eq:enmomuni2}
T_{0i}&=0,\\
\label{eq:enmomuni3}
T_{ij}&=0 \ (i\neq j),\\
\label{eq:enmomuni4}
T_{ii}&\equiv p_{\phi}= - \frac{1}{2}\dot{\phi}^2 - V(\phi).
\end{align}
\noindent
Параметр уравнения состояния для фантомного скалярного поля определяется как:
\begin{equation}
w_{\phi}\equiv\dfrac {p_{\phi}}{\rho_{\phi}}=\dfrac{-\dot{\phi}^2/2 - V(\phi)}{-\dot{\phi}^2/2 + V(\phi)}.
\end{equation}
Уравнение движения Клейна-Гордона для фантомного скалярного поля:
\begin{equation}
\ddot{\phi}+3H{\dot\phi}-\frac{\partial V(\phi)}{\partial \phi}=0.
\label{eq:KleinGordon1}
\end{equation}

В Таблице \ref{table:PP} приведен неполный список фантомных потенциалов:
\begin{table}[h!]
\begin{tabular}{|p{4.5cm}|p{7cm}|p{3.5cm}|}
\hline %\hline
\hspace{0.5 cm}\textrm{Название} & \hspace{3cm}\textrm{Форма} & \hspace{0.5 cm}\textrm{Ссылка} \\
\hline\hline
\hline
Fifth power & $V(\phi)=V_0\phi^5$  & \cite{Scherrer:2008be}\\
\hline
Inverse quadratic  & $V(\phi)=V_0\phi^{-2}$ & \cite{Scherrer:2008be}\\
\hline
Exponent & $V(\phi)=V_0\exp(\beta\phi)$, $\beta = {\rm const} >0$ & \cite{Scherrer:2008be} \\
\hline
Quadratic & $V(\phi)=V_0\phi^2$ & \cite{Dutta:2009dr}\\
\hline
Gaussian & $V(\phi)=V_0(1-\exp(\phi^2/\sigma^2))$, $\sigma={\rm const}$ & \cite{Dutta:2009dr}\\
\hline
pseudo-Nambu-Goldstone boson~(pNGb)& $V(\phi)=V_0(1-\cos(\phi/\kappa))$, $\kappa = {\rm const} >0$ & \cite{Frieman:1995pm}\\
\hline
Inverse hyperbolic cosine & $V(\phi)=V_0(\cosh(\psi\phi))^{-1}$, $\psi= {\rm const} >0$ &  \cite{Dutta:2009dr}\\
[0.2cm]
\hline
\end{tabular}
\caption{\rm Список фантомных потенциалов.}
\label{table:PP}
\end{table}
\section{Модели взаимодействия материи и темной энергии}
Как было указано ранее, одной из нерешенных проблем  современной космологии является проблема совпадения в стандартной $\Lambda$CDM модели. Ввиду того, что величина плотности темной энергии и величина плотности энергии
материи в современной вселенной имеют одинаковый порядок, можно предположить, что материя и темная энергия каким-то образом взаимодействуют друг с другом.

 В моделях взаимодействия (coupling models) рассматривается взаимодействие материи и темной энергии. В этих моделях темная энергия представлена в виде скалярного поля.
Взаимодействие между материей и темной энергией  описано модифицированными уравнениями непрерывности для материи и темной энергии, соответственно, как:
  \begin{eqnarray}
\dot{\rho}_{\rm m}+3H\rho_{\rm m}=\delta_{\rm couple},\label{eq:coupldm}\\
\dot{\rho}_{\rm \phi}+3H(\rho_{\rm \phi}+p_{\rm \phi})=-\delta_{\rm couple},\label{eq:couplde}
\end{eqnarray}
где $\rho_{\rm m}$ - плотность энергии материи; $\rho_{\rm \phi}$ и $p_{\rm \phi}$ плотность энергии и давление темной энергии, соответственно; $\delta_{\rm couple}$ - коэффициент взаимодействия между материей и темной энергией.

В моделях взаимодействия между материей и темной энергией применяются следующие формы  коэффициента взаимодействия, $\delta_{\rm couple}$, (\cite{Amendola:1999er}, \cite{Zimdahl:2001ar}):
 \begin{eqnarray}
\delta_{\rm couple}=nQ\rho_{\rm m}\dot{\phi},\label{eq:delta1}\\
\delta_{\rm couple}=\alpha H(\rho_{\rm m}+\rho_{\rm \phi}),\label{eq:delta2}
\end{eqnarray}
где $n=\sqrt{8\pi G}$; $\alpha$ и $Q$ - безразмерные константы. Согласно данным Planck 2015, (\cite{Ade:2015xua}), $Q<0.1$.

Модели взаимодействия материи и темной энергии подразделяются на два типа.
\subsection{Модели взаимодействия материи и темной энергии первого типа}
Модели взаимодействия материи и темной энергии первого типа характеризуются экспоненциальным потенциалом и линейным взаимодействием, определяемым коэффициентом взаимодействия, представленным в уравнении, Ур.~(\ref{eq:delta1}), (\cite{Amendola:1999er}).

Уравнение скалярного поля имеет вид:
 \begin{equation}
\ddot{\phi}+3H{\dot\phi}-\frac{\partial V(\phi)}{\partial \phi}=-nQ\rho_{\rm m}\dot{\phi},
\label{eq:scalcoup}
\end{equation}
где $V(\phi)=V_0e^{-n\lambda\phi}$ - потенциал скалярного поля и $\lambda$ - модельный параметр.

Уравнение непрерывности для темной энергии:
\begin{equation}
\dot{\rho}_{\rm \phi}+3H(\rho_{\rm \phi}+p_{\rm \phi})=-nQ\rho_{\rm m}\dot{\phi}.
\label{eq:deen}
\end{equation}
Плотность энергии материи эволюционирует как:
 \begin{equation}
\dot{\rho}_{\rm m}+3H\rho_{\rm m}=nQ\rho_{\rm m}~~\Rightarrow~~{\rho}_{\rm m}={\rho}_{\rm m0}a^{-3}e^{nQ\phi}.
\label{eq:maten}
\end{equation}

\subsection{Модели взаимодействия материи и темной энергии второго типа}
Для этого типа моделей взаимодействия материи и темной энергии, на основании выполнения требования, $\rho_{\rm m}/\rho_\phi=\rm const$, конструируется вид потенциала и динамика взаимодействия между материей и темной энергией, (\cite{Zimdahl:2001ar}).

Уравнение взаимодействия, Ур.~(\ref{eq:couplde}), эквивалентно соотношению:
 \begin{equation}
 \dot{\phi}\Bigr[\ddot{\phi}+3H{\dot\phi}-\frac{\partial V(\phi)}{\partial \phi}\Bigl]=-\delta_{\rm couple}.
\end{equation}
Коэффициент взаимодействия определен как:
\begin{equation}
\delta_{\rm couple}=-3H\Pi_{\rm m}=3H\Pi_{\phi},
\end{equation}
\begin{equation}
\Pi_{\rm m}=-\Pi_{\phi}=\dfrac{\rho_{\rm m}\rho_{\phi}}{\rho}(\gamma_{\phi}-1),
\end{equation}
где $\gamma_{\phi}=\frac{p_{\rm \phi}+\rho_{\phi}}{\rho_{\phi}}=\frac{\dot{\phi}^2}{\rho_{\phi}}$ и $\rho=\rho_{\rm m}+\rho_{\phi}$.

Уравнения непрерывности для материи и темной энергии, соответственно,  имеют вид:
  \begin{eqnarray}
\dot{\rho}_{\rm m}+3H(\rho_{\rm m}+\Pi_{\rm m})=0,\label{eq:coupldm1}\\
\dot{\rho}_{\rm \phi}+3H(\rho_{\rm \phi}+p_{\rm \phi}+\Pi_{\phi})=0.\label{eq:couplde1}
\end{eqnarray}
Вид потенциала скалярного поля конструируется следующим образом:
\begin{equation}
V(\phi)=\dfrac{1}{6\pi G}\Bigr(1-\dfrac{\gamma_{\phi}}{2}\Bigl)\dfrac{1+r}{(\gamma_{\phi}+r)^2}\dfrac{1}{t^2}~~\Rightarrow~~\dfrac{\partial V(\phi)}{\partial \phi}=-\lambda V(\phi),
\label{eq:potencoupl}
\end{equation}
где $r\equiv\dfrac{\rho_{\rm m}}{\rho_{\rm \phi}}=\rm const$ и $\lambda=\sqrt{\frac{24\pi G}{\gamma_{\phi}(1+r)}}$.

Из уравнений, Ур.~(\ref{eq:potencoupl}), следует, что вид потенциала имеет экспоненциальную форму:
\begin{equation}
V(\phi)=V_0e^{-\lambda(\phi-\phi_0)}.
\end{equation}
 Отсутствие убедительного объяснения причины начала взаимодействия темной энергии и материи в эпоху перехода от замедленного к ускоренному расширению вселенной является существенным недостатком этой модели.

\section{CPL параметризация параметра уравнения состояния}
В динамических моделях темной энергии, в которых плотность темной энергии зависит от времени, параметр уравнения состояния моделируется в виде: $p= w(a)\rho$. Такой тип параметризации называется $w$CDM параметризацией\footnote{Темная энергия иногда характеризуется только параметром уравнения состояния, а соответствующие космологические модели темной энергии называются $w$CDM моделями, (\cite{Barger:2006vc}).}. Эта параметризация не имеет физической мотивации. Применение $w$CDM параметризации
обычно используется в качестве анзаца в анализе данных для количественной оценки динамических моделей темной энергии.
Параметризация параметра уравнения состояния, $w(a)$, используется для того, чтобы отличить модели темной энергии между собой. В частности,
этот подход можно использовать для того, чтобы отличить $\Lambda$CDM модель от других моделей темной энергии в современную эпоху.

Зависящий от времени параметр уравнения состояния для моделей темной энергии часто характеризуется Chevallier-Polarsky-Linder (CPL) $w_0-w_a$
параметризацией, (\cite{cp01}, \cite{linder03}):
\begin{equation}
 w(a) = w_0 + w_a(1-a),
\label{eq:EOS}
\end{equation}
где $w_0=w(a=1)$ и $w_a=(dw/dz)|_{z=1}=-a^{-2}({\rm d} w/{\rm d}a)|_{a=1/2}$.
Несмотря на то, что эта параметризация очень проста, она достаточно гибкая, чтобы точно описывать
параметры уравнения состояния для большинства моделей темной энергии. CPL параметризация не может описать произвольные модели темной энергии с хорошей точностью (до нескольких процентов) в широком диапазоне величин красного смещения, (\cite{linder03}).

Первое уравнение Фридмана, в котором темная энергия аппроксимируется CPL $w_0-w_a$ параметризацией, имеет вид:
\begin{equation}
 E(a) = ({\Omega_{\rm r0}a^{-4} + \Omega_{\rm m0}a^{-3} + \Omega_\Lambda a^{-3(1+w_0+w_a)}e^{-3w_a(1-a)}})^{1/2}.
\label{eq:E_CPL}
\end{equation}

%%%%%%%%%%%%%%%%%%%%%%%%%%%%%%%%%%%%%%%%%%%%%%%%%%%%%%%%%%%%%%%%%%%%%%%%%%
%%%%%%%%%%%%%%%%%%%       Beginng of chapter 7   %%%%%%%%%%%%%%%%%%%%%%%
%%%%%%%%%%%%%%%%%%%%%%%%%%%%%%%%%%%%%%%%%%%%%%%%%%%%%%%%%%%%%%%%%%%%%%%%%%
\chapter{Динамика и функция темпа роста флуктуаций плотности материи в Ратра-Пиблс $\phi$CDM модели}\label{chapter:7}
Эта глава основана на исследованиях, представленных в статьях, (\cite{avs14}) и (\cite{avs15}).

В этой главе подробно исследуется обратно-степенной потенциал Ратра-Пиблса, $V(\phi)\propto1/\phi^{\alpha}$.
Впервые этот потенциал был рассмотрен Джимом Пиблсом и Бхарат Ратра в 1988, (\cite{Ratra:1987aj}, \cite{Ratra:1987rm}). Модель скалярного поля с потенциалом Ратра-Пиблса является простейшей $\phi$CDM моделью скалярного поля квинтэссенции замороженного типа. Эта модель была предложена с целью решения проблемы тонкой настройки в стандартной $\Lambda$CDM модели.

\section{Основные уравнения}\label{sec:one}
Потенциал Ратра-Пиблс имеет вид:
\begin{equation}
V(\phi)=\dfrac{\kappa}{2}M_{\rm pl}^2\phi^{-\alpha},
\label{eq:Potential1}
\end{equation}
\noindent
где $\alpha$ - модельный параметр, величина которого положительна, $\alpha>0$. Величина этого параметра влияет  на скорость изменения потенциала со временем, тем самым  определяет форму потенциала. В наших исследованиях мы рассматриваем величины параметра $\alpha$ в диапазоне $0 < \alpha \leq 0.7$. Этот диапазон величин соответствует современным космологическим наблюдениям, (\cite{Samushia:2009dd}). Для величины параметра, $\alpha=0$, Ратра-Пиблс $\phi$CDM модель сводится к $\Lambda$CDM модели. Модельный параметр $\kappa$ определяется через параметр $\alpha$\footnote{Вычисление параметра $\kappa$ представлено ниже.}. Модельный параметр $\kappa$ соотносится к шкале массы скалярных частиц, $M_\phi$, как:
\begin{equation}
M_\phi\sim\Bigl(\dfrac{\kappa M_{\rm pl}^2}{2} \Bigl)^{\frac{1}{\alpha+4}}.
\label{eq:kappa1}
\end{equation}
Мы рассматриваем плоскую и изотропную вселенную, которая описывается
FLRW метрикой пространства-времени:
\begin{equation}
ds^2=dt^2-a(t)^2d{\bf x}^2.
\end{equation}
Уравнение движения Клейна-Гордона в Ратра-Пиблс $\phi$CDM модели имеет вид:
\begin{equation}
\ddot{\phi}+3H{\dot\phi}-\frac{1}{2}\kappa\alpha M_\mathrm{pl}^2\phi^{-(\alpha+1)}=0.
\label{eq:KleinGordonRP}
\end{equation}
Плотность энергии, давление и параметр уравнения состояния в Ратра-Пиблс $\phi$CDM модели определены, соответственно, как:
\begin{eqnarray}
\rho_\phi & = &\dfrac{M_\mathrm{pl}^2} {32\pi} \Bigl(\dot{\phi}^2 + \kappa M_\mathrm{pl}^2\phi^{-\alpha} \Bigr),
\label{eq:Rho}  \\
p_\phi & = & \dfrac{M_{\mathrm{pl}}^2}{32\pi} \Bigl(\dot{\phi}^2 - \kappa M_\mathrm{pl}^2\phi^{-\alpha} \Bigr),
\label{eq:P}\\
w_{\rm \phi} & = & \dfrac{{\dot\phi}^2 - \kappa M_\mathrm{pl}^2\phi^{-\alpha}}{{\dot\phi}^2 + \kappa M_\mathrm{pl}^2\phi^{-\alpha}}.
\label{eq:w}
\end{eqnarray}
Из уравнения состояния, Ур.~(\ref{eq:w}), следует, что требование выполнения условия, $w_0\simeq-1$, накладывает следующее ограничение, $\dot{\phi}^2/2\ll V(\phi)$.
Ратра-Пиблс $\phi$CDM модель скалярного поля имеет следящее решение. Это означает, что плотность энергии скалярного поля, $\rho_\phi$, в ранние периоды развития вселенной сначала отслеживает плотность энергии излучения, а затем плотность энергии материи, оставаясь при этом субдоминантой. Только на  поздних этапах эволюции вселенной плотность энергии скалярного поля, $\rho_\phi$, становится доминирующей.

Величина параметра уравнения состояния в Ратра-Пиблс $\phi$CDM модели скалярного поля в эпоху доминирования излучения или материи может быть приблизительно определена как, (\cite{Zlatev:1998tr}):
\begin{equation}
 w_{\rm \phi}\approx\dfrac{\frac{\alpha}{2} w_{\rm bac}-1}{1+\frac{\alpha}{2}},
\label{eq:omegab}
\end{equation}
где $w_{\rm bac}$ - фоновый параметр уравнения состояния для доминирующих эпох излучения или материи. Для эпохи доминирования излучения, $w_{\rm bac}=1/3$, и для эпохи доминирования материи, $w_{\rm bac}=0$. Приближение, представленое в уравнении, Ур.~(\ref{eq:omegab}), является верным при $\rho_{\rm bac}\gg\rho_{\rm \phi}$, где $\rho_{\rm bac}$ - величина фоновой плотности энергии.

Модель скалярного поля с потенциалом Ратра-Пиблса имеет не только следящее решение, но и аттракторное решение\footnote{Аттрактор - это набор численных величин, к которым система стремится в результате своей эволюции, при широком разнообразии величин начальных условий.}. Это означает, что эволюция плотности энергии скалярного поля, $\rho_\phi$, в Ратра-Пиблс $\phi$CDM модели является нечувствительной к начальным условиям, ($\phi_{\rm in}$, $\dot{\phi}_{\rm in}$), и решения для широкого диапазона начальных условий сходятся в одно и то же общее решение в современную эпоху.

Параметр плотности энергии и первое уравнение Фридмана в Ратра-Пиблс $\phi$CDM модели определены, соответственно, как:
\begin{eqnarray}
\Omega_{\rm \phi}(a) & = & \dfrac{1}{12H_0^2}\Bigl(\dot{\phi}^2+\kappa M_{\rm pl}^2\phi^{-\alpha}\Bigr),
\label{eq:omegafi}\\
E(a) & = &\Bigl(\Omega_{\rm r0}a^{-4} + \Omega_{\rm m0} a^{-3}  + \frac{1}{12H_0^2}\Bigl(\dot{\phi}^2+\kappa M_{\rm pl}^2\phi^{-\alpha}\Bigr)\Bigr)^{1/2}.
\label{eq:FriedmannRP}
\end{eqnarray}
\subsection{Вычисление модельного параметра $\kappa$ и начальных условий}

Вычисления параметра $\kappa$ и  начальных условий основаны на исследованиях, представленных в статьях: (\cite{Farooq:2013syn}, Sec.~3.6.3,) и (\cite{avs14}, Appendix А).

В уравнении скалярного поля, Ур.~(\ref{eq:KleinGordonRP}), представим скалярный фактор, $a(t)$, и скалярное поле, $\phi(t)$, в виде степенного закона:
\begin{equation}
a(t)=a_\star \Bigl(\frac{t}{t_\star}\Bigl)^n,~~~~~~~\phi(t)= \phi_\star \Bigl(\frac{t}{t_\star}\Bigl)^p,
\label{eq:Power}
\end{equation}
где $a_\star \equiv a(t_\star)$ и $\phi_\star \equiv \phi(t_\star)$ - величины скалярного фактора и скалярного поля в момент времени, $t=t_\star$, соответственно. Параметр, $p$, связан с параметром, $\alpha$, следующим выражением, $p=2/(2+\alpha)$. В результате этих подстановок мы получим:
\begin{equation}
\phi_\star^{\alpha+2}=\frac{(\alpha+2)^2}{4(6n+3n\alpha-\alpha)}\kappa
\alpha M_{\rm pl}^2 t_\star^2.
 \label{eq:Scalmod2}
\end{equation}
Используя уравнения, Ур.~(\ref{eq:Rho}), Ур.~(\ref{eq:FriedmannRP}) - Ур.~(\ref{eq:Scalmod2}), мы найдем: %Ур.~(\ref{eq:Power})
\begin{eqnarray}
\rho & = &   \frac{3n}{8\pi}\Bigl(\frac{M_{\rm pl}}{t_\star} \Bigl)^2  \frac{\phi_\star^2}{\alpha(\alpha+2)}
\Bigl(\frac{t}{t_\star}\Bigl)^{\frac{-2\alpha}{\alpha+2}}
 \label{eq:Rho1},
\\
\Bigl(\frac{n}{t}\Bigl)^2  &= &\frac{8\pi}{3M_{\rm pl}^2}\rho,
\label{eq:Rho2}
\end{eqnarray}
где $\rho \equiv \rho_\phi$ является плотностью темной энергии, доминирующей во вселенной во время, $t<t_\star$. Предполагая $\rho(t) = \rho_\star ({t}/{t_\star})^\beta$, мы получим, $\beta = -2\alpha/(\alpha+2)$. С другой стороны, учитывая, что доминирующая компонента темной энергии представлена как, $\rho_\star$, в момент времени, $a=a_\star$:
\begin{equation}
\rho=\rho_{\star}\Bigl(\frac{a_\star}{a}\Bigl)^{\frac{2}{n}},
 \label{eq:Rho3}
\end{equation}
где $n=1/2$ и $n=2/3$ - величины параметра $n$ для эпохи доминирования излучения и материи, соответственно.

 Для того, чтобы получить выражение для $\phi_\star^2$, найдем чему равно выражение $1/t^2$ из уравнения, Ур.~(\ref{eq:Rho2}), подставим уравнение, Ур.~(\ref{eq:Rho3}),  в уравнение, Ур.~(\ref{eq:Rho1}), предполагая, что $a=a_\star$ и $\rho=\rho_{\star}$. Сравнивая полученный результат с уравнением, Ур.~(\ref{eq:Scalmod2}), мы найдем:
\begin{equation}
\kappa=\frac{32 \pi}{3nM_{\rm pl}^4}
\Bigl(\frac{6n+3n\alpha-\alpha}{\alpha+2}\Bigl)[n\alpha(\alpha+2)]^{\frac{\alpha}{2}}
\rho_{\star }.
 \label{eq:kappa}
\end{equation}
Подставляя уравнение, Ур.~(\ref{eq:kappa}), в уравнение, Ур.~(\ref{eq:Scalmod2}), и используя уравнение, Ур.~(\ref{eq:Rho2}), мы получим:
\begin{eqnarray}
\phi_\star &= &[n\alpha (\alpha+2)]^{\frac{1}{2}},
\label{eq:phi2}
\\
\phi&=&[n\alpha (\alpha+2)]^{\frac{1}{2}} \Bigl(\frac{a}{a_\star}\Bigl)^{\frac{2}{n(\alpha+2)}}.
\label{eq:phi3}
\end{eqnarray}
Подставляя в уравнение, Ур.~(\ref{eq:phi3}), величину, $n=1/2$, и  предполагая, $a_\star=a_0$, можно получить уравнения для начальных условий в эпоху доминирования излучения,  Ур.~(\ref{eq:phi0}) и Ур.~(\ref{eq:phi0div}).

Подставим уравнение, Ур.~(\ref{eq:phi3}), в уравнение, Ур.~(\ref{eq:KleinGordon}):
\begin{equation}
\kappa=\frac{4n}{M_{\rm pl}^2 t_\star^2}
\Bigl(\frac{6n+3n\alpha-\alpha}{\alpha+2}\Bigl)[n\alpha(\alpha+2)]^{\alpha/2}.
 \label{eq:kappa2}
\end{equation}
Так как уравнение, Ур.~(\ref{eq:kappa}), должно быть верным для произвольного момента времени, $t_\star$, мы полагаем $t_\star=M_{\rm pl}^{-1}$.

В результате, для величин $n=1/2$ и $n=2/3$, мы получим:
\begin{eqnarray}
\kappa(n=1/2) & = & \Bigl(\frac{\alpha+6}{\alpha+2}\Bigl)
\Bigl[\frac{1}{2}\alpha(\alpha+2)\Bigl]^{\alpha/2},
\label{eq:kappa3}
\\
\kappa(n=2/3) & = & \frac{8}{3} \Bigl(\frac{\alpha+4}{\alpha+2}\Bigl)
\Bigl[\frac{2}{3}\alpha(\alpha+2)\Bigl]^{\alpha/2}.
 \label{eq:kappa4}
\end{eqnarray}

\subsection{Начальные условия}
Мы численно проинтегрировали систему уравнений, Ур.~(\ref{eq:KleinGordonRP}) и Ур.~(\ref{eq:FriedmannRP}). Начальные условия были установлены в эпоху доминирования излучения, для момента,  $a_{\rm in}=5\cdot10^{-5}$. Расчеты проводились до современной эпохи, $a_0=1$.

 Несмотря на то, что потенциал Ратра-Пиблса имеет аттракторное решение, и широкий диапазон начальных условий может привести к одинаковому решению в современную эпоху, для лучшей численной сходимости мы вычислили величину модельного параметра $\kappa$  и начальные условия из уравнений, Ур.~(\ref{eq:phi2}) и Ур.~(\ref{eq:phi3}), для эпохи доминирования излучения, $n=1/2$. Для вычислений мы применили следующие начальные условия:
\begin{eqnarray}
 \phi_{\rm in} &=&\left[\frac{1}{2}\alpha(\alpha+2)\right]^{1/2}t_{\rm in}^{\frac{4}{\alpha+2}},
 \label{eq:phi0}
 \\
{\dot{\phi}}_{\rm in} &=&\Bigl(\frac{8\alpha}{\alpha+2}\Bigl)^{1/2}t_{\rm in}^{\frac{2-\alpha}{2+\alpha}}.
 \label{eq:phi0div}
\end{eqnarray}
\noindent
Величина модельного параметра $\kappa$ была определена из уравнения, Ур.~(\ref{eq:kappa3}). В соответствии с зависимостью скалярного фактора от физического времени в эпоху доминирования излучения, мы использовали выражение, $a_{\rm in} \propto t_{\rm in}^{1/2}$, представленное в уравнении, Eq.~(\ref{eq:DWR}).
В наших вычислениях  мы применяли величины параметров плотности энергии для материи, темной энергии и параметра Хаббла для современной эпохи: $\Omega_{\rm m0}=0.315$, $\Omega_{\rm
\phi0}=0.685$, $h=0.673$. Эти результаты были получены коллаборацией Planck 2013, (\cite{Ade:2013zuv}).

\section{Динамика и эволюция энергетических компонент во вселенной в Ратра-Пиблс $\phi$CDM модели}
\begin{figure}[h!]
\begin{center}
\includegraphics[width= \columnwidth]{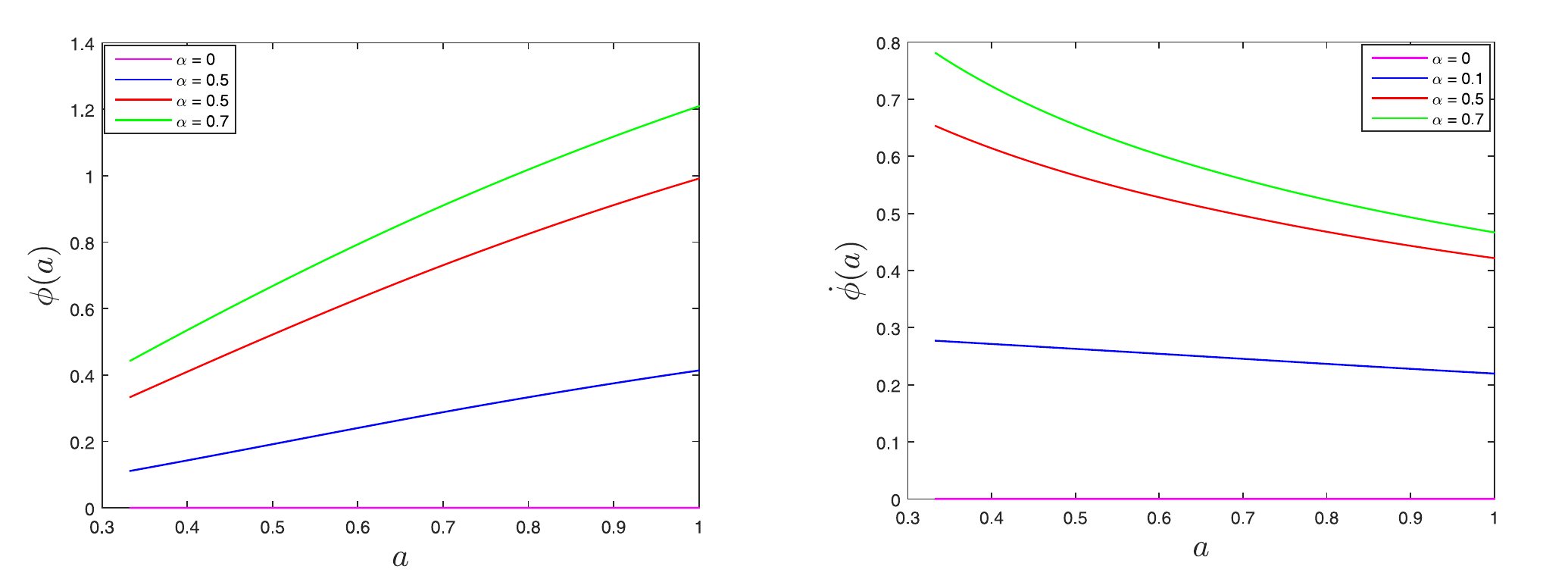}
\end{center}
\caption{Левая панель: зависимость амплитуды скалярного поля, $\phi(a)$, от величины параметра $\alpha$. Правая панель: зависимость производной по времени скалярного поля, $\dot{\phi}(a)$, от величины параметра $\alpha$.}
\label{fig:f23}
\end{figure}

\begin{figure}[h!]
\begin{center}
\psfig{file=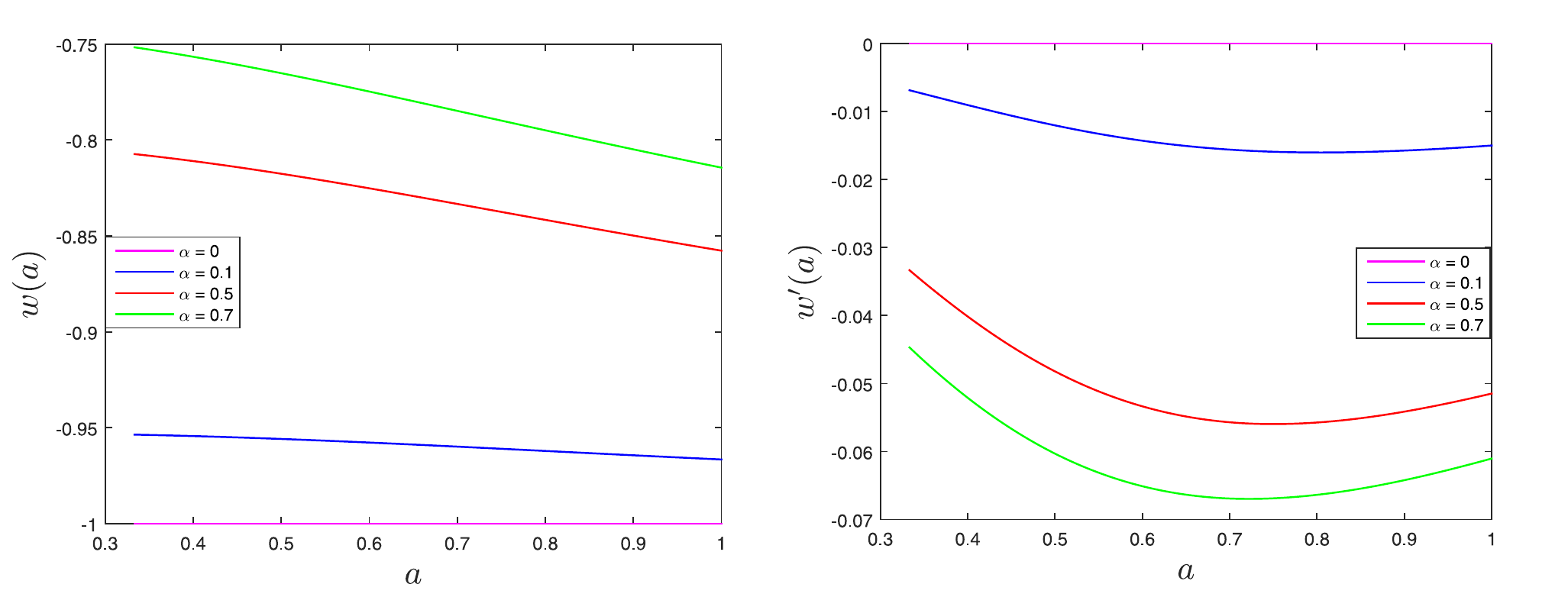,width= \columnwidth}
\end{center}
\caption{Левая панель: зависимость уравнения состояния, $w(a)$, от величины параметра $\alpha$. Правая панель: зависимость производной по скалярному фактору параметра уравнения состояния, $w'(a)$, от величины параметра $\alpha$.}
 \label{fig:f24}
\end{figure}

Мы проанализировали каким образом изменение величины модельного параметра $\alpha$ в Ратра-Пиблс $\phi$CDM модели влияет на величину скалярного поля, $\phi$, и на скорость изменения скалярного поля, $\dot{\phi}$. Результаты этого анализа представлены на Рис.~(\ref{fig:f23}) и Рис.~(\ref{fig:f24}).

В Ратра-Пиблс $\phi$CDM модели увеличение величины параметра $\alpha$ вызывает более сильную зависимость от времени скалярного поля, $\phi$, и ее производной по времени, $\dot{\phi}$, а также параметра уравнения состояния, $w$, и его производной по масштабному фактору, $dw/da$.

Как и ожидалось, для $\Lambda$CDM модели величина уравнения состояния, $w$, равна минус единице и величины $\phi$, $\dot{\phi}$ и $dw/da$, соответственно, равны нулю.

\begin{figure}[h!]
\begin{center}
\psfig{file=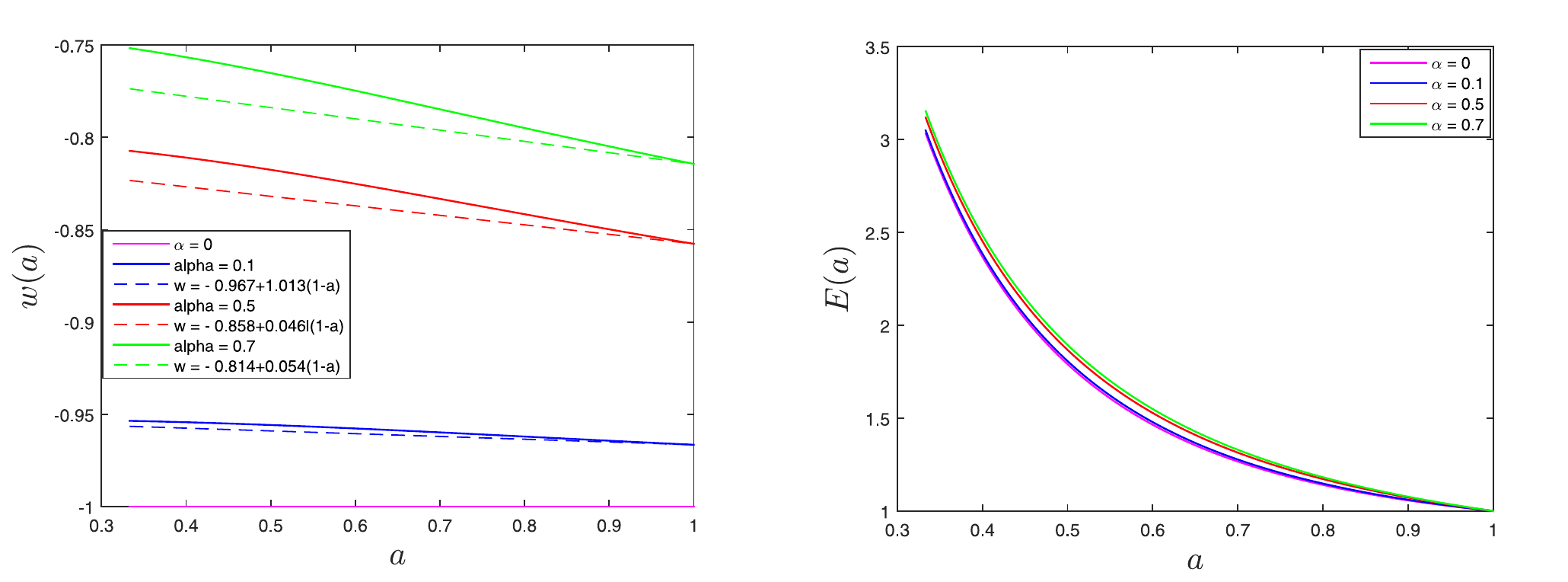,width=\columnwidth}
\end{center}
 \caption{Левая панель: параметр уравнения состояния, $w(a)$, для разных величин параметра $\alpha$ вместе с предсказаниями, вычисленными для CPL параметризации  с соответствующими оптимальными величинами $w_0$ и $w_a$. Правая панель: нормированный параметр Хаббла, $E(a)$, в зависимости от величины параметра $\alpha$.}
\label{fig:f25}
\end{figure}

Мы применили CPL параметризацию к эффективному параметру уравнения состояния, $w(a)$, в Ратра-Пиблс $\phi$CDM модели, Ур.~(\ref{eq:EOS}). Эта параметризация обеспечивает хорошую аппроксимацию эффективного параметра уравнения состояния в диапазоне величин масштабного фактора, $a\in(0.98;1)$, Рис.~(\ref{fig:f25})~(левая панель).

Мы изучали эволюцию нормированного параметра Хаббла, $E(a)$,
для разных величин параметра $\alpha$ в Ратра-Пиблс $\phi$CDM модели.
 \begin{figure}[h!]
\begin{center}
\psfig{file=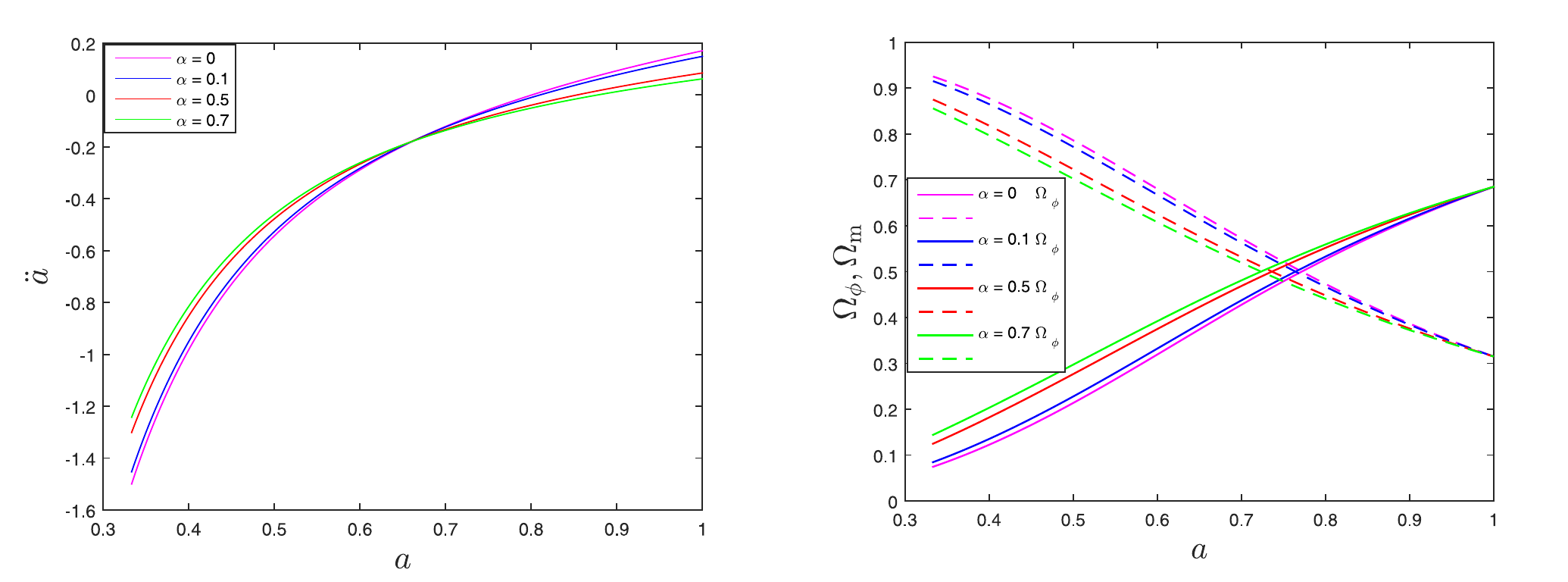,width=\columnwidth}
\end{center}
 \caption{Левая панель: величина второй производной скалярного фактора по времени, $\ddot{a}$, в зависимости от величины параметра $\alpha$. Правая панель: параметр плотности энергии материи, $\Omega_{\rm m}(a)$, (пунктирные линии)
 и параметр плотности энергии скалярного поля, $\Omega_{\rm \phi}(a)$, (сплошные линии) как функции
масштабного фактора для разных величин параметра $\alpha$.}
\label{fig:f26}
\end{figure}
Результаты этого исследования представлены на Рис.~(\ref{fig:f25})~(правая панель). С увеличением величины параметра  $\alpha$, расширение вселенной происходит быстрее. Самый медленный темп расширения соответствует $\Lambda$CDM модели.

Соотношение между динамикой и энергетическими компонентами во вселенной в Ратра-Пиблс $\phi$CDM модели показаны на Рис.~(\ref{fig:f26}). При одной и той же величине модельного параметра $\alpha$, динамическое доминирование темной энергии, Рис.~(\ref{fig:f26})~(левая панель),  начинается раньше чем энергетическое доминирование, Рис.~(\ref{fig:f26})~(правая панель). С увеличением величины параметра $\alpha$,  энергетическое доминирование темной энергии начинается раньше, Рис.~(\ref{fig:f26})~(правая панель).

\section{Эволюция крупномасштабной структуры в Ратра-Пиблс $\phi$CDM модели}
Эволюция флуктуаций плотности материи зависит от космологической модели темной энергии. Влияние темной энергии на развитие крупномасштабной структуры во вселенной происходит за счет влияния темной энергии на темп расширения вселенной. В свою очередь, темп расширения вселенной влияет на рост флуктуаций плотности материи.

\begin{figure}[h!]
\begin{center}
\psfig{file=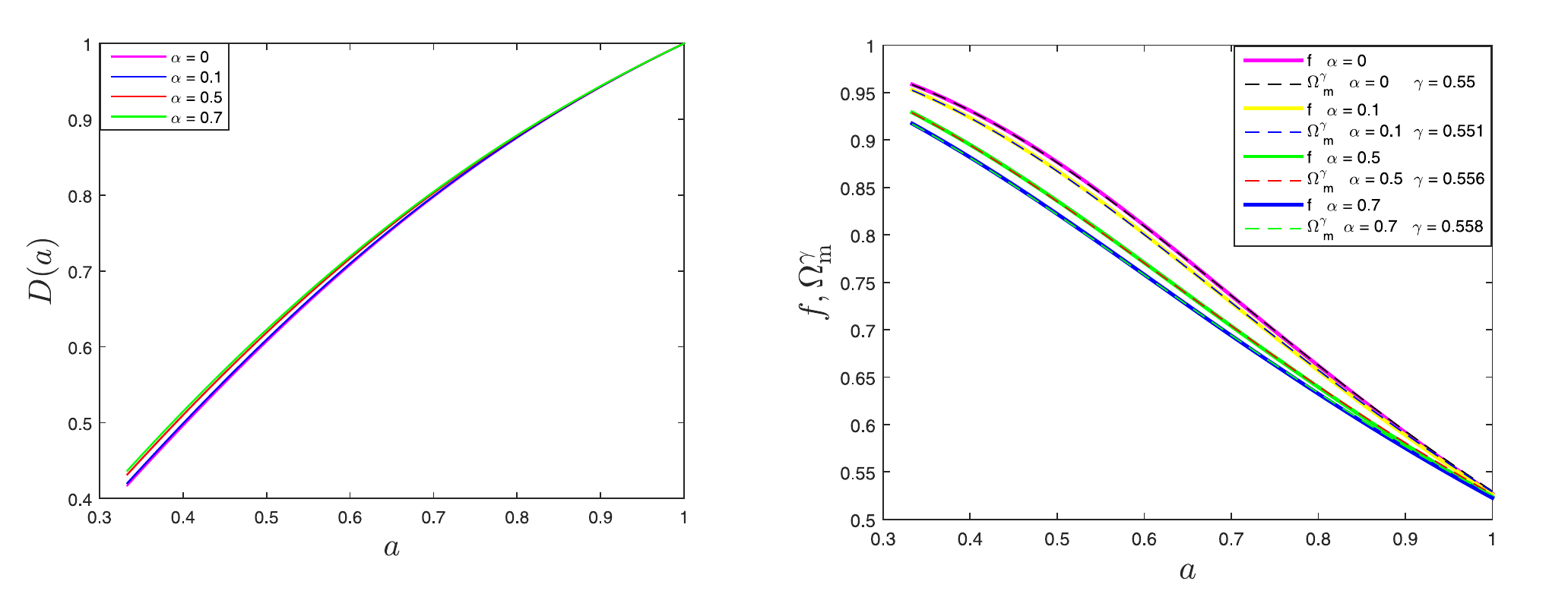,width=\columnwidth}
\end{center}
\caption{Левая панель: функция линейного фактора роста флуктуаций плотности материи, $D(a)$, для разных величин параметра $\alpha$. Правая панель: функция темпа роста флуктуаций плотности материи, $f(a)$, (сплошные линии) в зависимости от величины параметра $\alpha$ вместе с предсказаниями $\Omega_{\rm m}^{\gamma}$ (пунктирные линии), вычисленными для соответствующих наилучших величин параметра $\gamma$.}
\label{fig:f27}
\end{figure}

Мы исследовали эволюцию крупномасштабной структуры в расширяющейся вселенной в Ратра-Пиблс $\phi$CDM модели. Для вычисления величин флуктуаций плотности материи мы использовали уравнение,  Ур.~(\ref{eq:deltaeq}). На Рис.~(\ref{fig:f27})~(левая панель) показана эволюция функции линейного фактора роста флуктуаций плотности материи, $D(a)=\delta(a)/\delta(a_0)$, в зависимости от параметра $\alpha$. С увеличением величины параметра, $\alpha$, функция линейного фактора роста  флуктуаций плотности материи, $D(a)$, становится более зависимой от времени.

Как обсуждалось ранее, с увеличением величины параметра $\alpha$, расширение Хаббла происходит быстрее, Рис.~(\ref{fig:f25})~(правая панель), в то время как доминирование энергии скалярного поля начинается раньше, Рис.~(\ref{fig:f26})~(правая панель). Рост флуктуаций плотности материи происходит только в период эпохи доминирования материи, (\cite{Frieman:2008sn}), поэтому с увеличением величины параметра $\alpha$ остается  меньше времени для роста флуктуаций плотности материи. Для достижения той же амплитуды флуктуаций плотности материи, $\delta(a_0)$, в настоящий момент времени, в Ратра-Пиблс $\phi$CDM модели с увеличением величины параметра $\alpha$
требуется более большая величина начальной амплитуды флуктуаций плотности материи. Таким образом, скалярное поле при большей величине параметра $\alpha$ вызывает более сильные амплитуды флуктуаций плотности материи в начале их образования и во все последующие моменты времени вплоть до современной эпохи.
\section{Индекс роста в Ратра-Пиблс $\phi$CDM модели}
Мы исследовали насколько хорошо степенная параметризация функции темпа роста флуктуаций плотности материи, $f(a)$, и фракционной плотности материи, $\Omega_{\rm m}$(a), представленная в Ур.~(\ref{eq:f1f2}), может быть применена в Ратра-Пиблс $\phi$CDM модели. В этом случае вместо эффективного индекса роста $\gamma$(a), мы применили величину Линдер $\gamma$-параметризации, $\gamma$, которая определена в уравнении, Ур.~(\ref{eq:Lgamma}).

Результаты этих исследований показаны на Рис.~(\ref{fig:f27})~(правая панель).
Величина Линдер $\gamma$-параметризации, $\gamma$, в Ратра-Пиблс $\phi$CDM модели зависит от величины параметра $\alpha$, при этом величина Линдер $\gamma$-параметризации
увеличивается с увеличением величины параметра $\alpha$. Величина $\gamma$ в Ратра-Пиблс $\phi$CDM модели
немного выше величины $\gamma$ в модели $\Lambda$CDM, для которой $\gamma \approx 0.55$.
Темп роста флуктуаций плотности материи происходит медленнее с ростом величины параметра $\alpha$, Рис.~(\ref{fig:f27})~(правая панель). Это происходит в результате того, что расширение Хаббла и темп роста флуктуаций плотности материи являются взаимосвязанными и противоположно направленными процессами. Более быстрое расширение Хаббла, которое соответствует большей величине параметра $\alpha$, Рис.~(\ref{fig:f25})~(правая панель), приводит к большему подавлению темпа роста флуктуаций плотности материи.

\begin{figure}[h!]
\begin{center}
\psfig{file=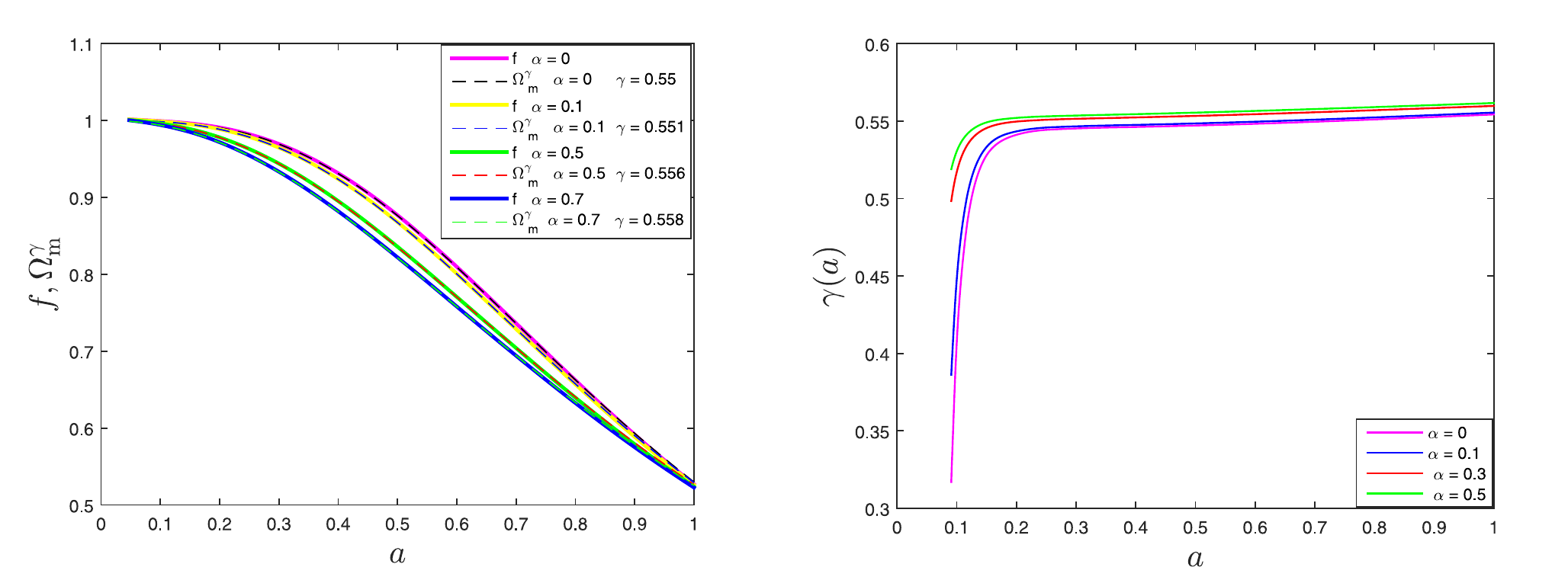,width=\columnwidth}
\end{center}
 \caption{Левая панель: функция темпа роста, $f(a)$, для разных величин параметра $\alpha$ (сплошные линии) вместе с предсказаниями $\Omega_{\rm m}^\gamma$ (пунктирные линии), вычисленными для соответствующих наилучших величин параметра $\gamma$ в диапазоне величин красного смещения, $z\in(0;10)$. Правая панель: функция индекса роста, $\gamma(a)$, для разных величин параметра $\alpha$ в диапазоне величин красного смещения, $z\in(0;10)$.}
\label{fig:f28}
\end{figure}

Мы исследовали применимость Линдер $\gamma$-параметризации для больших величин красного смещения.
Мы обнаружили, что эта параметризация может быть применимой в диапазоне величин красного смещения,
$z\in(0;5)$. Линдер $\gamma$-параметризация не применима для более больших величин красного смещения, Fig.~(\ref{fig:f28})~(левая панель).

Мы исследовали поведение функции эффективного индекса роста, $\gamma(a)$, представленного в Ур.~(\ref{eq:effgamma}), при больших величинах красного смещения, Рис.~(\ref{fig:f28})~(правая панель). Мы обнаружили, что на некотором диапазоне величин скалярного фактора, функция эффективного индекса роста, $\gamma(a)$, почти не зависит от величины скалярного фактора. При этом, слабая зависимость функции эффективного индекса роста от величины скалярного фактора происходит в диапазоне величин скалярного фактора: в $\Lambda$CDM модели, $a\in(0.25;1)$ (или $z\in(0;3)$); в Ратра-Пиблс $\phi$CDM модели, $a\in(0.18;1)$ (или $z\in(0;5)$), для $\alpha=0.5$.  Таким образом, с уменьшением величины параметра $\alpha$, режим слабой  зависимости функции эффективного индекса роста от величины скалярного фактора прекращается позже, Рис.~(\ref{fig:f28})~(правая панель).  Сравнивая Рис.~(\ref{fig:f28})~(левая панель) и  Рис.~(\ref{fig:f28})~(правая панель), мы видим, что прекращение применимости Линдер $\gamma$-параметризации для разных величин параметра $\alpha$ совпадает с моментом прекращения слабой зависимости функции эффективного индекса роста, $\gamma(a)$, от скалярного фактора. Таким образом, только в пределах диапазона величин скалярного фактора, при которых функция эффективного индекса роста слабо зависит от величины скалярного фактора, можно применять Линдер $\gamma$-параметризацию.

\section{Заключение}
Мы скрупулезно исследовали различные свойства Ратра-Пиблс $\phi$CDM модели в сравнении с $\Lambda$CDM моделью.
В частности, мы изучали динамику Ратра-Пиблс $\phi$CDM модели в зависимости от величины модельного параметра $\alpha$. С увеличением величины параметра $\alpha$, происходит увеличение скорости изменения потенциала в зависимости от времени. Это вызывает более сильную зависимость от времени скалярного поля, $\phi$, ее производной по времени, $\dot{\phi}$, а также параметра уравнения состояния, $w$, и его производной по масштабному фактору, $dw/da$.

Мы показали, что Ратра-Пиблс $\phi$CDM модель отличается от $\Lambda$CDM модели рядом характеристик. Эти характеристики являются общими для класса $\phi$CDM моделей квинтэссенции замороженного типа, и эти характеристики не зависят от величины модельного параметра $\alpha$:
\begin{itemize}
\item[$\bullet$]{\it В $\phi$CDM моделях величина темпа расширения вселенной, $E$(a), всегда больше чем величина темпа расширения вселенной для $\Lambda$CDM модели.}
\item[$\bullet$]{\it Момент доминирования темной энергии в $\phi$CDM моделях начинается раньше, чем в $\Lambda$CDM модели (при условии фиксирования других космологических модельных параметров)}.
\item[$\bullet$] {\it Ратра-Пиблс $\phi$CDM модель и $\Lambda$CDM модель отличаются в своих прогнозах для темпа роста флуктуаций плотности материи во вселенной: $\phi$CDM модель скалярного поля предсказывает более медленный темп роста флуктуации плотности материи чем $\Lambda$CDM модель.}
\item[$\bullet$] Мы изучили применимость Линдер $\gamma$-параметризации для Ратра-Пиблс $\phi$CDM модели. Мы обнаружили, что эта параметризация применима для этой модели. Величина индекса роста, $\gamma$, в Линдер $\gamma$-параметризации для Ратра-Пиблс $\phi$CDM модели возрастает с увеличением величины модельного параметра $\alpha$. {\it Величина индекса роста, $\gamma$, в  Линдер $\gamma$-параметризации для $\phi$CDM  модели немного больше чем для $\Lambda$CDM модели.}
    \item[$\bullet$] Мы определили границы применимости Линдер $\gamma$-параметризации в Ратра-Пиблс $\phi$CDM модели, $z \in(0; 5)$.
    {\it Применимость Линдер $\gamma$-параметризации прекращается позже для $\Lambda$CDM модели, чем для $\phi$CDM модели.}
\end{itemize}

 %%%%%%%%%%%%%%%%%%%%%%%%%%%%%%%%%%%%%%%%%%%%%%%%%%%%%%%%%%%%%%%%%%%%%%%%%%%
%%%%%%%%%%%%%%%%%%%       Beginng of chapter 8    %%%%%%%%%%%%%%%%%%%%%%%
%%%%%%%%%%%%%%%%%%%%%%%%%%%%%%%%%%%%%%%%%%%%%%%%%%%%%%%%%%%%%%%%%%%%%%%%%%
\chapter{Ограничения величин модельных параметров в Ратра-Пиблс $\phi$CDM модели}\label{chapter:8}
\section{Ограничения величин модельных параметров в Ратра-Пиблс $\phi$CDM модели по данным функции темпа роста флуктуаций плотности материи}
Мы провели ограничения на величины модельных параметров $\alpha$ и $\Omega_{\rm m}$ в Ратра-Пиблс $\phi$CDM модели, используя компиляцию наблюдательных данных функции темпа роста флуктуаций плотности материи, (\cite{Gupta2011}). Эти данные представлены в Таблице \ref{table:grd}.
\begin{table}[h!]
\begin{center}
\begin{tabular}{|p{1.5cm}|p{1.5cm}||p{1.5cm}|}
\hline
\hspace{0.2cm}\textrm{$f_\mathrm{\rm obs}$} & \hspace{0.2cm}\textrm{$z$} & \hspace{0.2cm}\textrm{$\sigma$} \\
\hline\hline
\hline
$0.510$ & $0.15$ & $0.11$\\
[0.1cm]
\hline
$0.600$ & $0.22$ & $0.10$\\
[0.1cm]
\hline
$0.654$ & $0.32$ & $0.18$\\
[0.1cm]
\hline
$0.700$ & $0.35$ & $0.18$\\
[0.1cm]
\hline
$0.700$ & $0.41$ & $0.07$\\
[0.1cm]
\hline
$0.750$ & $0.55$ & $0.18$\\
[0.1cm]
\hline
$0.730$ & $0.60$ & $0.07$\\
[0.1cm]
\hline
$0.910$ & $0.77$ & $0.36$\\
[0.1cm]
\hline
$0.700$ & $0.78$ & $0.08$\\
[0.1cm]
\hline
$0.900$ & $1.40$ & $0.24$\\
[0.1cm]
\hline
$1.460$ & $3.00$ & $0.29$\\
[0.1cm]
\hline
\end{tabular}
\caption{Наблюдательные данные функции темпа роста  флуктуаций плотности материи $f_\mathrm{\rm obs}$, величины красного смещения, $z$, величина неопределенности в 1$\sigma$.}
\label{table:grd}
\end{center}
\end{table}

\begin{figure}[h!]
\begin{center}
\psfig{file=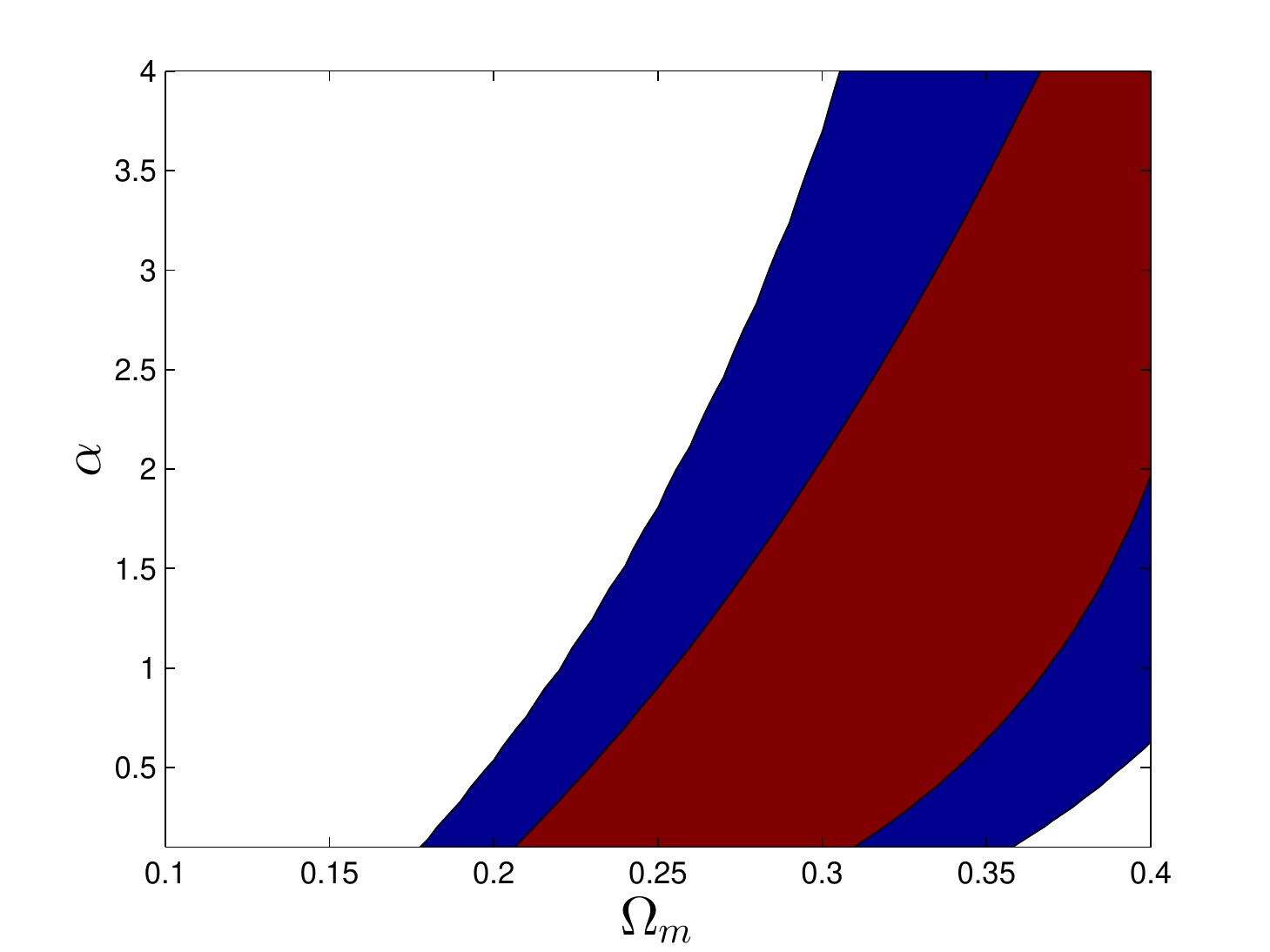,width= 0.8\columnwidth}
\end{center}
 \caption{ 1$\sigma$  и 2$\sigma$  контуры для параметров
$\Omega_{\rm m}$ и $\alpha$ в Ратра-Пиблс $\phi$CDM модели.
 Ограничения по наблюдательным данным функции темпа роста флуктуаций плотности материи, (\cite{Gupta2011}).}
\label{fig:f29}
\end{figure}

Для определения теоретических величин функции темпа роста флуктуаций плотности материи, $f_\mathrm{th}$, мы численно решили линейное уравнение флуктуаций плотности материи, Ур.~(\ref{eq:deltaeq}), для ряда величин $\alpha$ и $\Omega_\mathrm{m}$. Мы вычислили величины функции $\chi^2(\alpha,\Omega_\mathrm{obs})$ в предположении, что наблюдательные величины функции темпа роста флуктуаций плотности материи распределены в соответствии с  распределением Гаусса:
\begin{equation}
\chi^2(\alpha,\Omega_\mathrm{obs}) = \frac{[f_\mathrm{obs} - f_\mathrm{th}(\alpha,\Omega_\mathrm{m})]^2}{\sigma^2},
\end{equation}
\noindent
где $\sigma$ - среднеквадратичное отклонение наблюдательных данных функции темпа роста  флуктуаций плотности материи. Затем мы вычислили функцию вероятности, $\mathcal{L}^\mathrm{f}(\alpha,\Omega_\mathrm{m})$:
\begin{equation}
\mathcal{L}^\mathrm{f}(\alpha,\Omega_\mathrm{m})\propto \mathrm{exp}[-\chi^2(\alpha,\Omega_\mathrm{m})/2].
\end{equation}
Результаты этих вычислений представлены на Рис.~(\ref{fig:f29}).

Полученные 1$\sigma$ и 2$\sigma$ контуры в плоскости $\alpha$ - $\Omega_\mathrm{m}$ сильно вырождены относительно ограничениям на модельный параметр $\alpha$, Рис.~(\ref{fig:f29}). Таким образом, наблюдательные данные по функции темпа роста флуктуаций плотности материи сами по себе не могут одновременно ограничивать оба модельных параметра $\alpha$ и $\Omega_\mathrm{m}$ в Ратра-Пиблс $\phi$CDM модели.
Тем не менее, мы нашли ограничение на модельный параметр $\Omega_\mathrm{m}$ для $\Lambda$CDM модели и для Ратра-Пиблс $\phi$CDM модели, используя только наблюдательные величины функции темпа роста флуктуаций плотности материи. Если мы зафиксируем ординату с $\alpha=0$ на Рис.~(\ref{fig:f29}), которая соответствует пространственно-плоской $\Lambda$CDM модели, мы получим оптимальную величину $\Omega_\mathrm{m} = 0.278 \pm 0.03$, которая находится в пределах
1$\sigma$ данных Planck 2013,
\cite{Ade:2013zuv}. Для $\Lambda$CDM модели величины $0.18 \leq \Omega_{\rm m} \leq 0.36 $ содержатся в  $2\sigma$ уровня доверия, Рис.~(\ref{fig:f29}). Для Ратра-Пиблс $\phi$CDM модели величины $\Omega_{\rm m} < 0.18$ находятся за пределами $2\sigma$ уровня доверия, но величины $\Omega_{\rm m} \geq 0.36$ все еще допускаются при больших величинах параметра $\alpha$, Рис.~(\ref{fig:f29}).
\section{Ограничение на величины модельных параметров в Ратра-Пиблса $\phi$CDM модели по данным BAO}
Чтобы устранить вырождение между  модельными параметрами $\alpha$ и $\Omega_{\rm m}$, которое возникло в результате применения ограничений по наблюдательным данным функции темпа роста флуктуаций плотности материи, мы провели дополнительные ограничения этих параметров по BAO данным. В данном исследовании мы  руководствовались методом, представленным в статье (\cite{Giostri2012}).

 Мы рассчитали величины угловых расстоянй:
\begin{equation}
d_\mathrm{A}(z, \alpha, \Omega_\mathrm{m}, H_0) = \int_0^z \frac{dz'}{H(z', \alpha, \Omega_\mathrm{m}, H_0)}
\end{equation}
\noindent
и величины масштаба расширения:
\begin{equation}
D_\mathrm{V}(z, \alpha, \Omega_\mathrm{m}, H_0) =
 [d_\mathrm{A}^2(z, \alpha, \Omega_\mathrm{m}, H_0)z/H(z, \alpha, \Omega_\mathrm{m}, H_0)]^{1/3}.
 \end{equation}
\noindent
Мы сконструировали комбинацию величин угловых расстояний $d_\mathrm{A}(z_{\rm dec})$ и масштаба расширения, $D_\mathrm{V}(z_\mathrm{BAO})$, (\cite{Eisenstein:2005su}):
\begin{equation}
   \eta(z) \equiv d_\mathrm{A}(z_\mathrm{rec})/D_\mathrm{V}(z_\mathrm{BAO}).
   \label{eq:CMBBAO}
  \end{equation}

  Выражение, представленное в уравнении, Ур.~(\ref{eq:CMBBAO}), является
   BAO/CMBR ограничениями.

BАО и СМВR наблюдательные данные являются зависимыми друг от друга. Предполагая, что эти данные подчиняются закону Гаусса, мы вычислили величины функции $\chi^2_\mathrm{BАО}$, используя ковариантную обратную матрицу, $\bm{C}^{-1}$:
\begin{equation}
\chi^2_\mathrm{BАО} = \bm{X}^\mathrm{T}\bm{C}^{-1}\bm{X}.
\label{eq:Hibao}
\end{equation}
\noindent
Также мы вычислили величины функции вероятности, используя результаты уравнения, Ур.~(\ref{eq:Hibao}):
\begin{equation}
\mathcal{L}^\mathrm{BАО}(\alpha,\Omega_\mathrm{m},H_0) \propto \mathrm{exp}(-\chi^2_\mathrm{BАО}/2),
\end{equation}
\noindent
где $\bm{X} = \eta_\mathrm{th} - \eta_\mathrm{obs}$. Величина вектора, $\bm{X}$, вычисляется как:
\begin{equation}
{\bf X}=\left(
          \begin{array}{cccc}
          \displaystyle\frac{d_\mathrm{A}(z_{\rm rec})}{D_\mathrm{V}(0.106)} -30.95 \\[10pt]
            \displaystyle\frac{d_\mathrm{A}(z_{\rm rec})}{D_\mathrm{V}(0.2)} -17.55 \\[10pt]
            \displaystyle\frac{d_\mathrm{A}(z_{\rm rec})}{D_\mathrm{V}(0.35)} -10.11 \\[10pt]
             \displaystyle\frac{d_\mathrm{A}(z_{\rm rec})}{D_\mathrm{V}(0.44)} -8.44 \\[10pt]
             \displaystyle\frac{d_\mathrm{A}(z_{\rm rec})}{D_\mathrm{V}(0.6)} -6.69 \\[10pt]
              \displaystyle\frac{d_\mathrm{A}(z_{\rm rec})}{D_\mathrm{V}(0.73)} -5.45 \\
          \end{array}
        \right).
\end{equation}
Обратная ковариационная матрица наблюдательных данных, $\bf C^{-1}$,  определена как:
\begin{equation}
{\bf C^{-1}}=\left(
          \begin{array}{cccccc}
          0.48435 & -0.101383 &-0.164945 &-0.0305703 &-0.097874 & -0.106738\\
           -0.101383 & 3.2882 & -2.45497 & -0.0787898 & -0.252254 & -0.2751\\
           -0.164945 & -2.45497 & 9.55916 & -0.128187 & -0.410404 & -0.447574\\
           -0.0305703 & -0.0787898 & -0.128187 & 2.78728 & -2.75632 & 1.16437\\
          -0.097874 & -0.252254 & -0.410404 & -2.75632 & 14.9245 & -7.32441 \\
         -0.106738 & -0.2751 & -0.447574 & 1.16437 & -7.32441 & 14.5022  \\
          \end{array}
        \right).
\end{equation}
В распределении Гаусса мы использовали предварительную величину параметра Хаббла для современной эпохи, $H_0 = 74.3 \pm 2.1$, (\cite{fr12}).
Функция вероятности, полученная для функции темпа роста флуктуаций плотности материи, $\mathcal{L}^\mathrm{f}$, и функция вероятностей, полученная для BAO/CMBR ограничений, $\mathcal{L}^\mathrm{BАО}$, независимы друг от друга, поэтому, согласно уравнению, Ур.~(\ref{eq:HSM}), общая функция вероятности, $\mathcal{L}$, является результатом умножения этих функций друг на друга, т. е. $\mathcal{L}=\mathcal{L}^\mathrm{f}\cdot\mathcal{L}^\mathrm{BАО}$.

\begin{figure}[h!]
\begin{center}
\psfig{file=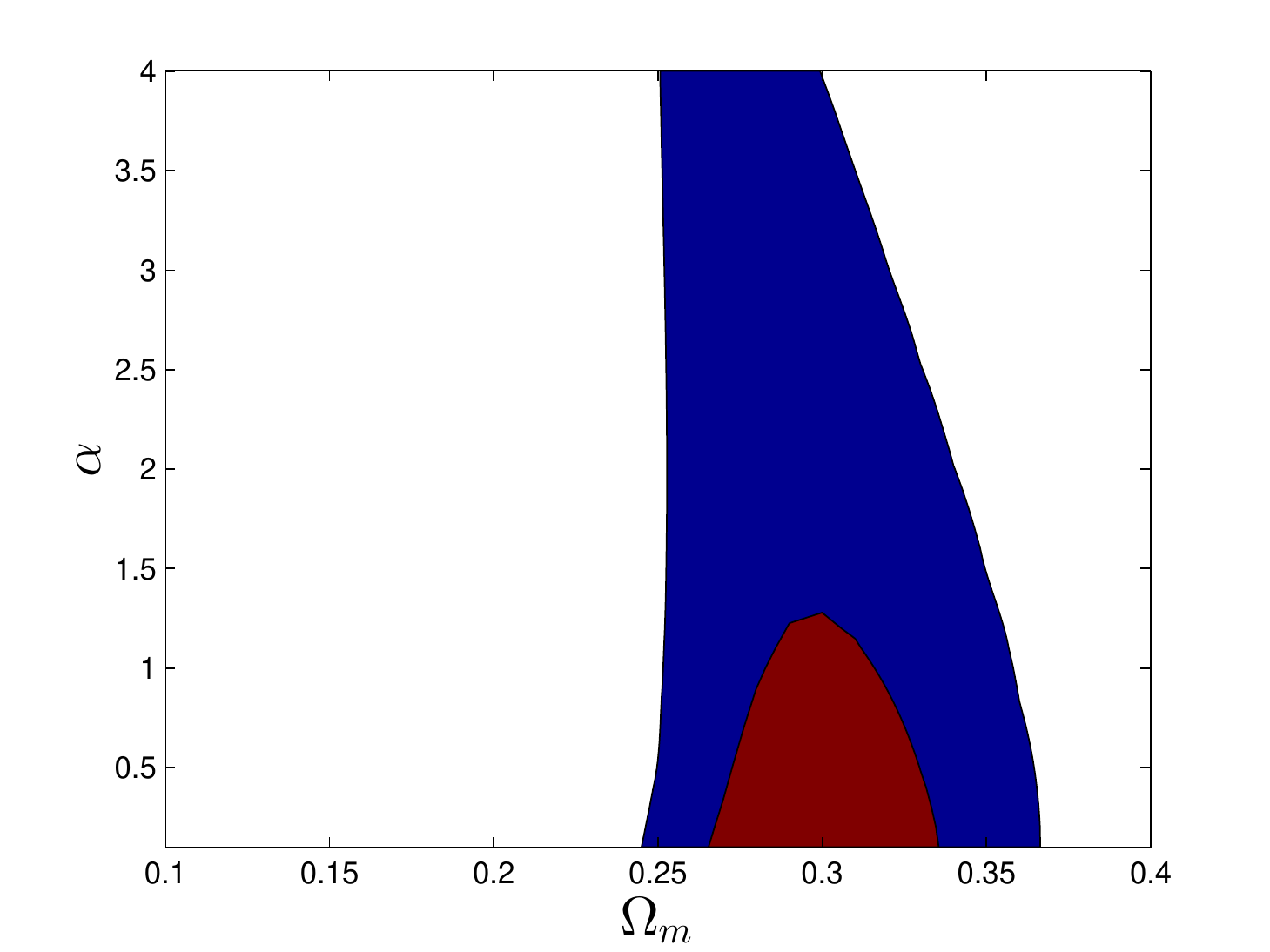,width= 0.8\columnwidth}
\end{center}
 \caption{1$\sigma$  и  2$\sigma$ контуры на параметры
$\Omega_{\rm m}$ и $\alpha$  в Ратра-Пиблс $\phi$CDM модели.
Ограничения, полученные после добавления
измерений  BAO/CMBR предварительных расстояний, (\cite{Giostri2012}).}
\label{fig:f30}
\end{figure}

Результаты наших вычислений представлены на Рис.~(\ref{fig:f30}).

После проведения BAO/CMBR анализа, мы получили новые ограничения на параметры $\Omega_{\rm m}$ и $\alpha$. Параметр $\Omega_{\rm m}$ ограничен в пределах величин $0.26 < \Omega_{\rm m} < 0.34$ для уровня доверия в 1$\sigma$. Для параметра $\alpha$ мы получили диапазон величин $0\leq\alpha\leq1.30$ для  уровня доверия в 1$\sigma$, Рис.~(\ref{fig:f30}).

\section{Заключение}
Для ограничения параметров в Ратра-Пиблс $\phi$CDM модели скалярного поля, мы применили компиляцию наблюдательных данных: функции темпа роста флуктуаций плотности материи, BAO и предварительного расстояния от CMBR.
При применении только наблюдательных данных функции темпа роста флуктуаций плотности материи происходит сильное вырождение между модельными параметрами
$\Omega_\mathrm {m} $ и $\alpha$. Это означает, что
допускаются более большие величины параметра $\alpha$
при увеличении величины параметра $\Omega_\mathrm{m}$.
Вырожденность устраняется
после объединения ограничений по наблюдательным данным функции темпа роста флуктуаций плотности материи и ограничений по наблюдательным данным от отношения расстояния-красного смещения данных BAO и предварительного расстояния от CMBR.

В результате, мы получили ограничения на модельные параметры в Ратра-Пиблс $\phi$CDM модели скалярного поля: $\Omega_\mathrm{m} = 0.30 \pm 0.04$ и $0\leq\alpha\leq1.30$ для  уровня доверия в 1$\sigma$. Оптимальная величина для параметра $\alpha$, $\alpha = 0.00$.

%%%%%%%%%%%%%%%%%%%%%%%%%%%%%%%%%%%%%%%%%%%%%%%%%%%%%%%%%%%%%%%%%%%%%%%%%%
%%%%%%%%%%%%%%%%%%%       Beginning of chapter 9     %%%%%%%%%%%%%%%%%%%%%%%
%%%%%%%%%%%%%%%%%%%%%%%%%%%%%%%%%%%%%%%%%%%%%%%%%%%%%%%%%%%%%%%%%%%%%%%%%%
%%%%%%%%%%%%%%%%%%%%%%%%%%%%%%%%%%%%%%%%%%%%%%%%%%%%%%%%%%%%%%%%%%%%%%%%%%
\chapter{Ограничения величин модельных параметров в квинтэссенциальных и фантомных $\phi$CDM моделях по наблюдательным данным}\label{chapter:9}

Эта глава основана на исследованиях, представленных в статье, (\cite{Avsajanishvili:2017zoj}).

Мы изучали квинтэссенциальные (канонические скалярные поля) и фантомные (не канонические скалярные поля) модели темной энергии для плоской вселенной. До сих пор нет окончательного решения относительно того, какая из этих моделей предпочтительна для объяснения темной энергии на основе результатов различных наблюдений, (\cite{Suzuki:2011hu}, \cite{Novosyadlyj:2012qu}, \cite{Ade:2013zuv}, \cite{Betoule:2014frx},  \cite{Ade:2015xua}). Мы применили спрогнозированные данные, которые были вычислены  для предстоящего эксперимента Dark Energy Spectroscopic Instrument (DESI) и изучили $\phi$CDM модели в сравнении со стандартной $\Lambda$CDM моделью. Наше исследование основано на сравнении данных: темпа расширения вселенной, функции темпа роста флуктуаций плотности материи и угловых расстояний, которые будут получены из эксперимента DESI.
\section{Метод определения модельных параметров и начальных условий}
Мы изучали $\phi$CDM модели с 10 квинтэссенциальными и 7 фантомными потенциалами, список которых представлен в Таблице \ref{table:QP} и в Таблице \ref{table:PP}. Все эти модели имеют одинаковые параметры $\Omega_\mathrm{m0}$ и $H_\mathrm{0}$. Кроме этих параметров каждая $\phi$CDM модель имеет свой набор дополнительных модельных параметров, определяющих форму потенциала, $V(\phi)$.

Для каждого потенциала, для широкого диапазона величин свободных параметром и начальных условий,  ($\phi_0$, $\dot{\phi_0}$), мы численно решили систему дифференциальных уравнений: уравнение движения Клейн-Гордона для квинтэссенциальных/фантомных моделей, соответственно, Ур.~(\ref{eq:KleinGordon})/ Ур.~(\ref{eq:KleinGordon1}), и первое уранение Фридмана, Ур.~(\ref{eq:freqphi}), а затем линейное уравнение флуктуаций плотности материи, Ур.~(\ref{eq:deltaeq}).
   В связи с тем, что для всех потенциалов диапазоны величин начальных условий и модельных параметров точно не известны, мы разработали метод определения этих диапазонов.
    Для каждого потенциала мы нашли достоверные решения, для которых одновременно должны выполняться следующие три критерия:
\begin{enumerate}[1.]
 \item
Момент равенства между плотностью материи и плотностью темной энергии,
 $\Omega_\mathrm{m}=\Omega_\mathrm{\phi}$, происходит относительно недавно, при $a\in(0.6;0.8)$, Рис.~(\ref{fig:f26})~(правая панель).

\item
Функция темпа роста флуктуаций плотности материи, $f(a)$, и фракционная плотность материи, $\Omega_{\rm m}(a)$, параметризированы Линдер $\gamma$-параметризацией, $\gamma$, которая описывается уравнением, Ур.~(\ref{eq:Lgamma}).

\item
Выбранные величины параметра уравнения состояния в современную эпоху должны соответствовать ожидаемым величинам параметра уравнения состояния в современную эпоху для этих моделей: для фантомных моделей  $w_0<-1$; для моделей квинтэссенции  $-1<w_0<-0.75$: для моделей замерзания: $w_a<0$ и для моделей таяния: $w_a>0$.
\end{enumerate}
Несмотря на то что потенциал Ратра-Пиблса имеет аттракторное решение, для лучшей численной сходимости мы выбрали определенное решение со следующими начальными условиями, ($\phi_0$, $\dot{\phi_0}$), и модельным параметром, $V_0$,
(\cite{Ratra:1987aj}, \cite{Farooq:2013syn}, \cite{avs14}):
\begin{align}\label{eq:Incond}
V_0 &= \frac{8}{3} \left(\frac{\alpha+4}{\alpha+2}\right)
\Bigl[\frac{2}{3}\alpha(\alpha+2)\Bigl]^{\alpha/2},\\
\phi_{\rm in} &= \left[\frac{2}{3}\alpha(\alpha + 2)\right]^{1/2}t_{\rm in}^{\frac{3}{\alpha + 2}},\\
\dot{\phi}_{\rm in} &= \left[\dfrac{6\alpha}{\alpha + 2}\right]^{1/2}t_{\rm in}^{\frac{1-\alpha}{2+\alpha}}.
\end{align}
Начальная величина скалярного фактора, $a_{\rm in} \propto t_{\rm in}^{2/3}$, была выбрана для эпохи доминирования материи, Ур.~(\ref{eq:DWR}). В наших вычислениях мы применяли величины  модельного параметра $\alpha$ в диапазоне, $0<\alpha\leq0.7$, (\cite{Samushia:2009dd}).

В результате применения феноменологического метода, описанного выше, для каждого потенциала мы нашли диапазоны: возможных начальных условий и свободных модельных параметров, описывающих форму потенциала, $V(\phi)$. Эти величины, совместно с общими свободными модельными параметрами $\Omega_\mathrm{m0}$ and $H_\mathrm{0}$ представлены в Таблице \ref{table:QP} и Таблице \ref{table:PP}.
Мы использовали эти данные как начальные условия (предварительные диапазоны) в MCMC анализе для каждой $\phi$CDM модели.
\begin{table*}[h!]
\begin{tabular}{|m{7cm}|m{3cm} m{3cm}|}
%\begin{tabular}{|p{4cm}|p{12.5cm}|}
\hline %\hline
\multicolumn{1}{|C{7cm}|}{Потенциалы квинтэссенции} & \multicolumn{2}{C{8cm}|}{Свободные параметры} \\
\hline\hline
\hline
 $V(\phi)=V_0M_\mathrm{pl}^2\phi^{-\alpha}$  &   $H_\mathrm{0}(50\div90)$\par
                                                 $\Omega_\mathrm{m0}(0.25\div0.32)$ &
                                                 $V_0(3\div5)$ \par
                                                 $\alpha(10^{-6}\div0.7)$ \\

\hline
$V(\phi)=V_0\exp(-\lambda\phi/M_{\rm pl})$ & $H_\mathrm{0}(50\div90)$\par $\Omega_\mathrm{m0}(0.25\div0.32)$ \par $V_0(10\div10^3)$ & $\lambda(10^{-7}\div10^{-3})$ \par  $\phi_0(0.2\div1.6)$ \par $\dot{\phi_0}(79.8\div338.9)$\\

\hline
$V(\phi)=V_0(\exp({M_{\rm pl}/\phi})-1)$  &   $H_\mathrm{0}(50\div90)$\par $\Omega_\mathrm{m0}(0.25\div0.32)$ \par  $V_0(10\div10^2)$ & $\phi_0(1.5\div10)$ \par $\dot{\phi_0}(350\div850)$\\

\hline
$V(\phi)=V_0\phi^{-\chi}\exp(\gamma\phi^2/M_{\rm pl}^2)$   & $H_\mathrm{0}(50\div90)$\par $\Omega_\mathrm{m0}(0.25\div0.32)$ \par $V_0(10^{-2}\div10^{-1})$ \par $\chi(4\div8)$ & $\gamma(6.5\div7)$ \par  $\phi_0(5.78\div10.55)$ \par $\dot{\phi_0}(680.6\div879)$\\

\hline
$V(\phi)=V_0(\cosh(\varsigma\phi)-1)^g$ &  $H_\mathrm{0}(50\div90)$\par $\Omega_\mathrm{m0}(0.25\div0.32)$ \par   $V_0(5\div8)$ \par  $\varsigma(0.15\div1)$ &  $g(0.1\div0.49)$ \par  $\phi_0(1.8\div5.8)$ \par $\dot{\phi_0}(360\div685)$\\

\hline
$V(\phi)=V_0(\exp(\nu\phi) + \exp(\upsilon\phi))$ & $H_\mathrm{0}(50\div90)$\par $\Omega_\mathrm{m0}(0.25\div0.32)$ \par $V_0(1\div12)$ & $\nu(6\div12)$ \par  $\phi_0(0.014\div1.4)$ \par $\dot{\phi_0}(9.4\div311)$\\

\hline
$V(\phi)=V_0((\phi-B)^2 + A)\exp(-\mu\phi)$  & $H_\mathrm{0}(50\div90)$\par $\Omega_\mathrm{m0}(0.25\div0.32)$ \par $V_0(40\div70)$ \par $A(1\div40)$ & $B(1\div60)$ \par $\mu(0.2\div0.9)$ \par $\phi_0(5.8\div8.45)$ \par  $\dot{\phi_0}(681\div804.5)$\\

\hline
$V(\phi)=V_0\sinh^m(\xi M_\mathrm{pl}\phi)$ & $H_\mathrm{0}(50\div90)$\par $\Omega_\mathrm{m0}(0.25\div0.32)$ \par $V_0(1\div10)$ \par $m(-0.1\div-0.3)$  & $\xi(10^{-2}\div1)$ \par $ \phi_0(0.5\div2.5)$ \par $\dot{\phi_0}(190\div367)$\\

\hline
$V(\phi)=V_0\exp({M_{\rm pl}/\phi})$ & $H_\mathrm{0}(50\div90)$\par $\Omega_\mathrm{m0}(0.25\div0.32)$ \par $V_0(10^2\div10^3)$ & $ \phi_0(5.78\div10.55)$ \par $\dot{\phi_0}(680.6\div879)$\\

\hline
$V(\phi)=V_0(1+\exp(-\tau\phi))$ & $H_\mathrm{0}(50\div90)$\par $\Omega_\mathrm{m0}(0.25\div0.32)$ \par $V_0(1\div10^2)$ & $\tau(10\div10^2)$ \par $\phi_0(0.01\div0.075)$ \par $\dot{\phi_0}(9.4\div32)$\\

\hline
\end{tabular}
\caption{\rm Список потенциалов квинтэссенции и свободных параметров.}
\label{table:QP1}
\end{table*}
%\bigskip
\begin{table*}[h!]
\begin{tabular}{|m{7cm}|m{3cm} m{3cm}|}
\hline %\hline
\multicolumn{1}{|C{7cm}|}{Фантомные потенциалы} & \multicolumn{2}{C{8cm}|}{Свободные параметры} \\
\hline\hline
\hline
$V(\phi)=V_0\phi^5$ &  $H_\mathrm{0}(50\div90)$\par $\Omega_\mathrm{m0}(0.25\div0.32)$ \par $V_0(10^{-3}\div10^{-2})$ & $\phi_0(3.37\div3.94)$ \par $\dot{\phi_0}(523\div563.6)$\\

\hline
 $V(\phi)=V_0\phi^{-2}$ & $H_\mathrm{0}(50\div90)$\par $\Omega_\mathrm{m0}(0.25\div0.32)$ \par $V_0(30\div50)$ & $\phi_0(2.83\div5.15)$ \par $\dot{\phi_0}(471.4\div600)$\\

\hline
$V(\phi)=V_0\exp(\beta\phi)$ &   $H_\mathrm{0}(50\div90)$\par $\Omega_\mathrm{m0}(0.25\div0.32)$ \par $V_0(1\div20)$ & $\beta(0.08\div0.3)$ \par $ \phi_0(0.2\div9.14)$ \par $\dot{\phi_0}(79.8\div830.9)$\\

\hline
$V(\phi)=V_0\phi^2$ &  $H_\mathrm{0}(50\div90)$\par $ \Omega_\mathrm{m0}(0.25\div0.32)$ \par $V_0(1\div20)$ & $\phi_0(0.67\div2.8)$ \par $\dot{\phi_0}(191\div450)$\\

\hline
$V(\phi)=V_0(1-\exp(\phi^2/\sigma^2))$ &   $H_\mathrm{0}(50\div90)$\par $ \Omega_\mathrm{m0}(0.25\div0.32)$ \par $V_0(5\div30)$ & $\sigma(5\div30)$ \par $\phi_0(0.67\div2.8)$ \par $\dot{\phi_0}(191\div450)$\\

\hline
$V(\phi)=V_0(1-\cos(\phi/\kappa))$ &  $H_\mathrm{0}(50\div90)$\par $ \Omega_\mathrm{m0} (0.25\div0.32)$ \par $V_0(1\div4)$ & $\kappa(1.1\div2)$ \par $ \phi_0(2.3\div3.37)$ \par $ \dot{\phi_0}(420\div500)$\\

\hline
 $V(\phi)=V_0(\cosh(\psi \phi))^{-1}$ & $H_\mathrm{0}(50\div90)$\par $ \Omega_\mathrm{m0}(0.25\div0.32)$ \par $V_0(10^{-3}\div10^2)$ & $ \psi(10^{-3}\div1)$ \par $  \phi_0(1.4\div2.3)$ \par $\dot{\phi_0}(310\div420.7)$\\

\hline
\end{tabular}
\caption{\rm Список фантомных потенциалов и свободных параметров.}
\label{table:PP1}
\end{table*}
\section{МCMC анализ для исследования моделей темной энергии}
Для всех моделей темной энергии мы вычислили величины нормированного параметра Хаббла, углового расстояния и функции темпа роста флуктуаций плотности материи в диапазоне величин красного смещения, $z\in(0.15;1.85)$:
\begin{itemize}
\item[$\bullet$]{\it Нормированный параметр Хаббла, $E$(z)}\\
Для вычисления величины нормированного параметра Хаббла, $E(z)$, мы использовали уравнение, Ур.~(\ref{eq:freqphi}).
\item[$\bullet$]{\it Угловое расстояние, $d_A$(z)}\\
Мы вычисляли угловые расстояния, применяя уравнение:
\begin{equation}
d_A(z)=\frac{1}{H_\mathrm{0}(1+z)}\int_0^z\frac{dz'}{E(z')}.
\label{eq:ADD5}
\end{equation}
Это уравнение является частным случаем для пространственно-плоской вселенной, оно было получено из уравнения, Ур.~(\ref{eq:ADD2}).
\item[$\bullet$]{\it Комбинация функции темпа роста флуктуаций плотности материи и амплитуды спектра мощности, $f(a)\sigma_8$(a)}\\
Вычисление величины функции темпа роста флуктуаций плотности материи, $f(a)$, проводилось с использованием уравнениия, Ур.~(\ref{eq:GR}). Амплитуду спектра мощности материи можно определить через функцию, $\sigma_8(a)$. Функция $\sigma_8(a)\equiv D(a)\sigma_8$, где $\sigma_8 \equiv \sigma_8(a_0)$,  является среднеквадратичными линейными флуктуациями в распределении плотности материи на масштабах $8h^{-1}$ Mпс. Мы зафиксировали величину $\sigma_8=0.815$, которая наилучшим образом соответствует $\Lambda$CDM модели, (\cite{Ade:2015xua}).
\end{itemize}
Ввиду того, что наблюдательные данные для темпа расширения вселенной, $H(z)$, функции темпа роста флуктуаций плотности материи, $f\sigma_8(a)$, и  угловых расстояний, $D_A(z)$, являются зависимыми друг от друга, мы вычислили ковариантные матрицы для этих измерений. Для этих расчетов мы следовали стандартному подходу, применяемому при вычислении матриц Фишера, (\cite{Font-Ribera:2013rwa}).
Мы предположили покрытие в 14000 кв. градусов неба, при этом величина волновых чисел достигает величины $k_\mathrm{max} = 0.2\ \mathrm{Mpc}/h$. В наших расчетах мы применяли величины среднеквадратических отклонений в соответствии с данными, представленными в Таблице V, (\cite{Font-Ribera:2013rwa}). Мы также учитывали ковариации между измерениями в одном и том же диапазоне величин красного смещения. Измерения $D_A(z)$ и $H(z)$ коррелируют между собой с отрицательным знаком примерно на $40\%$, а величина корреляции измерений $D_A(z)$ и $H(z)$ с измерениями $f\sigma_8(a)$ меньше чем на $10\%$ для всех диапазонов величин красного смещения.

\begin{figure}[h!]
\begin{center}
\includegraphics[width=\columnwidth]{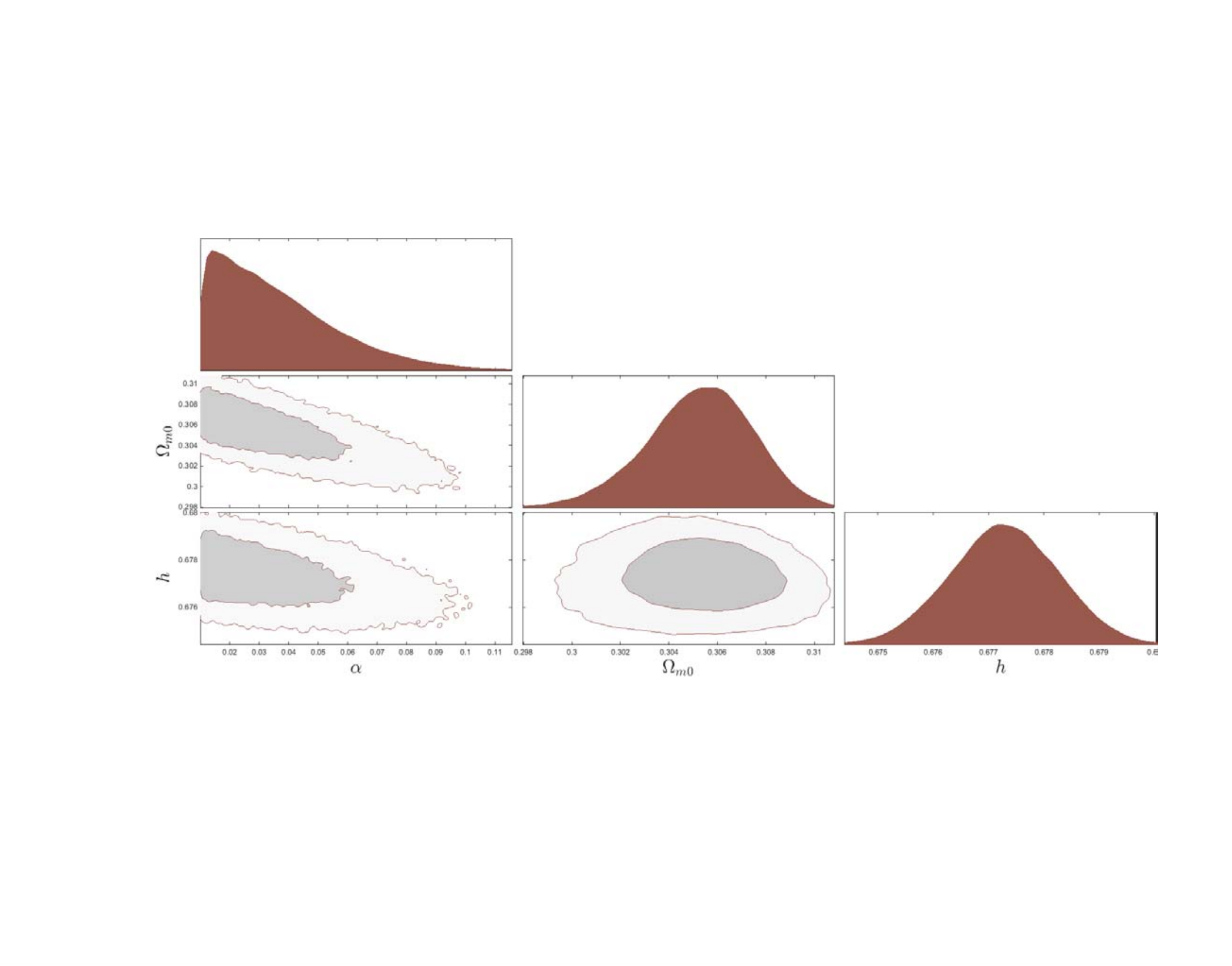}
\end{center}
 \caption{Контуры с 2$\sigma$ уровнем доверия для различных пар свободных параметров  ($\alpha$, $\Omega_{\rm m0}$, $h$), для которых
$\phi$CDM модель с потенциалом Ратра-Пиблса
$V(\phi)=V_0M_\mathrm{pl}^2\phi^{-\alpha}$ наилучшим образом соответствует $\Lambda$CDM модели.}
\label{fig:f31}
\end{figure}

\begin{figure}[h!]
\begin{center}
\includegraphics[width=1\columnwidth]{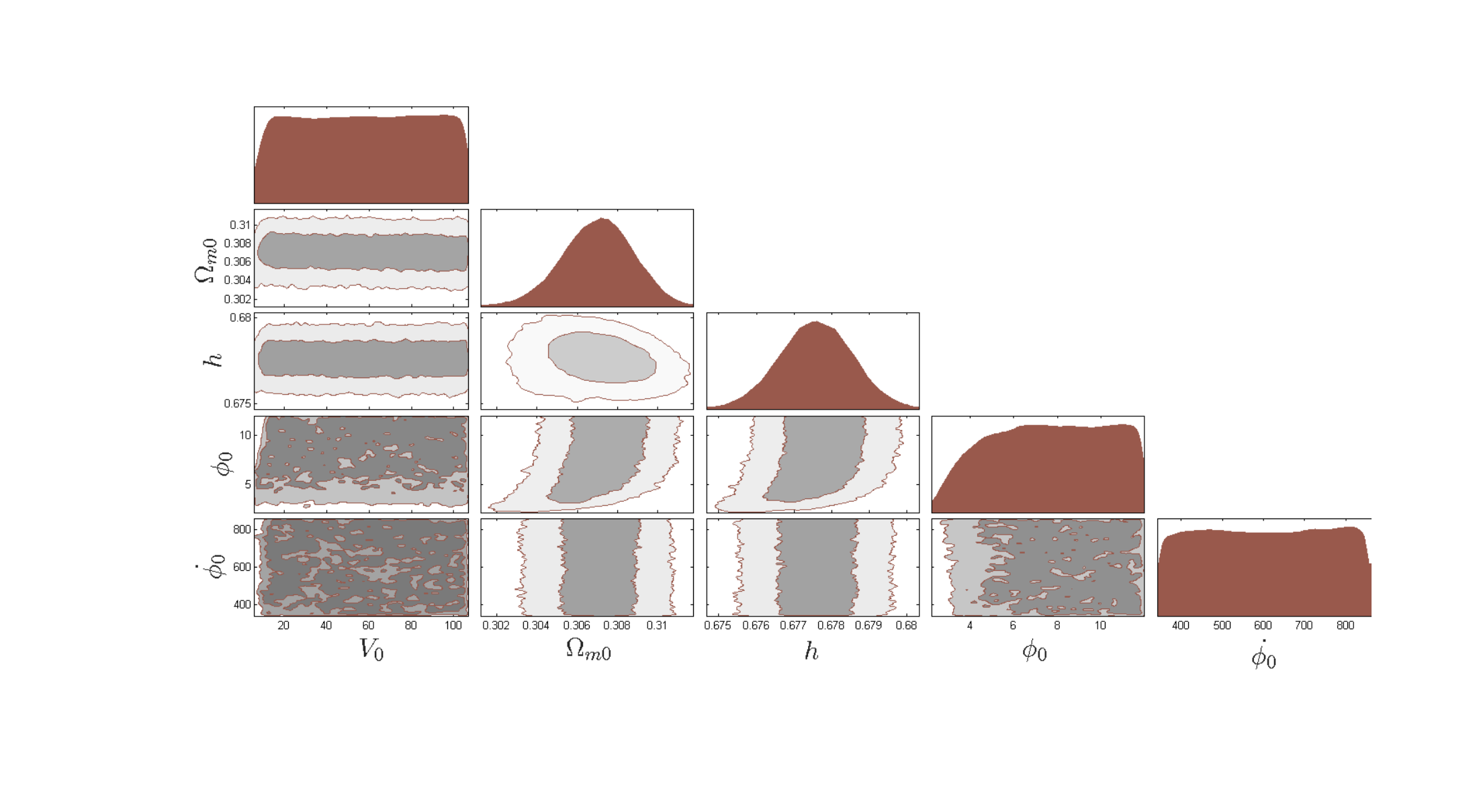}
\end{center}
 \caption {Контуры с 2$\sigma$ уровнем доверия  для различных пар свободных параметров ($V_0$, $\Omega_{\rm m0}$, $h$, $\phi_0$, $\dot{\phi_0}$), для которых
$\phi$CDM модель с Златев-Ванг-Стейнхардт (Zlatev-Wang-Steinhardt) потенциалом  $V(\phi)=V_0(\exp({M_{\rm pl}/\phi})-1)$ наилучшим образом соответствует $\Lambda$CDM модели.}
\label{fig:f32}
\end{figure}

\begin{figure}[h!]
\psfig{file=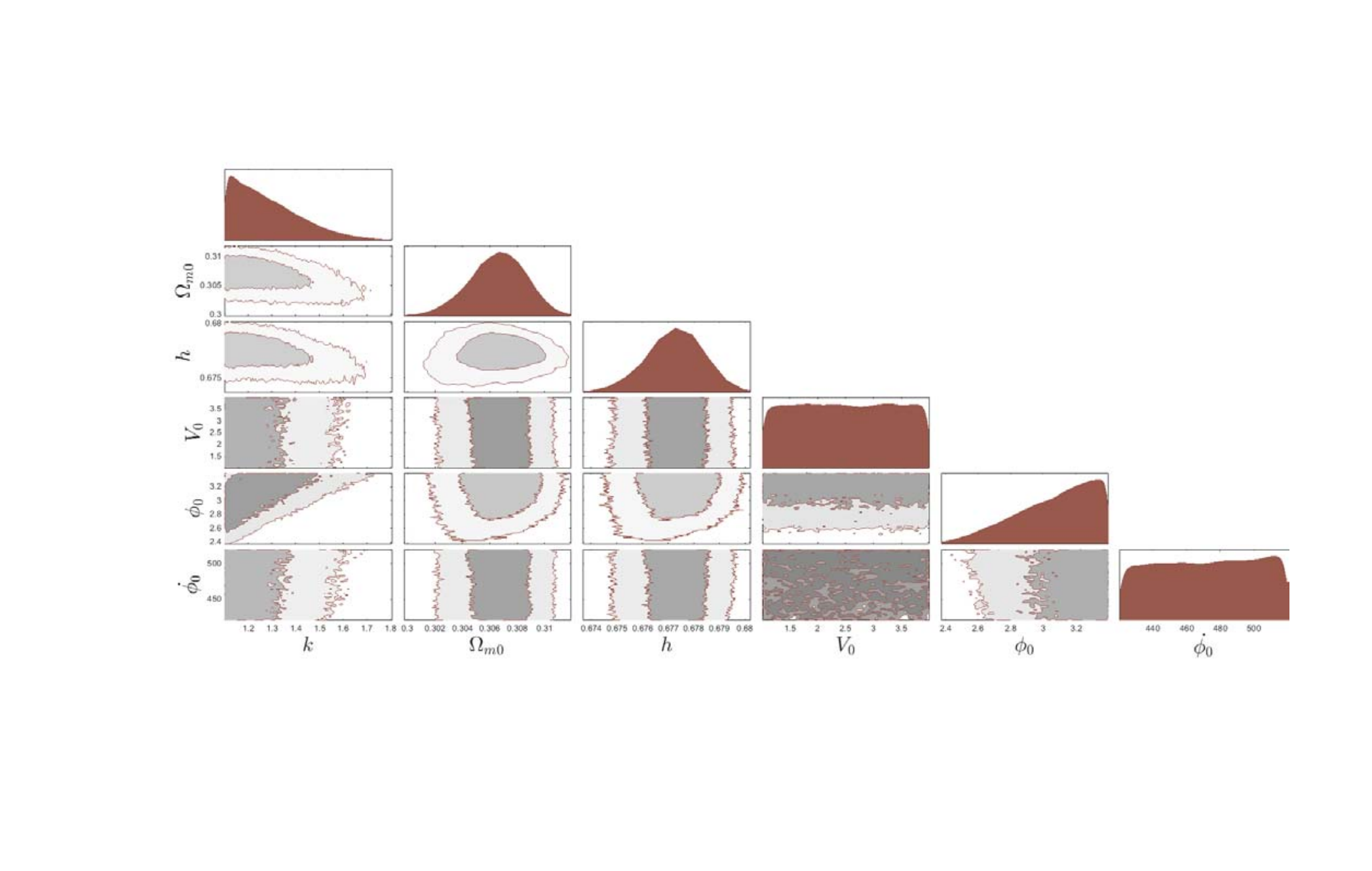,width= 1\columnwidth}
 \caption{Контуры с 2$\sigma$ уровнем доверия для различных пар свободных параметров ($k$, $\Omega_{\rm m0}$, $h$, $V_0$, $\phi_0$, $\dot{\phi_0}$), для которых
$\phi$CDM модель с фантомным Псевдо-Намбу-Голдстоун бозон (pNGb) потенциалом $V(\phi)=V_0(1-\cos(\phi/\kappa))$ наилучшим образом соответствует $\Lambda$CDM модели.}
\label{fig:f33}
\end{figure}

 Проведя MCMC анализ, мы обнаружили, что величины модельных параметров, которым соответствует максимальная вероятность (апостериорные диапазоны), находятся в пределах предварительных диапазонов величин этих модельных параметров, представленных в Таблице \ref{table:QP1} и Таблице \ref{table:PP1}. Таким образом, предварительные диапазоны величин модельных параметров не нуждаются в корректировке.
Примеры ограничений, полученных в результате проведения MCMC анализа, для квинтэссенциальных потенциалов Ратра-Пиблса (Ratra-Peebles),  Златев-Вэнг-Стейнхардта (Zlatev-Wang-Steinhardt), а также для фантомного потенциала Псевдо-Намбу-Голдстоун бозон (pNGb) показаны  на Рис.~(\ref{fig:f31} - \ref{fig:f33}).

\section{Статистика Байеса}
Для качественной оценки моделей, для возможности различения этих моделей друг от друга, а также для определения более предпочтительных моделей в сравнении с $\Lambda$CDM моделью, мы провели статистический анализ Байеса. С этой целью мы вычислили Акайке (Akaike) ($AIC$), (\cite{Akaike1974}), и Байесовский (Bayesian) ($BIC$), (\cite{Schwarz1978}), статистические информационные критерии, а также доказательства Байеса.

Статистические информационные критерии $AIC$ и $BIC$ являются функциями количества модельных параметров, $N$. Информация, полученная при вычислении величин $AIC$ и $BIC$ дополняет друг друга.
Статистические информационные критерии $AIC$ и $BIC$ определяются, соответственно, как:
\begin{equation}
AIC = -2\ln{\mathcal{L}_{\rm max}} + 2k
\end{equation}
и
\begin{equation}
BIC = -2\ln{\mathcal{L}_{\rm max}} + k\ln{N},
\end{equation}
где $\mathcal{L}_{\rm max}\propto \mathrm{exp}(-\chi^2_\mathrm{min}/2)$ - максимальная величина функции вероятности, $k$ - количество данных.

Доказательства Байеса для модели с набором параметров, $\bm{p}$, определяются интегралом:
\begin{equation}
\mathcal{E} = \int \mathrm{d}^3\bm{p} \mathcal{P}(\bm{p}),
 \label{eq:evidence}
\end{equation}
где $\mathcal{P}$ - величины апостериорных вероятностей, которые прямопропорциональны локальной плотности точек в MCMC анализе. Величины границ интегрирования определены из предварительно найденных диапазонов величин модельных параметров из Таблицы \ref{table:QP} и из Таблицы \ref{table:PP}. Модели с более высокой величиной доказательства Байеса предпочительнее чем модели с более низкой величиной доказательства Байеса.

Мы исследовали, насколько узки должны быть предварительные диапазоны величин дополнительных параметров для конкурентоспособности (в смысле
доказательств Байеса) со стандартной $\Lambda$CDM моделью.
Мы проверили, что предварительные диапазоны дополнительных параметров включают в себя большую часть величин из апостериорных диапазонов. Для всех моделей мы численно проинтегрировали апостериорную вероятность. Результаты интегрирования представлены в Таблице \ref{table:bicq} и в Таблице \ref{table:bicp}.
Все эти числа нормированы относительно $\Lambda$CDM модели.
\begin{table*}[t]
\begin{center}
\begin{tabular}{|l|l|l|l|}
\hline
\hspace{0.5 cm}\textrm{Потенциалы квинтэссенции} & \textrm{AIC}  & \textrm{BIC} & \textrm{Байес фактор}\\
\hline\hline
\hline
 $V(\phi)=V_0M_\mathrm{pl}^2\phi^{-\alpha}$&10&18.7&0.5293\\
[0.2cm]
\hline
$V(\phi)=V_0\exp(-\lambda\phi/M_{\rm pl})$&12&22.4&0.0059\\
[0.2cm]
\hline
$V(\phi)=V_0(\exp({M_{\rm pl}/\phi})-1)$&10&18.7&0.0067\\
[0.2cm]
\hline
$V(\phi)=V_0\phi^{-\chi}\exp(\gamma\phi^2/M_{\rm pl}^2)$&14&26.2&0.0016\\
[0.2cm]
\hline
$V(\phi)=V_0(\cosh(\varsigma\phi)-1)^g$&14&26.2&0.0012\\
[0.2cm]
\hline
$V(\phi)=V_0(\exp(\nu\phi) + \exp(\upsilon\phi))$&14&26.2&0.0053\\
[0.2cm]
\hline
$V(\phi)=V_0((\phi-B)^2 + A)\exp(-\mu\phi)$&16&29.9&0.0034\\
[0.2cm]
\hline
$V(\phi)=V_0\sinh^m(\xi M_\mathrm{pl}\phi)$&14&26.2&0.0014\\
[0.2cm]
\hline
$V(\phi)=V_0\exp({M_{\rm pl}/\phi})$&10&18.7&0.0077\\
[0.2cm]
\hline
$V(\phi)=V_0(1+\exp(-\tau\phi))$&12&22.4&0.0024\\
[0.2cm]
\hline
\end{tabular}
\caption{\rm Список потенциалов  квинтэссенции с соответствующими величинами $AIC$, $BIC$ и фактора Байеса.}
\label{table:bicq}
\end{center}
\end{table*}

\begin{table*}[t]
\begin{center}
\begin{tabular}{|l|l|l|l|}
\hline %\hline
\hspace{0.2 cm}\textrm{Фантомные потенциалы} &\textrm{AIC}  &\textrm{BIC} & \textrm{Байес фактор}\\
\hline\hline
\hline
$V(\phi)=V_0\phi^5$ &10.0 &18.7 &0.0921\\
[0.2cm]
\hline
$V(\phi)=V_0\phi^{-2}$  &10.0 &18.7 &0.0142\\
[0.2cm]
\hline
$V(\phi)=V_0\exp(\beta\phi)$ &22.4 &12.0  &0.0024\\
[0.2cm]
\hline
$V(\phi)=V_0\phi^2$  &10.0& 18.7& 0.0808\\
[0.2cm]
\hline
$V(\phi)=V_0(1-\exp(\phi^2/\sigma^2))$  &12.0& 22.4& 0.0113\\
[0.2cm]
\hline
$V(\phi)=V_0(1-\cos(\phi/\kappa))$  &12.0& 22.4 &0.0061\\
[0.2cm]
\hline
 $V(\phi)=V_0(\cosh(\psi \phi))^{-1}$  &12.0& 22.4 &0.0056\\
[0.2cm]
\hline
\end{tabular}
\caption{\rm Список фантомных потенциалов с соответственными величинами $AIC$, $BIC$ и фактора Байеса.}
\label{table:bicp}
\end{center}
\end{table*}
\section{$\phi$CDM модели в CPL фазовом пространстве}
Чтобы проверить насколько хорошо $\phi$CDM модели согласуются с $\Lambda$CDM моделью и как они отличаются друг от друга, мы представили набор возможных величин параметров уравнения состояния  $w_0$ и $w_a$ для каждого потенциала темной энергии в CPL-$\Lambda$CDM фазовом пространстве.

\begin{figure}[h!]
\begin{center}
\includegraphics[width=\columnwidth]{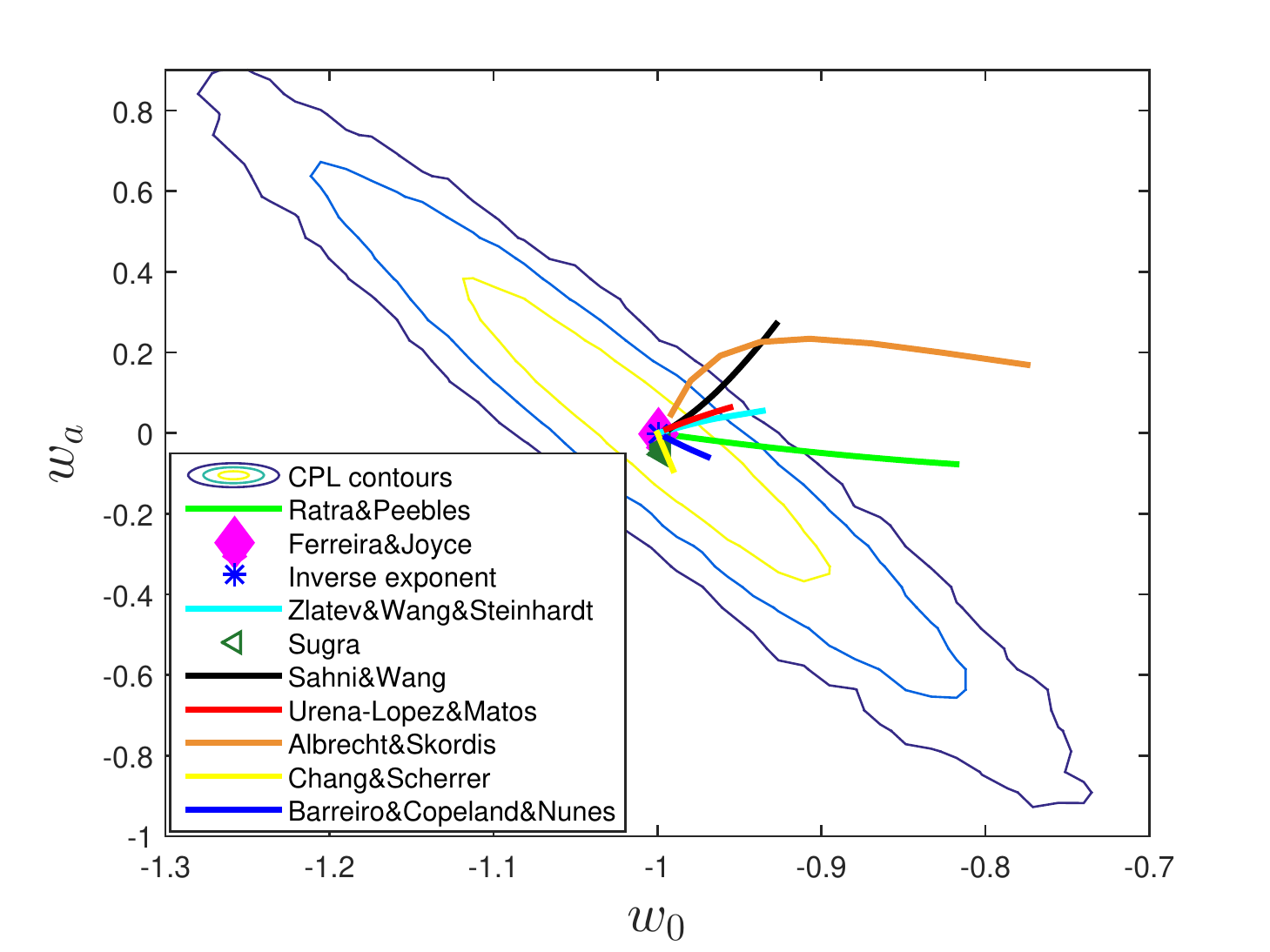}
\end{center}
 \caption{Сравнение возможных $w_0$ и $w_a$ величин для различных потенциалов $\phi$CDM моделей  квинтэссенции скалярного поля с 3$\sigma$ уровня доверия CPL-$\Lambda$CDM  контурами.}
 \label{fig:f34}
\end{figure}

\begin{figure}[h!]
\begin{center}
\includegraphics[width=\columnwidth]{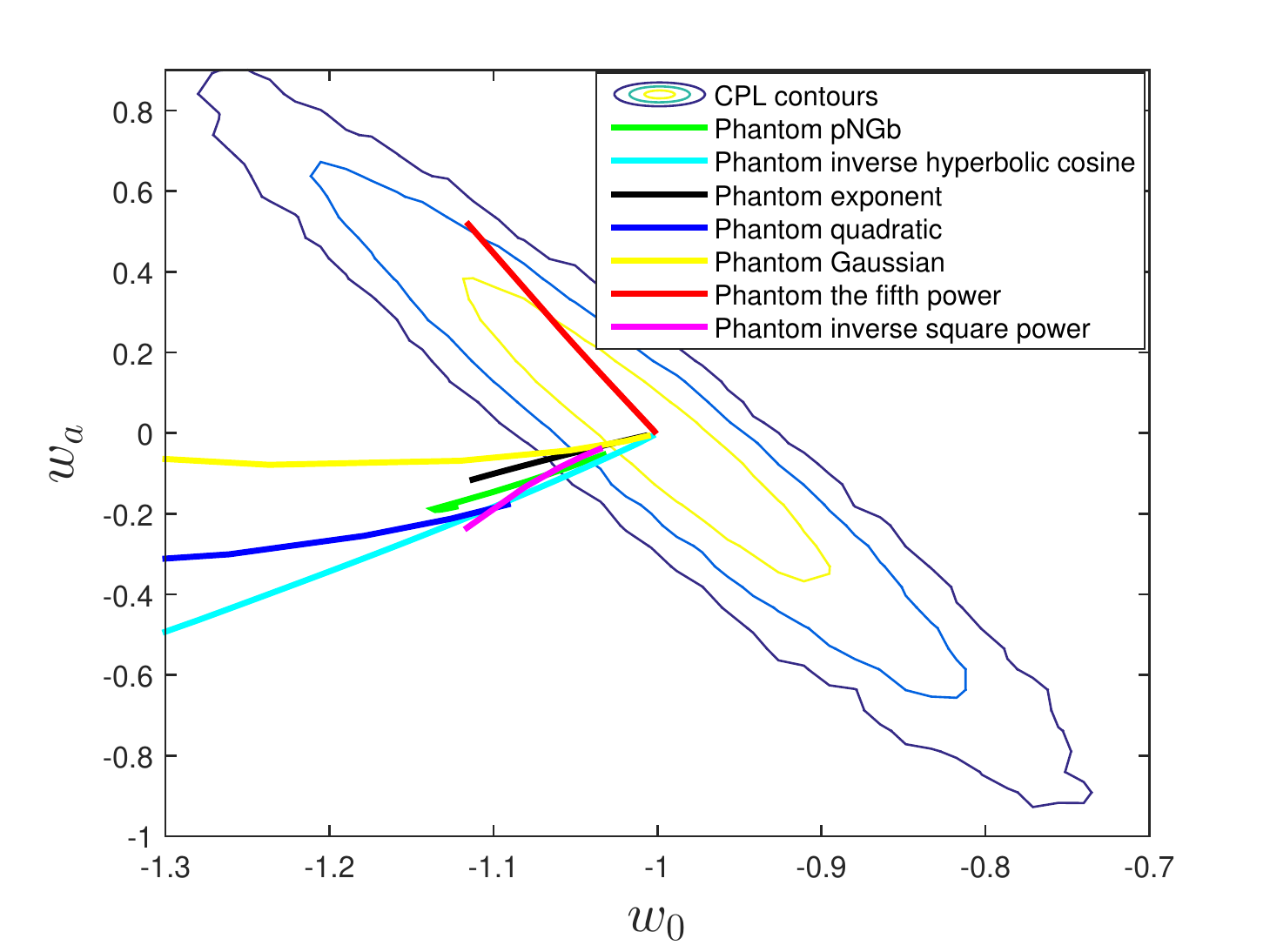}
\end{center}
 \caption {Сравнение возможных $w_0$ и $w_a$ величин для различных потенциалов $\phi$CDM фантомных моделей скалярного поля с 3$\sigma$ уровня доверия CPL-$\Lambda$CDM  контурами.}
 \label{fig:f35}
 \end{figure}

Отображение $\phi$CDM моделей на плоскости $w_0-w_a$ показаны на Рис.~(\ref{fig:f34}) для моделей квинтэссенции и на Рис.~(\ref{fig:f35}) для фантомных моделей. Эти кривые представляют собой максимальный диапазоны эффективных
величин параметра уравнения состояния, $w(a)$, для каждого потенциала на плоскости $w_0-w_a$. Эти диапазоны были получены для различных величин параметров модели или начальных условий из предварительных диапазонов этих величин.
Контуры CPL-$\Lambda$CDM с 1$\sigma$, 2$\sigma$ и 3$\sigma$ интервалами уровня доверия получены при аппроксимации CPL параметризации данными, $H(z)$, $d_A(z)$ и $f(a)\sigma_8(a)$, для каждой исследуемой модели и для $\Lambda$CDM модели.

На Рис.~(\ref {fig:f34}) показано, что
диапазон изменений величин параметра уравнения состояния для потенциалов: Феррейра-Джойса (Ferreira-Joyce), обратной экспоненты (inverse exponent) и Сугра (Sugra) очень мал, он почти совпадает с параметром уравнения состояния для $\Lambda$CDM
модели ($w_0=-1, w_a=0$). Величины параметра уравнения состояния для потенциалов:
 Чанг-Шерер (Chang-Scherrer), Урена-Лопес-Матос (Ur\~{e}na-L\'{o}pez-Matos), Баррейро-Копеланд-Нюнес (Barreiro-Copeland-Nunes) находятся в пределах 3$\sigma$ уровня доверительного интервала CPL-$\Lambda$CDM контуров. Таким образом, эти потенциалы не различимы от стандартной $\Lambda$CDM модели. Величины параметра уравнения состояния для Ратра-Пиблс (Ratra-Peebles), Златев-Ванг-Стейнхардт (Zlatev-Wang-Steinhardt), Албрехт-Скордис (Albrecht-Skordis), Сахни-Ванг (Sahni-Wang) потенциалов находятся за пределами 3$\sigma$ уровня доверительного интервала. Это означает, что в зависимости от величины параметра уравнения состояния в современную эпоху, эти модели могут быть различимы или же они могут быть неразличимы от $\Lambda$CDM модели.

Полученные результаты для фантомных потенциалов представлены на Рис.~(\ref{fig:f35}). Очевидно, что величины параметра уравнения состояния для фантомного квадратичного (quadratic) потенциала находятся за пределами 3$\sigma$ интервала уровня доверия CPL-$\Lambda$CDM контуров, т. е.  этот потенциал не может имитировать $\Lambda$CDM модель в современную эпоху. Кривые параметра уравнения состояния для Псевдо-Намбу-Голдстоун бозон (pNGb), обратного гиперболический косинуса (inverse hyperbolic cosine), экспоненциального (exponent), Гауссианы (Gaussian), обратно-квадратичного (inverse square power) потенциалов  расположены частично в пределах 3$\sigma$ интервала уровня доверия CPL-$\Lambda$CDM контуров и частично вне этих границ. Это означает, что в современную эпоху эти модели могут имитировать $\Lambda$CDM модель, т. е. они могут быть неразличимы с $\Lambda$CDM моделью, или же они могут быть различимыми с $\Lambda$CDM моделью. Кривая параметра уравнения состояния для фантомного потенциала пятой степени (fifth power) находится в пределах 3$\sigma$ интервала уровня доверия CPL - $\Lambda$CDM контуров, поэтому в современную эпоху эту модель нельзя отличить от $\Lambda$CDM модели.

Для каждого потенциала мы исследовали, может ли изменение величины одного из модельных параметров (при условии, что величины других модельных параметров и величины начальных условий фиксированы) или изменение величин начальных условий (при условии, что величины модельных параметров фиксированы) привести к максимальному диапазону величин параметра уравнения состояния (при условии, что остальные параметры фиксированы). Результом этого исследования является тот факт, что мы можем разделить все рассматриваемые потенциалы на два типа: (i) на потенциалы, эволюция которых зависит от величин начальных условий; (ii) на потенциалы, эволюция которых не зависит от величин начальных условий (такие потенциалы имеют аттракторное решение).
К первому типу относятся следующие потенциалы квинтэссенции: Златев-Ванг-Стейнхардт (Zlatev-Wang-Steinhardt), Сахни-Ванг (Sahni-Wang), а также фантомные потенциалы: квадратичный (quadratic), Гауссиана (Gaussian), пятой степени (fifth power), обратно-квадратичный (inverse square).
Ко второму типу относятся следующие потенциалы квинтэссенции\footnote{Потенциал Ратра-Пиблс (Ratra-Peebles) находится в привилегированном положении в сравнении с другими потенциалами, так как для него мы рассматривали решение с фиксированными начальными условиями, Ур.~(\ref{eq:Incond}). Таким образом, этот потенциал не рассматривался в данном исследовании.}: Сугра (Sugra), Урена-Лопес-Матос (Ur\~{e}na-L\'{o}pez-Matos), Албрехт-Скордис (Albrecht-Skordis), Чанг-Шерер (Chang-Scherrer), Баррейро-Копеланд-Нюнес (Barreiro-Copeland-Nunes), а также фантомные потенциалы: Псевдо-Намбу-Голдстоун бозон (pNGb), обратный гиперболический косинус (inverse hyperbolic  cosine), экспоненциальный (exponent).

\section{Заключение}

Используя разработанный нами феноменологический метод, мы реконструировали $\phi$CDM модели скалярного поля, представленные в Таблице \ref{table:QP} и в Таблице \ref{table:PP}. Таким образом, мы нашли предварительные диапазоны величин для начальных условий и параметров моделей темной энергии, эти результаты представлены в Таблице \ref{table:QP1} и в Таблице \ref{table:PP1}.
Мы получили ограничения на $\phi$CDM модели при сравнении данных, $H(z)$, $d_A(z)$, $f(a)\sigma_8(a)$, с соответственными данными, которые были сгенерированы для фидуциарной $\Lambda$CDM модели. На Рис.~(\ref{fig:f31} - \ref{fig:f33}) показаны примеры ограничений, полученных при применении MCMC анализа, для Ратра-Пиблс (Ratra-Peebles), Златев-Ванг-Стейнхардт (Zlatev-Wang-Steinhardt) потенциалов квинтэссенции и для обратного гиперболического косинуса (inverse hyperbolic cosine) фантомного потенциала.

Мы применяли статистические критерии Байеса для сравнения моделей, такие как Байес фактор, а также $AIC$ и $BIC$ статистические информационные критерии.
С этой целью мы проинтегрировали уравнение, Ур.~(\ref{eq:evidence}), в пределах, соответствующих предварительно найденным диапазонам величин модельных параметров, указанных в Таблице \ref{table:QP1} и в Таблице \ref{table:PP1}. Вычисленные величины $AIC$, $BIC$ и фактора Байеса для всех $\phi$CDM моделей суммированы в Таблице \ref{table:bicq} и в Таблице  \ref{table:bicp}. Эти цифры наглядно демонстрируют, что если $\Lambda$CDM модель является истинной моделью темной энергии, то возможность использования данных DESI для того чтобы однозначно различить между собой $\phi$CDM модели в настоящую эпоху находится под сомнением.

Мы исследовали, как $\phi$CDM модели отображаются на $w_0-w_a$ фазовой поверхности контуров CPL - $\Lambda$CDM, Рис.~(\ref {fig:f34}) и Рис.~(\ref {fig:f35}).
Мы обнаружили, что модели квинтэссенции Феррейра-Джойс (Ferreira-Joyce), обратно-экспоненциального (inverse exponent),
 Сугра (Sugra), Чанг-Шерер (Chang-Scherrer), Урена-Лопес-Матос (Ur\~{e}na-L\'{o}pez-Matos), Баррейро-Копеланд-Нюнес (Barreiro-Copeland-Nunes) и фантомная модель пятой степени (fifth power)  модели  не могут быть различимы c $\Lambda$CDM моделью в современную эпоху. Модели квинтэссенции: Ратра-Пиблс (Ratra-Peebles), Златев-Ванг-Стейнхардт (Zlatev-Wang-Steinhardt), Албрехт-Скордис (Albrecht-Skordis), Сахни-Ванг (Sahni-Wang) и фантомные модели: Псевдо-Намбу-Голдстоун бозон (pNGb), обратный гиперболический косинус (inverse hyperbolic cosine), экспоненциальный (exponent), Гауссиана (Gaussian), обратно-квадратичный (inverse square power) могут проявить себя как различимыми так и не различимыми с $\Lambda$CDM моделью. Квадратичная (quadratic) фантомная модель абсолютно различима с $\Lambda$CDM моделью.

Все исследуемые модели можно разделить на два типа: (i) на модели, эволюция которых не зависит от начальных условий; (ii) на модели, эволюция которых зависит от величины начальных условий в некоторый начальный момент времени в прошлом (такие потенциалы имеют аттракторное решение). К первому типу относятся квинтэссенциальные модели: Златев-Ванг-Стейнхардт (Zlatev-Wang-Steinhardt), Сахни-Ванг (Sahni-Wang), а также фантомные модели: квадратичная (quadratic), Гауссиана (Gaussian), пятой степени (fifth power), обратно-квадратичная (inverse square power).
К второму типу относятся модели квинтэссенции:
Сугра (Sugra), Чанг-Шерер (Chang-Scherrer), Албрехт-Скордис (Albrecht-Skordis), Урена-Лопес-Матос (Ur\~{e}na-L\'{o}pez-Matos), Баррейро-Копеланд-Нюнес (Barreiro-Copeland-Nunes), а также фантомные модели: Псевдо-Намбу-Голдстоун бозон (pNGb), экспоненциальная (exponent), обратно-квадратичная (inverse square power).

%%%%%%%%%%%%%%%%%%%%%%%%%%%%%%%%%%%%%%%%%%%%%%%%%%%%%%%%%%%%%%%%%%%%%%%%%%
%%%%%%%%%%%%%%%%%%%       Beginning of chapter 10   %%%%%%%%%%%%%%%%%%%%%%%
%%%%%%%%%%%%%%%%%%%%%%%%%%%%%%%%%%%%%%%%%%%%%%%%%%%%%%%%%%%%%%%%%%%%%%%%%%
%%%%%%%%%%%%%%%%%%%%%%%%%%%%%%%%%%%%%%%%%%%%%%%%%%%%%%%%%%%%%%%%%%%%%%%%%%
\chapter{Заключение}\label{chapter:11}

Данная диссертация посвящена изучению $\phi$CDM моделей. Детальное описание этих исследований представлено ниже.

\begin{enumerate}[I.]
\item
Мы исследовали различные свойства Ратра-Пиблс $\phi$CDM модели в сравнении  с $\Lambda$CDM моделью:
\begin{enumerate}[1.]
\item
Мы изучали динамику Ратра-Пиблс $\phi$CDM модели в зависимости от величины модельного параметра $\alpha$. Увеличение величины параметра $\alpha$  вызывает более сильную зависимость от времени скалярного поля, $\phi$, производной по времени скалярного поля, $\dot{\phi}$, а также параметра уравнения состояния, $w$, и его производной по масштабному фактору, $dw/da$.
\item
 Мы установили, что Ратра-Пиблс $\phi$CDM модель отличается от $\Lambda$CDM модели рядом характеристик, которые не зависят от величины модельного параметра $\alpha$. Эти характеристики являются общими для класса $\phi$CDM моделей скалярного поля квинтэссенции замороженного типа:
\begin{enumerate}[a)]
\item
В $\phi$CDM моделях величина темпа расширения вселенной всегда больше, чем величина темпа расширения вселенной для $\Lambda$CDM модели.
\item
Момент доминирования эпохи темной энергии для $\phi$CDM моделей начинается раньше чем для $\Lambda$CDM модели (при условии фиксирования других космологических модельных параметров).
\item
 $\phi$CDM модели и $\Lambda$CDM модель отличаются в своих прогнозах для темпа роста флуктуаций плотности материи во вселенной: $\phi$CDM модели предсказывают более медленный темп роста флуктуаций плотности материи, чем $\Lambda$CDM модель.
\item  Величина Линдер $\gamma$-параметризации для $\phi$CDM  модели возрастает с увеличением величины модельного параметра $\alpha$. Величина Линдер $\gamma$-параметризации для $\phi$CDM модели больше, чем для $\Lambda$CDM модели.
    \item Мы определили границы применимости Линдер $\gamma$-параметризации для Ратра-Пиблс $\phi$CDM модели, $z \in(0; 5)$.
     Применимость Линдер $\gamma$-параметризации прекращается позже для $\Lambda$CDM модели, чем для Ратра-Пиблс $\phi$CDM модели.
\end{enumerate}
\end{enumerate}
%---------------------------------------
\item
Мы провели ограничение модельных параметров $\Omega_\mathrm {m}$ и $\alpha$ в Ратра-Пиблс $\phi$CDM модели, используя различные наблюдательные данные:
\begin{enumerate}[a)]
\item
 При применении только наблюдательных данных функции темпа роста флуктуаций плотности материи происходит сильное вырождение между модельными параметрами
$\Omega_\mathrm {m} $ и $\alpha$. Это означает, что
допускаются более большие величины параметра $\alpha$
при увеличении величины параметра $\Omega_\mathrm{m}$. При этом невозможно найти ограничение на величину параметра $\alpha$.
\item Вырожденность устраняется
после объединения ограничений по наблюдательным данным функции темпа роста флуктуаций плотности материи и ограничений по наблюдательным данным от отношения расстояние-красное смещение
 данных BAO и предварительных расстояний до CMBR.
\item
В результате мы получили ограничения на модельные параметры Ратра-Пиблс $\phi$CDM модели скалярного поля: $\Omega_\mathrm{m} = 0.30 \pm 0.04$ и $0\leq\alpha\leq1.30$ для  уровня доверия в 1$\sigma$. Оптимальная величина для параметра $\alpha$, $\alpha = 0.00$.
\end{enumerate}

%-----------------------------------
\item
Мы провели исследования десяти моделей квинтэссенции и семи фантомных $\phi$CDM моделей скалярного поля:
\begin{enumerate}[1.]
\item
Мы восстановили эти модели, используя разработанный нами феноменологический метод. В результате этого исследования мы нашли диапазоны величин параметров для потенциалов $\phi$CDM моделей скалярного поля, при которых эти модели могут проявлять себя. Для каждой $\phi$CDM модели были найдены  диапазоны начальных условий для решения дифференциальных уравнений, описывающих динамику вcеленной.
\item
Применяя MCMC анализ, мы получили ограничения на $\phi$CDM модели скалярного поля, сравнивая наблюдательные данные для: темпа расширения вселенной, углового расстояния и функции темпа роста флуктуаций плотности материи с соответственными данными, сгенерированными для $\Lambda$CDM модели.
\item
Для определения более предпочтительных моделей в сравнении с $\Lambda$CDM моделью, основываясь на предсказанных данных DESI, мы провели статистический анализ Байеса. С этой целью мы вычислили для каждой модели величину фактора Байеса, а также $AIC$ и $BIC$ информационные критерии. Согласно результатам статистического анализа Байеса, мы не можем однозначно идентифицировать предпочтительные модели в сравнении с фидуциарной $\Lambda$CDM моделью на основе предсказанных данных DESI.
\item
Мы исследовали $\phi$CDM модели скалярного поля в $w_0-w_a$ фазовом пространстве контуров CPL - $\Lambda$CDM. Мы выявили подклассы квинтэссенциальных и фантомных $\phi$CDM моделей скалярного поля, которые в современную эпоху: (i) могут быть различимы с $\Lambda$CDM моделью, (ii) не могут быть различимы с $\Lambda$CDM моделью, (iii) могут вести себя как различимые так и не различимые с $\Lambda$CDM моделью.
\item
Более того, мы обнаружили, что все исследуемые модели можно разделить на два класса: (i) на модели, которые имеют  аттракторное решение; (ii) на модели, эволюция которых зависит от начальных условий.
\end{enumerate}

\end{enumerate}
%%%%%%%%%%%%%%%%%%%%%%%%%%%%%%%%%%%%%%%%%%%%%%%%%%%%%%%%%%%%%%%%%%%%%%%%%%
%%%%%%%%%%%%%%%%%%%       Beginning of chapter 11     %%%%%%%%%%%%%%%%%%%%%%%
%%%%%%%%%%%%%%%%%%%%%%%%%%%%%%%%%%%%%%%%%%%%%%%%%%%%%%%%%%%%%%%%%%%%%%%%%%
%%%%%%%%%%%%%%%%%%%%%%%%%%%%%%%%%%%%%%%%%%%%%%%%%%%%%%%%%%%%%%%%%%%%%%%%%%
\chapter{Будущие проекты}\label{chapter:11}
В планы будущих проектов входит:
\begin{enumerate}[1.]
\item
Исследование влияния нейтрино на формирование крупномасштабной структуры во вселенной в модели Меняющейся Массы Нейтрино (MaVaN). Исследование кластеризации нейтрино в MaVaN модели при взаимодействии нейтрино со скалярным полем.\\
\item
Изучение инфляционных моделей динамических $\phi$CDM моделей скалярного поля с ненулевой пространственной кривизной, (\cite{Ratra:1994vw}, \cite{Ratra:2017ezv}). Проведение Фишер матричного анализа и более расширенного матричного анализа Дали для этих моделей.\\
\item
Изучение моделей модифицированной гравитации.\\
\item
Изучение крупномасштабной структуры во вселенной в моделях модифицированной гравитации.
\end{enumerate}

%%%%%%%%%%%%%%%%%%%%%%%%%%%%%%%%%%%%%%%%%%%%%%%%%%%%%%%%%%%%%%%%%%%%%%%%%%%%%%%%%%%%

\end{document}